\documentclass[12pt,a4paper,twoside,openany,titlepage]{book}
\usepackage{fullpage}
\usepackage{array}
\usepackage{graphicx}
\usepackage[utf8]{inputenc}
\usepackage[english]{babel}
\usepackage{csquotes}
\usepackage[T1]{fontenc}
\usepackage{amsmath}
\usepackage[scaled=.95]{helvet}
\usepackage{caption}
\usepackage{keyval}
\usepackage{subfigure}
\usepackage{multirow}
\usepackage{float}
\usepackage{multicol}
\usepackage{changepage}
\usepackage[margin=1.2in,headsep=1.5cm]{geometry}
\usepackage{setspace}
\usepackage{afterpage}
\usepackage{amsmath}
\usepackage{amssymb}
\usepackage{hyperref}
\usepackage{textcomp}
\usepackage[toc,page]{appendix}
\usepackage{pdfpages}
\usepackage{bm}
\usepackage[backend=biber, style=phys, biblabel=brackets, sorting=none, hyperref = true]{biblatex}
\usepackage{fancyhdr}
\allowdisplaybreaks
\usepackage{cancel}
\newcommand\numberthis{\addtocounter{equation}{1}\tag{\theequation}}

\title{Titolo della tesi}
\author{Luigi Zanovello}
\date{Anno Accademico 2017-2018}

\begin{document}

\begin{titlepage}

\begin{center}

\vspace{-15mm}
 
\hspace*{-0.5cm}\includegraphics[scale=0.25]{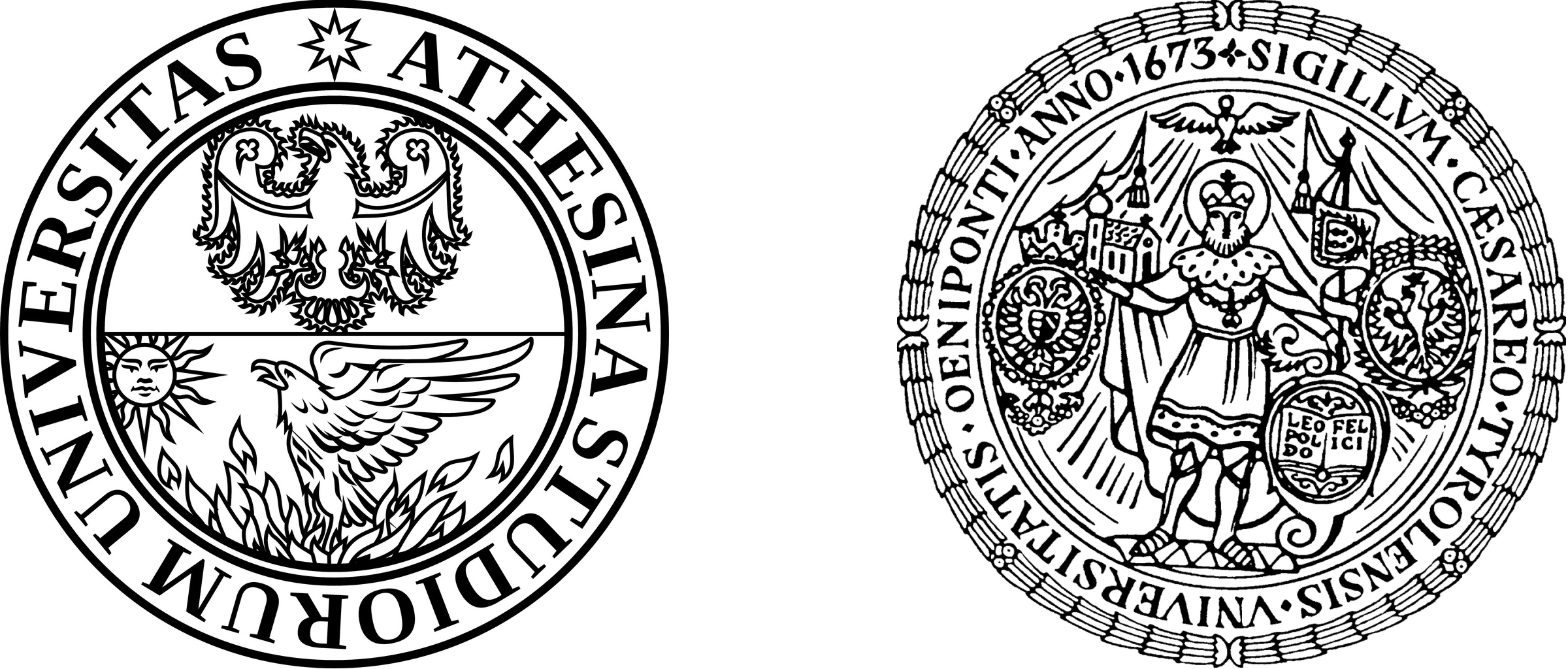}

\vspace{15mm}

\textsc{\LARGE Università degli Studi di Trento}\\[0.2cm]

\textsc{\Large Dipartimento di Fisica}\\[0.8cm]

\textsc{\LARGE Universit\"at Innsbruck}\\[0.2cm]

\textsc{\Large Institut f\"ur Theoretische Physik}\\[0.8cm]

\Large Thesis submitted for the Degree of\\

\LARGE Doctor of Philosophy

\vspace{0.5cm}
  
{\Huge \doublespacing \bfseries \begin{spacing}{1}{Target search of active particles in complex environments}\end{spacing}}

\vspace{0.5cm}

\vfill
 
\begin{minipage}{0.4\textwidth}
\begin{flushleft} \large
\emph{Supervisor:} \\
Prof. Pietro \textsc{Faccioli} \\
\vspace{0.5cm}
\emph{Supervisor:} \\
Univ.-Prof. Dr. Thomas \textsc{Franosch}
\end{flushleft}
\end{minipage}
\begin{minipage}{0.4\textwidth}
\begin{flushright} \large
\emph{Author:}\\
Luigi \textsc{Zanovello}
\end{flushright}
\end{minipage}
 
\vfill
 
{\large Academic year 2021/2022}
 
\end{center}
 
\end{titlepage}

\afterpage{\null\newpage}

\frontmatter

\pagestyle{plain}

\vspace*{\fill}
\noindent
This thesis was defended in Trento on the 2$^{\text{nd}}$ of May 2022.\\
Thesis defense committee:\\
Prof. Christoph \textsc{Dellago} (First referee)\\
Prof. Cristian \textsc{Micheletti} (Second referee)\\
Prof. Raffaello \textsc{Potestio}\\
Prof. Holger \textsc{Stark} (Third referee)\\

\afterpage{\null\newpage}

\frontmatter

\pagestyle{plain}

\chapter*{Abstract}
\addcontentsline{toc}{chapter}{Abstract}

Active particle is a general term used to label a large set of different systems, spanning from a flock of birds flying in a coordinated pattern to a school of fish abruptly changing its direction or to a bacterium self-propelling itself while foraging nourishment.
The common property shared by these systems is that their constituent agents, \textit{e.g.} birds, fishes, or bacteria, are capable of harvesting energy from the surrounding environment and converting it into self-propulsion and directed motion.
This peculiar feature characterizes them as out-of-equilibrium systems, in fact, the process of energy consumption and dissipation generates microscopically irreversible dynamics and drives them far from thermal equilibrium.
Thanks to their intrinsic out-of-equilibrium nature, active particle systems often display characteristic patterns and behaviors that are not observed in equilibrium physics systems, such as collective motion or motility-induced phase separation.
These features prompted the development of theories and algorithms to simulate and study active particles, giving rise to paradigmatic models capable of describing these phenomena, such as the Vicsek model for collective motion, the run-and-tumble model, or the active Brownian particle model.
At the same time, synthetic agents have been designed to reproduce the behaviors of these natural active particle systems, and their evolution could play a fundamental role in the nanotechnology of the $21^{\text{st}}$ century and the development of novel medical treatments, in particular controlled drug delivery.

A specific type of active particle that uses its directed motion to move at the microscale is called a microswimmer.
Examples of these agents are bacteria exploring their surroundings while searching for food or escaping external threats, spermatozoa looking for the egg, or artificial Janus particles designed for specific tasks.
Active agents at these scales use different swimming mechanisms, such as rotating flagella or phoretic motion along chemical gradients that they can create.
The outcome of their efforts is determined by the interplay of the translational diffusion intrinsic to the dynamics at these scales and the persistent motion that characterizes their self-propulsion.

The problem of finding a specific target in a complex environment is essential for microswimmers and active agents in general.
Target search is employed by animals and microorganisms for a variety of purposes, from foraging nourishment to escaping potential threats, such as in the case of bacterial chemotaxis.
The study of this process is therefore fundamental to characterize the behavior of these systems in nature.
Its complete description could then be employed in designing synthetic microswimmers for addressing specific problems, such as the aforementioned targeted drug delivery and the environmental cleansing of soil and polluted water.

Here, we provide a detailed study of the target search process for microswimmers exploring complex environments.
To this end, we generalize Transition Path Theory, the rigorous statistical mechanics description of transition processes, to the target-search problem.

The most general way of modeling a complex environment that the microswimmer has to navigate is through an external potential.
This potential can be characterized by high barriers separating metastable states in the system or by the presence of confining boundaries.
If a high energy barrier is located between the initial position of the microswimmer and its target, the target search becomes a rare event.
Rare events have been thoroughly investigated in equilibrium physics, and several algorithms have been designed to cope with the separation of timescales intrinsic to these problems and enable their investigation via efficient computer simulations.
Despite the large set of tools developed for studying passive particles performing rare transitions, the characterization of this process for non-equilibrium systems, such as active particles, is still lacking.

One of the main results of this thesis is the generalization to non-equilibrium systems of the Transition Path Sampling (TPS) algorithm, which was originally designed to simulate rare transitions in passive systems.
This algorithm relies on the generation of productive trajectories, \textit{i.e.} trajectories linking the initial state  of the particle to the target state, via a Monte Carlo procedure, without the need of simulating long thermal oscillations in metastable states.
These trajectories are then accepted according to a Metropolis criterion and are subsequently used to obtain the transition path ensemble, \textit{i.e.} the ensemble of all reactive paths that completely characterizes the process.
The TPS algorithm relies on microscopic reversibility to generate the productive trajectories, therefore its generalization to out-of-equilibrium systems lacking detailed balance and microscopic reversibility has remained a major challenge.
Within this work, after deriving a path integral representation for active Brownian particles, we provide a new rule for the generation and acceptance of productive non-equilibrium trajectories, which reduces to the usual expression for passive particles when the activity of the microswimmer is set to zero.
This new rule allows us to generalize the TPS algorithm to the case of active Brownian particles and to obtain a first insight into the counterintuitive target-search pathways explored by these particles.
In fact, while passive particles perform barrier crossing following the minimum energy path linking the initial state to the target state, we found that active particles, thanks to their activity and persistence of motion, can reach the target more often by surfing higher energy regions of the landscape that lie far from the minimum energy path.

The second result of this thesis is a systematic characterization of the target-search path ensemble for an active particle exploring an energy landscape.
We do so by analyzing the system's response to changes in the two adimensional parameters that define the parameter space of the model: the P{\'e}clet number and the persistence of the active particle.
Our findings show that active Brownian particles can increase their target-finding rates by tuning their P{\'e}clet number and their persistence according to the shape and characteristics of the external landscape.
We perform this analysis in two different landscapes, namely a double-well potential and the Brown-M\"uller potential, finding robust features in the target-search patterns.
In contrast, other observables of the system, \textit{e.g.} the target-finding rates, are more responsive to the features of the external environment.
Interestingly, our results suggest that, differently from what happens for passive particles, the presence of additional metastable states in the system does not hinder the target-search dynamics of active particles.

The third original contribution of this Ph.D. thesis is the generalization of the concept of the committor function to target-search problems.
The committor function was first introduced in the framework of Transition Path Theory to study reaction processes.
If a definition for a reactant and a product state embedded in the configuration space of the system is provided, the committor function quantifies the probability that a trajectory starting in a given configuration reaches the product state before it can enter the reactant.
For this reason, it has been proven to be pivotal for a complete characterization of these events and it is often regarded as the optimal reaction coordinate for thermally activated transitions.

The target search problem shares many similarities with transition processes since it is characterized by an initial state from which the agent begins its journey and a target state that the particle is aiming to reach, and often some barriers or obstacles separate the two.
Exploiting these similarities, we take advantage of the concept of the committor function to fully characterize a target-search process performed by an active agent.
First, we derive the Fokker-Planck equation for an active Brownian particle subject to an external potential, and we use its associated probability current to define the committor function for an active agent.
Then, we prove that the active committor satisfies the Backward-Kolmogorov equation analogously to the committor for passive particles.
We take advantage of this property to efficiently compute the committor function using a finite-difference algorithm, validating it with brute-force simulations.

Finally, we further validate our theory with experiments of a camphor self-propelled disk.
This self-propelled disk is capable of moving on a water surface and is studied during its exploration of a circular confining environment.
We start by analyzing long recorded trajectories of such a disk moving in a Petri dish, and, after defining a reactant and a product region in the system, we proceed to compute the committor function in three different regions contained in the dish.
We analyze all the trajectory slices passing through those regions and we measure how many of them hit the product region and how many hit instead the reactant first, and we obtain the committor in the three regions as a function of the angle.
Finally, we simulate a long trajectory of an active Brownian particle exploring a circular confining environment, and we compare the committor as an angular function obtained from brute-force simulations with the committor estimated from experimental data.

\afterpage{\null\newpage}

\pagestyle{plain}

\chapter*{List of Publications}
\addcontentsline{toc}{chapter}{List of Publications}

This dissertation is based on the following publications and manuscripts:

\vspace{0.8cm}

\noindent
\textsc{Luigi Zanovello, Michele Caraglio, Thomas Franosch, and Pietro Faccioli}\\
\textit{Target Search of Active Agents Crossing High Energy Barriers}\\
Physical Review Letters \textbf{126}, 018001 (2021). doi: \href{https://journals.aps.org/prl/abstract/10.1103/PhysRevLett.126.018001}{10.1103/PhysRevLett.126.018001}

\vspace{0.6cm}

\noindent
\textsc{Luigi Zanovello, Pietro Faccioli, Thomas Franosch, and Michele Caraglio}\\
\textit{Optimal navigation strategy of active Brownian particles in target-search problems}\\
The Journal of Chemical Physics \textbf{155}, 084901 (2021). doi: \href{https://aip.scitation.org/doi/10.1063/5.0064007}{10.1063/5.0064007}

\vspace{0.6cm}

\noindent
\textsc{Luigi Zanovello, Richard J.G. L\"offler, Michele Caraglio, Thomas Franosch, Martin M. Hanczyc, and Pietro Faccioli}\\
\textit{Survival strategies of artificial active agents}\\
Scientific Reports \textbf{13}, 5616 (2023). doi: \href{https://www.nature.com/articles/s41598-023-32267-3}{10.1038/s41598-023-32267-3}

\afterpage{\null\newpage}

\pagestyle{plain}

\tableofcontents

\afterpage{\null\newpage}

\pagestyle{plain}

\mainmatter

\chapter{Introduction}
\pagestyle{fancy}

Active particle is a term used to define agents that are capable of self-propelling and performing directed motion~\cite{bech2016}.
These agents are ubiquitous in nature and they span several length scales, from the microscopic to the macroscopic world.
These systems are manifold, and they include (but are not limited to) spermatozoa, bacteria, amoebae or other similar microorganisms, up to swarms of insects, flocks of birds, schools of fish, or pack of animals~\cite{Vicsek2012}.
Systems characterized by assemblies of interacting self-propelling agents are usually called \emph{Active matter} systems, and they display a plethora of interesting phenomena, such as collective motion~\cite{Vicsek2012}, Motility induced phase separation (MIPS)~\cite{Cates2015}, active turbulence~\cite{Mendelson1999,Riedel2005}, accumulation at the system's boundary~\cite{Li2009,Volpe2014} or activation of ratchet motors~\cite{DiLeonardo2010}.

However, differently from many other systems, they cannot be fully characterized by the laws of equilibrium statistical mechanics.
The mechanism that allows them to achieve directed motion is based on a process of energy consumption and dissipation that drives them far from thermal equilibrium~\cite{bech2016}.
Consequently, their dynamics is often time irreversible, and usual models and theories that are based on microscopic reversibility or detailed balance are not suitable for the description of these systems.
Therefore new theories are needed and new physics can be unveiled.
Additionally, the question arises of whether the dynamic laws of these agents are system-specific or if they can be considered more general and suitable to characterize this whole class of systems.
These queries originated a growing interest from the statistical physics community in the investigation of active particles and consequently moved the study of these systems from a mere biological standpoint to an interdisciplinary one.

In the past few decades, several models and theories were developed from the statistical physics community to describe the behavior and evolution of these systems.
Starting from the renowned Vicsek model~\cite{Vicsek1995}, a large class of models and algorithms has been designed to study the evolution of collective systems, using different alignment or particle interaction rules.
At the same time, a different set of models has been envisaged to describe the behavior of single self-propelled particles.
Among these, the active Brownian particle model~\cite{bech2016}, the Brownian circle swimmer~\cite{bech2016} and the Run-and-tumble model~\cite{Tailleur2008} have been pertinent to characterize the phenomenology of microorganisms or artificial particles moving at the microscopic scale.

This thesis will focus on the study and characterization of the dynamics of self-propelled particles at the microscopic scale, which are often called microswimmers.
In particular, this thesis aims at the description of the process of target search that is performed by many particles at that scale, and will involve the development of new algorithms and theories to tackle this problem.

\section{Microswimmers and self-propelled particles}

Microswimmers are self-propelled particles able to move in liquids at the micrometer scale.
They do so using a variety of mechanisms, but the general feature shared by all of them is the fact that energy needs to be consumed from the microswimmer in order to achieve self-propulsion.
Movement is needed for a variety of purposes, from exploring the surrounding environment for harvesting nourishment, to escaping threats.
These microswimmers have been first observed in $1676$ by Antonie van Leeuwenhoek, who referred to them as ‘‘animalcules''~\cite{Lane2015}, and which are today known as microorganisms.
However, it was only in the $19^{\text{th}}$ century that the biology community started developing an increasing interest in the observation and study of these organisms and of their self-propulsion mechanisms.

In particular, in the $1970$s Purcell held a talk, entitled ‘‘Life at low Reynolds numbers''~\cite{Purcell1977}, which would have lately become the theoretical basis for a physical description of these systems.
There, he provided for the first time a physically accurate description of the swimming mechanisms of the microswimmers in a world where the viscous forces dominates over the inertial ones, \textit{i.e.} in the regime of small Reynolds number.
Microorganisms in that regime usually deform their bodies in a cyclic manner to achieve self-propulsion, so that the final state coincides with the initial one and they can keep swimming.
However, not all cyclic deformations work for moving at low Reynolds numbers.
In fact, the motion of the microswimmer needs to be non-reciprocal, meaning that the cyclic deformation cannot follow the same steps forward and backward in time, otherwise the microswimmer would just move forward and then go back to its initial position due to the little-to-no effect of inertial forces.
A useful example for this phenomenon is Purcell's \emph{scallop theorem}.
While at the macroscopic scale a scallop moves by quickly closing its shell and then slowly opening it again, Purcell illustrates how a scallop at the microscopic scale wouldn't be able to obtain net movement with such a mechanism: since at low Reynolds number the inertial forces are dominated by the viscous forces, the motion of the scallop would end in the same place where it started.
For this reason, many motile microorganisms use non-reciprocal movements, such as the use of rotating flagella.

\subsection{Natural microswimmers}

Microswimmers can be found in a variety of forms in nature, featuring a plethora of different swimming mechanisms, such as the use of rotating flagella, cilia, or the peculiar amoeboid motion used by eukaryotic cells to crawl or swim.
Most of the times these mechanisms are not even optimized in efficiency since the energy consumed for moving is just a small amount compared to the one they need for their vital processes, therefore an efficient mechanism is not a fundamental aspect for their survival as long as it allows them to move and explore the surrounding environment~\cite{Purcell1977}.
In fact, the ability to self-propel and achieve directed motion is pivotal for their survival, allowing them to scout the environment for harvesting nourishment (for example by following external gradients as it happens during bacterial chemotaxis~\cite{Berg1972,Berg1990}) or to escape predators and dangerous threats.

\begin{figure}[H]
\centering
\subfigure{\label{fig:side}\includegraphics[height=55.8mm]{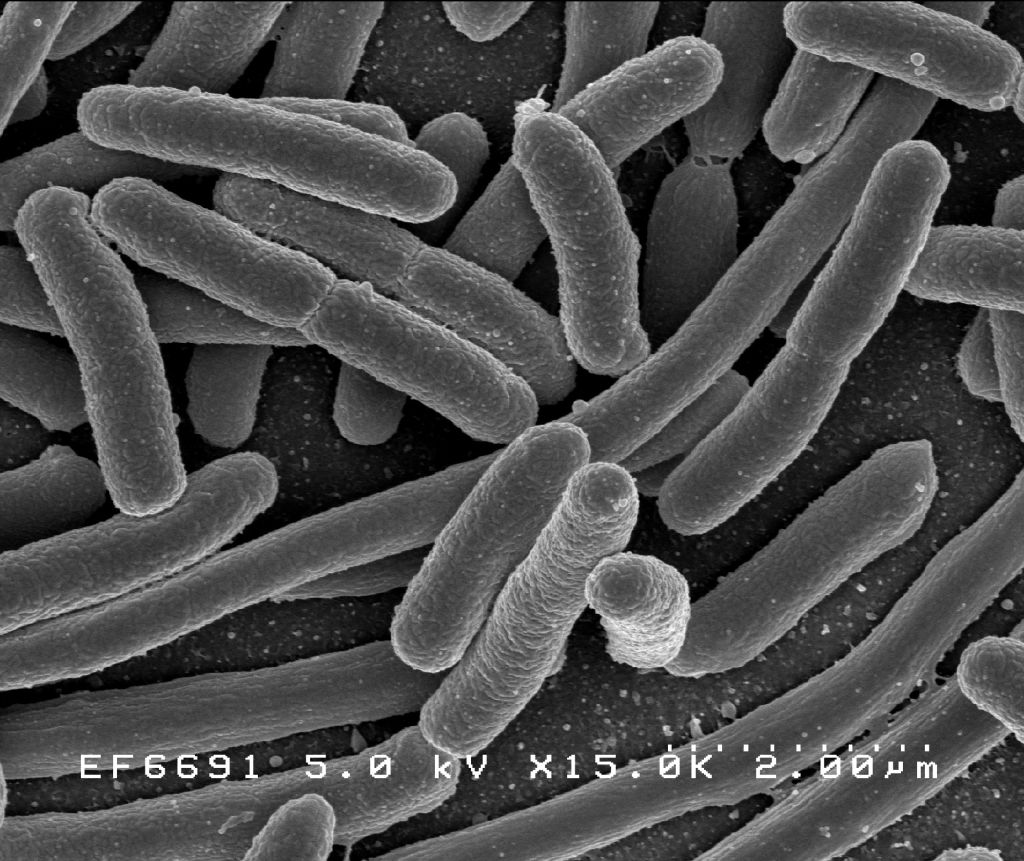}}
\subfigure{\label{fig:top}\includegraphics[height=55.8mm]{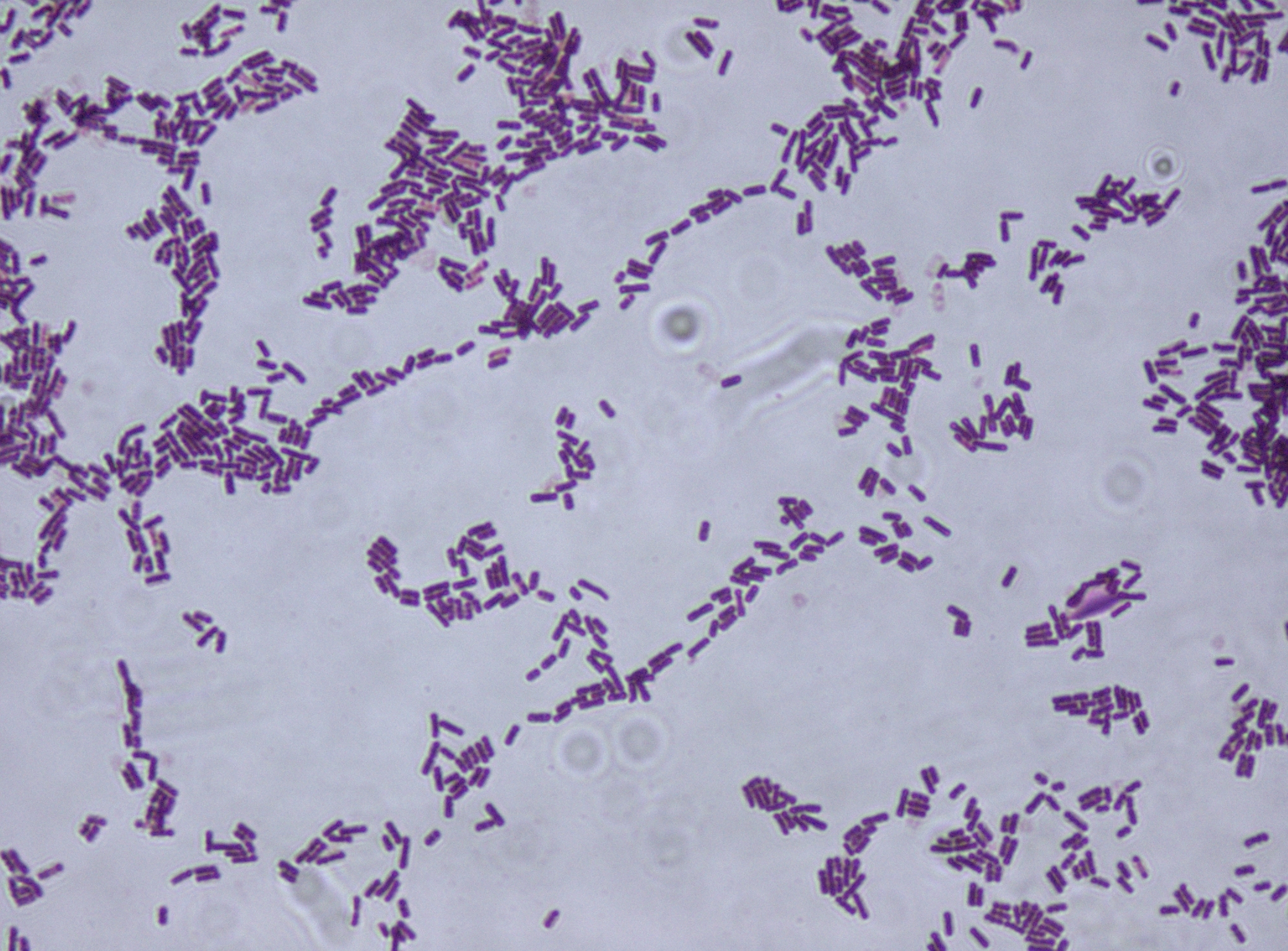}}
\subfigure{\label{fig:top}\includegraphics[height=50mm]{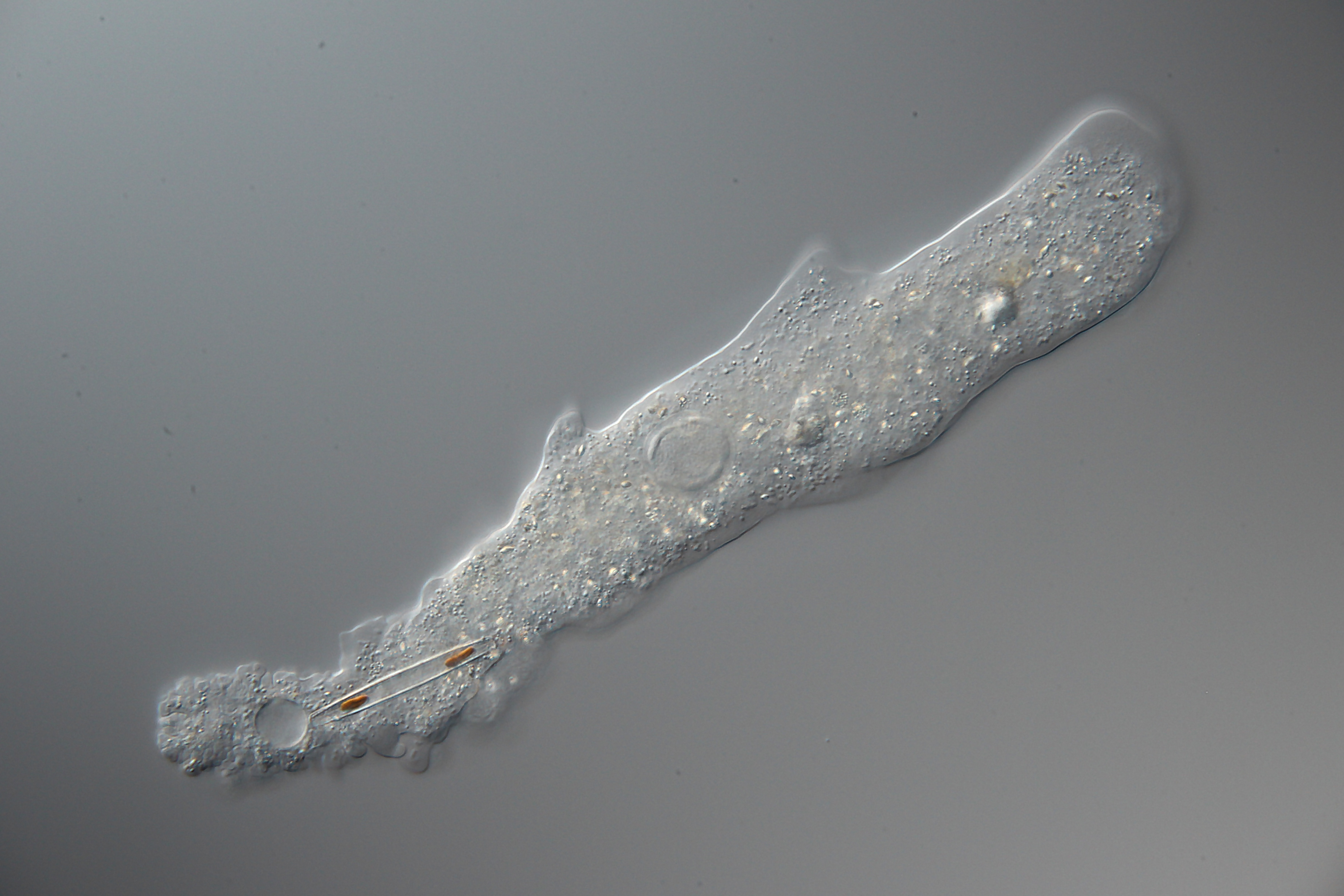}}
\subfigure{\label{fig:side}\includegraphics[height=50mm]{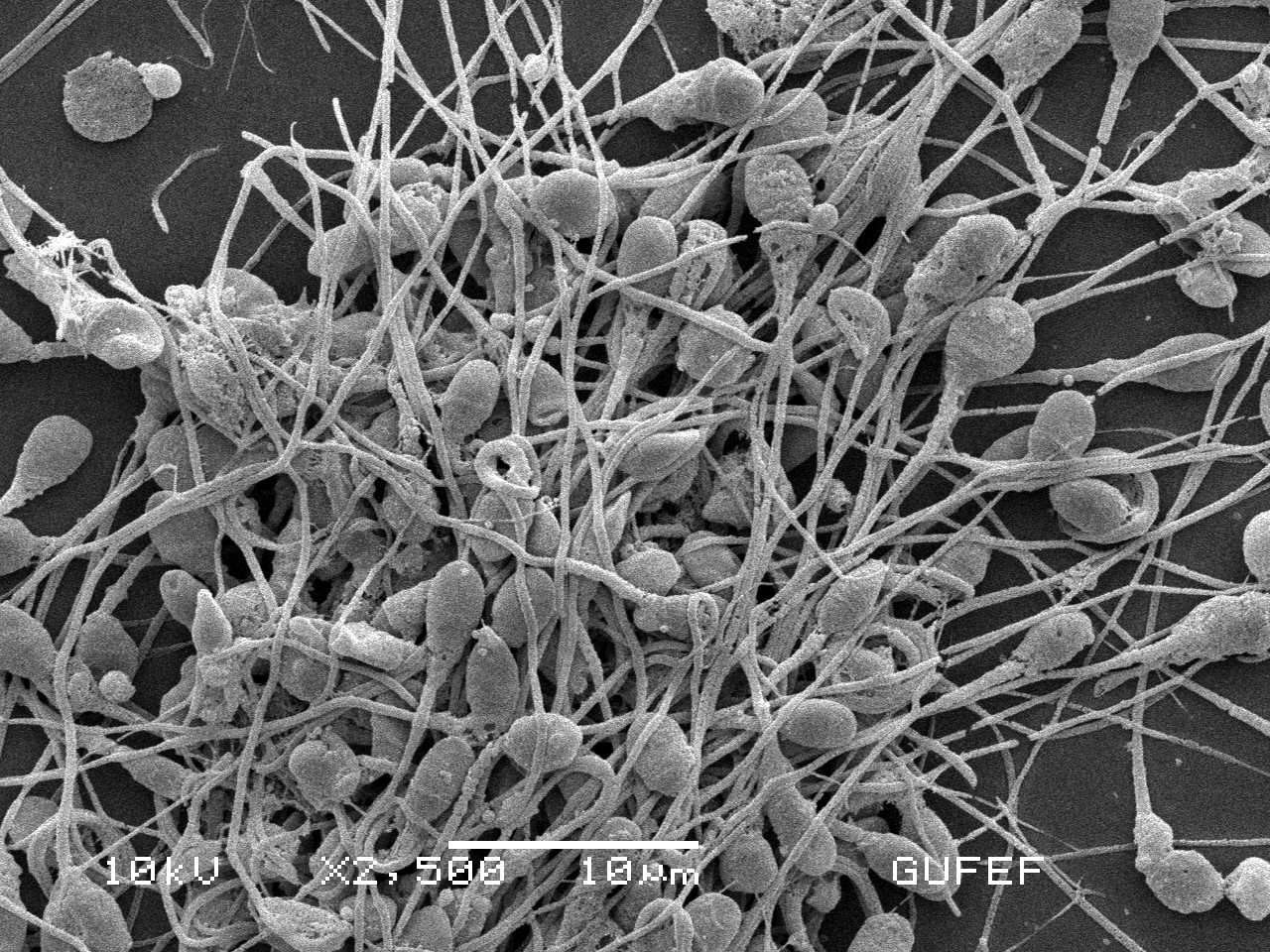}}
\caption{Natural microswimmers. \textit{Upper left panel}: Scanning electron microscopy image of \emph{Escherichia coli}. \href{https://commons.wikimedia.org/wiki/File:EscherichiaColi_NIAID.jpg}{Source: Wikimedia Commons: Rocky Mountain Laboratories, NIAID, NIH, public domain}. \textit{Upper right panel}: \emph{Bacillus subtilis} Gram stain. \href{https://commons.wikimedia.org/wiki/File:Bacillus_subtilis_Gram_stain.jpg}{Wikimedia Commons: Riraq25} \href{https://creativecommons.org/licenses/by-sa/3.0/}{CC BY-SA 3.0}. \textit{Lower left panel}: \emph{Amoeba proteus} locomotive form. \href{https://commons.wikimedia.org/wiki/File:Amoeba_proteus_locomotive_form.jpg}{Wikimedia Commons: SmallRex} \href{https://creativecommons.org/licenses/by-sa/4.0/}{CC BY-SA 4.0}. \textit{Lower right panel}: Scanning electron microscopy image of human \emph{spermatozoa}. \href{https://commons.wikimedia.org/wiki/File:Human_Spermatozoa,_Scanning_Electron_Micrograph.jpg}{Wikimedia Commons: Enver Kerem Dirican} \href{https://creativecommons.org/licenses/by-sa/4.0/}{CC BY-SA 4.0}. }
\label{fig:nat_microswimmers}
\end{figure}

Examples of these natural microswimmers are manifold.
Bacteria (such as \emph{Escherichia coli} (see Fig.~\ref{fig:nat_microswimmers} \textit{Upper left panel})~\cite{Berg2003,Bianchi2015,Chattopadhyay2006,Hill2007,DiLuzio2005}, \emph{Serratia marcescens}~\cite{Chen2015}, \emph{Bacillus subtilis} (see Fig.~\ref{fig:nat_microswimmers} \textit{Upper right panel})~\cite{Sokolov2010,Sokolov2007} \emph{Thiovulum majus}~\cite{Thar2001} or \emph{Vibrio Cholerae}~\cite{Utada2014}) have been extensively studied to analyze their swimming behavior, but also several eukaryotes (\textit{e.g.} protists such as \emph{Salpingoeca rosetta}~\cite{Kirkegaard2016}, amoebae (see Fig.~\ref{fig:nat_microswimmers} \textit{Lower left panel})~\cite{Wu2016} or algae like \emph{Chlamydomonas reinhardtii}~\cite{Kantsler2013}) and eucaryotic cells (\textit{e.g.} human \emph{spermatozoa} (see Fig.~\ref{fig:nat_microswimmers} \textit{Lower right panel})~\cite{Eisenbach2006,Friedrich2008,Brokaw1958} and phagocytes~\cite{Devreotes1988,deOliveira2016}) have been found to use similar self-propulsion mechanisms.

\subsection{Artificial microswimmers}
\label{sec:art_micr}

\begin{figure}[H]
\centering
\subfigure{\label{fig:side}\includegraphics[height=90mm]{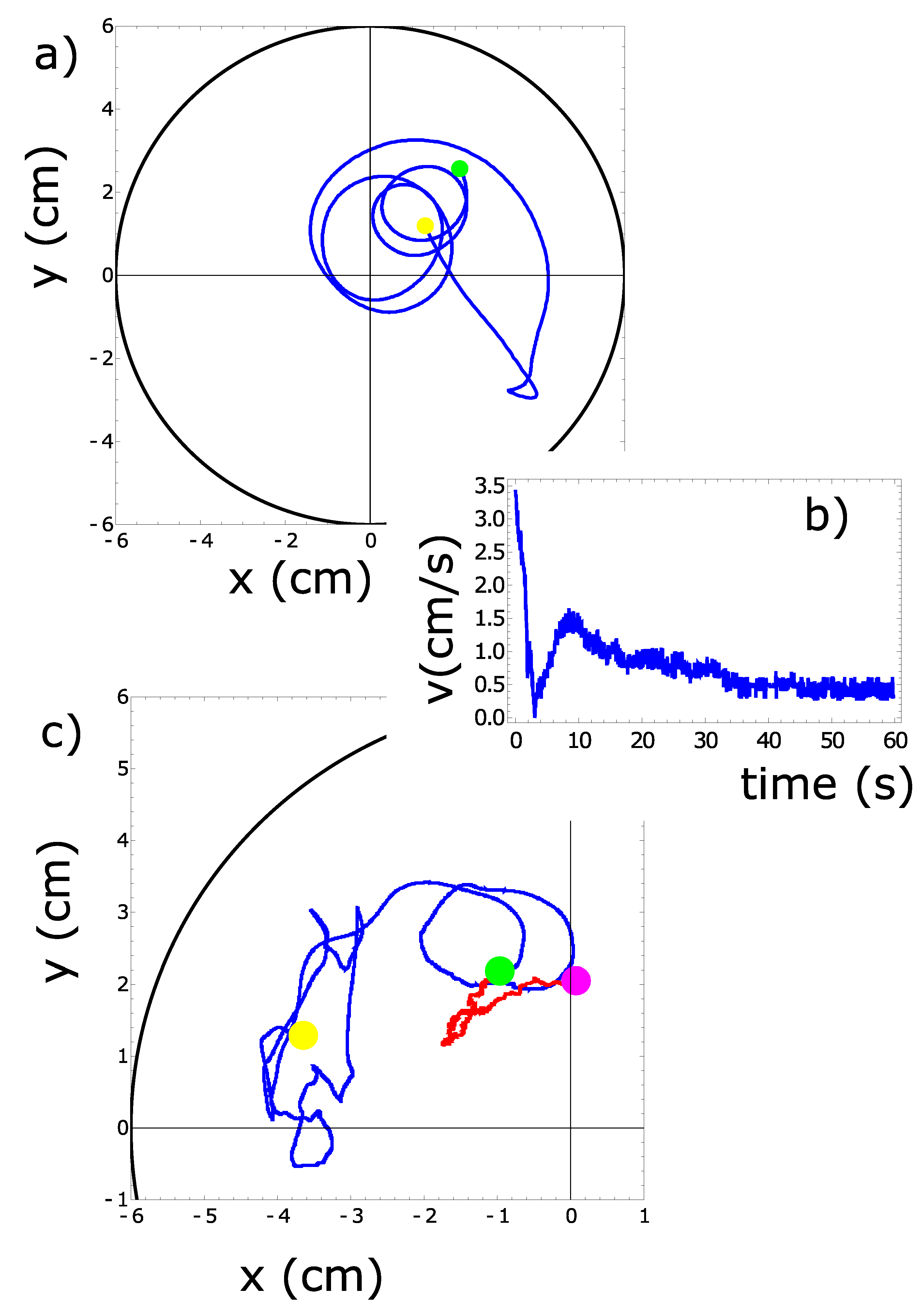}}
\subfigure{\label{fig:top}\includegraphics[height=90mm]{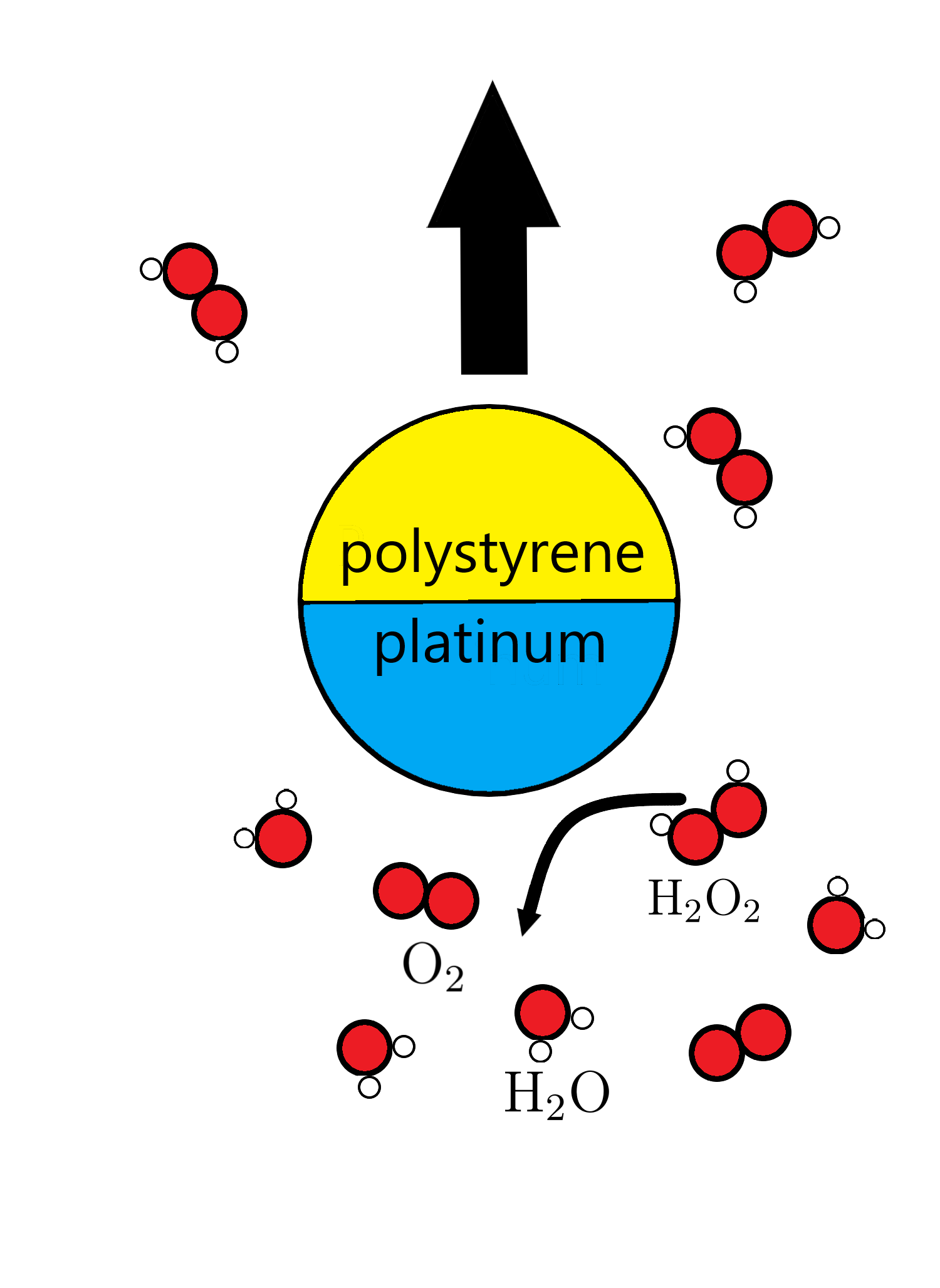}}
\caption{Synthetic microswimmers. \textit{Left panel}: (a) Short-time trajectory of a $4$ mm self-propelled camphene/polypropylene pill, at the beginning of the experiment. Colored dots mark different positions in time. (b) disk speed during the trajectory in (a). (c) Long-time trajectory of the pill at a later stage of the experiment. Different colors are used for different parts of the trajectory, the colored dots represent the positions of the disk at different times. The figure has been reproduced from Ref.~\cite{Löffler2021} under the license \href{https://creativecommons.org/licenses/by/4.0/}{CC BY 4.0}. \textit{Right panel}: representation of a polystyrene/platinum \emph{Janus particle} immersed in a hydrogen peroxide ($\text{H}_{2} \text{O}_{2}$) solution. Hydrogen peroxide decompose to oxygen and water, generating a concentration gradient surrounding the particle.}
\label{fig:synth_microswimmers}
\end{figure}

In the past few decades, the scientific community has been intrigued by the possibility of designing artificial self-propelled particles capable of mimicking the behavior of natural microswimmers.
Such a result would not only increase the level of understanding of these systems, but it would also allow expanding their potential applications significantly.
With the design of such biomimetic agents, several open problems could foresee a manageable solution in the next few years.
For example, the possibility of assembling microscopic particles that can be externally guided could provide a solution to the problem of targeted drug delivery~\cite{Naah2013,Patr2013,Chea2014}, one of the focal point of modern medicine, which would allow a drastic reduction in the side effects generated by the application of drugs during clinical treatments.
Another fundamental application could be the use of these synthetic agents for the decontamination of polluted water or soil~\cite{Vilela2017,Guo2020,Gao2014}, a problem becoming more and more pressing as the climate crisis intensifies.

For these reasons, the scientific community started developing the design and assembly of artificial microswimmers, coming in various shapes and featuring different swimming mechanisms.
The first attempt in this direction~\cite{bech2016} can be dated back to $2002$~\cite{Ismagilov2002}, when millimeters-sized artificial swimmers were built with a self-propulsion generated by chemical reactions.
Subsequent works involved the fabrication of self-propelling rods~\cite{Paxton2004}, particles with synthetic flagella~\cite{Dreyfus2005}, or self-propelled camphor and camphene objects~\cite{Nakata1997,Löffler2019}.
Following these first attempts on synthesizing artificial microswimmers, a famous class of artificial self-propelled particles was conceived, that is often referred to as \emph{Janus particle}~\cite{Golestanian2005,Howse2007}.
The name of these particles comes from the two-faced Roman god since they are usually polystyrene or silica particles partially coated with platinum (Pt) or palladium (Pd)~\cite{bech2016}, so that the final object has an asymmetric composition.
Thanks to this asymmetry, when these particles are inserted in liquid solutions of $\text{H}_{2} \text{O}_{2}$, the Pt or Pd coating the particle act as a catalyzer for a chemical reaction that dissociates the $\text{H}_{2} \text{O}_{2}$ molecules into $\text{H}_{2} \text{O}$ and $\text{O}_{2}$.
This produces a concentration gradient surrounding the particle that generates self-diffusiophoresis.

\section{The problem of target search}

\begin{figure}[H]
\centering
\includegraphics[height=45mm]{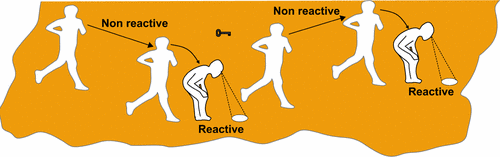}
\caption{An example of intermittent search strategy. An agent is looking for a target, and it performs an intermittent search. The agent alternates non-reactive phases characterized by fast relocation movements, which bring him to previously unexplored places, with reactive phases during which it actively searches for the target. The figure has been reproduced from Ref.~\cite{beni2011}.}
\label{fig:intermittent_strategies}
\end{figure}

The problem of finding a target is ubiquitous in nature and affects our everyday lives.
A bacterium looking for food, an animal trying to find a partner to mate or a human looking for his lost keys all face the same challenge, namely the need to find a specific target in a vast (and sometimes unknown) space.
This process can be energy consuming, and sometimes can be made more difficult by a limited amount of available time or resources, or by the presence of a complex environment, which hinders the target-finding process.

Target search in a complex or crowded environment is fundamental for many biological processes~\cite{beni2011}, encompassing macroscopic agents trying to find nourishment or escaping predators~\cite{Edwards2007,Benichou2006,Viswanathan1999}, the navigation of microswimmers trying to survive at the microscopic scale~\cite{Berg1972,Berg1990}, spermatozoa seeking a fertile egg~\cite{Eisenbach2006} and also the process of finding the needed compounds in chemical reactions~\cite{Mirny2008,Gorman2008}.
Notwithstanding its importance, however, the target-search problem for self-propelled particles has been investigated in relatively few studies, and only the concept of intermittent search strategy has been addressed in detail by the physics community~\cite{beni2011}.

The process of finding a target with unknown location through an intermittent search strategy relies on alternating phases of fast movements in the surrounding environment, during which the search process cannot be carried out, with slow phases with a careful exploration of the surroundings and the possibility of locating the target (see Fig.~\ref{fig:intermittent_strategies}).
The advantage of this type of strategy resides in the fact that the fast movements allow the exploration of different regions far from the initial one, which would be visited after long times if the agent would use a process based only on the careful and gradual exploration of the surrounding space.
The disadvantage of this strategy, however, is the fact that during these fast movements the agent is unable to locate the target, and consequently is sacrificing some time without the possibility for any immediate reward.

Intermittent searches are widely-used in nature, and their efficiency depends on the average duration of the two different phases~\cite{Benichou2007}.
This allows for a tuning in the model, so that the effectiveness of the search can be optimized by acting on the duration of the phases~\cite{Benichou2006}.
Furthermore, it is found that an intermittent strategy is often advantageous compared to a purely diffusive search~\cite{beni2011}.
In fact, intermittent strategies have been proven to yield shorter average search times than random walks for agents exploring one or two-dimensional spaces, while instead their gain reduces at higher dimensions.
The reason for this tendency is that if the searcher is found at a distance $d$ from the target, the higher is the dimensionality of the system and the smaller are the odds of performing a displacement in the correct direction to find the target.
While in the case of 1D a displacement of a distance $d$ can only be done in two possible directions, one of which will take the searcher to the target, as the dimension of the system increases also the number of possible moves of amplitude $d$ in the surrounding space does.
Instead, the number of moves taking the searcher to the target will remain constant to $1$, therefore decreasing the overall probability of finding the target using an intermittent pattern.
Additionally, it has been found that this efficiency gain over a diffusive search increases the smaller the density of the targets becomes, and that the optimal duration of the fast phase to achieve the best efficiency depends only on the dimension of the space in which the agent is moving.
For a generic model of intermittent search, this optimal duration can be found in a robust manner, independent of the modeling of the search phase.

Intermittent search strategies have therefore provided a useful tool to characterize the process of target search performed by active agents, and have paved the way for other target-search studies relying on different models.
This is the case for bacteria performing target-search using the run-and-tumble motion, which is characterized by alternating straight runs with constant speed and fixed orientation, with tumbling phases, during which the microswimmer can reorient and quickly changes its direction.
Bacteria performing run-and-tumble motion (such as \textit{E. coli}~\cite{Tailleur2009}) are able to efficiently follow gradients of chemical substances (\textit{e.g.} nourishment) by adjusting their tumbling rate according to the external gradient.
In this way they can increase their tumbling rate when the external environment is becoming less suitable, and decrease it when the concentration of food in the surrounding increases~\cite{Wadhams2004}.
Studies on these systems have revealed that also the external environment plays a fundamental role on the target-search process carried out by these active agents.
In particular, the presence of rigid boundaries, obstacles or barriers  heavily influences the target-search strategy performed by these bacteria.
In fact, differently from the case without obstacles, the presence of an heterogeneous environment shifts the optimal target-search towards a less ballistic and more diffusive motion~\cite{Volpe2017}.

Although intermittent strategies have been useful to characterize the target-search process in a large variety of systems, this class of models lacks the generality needed to be applied to all active particle systems.
In particular, particles performing directed motion characterized by a rotational diffusion on their swimming orientation constitute a large class of microswimmers that cannot be described by run-and-tumble models.
Similarly, intermittent search strategies cannot be applied to the target search performed by these particles, and their characterization has remained, up to date, virtually unexplored.

\section{The active Brownian particle model}
\label{sec:ABP_model}

In the past few decades several active particle models have been developed to describe the behavior of self-propelled agents.
Among these, the active Brownian particle (ABP) model~\cite{bech2016} is one of the simplest and more general.
The active Brownian particle model has been designed to describe the behavior self-propelled particles at the microscopic scale, whose motion evolves according to the interplay between the translational diffusion typical at this scale and the directed motion subject to a process of rotational diffusion that characterizes them.
Thanks to its generality, this model can be applied to many active particle systems, both for the case of biological microswimmers (\textit{e.g.} sperm cells~\cite{Woolley2003}) and also for artificial ones (\textit{e.g.} Janus particles~\cite{Jiang2010}).

\begin{figure}[H]
\centering
\includegraphics[height=90mm]{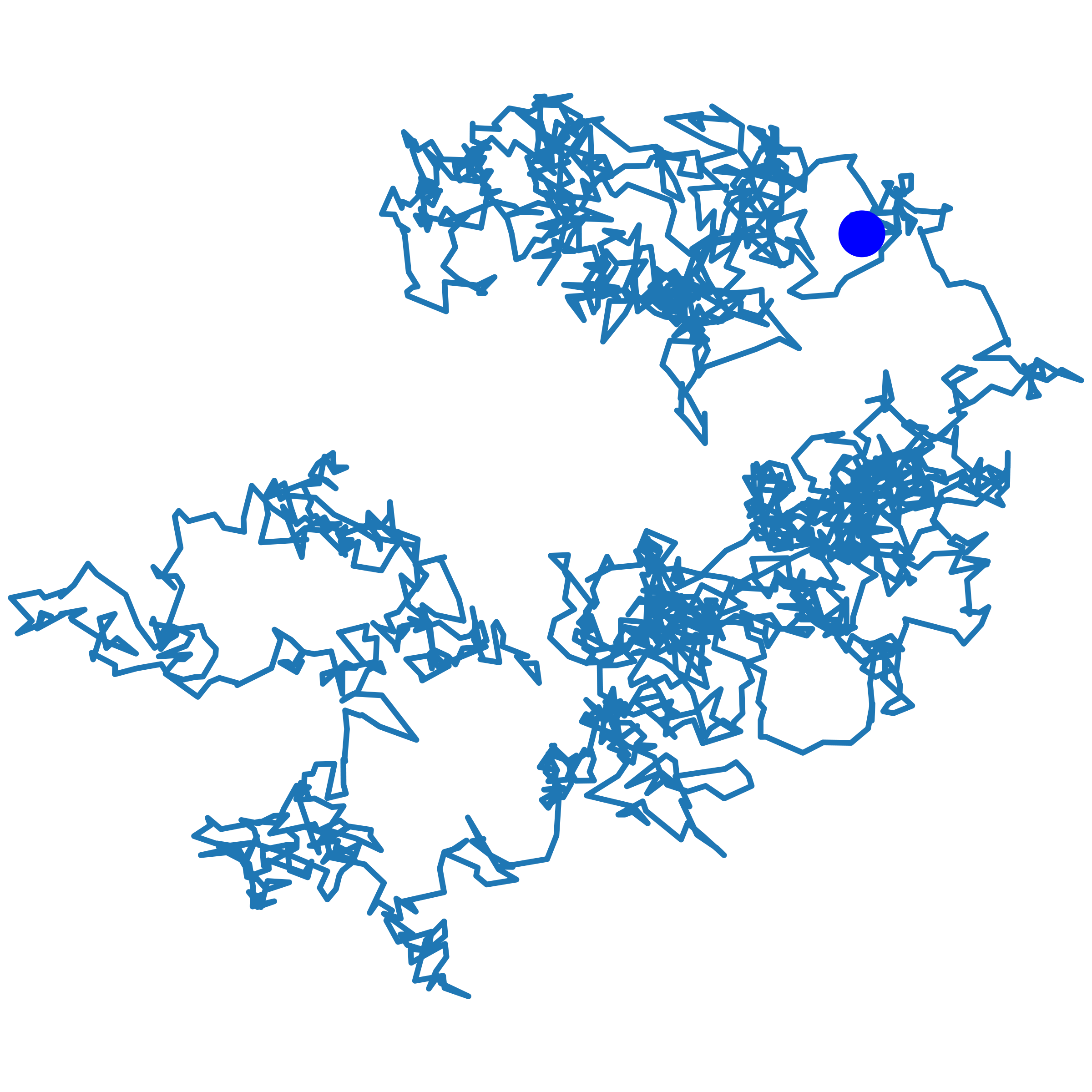}
\caption{Sketch of the trajectory of a passive particle on a two-dimensional plane.}
\label{fig:passive_traj}
\end{figure}

As anticipated, motion at the microscale is determined by translational diffusion. This effect generates a peculiar motion for particles in this regime, which is referred to as Brownian motion, named after the Scottish botanist Robert Brown who first observed this phenomenon in 1827~\cite{Brown1827}.
The trajectory of a passive particle moving at this scale is determined by the random collisions with surrounding particles, which generate erratic movements defined as diffusive motion, and is usually modeled through stochastic Wiener processes (a sketch of a passive particle trajectory can be seen in Fig.~\ref{fig:passive_traj}).
The extent of these fluctuations is determined by the translational diffusion coefficient typical for the considered object, which depends on the dimension of the particle and on the viscosity of the fluid where the object is immersed.
In the case of a spherical particle, the translational diffusion coefficient can be computed via the famous Stokes-Einstein relation:
\begin{equation}
D = \frac{k_{B}T}{6 \pi \eta R} \; ,
\end{equation}
where $k_{B}$ is the Boltzmann constant, $T$ is the temperature of the external bath, $\eta$ is the viscosity of the external fluid and $R$ is the radius of the particle.

\begin{figure}[H]
\centering
\includegraphics[height=90mm]{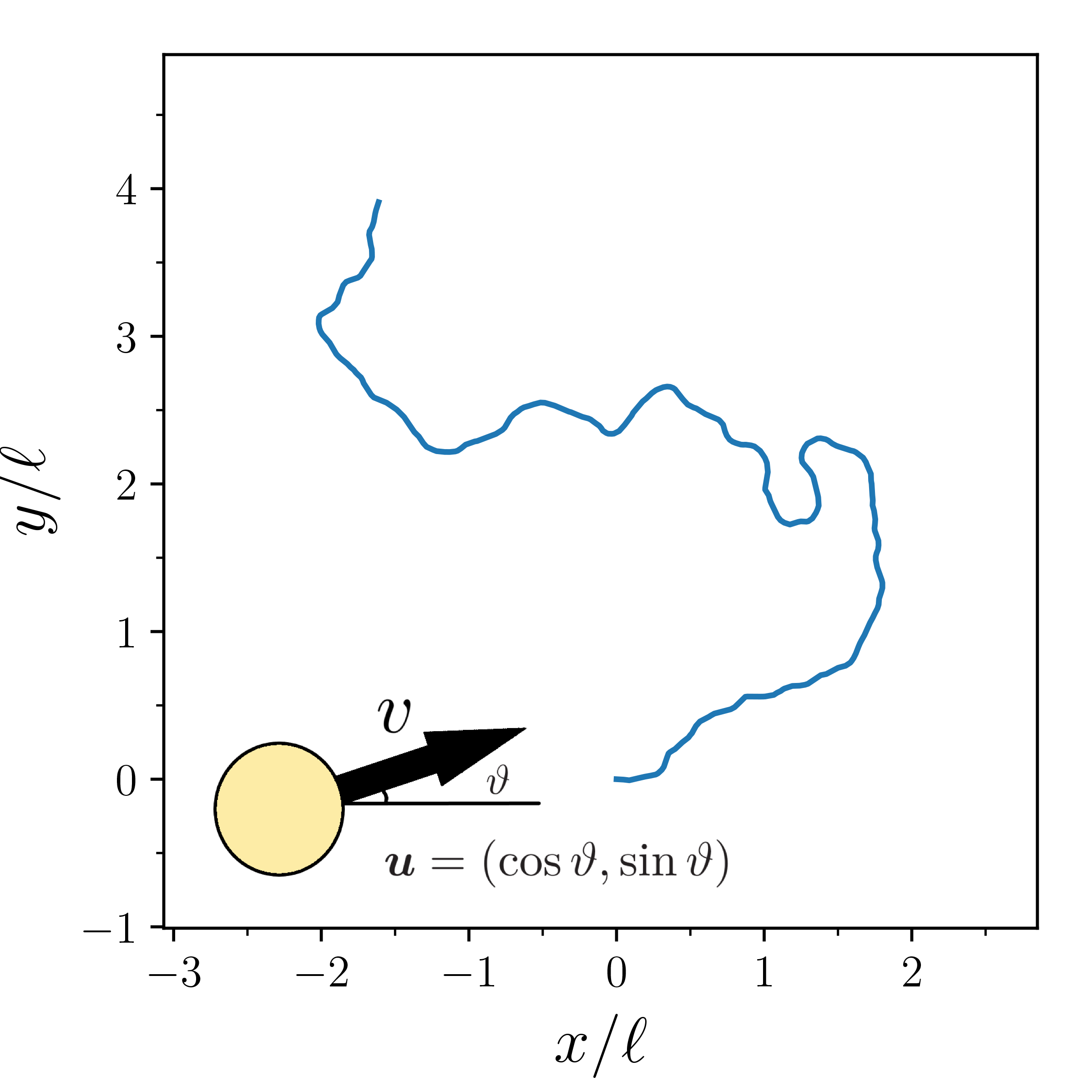}
\caption{Example of a trajectory of an ABP on the $xy$ plane, with negligible translational diffusion. $\ell$ identifies the persistence length of the particle. \textit{Lower left corner}: sketch of an ABP.}
\label{fig:ABP_traj}
\end{figure}

However, if the particle is active, the outcome of its motion is not determined exclusively by translational diffusion, because the particle also possesses the ability of self-propelling.
In this case, the trajectory of the particle will be determined by both effects, and, depending on the activity of the particle, it is also possible to achieve a regime where the activity dominates over the translational motion (for this reason some works use a reduced model where the translational diffusion is neglected, see~\cite{Chepizhko2013} as an example).
The activity of the particle is usually quantified using the \emph{P{\'e}clet number} $Pe \propto v$ (which in literature has been defined in many ways inconsistent with each other, hence the proportionality relation).
The regime in which the self-propulsion dominates over the diffusion processes is the regime of large $\text{Pe}$, and the case with negligible translational diffusion corresponds to the limit $\text{Pe} \rightarrow \infty$.
However, also the process of self-propulsion is influenced by the repeated interactions with the solvent molecules, and for this reason the active particle usually does not perform ballistic runs.
Instead, the active particle typically follows a trajectory with a swimming direction that changes with time: the orientation of the particle evolves according to a process of rotational diffusion, which describes the changes in the self-propulsion direction caused by the impacts with the solvent molecules.
The magnitude of these angular fluctuations is encoded in the rotational diffusion coefficient of the particle, which, similarly to the translational diffusion coefficient, also depends on the dimension of the particle and on the viscosity of the surrounding fluid.
For a spherical particle it reads~\cite{bech2016}:
\begin{equation}
D_{\vartheta} = \frac{k_{B}T}{8 \pi \eta R^{3}} \; .
\end{equation}
Finally, an important measure to define the self-propulsion of the particle compared to its rotational diffusion coefficient is the \emph{persistence length}, defined as $\ell := v/D_{\vartheta}$, that represents the typical length on which the trajectories of an ABP moving in free space look straight.
An example of an ABP trajectory can be seen in Fig.~\ref{fig:ABP_traj}.

The active Brownian particle model relies on a stochastic description of the diffusion processes to obtain the time evolution of the position of an active particle.
In the most general case of a two-dimensional system, the ABP equations of motion can be represented by a set of coupled Langevin equations in It\^{o} discretization, which read:

\begin{subequations}
\begin{eqnarray}\label{eom}
\bm{r}_{i\!+\!1} &=& \bm{r}_{i} + v\, \bm{u}_{i} \, \Delta t - \mu \bm{\nabla} U(\bm{r}_{i}) \Delta t + \sqrt{2D\Delta t} \, \bm{\xi}_i\;,\\ \label{eom2}
\vartheta_{i\!+\!1} &=& \vartheta_{i} + \sqrt{2D_{\vartheta}\Delta t} \, \eta_i\;.
\end{eqnarray}
\end{subequations}

Here, $\Delta t$ is a discrete time step, $\bm{r}_{i} = (x_{i},y_{i})$ denotes the position of the particle on the 2D plane at time $i\Delta t$, $\mu$ is the mobility of the particle, and $U(\bm{r})$ represents a general external potential acting on the particle.
$\sqrt{2D\Delta t} \, \bm{\xi}_i$ represents the contribution to the equations of motion generated by the translational diffusion process shared also by passive Brownian particles.
Here $D$ is the translational diffusion coefficient and $\bm{\xi}_i = (\xi_{x,i},\xi_{y,i})$ is a vector whose components are independent centered Gaussian random variables with unit variance.
The difference to the dynamics of a passive particle is given by the active self-propulsion, which is represented by the term $v\, \bm{u}_{i} \, \Delta t$, where $v$ is the modulus of the self-propulsion speed of the agent, which is constant in time, and $\bm{u}_{i} = \big(\cos(\vartheta_{i}),\sin(\vartheta_{i})\big)$ is its instantaneous orientation.
The orientation of the self-propulsion evolves in time due to the process of rotational diffusion, encoded in Eq.~\ref{eom2}.
$D_{\vartheta}$ is then the rotational diffusion coefficient and $\eta_{i}$ is another centered Gaussian random variable with unitary variance.

\begin{figure}[H]
\centering
\includegraphics[height=70mm]{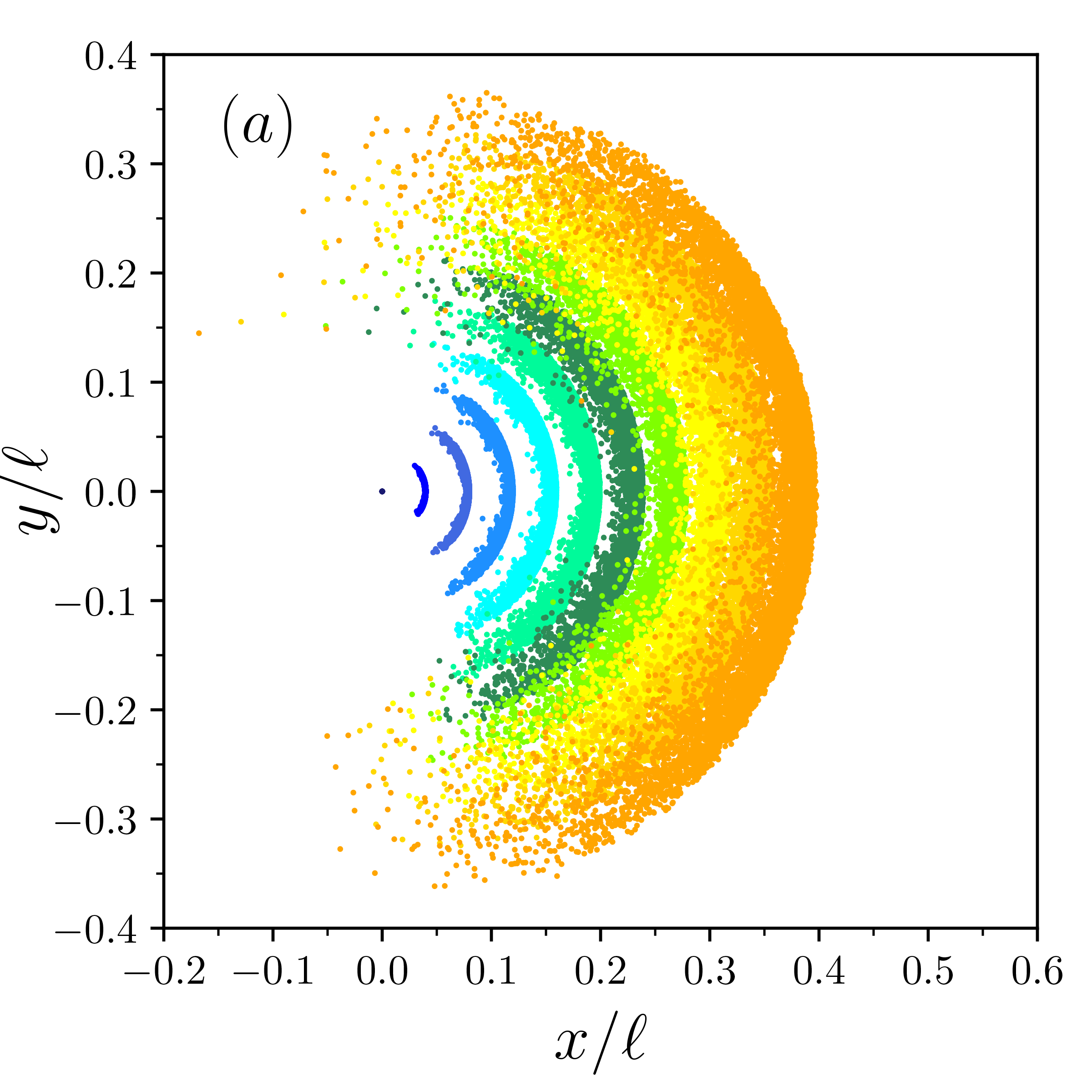}
\includegraphics[height=70mm]{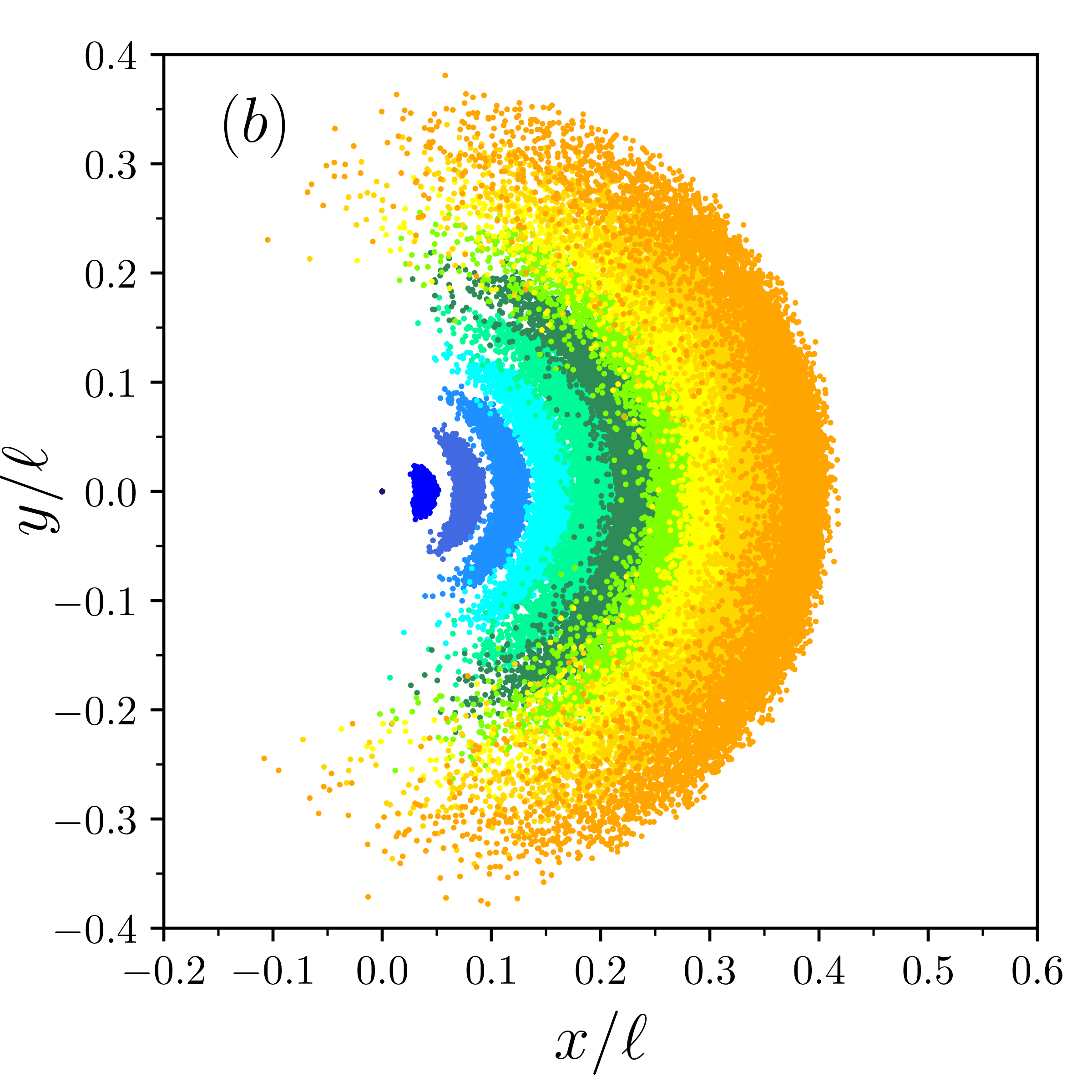}
\caption{(a) time evolution of an ensemble of $10^{4}$ non-interacting ABPs starting from the origin and with initial orientation directed along the $x$ axis, in the case with $D = 0$. Different color waves represent subsequent time steps during the simulations. The direction of motion of the ABPs gets gradually randomized from the rotational diffusion process. (b) time evolution of an ensemble of $10^{4}$ non-interacting ABPs starting from the origin and with initial orientation directed along the $x$ axis, in the case with $D > 0$. The color waves are more blurred by the presence of translational diffusion.}
\label{fig:ABP_evo}
\end{figure}

Note that if $v$ is set to zero, the ABP model reduces to the equations of motion for a passive particle, and the evolution of the position and orientation of the particle decouples.
Finally, while in the case of a passive particle the energy scale for the problem can be defined through the Einstein-Smoluchowski relation, $k_{B}T = D/\mu$ (for more on this topic see Appendix~\ref{app:LD}), in the case of an out-of-equilibrium setting such as the active particle this equation does not hold anymore.
However, it is still possible to define an effective thermal energy $k_{B}T_{\text{eff}}= D/\mu$, where $T_{\text{eff}}$ is the temperature of the bath in the passive limit, as a unit for the energy scale.
The time integration of Eqs.~(\ref{eom},\ref{eom2}) can provide the spatiotemporal evolution of an active self-propelling particle, and for this reason these equations are used as starting point for many studies and theories.
In Fig.~\ref{fig:ABP_evo}, the evolution of the positions of a set of $10^{4}$ ABPs obtained with the integration of Eqs.~(\ref{eom},\ref{eom2}) is displayed.

\subsection{ABPs in complex environments}
As already introduced, the process of how microswimmers move and perform target-search cannot be analyzed without accounting for their interactions with complex environments.
Microswimmers often need to explore environments that are crowded, limited by the presence of confining boundaries, or filled by obstacles or barriers, which hinder their navigation and target-search processes~\cite{bech2016}.
For this reason, in the past few years, huge efforts have been devoted by the active matter community to investigate the effects and mechanisms of microswimmers interacting with rigid boundaries, obstacles inserted in the system, passive particles, and with other microswimmers as well.

The study of how self-propelled agents interact with rigid boundaries has revealed that these particles do not visit configurations distributed according to a Gibbs-Boltzmann ensemble, differently to what happens with passive particles and equilibrium systems.
Instead, when their environment is limited by rigid boundaries they usually accumulate there, and tend to spend most of their navigation time close to the boundaries~\cite{Fily2015,Ezhilan2015}.
During this time, they can form complex structures caused by the accumulation of particles, such as vortexes~\cite{Bricard2015} or biofilms~\cite{Schaar2015}.

\begin{figure}[H]
\centering
\includegraphics[height=40mm]{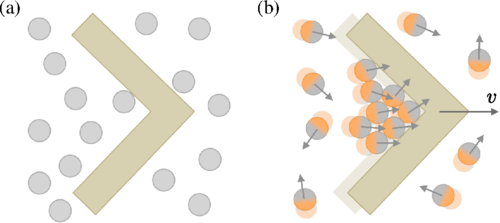}
\caption{(a) wedge inserted in a bath of passive particles. The pressure exerted by the bath on the wedge is equal on all sides, so no directed motion is observed. (b) wedge inserted in an active bath. The active agents accumulate on the inner side of the wedge and their trapping generates directed motion of the wedge. The figure has been reproduced from Ref.~\cite{bech2016}.}
\label{fig:wedge}
\end{figure}

This trapping mechanism close to the walls is generated by the self-propulsion, in fact, every time an active particle collides with the confining boundaries it keeps sliding along them until the rotational diffusion process is able to shift the swimming orientation away from the wall.
This causes the emergence of a detention time typical of the interaction of these particles with the confining boundaries, and it has been shown that the distribution of these detention times strongly depends on the swimming mechanism adopted by the self-propelling agent~\cite{Schaar2015}.

\begin{figure}[H]
\centering
\includegraphics[height=45mm]{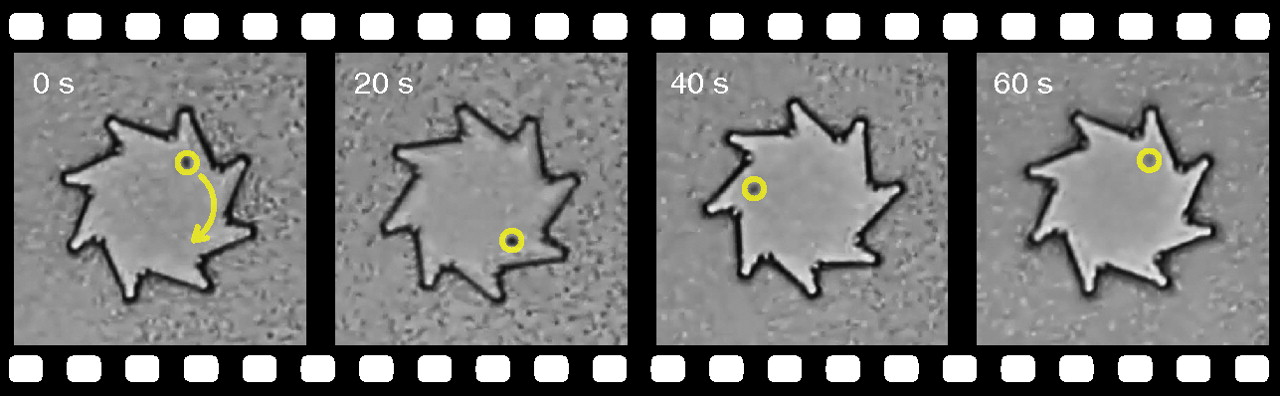}
\caption{Ratchet micromotor driven by bacteria. An asymmetric gear with a diameter of $48$ \textmu m and thickness of $10$ \textmu m is inserted in a bath of \emph{E. coli} bacteria, which can be seen in the background. The ratchet rotates clockwise thanks to the self-propulsion of trapped bacteria. The yellow point can be used to track the rotation process. The figure has been reproduced from Ref.~\cite{DiLeonardo2010}.}
\label{fig:bacterial_ratchet}
\end{figure}

The presence of rigid obstacles in the system introduces another degree of complexity, and the dynamics of the self-propelled particles in such an environment will very much vary depending on the characteristics of the microswimmer or on the shape and density of the obstacles~\cite{bech2016}.
For example, if a wedged-shaped object is inserted in an active bath, a totally different behavior will emerge if compared to the same wedge inserted in a passive bath~\cite{Kaiser2012}.
In fact, if the wedge is inserted in a passive bath it will only fluctuate without net motion, because the pressure exerted from the collisions with the passive particles will be equal on all the sides of the object.
If instead the wedge is inserted in a bacterial bath, the self-propelled particles will tend to accumulate on the inner side of the wedge because of the persistence in their swimming direction and the reduced escape angle of the inner cusp.
This accumulation of microswimmers will cause the wedge to move on average, and it has been shown that, depending on the density of the active particles in the surrounding bath, the speed of the wedge can be optimized.
Similarly, if a ratchet is inserted in a bath of bacteria, the microorganisms will accumulate on the inner vertexes of the ratchet and will cause its rotation~\cite{DiLeonardo2010}, which would be impossible if the ratchet was inserted in a passive bath (see Fig.~\ref{fig:bacterial_ratchet}).
Finally, the motion of a microswimmer placed in an array of obstacles can change from diffusive to subdiffusive with spontaneous trapping of the microswimmers within obstacles clusters depending on the density of the obstacles~\cite{Chepizhko2013}.

When instead active particles are studied in environments crowded by other active or passive agents, further interesting phenomena arise.
Active particles characterized by purely repulsive interactions can undergo motility-induced phase separation (MIPS)~\cite{Cates2015}, a state of the system characterized by a phase separation between liquid and gas phases, generated by a local increase in the particle density with a consequent decrease of the particles' speeds.
More generally, when repulsive active particles interact they can form small aggregates of particles that are not able to separate due to an unfavorable set of the particles' orientations.
This leads to the emergence of active jamming~\cite{Berthier2014,Reichhardt2015} or the creation of living crystals~\cite{Buttinoni2013,Palacci2013}.
If few active particles are instead suspended in a bath of passive particles, they can promote the formation of hard sphere glasses through a phenomenon called active doping~\cite{Ni2013}, or they can induce particle separations~\cite{Yang2012}.

Finally, the topic of target search for active particles in heterogeneous environments has been recently tackled by few studies, limited to the cases of a target located at the center or at the boundary of environments with rigid boundaries or obstacles.
In these cases, it has been found that the mean search time can be minimized acting on the speed of the agent and on its rotational diffusion process, and that the location of the target plays an important role in determining how effective will be the target search processes~\cite{Wang2016}.
Furthermore, it has been observed how the presence of an heterogeneous environment, such as in the case of a landscape filled with obstacles, can heavily influence the target-search dynamics~\cite{Wang2017}.
Notwithstanding these specific studies on the topic, a more general characterization of active Brownian particles looking for targets in external landscapes is up to date still lacking.

\subsection{ABPs in external potentials}
\label{sec:ABPs_ext_pot}
Not all complex environments can be reproduced by introducing rigid boundaries or obstacles in the system.
A more general approach in modeling these systems relies on the use of external potentials acting on the particle and mimicking the effect of an external environment.
For this reason the active matter community performed several theoretical and experimental studies to analyze how self-propelled particles behave when inserted in simple energy landscapes or how these agents are able to escape a potential well.
\begin{figure}[H]
\centering
\includegraphics[height=100mm]{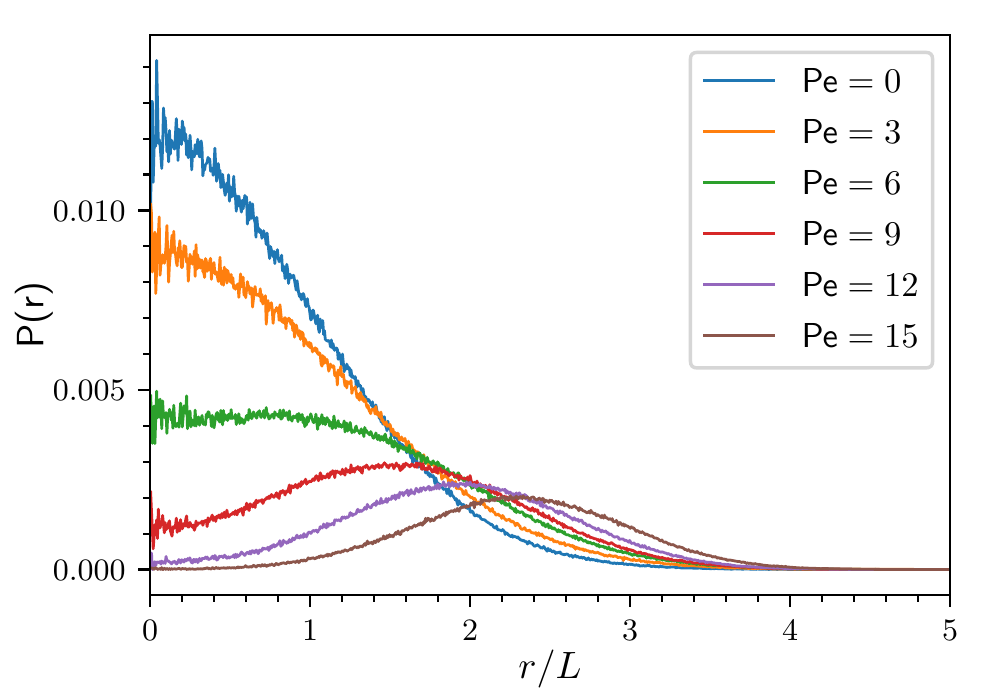}
\caption{Normalized radial distribution function for particles exploring a harmonic well, defined by $U(\bm{r})=k_{e}\bm{r}^{2}/2$. The P{\'e}clet number, in this case defined as $\text{Pe} = v\sqrt{3}/2\sqrt{DD_{\vartheta}}$, quantifies the activity of the particle. For $\text{Pe}=0$ the particle is passive and its positions are distributed according to the Boltzmann distribution, which, in the case of a harmonic well, is a Gaussian distribution. The more $\text{Pe}$ increases, the more the steady-state distribution becomes different from the equilibrium distribution. At large $\text{Pe}$ the particle visits almost exclusively regions of the well far from the center, accumulating along the boundaries. $L$ identifies a characteristic length scale for the problem, $L=\sqrt{k_{B}T/k_{e}}$.}
\label{fig:radial}
\end{figure}
In particular, it has been found that active particles behave once again totally different from passive particles when inserted in a harmonic well.
While passive particles visit configurations according to a Boltzmann distribution, with a peak at the bottom of the well, active particles tend once again to accumulate at the boundaries of the system, trying to climb up the potential walls as much as their self-propulsion and persistence of motion allows (see Fig.~\ref{fig:radial}).
Active particles cross quickly the lower energy regions of the landscape, and then slide for long times along the energy walls trying to climb them.
As soon as the rotational diffusion process is able to turn their direction of motion away from the walls, the particle will move quickly through the central part of the well eventually meeting the potential walls on the opposite sides.
Intuitively, this behavior strengthens with increasing activity of the particle and with decreasing rotational diffusion process as it has been shown from recent works~\cite{Pototsky2012}.

Studies on these topics have provided a mathematical description for the steady state distribution of an active particle in a harmonic trap~\cite{Malakar2020}, finding also how an increased steepness of the harmonic trap can induce a passive-like behavior in the active agent, forcing it to visit configurations at the bottom of the well.

Experimental research has revealed how this behavior can be observed also in the case of microswimmers trapped with an acoustic tweezer~\cite{Takatori2016}, in fact, depending on the confinement strength the probability distribution of the configuration can turn from a passive-like to a distribution peaked at the boundaries of the system.

Finally, the study of active particles exploring energy landscapes has led to the problem of an active agent escaping from a potential well.
Also in this case it has been found that active particles display striking differences compared to the passive particle behavior.
In fact, differently from passive particles that follow Kramers law for rare events~\cite{Hanggi1990}, active particles show enhanced escape rates from potential wells.
The escape times exponentially reduce with the activity of the particle due to an effective reduction of the height of the potential barrier that activity provides~\cite{Ebeling2005}.
Additionally, it has been discovered that the shape of the potential barrier influences the escape odds of an active agent.
While for a passive particle the only limiting factor for the transition is the barrier height, for active particles the shape of the barrier plays a fundamental role, and it might happen that active particles take transitions over higher barriers more often than over smaller ones, in a fashion that is extremely counterituitive for passive particles~\cite{Woillez2019}.

Notwithstanding these important results discovered in the past few years, to date the general problem of target search approached by modeling the surrounding environment through an external potential has not yet been addressed.

\section{Transition Path Theory}

To study the process of target search in a systematic way, a theoretical background is required to correctly formulate the problem.
Therefore, we generalize to active motion what is known in the literature as Transition Path Theory~\cite{Vanden-Eijnden2006,Weinan2010,Vanden-Eijnden2014,Metzner2006}, which provides a rigorous framework for studying the problem of kinetically activated transition events between metastable states in an ergodic system.
Transition Path Theory can be easily applied to the target-search problem, and for this reason will be used throughout this entire thesis.

Thermally activated transition processes are ubiquitous in nature, spanning from biology (\textit{e.g.} protein folding~\cite{Dill2008,Dill2012}, the process in which a protein has to find a specific metastable state in its free energy landscape, and has to reach it by overcoming several energy barriers) to chemistry (\textit{e.g.} chemical reactions~\cite{Hanggi1990}, which often require some activation energy in order to take place, allowing the system to jump over energy barriers separating the metastable states of the reaction).
Transition Path Theory allows for a mathematical description of these events: by providing a definition of an initial state for the system and of the final state of the process, a Transition Path Theory formulation provides a complete description of how the transition event occurs.
It does so through the use of two observables, the \emph{transition probability density}  and the \emph{transition current}, which provide complementary and complete information on the problem.
Additionally, Transition Path Theory defines the concept of \emph{committor function}, which is often regarded as the optimal reaction coordinate for the process~\cite{Krivov2018,Elber2017} and can be used to obtain the transition probability density and transition current.

Before diving in the details of these observables, we proceed with providing a general description of the transition process itself.
This description will then be followed by the characterization and derivation of the observables introduced by Transition Path Theory, while in the final part of the section we will introduce the concept of \emph{transition rates} and \emph{Transition Path Times}.
The transition rates are quantities that were inherited from the chemistry community and that were usually employed as one of the main quantities to characterize these processes.
In this last section we will also discuss how Transition Path Theory can provide complementary temporal information for these processes through the use of the Transition Path Times.

\subsection{Characterizing transition processes}

Consider a generic transition in which a particle is found in an initial state and has to reach a specific target state.
In this problem, a definition for the initial and final state is required.
These states are usually referred to as \emph{reactant} and \emph{product} states since the nomenclatures are historically inherited from the chemistry community.
They can be defined as follows.

Given a configuration space $\Omega$, which represents the space where the transition occurs, the reactant state R and the product state P will be identified as two non-overlapping regions embedded in $\Omega$, therefore $\text{R} \subset \Omega$, $\text{P} \subset \Omega$, and $\text{R} \cap \text{P} = \emptyset$.
The remainder of the configuration space $\Omega$ is identified as the \emph{transition region}, $\Omega_{\text{T}} = \Omega \setminus (\text{R} \cup \text{P})$, where the trajectories linking the two states are found.

If we consider an ergodic and infinitely long trajectory $x$ visiting configurations in $\Omega$,
\begin{equation}
x(t) \in \Omega, \; \forall t \in ]-\infty,\infty[,
\end{equation}
this trajectory will visit every configuration in $\Omega$ and will move from R to P and vice versa infinitely many times.

Since we are interested in the study of transition processes, we need a proper definition for the trajectory slices relevant for the problem, \textit{i.e.} parts of the trajectory $x$ starting in R and ending up in P.
Once a proper definition of these slices is set, their study can then be used to completely characterize the transition process.
Therefore, we introduce the concept of \emph{reactive trajectory}, which are trajectories that enter the transition region by exiting from the reactant state and exit the transition region by entering in the product state.

\begin{figure}[H]
\centering
\includegraphics[height=60mm]{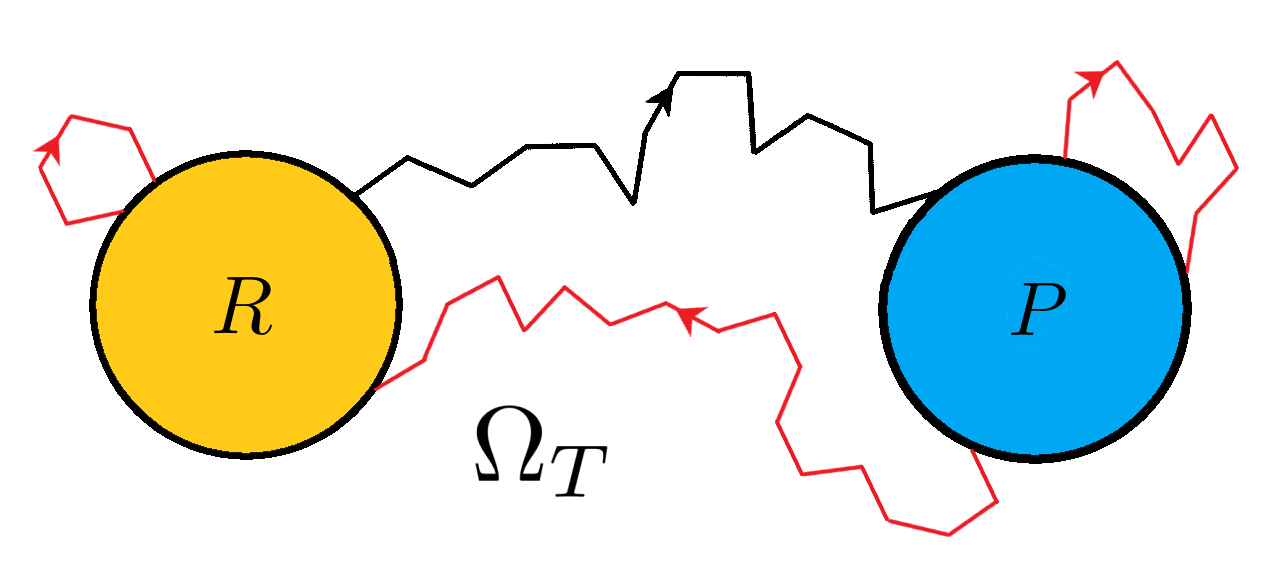}
\caption{Example of trajectory slices of $x(t)$. R and P indicate the reactant and product states respectively, $\Omega_{\text{T}}$ the transition region. In black an example of a reactive trajectory for the process, which enters the transition region by exiting from the R state and exits the transition region by entering the P state. In red examples of non-reactive trajectories.}
\label{fig:reactive}
\end{figure}

This definition can be formalized by introducing two characteristic functions, $t^{+}(t)$ and $t^{-}(t)$.
$t^{+}(t)$ defines the smallest time $t' \geq t$ for which $x(t') \in (\text{R} \cup \text{P})$ (so the smallest time after $t$ whose corresponding visited configuration was lying in the reactant or product state), while $t^{-}(t)$ provides instead the largest time $t' \leq t$ for which $x(t') \in (\text{R} \cup \text{P})$ (so the larger time smaller than $t$ whose visited configuration was found in the reactant or product).
Subsequently, a reactive trajectory can be defined as the continuous slice $x(t)$ of an ergodic trajectory that lies entirely in $\Omega_{\text{T}}$ and for which $x[t^{-}(t)] \in \text{R}$ and $x[t^{+}(t)] \in \text{P}$.
Instead, trajectories exiting from P and reaching R as well as trajectories exiting and reentering the same basin without reaching the other are non-reactive trajectories.
An example of a reactive as well as some typical non-reactive trajectories are displayed in Fig.~\ref{fig:reactive}.

\subsection{The committor function}
\label{sec:comm_func_pass}

Given a phase space $\Omega$ in which a reactant and a product state for a Markovian process are identified, a committor function can be defined~\cite{Metzner2006}.
The forward committor $q^{+}_{\text{P}}(\bm{x}_{i})$ is a function that evaluates the probability that a trajectory initiated at some point $\bm{x}_{i} \in \Omega$ will enter the product state P before entering in the reactant state R, and which is therefore a fundamental quantity to characterize the transition process.\footnote{Note that the system Markovianity is required to ensure that the committor depends only on the configuration $\bm{x}_{i}$ and not on the previous history of the system.}
Provided the ergodic assumption and since R and P are disjoint, the probability that a trajectory starting in $\bm{x}_{i}$ will instead enter R before entering P can consequently be obtained as $q^{+}_{\text{R}}(\bm{x}_{i}) = 1-q^{+}_{\text{P}}(\bm{x}_{i})$.
A backward committor $q^{-}_{\text{P}}(\bm{x}_{i})$ can then be defined as a function that provides the probability that a trajectory ending in $\bm{x}_{i}$ reached that final configuration by coming from P and not from R.
Analogously to the case of a forward committor, the probability that a trajectory ending in $\bm{x}_{i}$ was coming from R rather than P is obtained as $q^{-}_{\text{R}}(\bm{x}_{i}) = 1-q^{-}_{\text{P}}(\bm{x}_{i})$, provided the system Markovianity and the fact that the R and P regions are disjoint.
The probability for a path visiting state $\bm{x}_{i}$ of being reactive then reads:
\begin{equation}
P_{r}(\bm{x}_{i}) = q^{+}_{\text{P}}(\bm{x}_{i})q^{-}_{\text{R}}(\bm{x}_{i}) = q^{+}_{\text{P}}(\bm{x}_{i})(1-q^{-}_{\text{P}}(\bm{x}_{i})) \; .
\end{equation}
Now, if the dynamics of the system is microscopically reversible, the forward and backward committor are indistinguishable and the probability of a path of being reactive reduces to:
\begin{equation}
\label{eq_preact}
P_{r}(\bm{x}_{i}) = q(\bm{x}_{i})(1-q(\bm{x}_{i})) \; ,
\end{equation}
where we denote $q(\bm{x}_{i}) = q^{+}_{\text{P}}(\bm{x}_{i})$ for simplicity, and which will be called committor function.

The committor function can be obtained from the knowledge of the probability current of the system, which in turn can be calculated by solving the Fokker-Planck equation for the considered system (for more on the Fokker-Planck equation in continuity form see Appendix~\ref{app:FP}).

In particular, starting from the propagator of the system $p(\bm{x}_{\text{f}},t|\bm{x}_{0},0)$ (which represents the probability that a trajectory started in $\bm{x}_{0}$ at time $t_{0} = 0$ will reach the configuration $\bm{x}_{\text{f}}$ at time $t$), one can define a modified propagator $p^{*}_{\partial \text{W}}(\bm{x}_{\text{f}},t|\bm{x}_{0},0)$, where $\partial \text{W} = \partial \text{R} \cup \partial \text{P}$, which accounts for the probability that a trajectory started in $\bm{x}_{0}$ at time $t_{0} = 0$ will reach $\bm{x}_{\text{f}}$ at time $t$ without touching the boundaries of R and P.
This modified propagator can be obtained from $p(\bm{x}_{\text{f}},t|\bm{x}_{0},0)$ through the use of a characteristic function that suppresses all contributions to $p(\bm{x}_{\text{f}},t|\bm{x}_{0},0)$ coming from the trajectories that reach the boundary of R or P at a time $t'$, with $0 \leq t' \leq t$.

Since the propagator for a passive particle obeying the overdamped Langevin equations (for more on the Langevin overdamped dynamics see Appendix~\ref{app:LD}) can be written in path integral form as:
\begin{equation}
p(\bm{x}_{\text{f}},t|\bm{x}_{0},0) = \mathcal{N} \int \mathcal{D}\bm{x} \; e^{-S_{OM}[\bm{x}]} \; ,
\end{equation}
where $\mathcal{D}\bm{x}$ is a shorthand notation for $\mathcal{D}\bm{x}(\tau)$ and $S_{OM}[\bm{x}]$ is the Onsager-Machlup functional~\cite{Onsager1953,Machlup1953}:
\begin{equation}
S_{OM}[\bm{x}] = \frac{1}{4D} \int_{0}^{t} d\tau \bigg(\bm{\dot{x}}(\tau) + \frac{D}{k_{B}T} \bm{\nabla} U \big[ \bm{x}(\tau) \big] \bigg)^{2} \; ,
\end{equation}
(for a complete derivation see Appendix~\ref{app:PI}), the modified propagator $p^{*}_{\partial \text{W}}(\bm{x}_{\text{f}},t|\bm{x}_{0},0)$ can be written as:
\begin{equation}
p^{*}_{\partial \text{W}}(\bm{x}_{\text{f}},t|\bm{x}_{0},0) = \mathcal{N} \int \mathcal{D}\bm{x} \; e^{- \int_{0}^{t} d\tau \big[ \frac{1}{4D} \big(\bm{\dot{x}}(\tau) + \frac{D}{k_{B}T} \bm{\nabla} U [\bm{x}(\tau)] \big)^{2} + \Omega_{\text{W}}[\bm{x}(\tau)]\big]} \; ,
\end{equation}
where $\Omega_{\text{W}}(\bm{x})$ is a characteristic function that assumes value $0$ if the particle is outside R and P and is very large inside the two regions, suppressing the contributions to the path integral of the paths touching the two regions.
Then, by imposing that $\Omega_{\text{W}}(\bm{x}_{\text{f}})p^{*}_{\partial \text{W}}(\bm{x}_{\text{f}},t|\bm{x}_{0},0) = 0$, it follows that $p(\bm{x}_{\text{f}},t|\bm{x}_{0},0)$ and $p^{*}_{\partial \text{W}}(\bm{x}_{\text{f}},t|\bm{x}_{0},0)$ are solution to the same Fokker-Planck equation for the system.
Indeed, the two probabilities differ only from the boundary conditions in path space imposed by $\Omega_{\text{W}}(\bm{x})$, which does not explicitly depend on time.
Therefore, in the case of a passive particle obeying the overdamped Langevin equation:
\begin{equation}
    \frac{\partial}{\partial t} p^{*}_{\partial \text{W}}(\bm{x}_{\text{f}},t|\bm{x}_{0},0) = D \bm{\nabla} \cdot [\bm{\nabla} + \beta \bm{\nabla} U(\bm{x}(t))]p^{*}_{\partial \text{W}}(\bm{x}_{\text{f}},t|\bm{x}_{0},0).
    \label{eq:fpwp}
\end{equation}
This equation can be written as a continuity equation, giving:
\begin{equation}
     \frac{\partial}{\partial t} p^{*}_{\partial \text{W}}(\bm{x}_{\text{f}},t|\bm{x}_{0},0) = - \bm{\nabla} \cdot \bm{J}^{*}_{\partial \text{W}}(\bm{x}_{\text{f}},t|\bm{x}_{0},0) \; ,
\end{equation}
where:
\begin{equation}
\bm{J}^{*}_{\partial \text{W}}(\bm{x}_{\text{f}},t|\bm{x}_{0},0) = - D [\bm{\nabla} + \beta \bm{\nabla} U(\bm{x})] p^{*}_{\partial \text{W}}(\bm{x}_{\text{f}},t|\bm{x}_{0},0) \; .
\end{equation}
Since the committor function $q(\bm{x}_{i})$ is by definition related to the flux of the associated probability current $\bm{J}^{*}_{\partial \text{W}}(\bm{x}_{\text{f}},t|\bm{x}_{0},0)$ through the boundary of the product region, the expression of the committor in point $\bm{x}_{i}$ can be obtained as:
\begin{equation}
    q(\bm{x}_{i}) = - \int_{0}^{\infty} dt \int_{\partial \text{P}} d\bm{\sigma}' \cdot \bm{J}^{*}_{\partial \text{W}}(\bm{x}',t|\bm{x}_{i},0) \; ,
    \label{eq:committorp}
\end{equation}
where $d\bm{\sigma}'$ is an infinitesimal surface element on the boundary of the product region directed inwards.
The committor $q(\bm{x}_{i})$ will then provide the probability of having reached the product state P at any time without having touched the boundary of R.

We point out that iso-committor surfaces (hypersurfaces formed by a set of configuration space points $\bm{x}_{j}$ for which $q(\bm{x}_{j}) = \bar{q}$) provide a convenient foliation of the configuration space, which can be used to characterize the transition process.
Finally, an important property of the committor is that, provided that the boundary conditions $q(\bm{x}_{i})|_{\partial \text{R}}\!=\!0$ and $q(\bm{x}_{i})|_{\partial \text{P}}\!=\!1$ are satisfied, it is a solution to the backward Kolmogorov equation:
\begin{equation}
\hat{H}^{\dag}_{\text{FP}} q(\bm{x}_{i}) = 0 \; ,
\end{equation}
where $\hat{H}^{\dag}_{\text{FP}} = -D \big(\bm{\nabla}^{2} - \beta \bm{\nabla} U(\bm{x}(t)) \bm{\nabla} \big)$ is the adjoint of the Fokker-Planck operator, called backward Kolmogorov operator. For a complete proof see Appendix~\ref{app:BK}.

\subsection{Transition Probability Density and Transition Current}

Two fundamental observables provided by Transition Path Theory are the transition probability density and the transition current.

On the basis of Transition Path Theory's ergodicity assumption, we can define a time-independent distribution $m(\bm{x})$ which quantifies the probability that a reactive path visits the configuration $\bm{x}$.

The ergodicity assumption allows us to interchange time averages with ensemble averages.
Subsequently, the average value of an observable $A$ along a trajectory $x$ can be found equivalently as:
\begin{equation}
\underset{T \rightarrow \infty}{\lim} \frac{1}{2T} \int_{-T}^{T} dt A\big[ x(t) \big] = \int_{\Omega} d\bm{x} \; m_{\Omega}(\bm{x}) A(\bm{x}) \; ,
\end{equation}
where $m_{\Omega}(\bm{x})$ is an equilibrium probability distribution for the system.
Under this assumption, the time-independent transition probability density can be obtained as the distribution $m(\bm{x})$ that satisfies:
\begin{equation}
\underset{T \rightarrow \infty}{\lim} \frac{\int_{-T}^{T}dt A\big[ x(t) \big] h_{\text{T}}\big[ x(t) \big]}{\int_{-T}^{T}dt h_{\text{T}}\big[ x(t) \big]} = \int_{\Omega_{\text{T}}} d\bm{x} \; m(\bm{x}) A(\bm{x}) \; ,
\end{equation}
where $h_{\text{T}}\big[ x(t) \big]$ is a characteristic function that ensures that the considered trajectory $x$ is reactive,
\begin{equation}
h_{\text{T}}\big[ x(t) \big] = h_{\Omega_{\text{T}}}\big[ x(t) \big] h_{\text{R}}\big[ x\big[t^{-}(t)\big] \big] h_{\text{P}}\big[ x\big[t^{+}(t)\big] \big] \; ,
\end{equation}
with $h_{\text{R}}$ a characteristic function for the reactant state, $h_{\text{P}}$ one for the product state, and $h_{\Omega_{\text{T}}}$ one for the transition region, which are $1$ if the argument of the function belongs to the phase space region in the function subscript and are $0$ otherwise.

For an equilibrium system, the probability that any trajectory slice visits a configuration $\bm{x}_{i}$ at time $t$ is given by the Boltzmann distribution, and since the probability of a trajectory being reactive can be obtained from Eq.~\ref{eq_preact} (which is assuming microscopic reversibility and stating that a reactive trajectory that visits $\bm{x}_{i}$ at time $t$ must have one end reaching P and the other reaching R), the transition probability density $m(\bm{x})$ can be obtained as:
\begin{equation}
m(\bm{x}) = \frac{1}{Z_{\text{T}}} e^{-\beta U(\bm{x})} q(\bm{x}) \big(1 - q(\bm{x}) \big) \; ,
\end{equation}
with $Z_{\text{T}}$ the normalization factor
\begin{equation}
Z_{\text{T}} = \int_{\Omega_{\text{T}}} d\bm{x} \; e^{-\beta U(\bm{x})} q(\bm{x}) \big(1 - q(\bm{x}) \big) \; .
\end{equation}

Complementary information on the reaction process is provided by the transition current $\bm{J}_{\text{T}}(\bm{x})$, which yields the information on the flux of reactive paths across hypersurfaces enclosing an arbitrary region N of the system, and which can be used to find which reactive paths are more likely to be undergone during the transition.
Analogously to the case of the transition probability density, the transition current $\bm{J}_{\text{T}}(\bm{x})$ can be defined as:
\begin{equation}
\begin{aligned}[t]
\underset{\Delta \tau \rightarrow 0}{\lim} \underset{T \rightarrow \infty}{\lim} \frac{1}{2T}\int_{-T}^{T}d\tau \big( h_{\text{N}}&\big[ x(\tau) \big] h_{\Omega \setminus \text{N}}\big[ x(\tau + \Delta \tau) \big] - h_{\Omega \setminus \text{N}}\big[ x(\tau) \big] h_{\text{N}}\big[ x(\tau + \Delta \tau) \big] \big) \cdot \\ & \cdot h_{\text{T}}\big[ x(t) \big] \equiv \int_{\partial \text{N}} d\sigma \; \hat{\bm{n}}(\bm{x}) \bm{J}_{\text{T}}(\bm{x}) \; ,
\end{aligned}
\end{equation}
where $d\bm{\sigma} = d\sigma \; \hat{\bm{n}}(\bm{x})$ is an infinitesimal surface element directed outwards the surface of N.
Similarly to the transition probability density  also the transition current can be written as a function of the committor:
\begin{equation}
\bm{J}_{\text{T}}(\bm{x}) = D\bm{\nabla} q^{+}(\bm{x})\frac{e^{-\beta U(\bm{x})}}{Z} \; .
\end{equation}

\subsection{Other observables: transition rates and Transition Path Times}

In the study of transition processes, there are two fundamental observables that provide important information on the system: the transition rates and the Transition Path Times (TPT).

\begin{figure}[H]
\centering
\includegraphics[height=60mm]{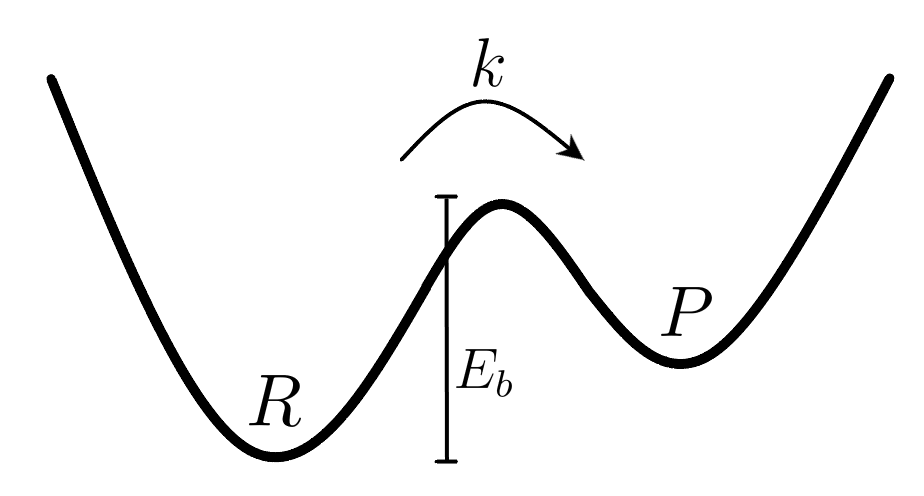}
\caption{Sketch of a transition process involving the jump over an energy barrier. The two metastable states R and P are separated by an energy barrier with height $E_{b}$. From Kramers theory, the transition rate $k$ and consequently the crossing timescale will depend exponentially from the magnitude of $E_{b}$.}
\label{fig:kramers}
\end{figure}

The transition rates are a useful tool to understand how likely and frequent is the transition under consideration.
In particular, they have been historically fundamental to characterize chemical reactions and transitions that involve jumps over high-energy barriers that separate the reactant and the product basin\footnote{Note that here we used the term ‘‘basin'' rather than ‘‘state'' or ‘‘region'' adopted before. We use the three terms with an equivalent meaning: ‘‘state'' or ‘‘region'' are in fact more useful to discuss a general phase-space formulation of the problem, while ‘‘basin'' is more fitting the the barrier-hopping scenarios, but a reactant state or a reactant basin have the same meaning.} (see Fig.~\ref{fig:kramers}), and they provide an insight complementary to the information obtained from Transition Path Theory.
These barrier-hopping transitions are usually dominated by rare events and are therefore characterized by small transition rates~\cite{Hanggi1990}.
In fact, when a high-energy barrier separates the two basins, the particle performing the transition has to climb on top of the barrier solely relying on a fortunate series of thermal fluctuations.
When the energy associated with these thermal fluctuations is small compared to the height of the energy barrier ($E_{b} \gg k_{B}T$), this condition can require a time much longer than the typical time associated with the sampling of the metastable states in the system (the basins R and P).
This leads to a separation of timescales in the system: the timescale associated with a local sampling of the metastable states $\tau_{l}$ will be much smaller than the timescale associated with crossing events over the barriers, $\tau_{c}$.

In particular, the transition rates were originally estimated in the case of chemical reactions using the Van't Hoff-Arrhenius law.
However, it was only with Kramers work on barrier-crossing events~\cite{Kramers1940} that it was possible to obtain a deeper understanding of these transition processes.
In his work he provided a way to obtain the transition rates for a general process obeying the overdamped Langevin dynamics and involving the overcome of an energy barrier.
In this case, these rates can be obtained using Kramers law:
\begin{equation}
k = C e^{-\beta E_{b}}
\end{equation}
where $k$ is the transition rate, $C$ is a prefactor physically interpreted as an attempt frequency, $\beta = 1/k_{B}T$ and $E_{b}$ is the height of the energy barrier separating the two metastable states.
From this equation emerges how the transition rates in barrier-hopping events depend exponentially on the height of the energy barrier (\textit{e.g.} the activation energy in a chemical reaction).
The timescales associated with these rare events can be obtained once the rates are estimated, in fact it is found that $\tau_{c} = k^{-1}$.

Another important observable to characterize the transition process is the Transition Path Time, which has received increasing interest in the past few years~\cite{Sega2007,Zhang2007,Caraglio2020,Chung2009,Neupane2016}.
Assuming that one can measure the time required for a transition to occur, and one can estimate its transition rate, one can still wonder how much of the measured time was required to exit the reactant basin and find a reactive trajectory, and how much of it was instead needed to actually follow this reactive path.

To solve this problem, it is once again useful to rely on Transition Path Theory and on the concept of reactive paths.
In particular, it is convenient to define the Transition Path Time, $t_{\text{TPT}}$, which is the time required to follow a reactive path.
It provides an information complementary to the time estimated from the transition rates.
This time in fact differs from the time associated with the complete transition because it does not account for the time spent in local oscillations in the reactant state nor for the time spent in unproductive trajectories, which exit and reenter R without reaching P.

\section{Thesis purpose and content}

The main objective of this thesis is the complete and general characterization of target-search processes performed by active agents exploring a complex external environment.
This will be achieved by designing a set of tools and methodologies that will be verified for simple yet paradigmatic models and that are suitable to the application to more sophisticated systems.

We will choose the equations of motion of the active Brownian particle model as a starting point for our study (Eqs.~(\ref{eom},\ref{eom2})).
These equations will be integrated by computer simulations to obtain the spatiotemporal evolution of such a particle exploring an external landscape, which will be modeled using an external potential in the equations of motion.
The landscape will be shaped to enforce a non-trivial exploration of the system, and will involve the use of confining potentials and the presence of high-energy barriers.
These energy barriers will separate metastable states in the system, and by selecting as initial (reactant) state and as target (product) state two metastable states in the landscape, a rare target-search dynamics will be induced.
The typical target-search paths will have to climb over these energy barriers, resulting in low target-finding rates.
For this reason, the development of algorithms for simulating rare events in active particle systems will be needed to efficiently harvest the required transition statistics, which will then be analyzed using the Transition Path Theory framework generalized to the non-equilibrium setting of the active particle systems.

The content of this thesis is organized as follows.
The first chapter is devoted to the development of algorithms for simulating rare events in active particle systems.
First, we will derive a path-integral representation for the propagator of an active Brownian particle.
This path-integral representation of the system will then be used to obtain a generalized version of Transition Path Sampling (TPS), a famous and reliable Monte Carlo algorithm designed to simulate rare transitions in equilibrium systems.
This generalized version will be able to simulate rare transition using the TPS scheme in active systems, and more in general in non-equilibrium systems, for the first time.
We will then validate this new version of the algorithm against brute-force simulations by studying the target-search problem of an active Brownian particle in a double-well potential, which will involve the climb of an energy barrier.

The second chapter of this thesis focuses on a systematic characterization of the main features of an active target-search process in an external energy landscape.
This will be achieved by providing an analysis of the parameter space of the model and of the robust features displayed by the target-search paths.
We will then perform this analysis in two different confining energy landscapes displaying important differences, so that the robustness of the target-search patterns can be tested in different scenarios.

The third chapter is centered on the generalization of the committor function to a target-search problem performed by an active Brownian particle and on the development of an efficient way to compute it.
The chapter will open with the derivation of the Fokker-Planck equation for an active Brownian particle moving in an external energy landscape, and will be followed by the derivation of an expression for the committor function for the agent.
Next, we will prove that the active committor function satisfies the backward-Kolmogorov equation for an active agent, in a similar fashion to a passive system, and we will take advantage of this property to develop an efficient way to compute the committor.
In particular, the solution of the backward-Kolmogorov equation can be numerically estimated using a finite-differences algorithm on a lattice.
Finally, we will validate the committor obtained as a solution of the backward-Kolmogorov equation with brute-force simulations in a simple double-well potential.

The fourth and final chapter of this thesis is devoted to an experimental validation of the active committor by means of a self-propelled camphor-camphene sphere.
We will compare the experimental results of the motion of such a self-propelled sphere in a Petri dish with a brute-force simulation of an active Brownian particle with a negligible translational diffusion process exploring a circular environment with rigid boundaries.
After defining a reactant and a product state for the system, we will compare the theoretical prediction of the committor obtained from the simulation with the experimental estimate of the same quantity.

\afterpage{\null\newpage}

\pagestyle{plain}

\chapter{Simulating rare transitions in out-of-equilibrium settings}
\label{ch:prl_tps_abps}
\pagestyle{fancy}

\section{Enhanced sampling algorithms}

As we discussed in the introduction, rare transitions are characterized by a separation of timescales between the short times required to locally sample a metastable state of the system and the longer times required to migrate to a different metastable state.
Consequently, these transitions constitute rare events, which usually require waiting times much larger than the typical time scale at which the system is studied before a single event can be observed.
For this reason, studying these systems through computer simulations and direct integration of their equations of motion can become cumbersome in the best cases, and totally unfeasible in others.
In fact, simulating a complex system consisting of many degrees of freedom for a considerable amount of time will require computational resources which, in most cases, are not available yet.
Even though the supercomputing era is steadily increasing the availability of the resources needed to simulate these systems, a complete description of some processes will probably remain out of reach for the upcoming years.
For example, the simulation of processes such as protein folding at the atomistic resolution can be currently studied only using specific-purpose supercomputers and only for the case of small proteins, consisting of relatively short ammino acids chains~\cite{Shaw2010}, while longer chains and biologically relevant timescales still remain out of reach.
For this reason, a different approach has been developed by the statistical physics community to study these systems in the past few decades, which is known as \emph{enhanced sampling}.
Enhanced sampling relies on finding smart shortcuts to study the relevant parts of these processes, without the need of investing long simulation times to investigate the complete evolution of the system.
Several enhanced-sampling techniques have been developed in the past few decades, relying on different schemes and ideas.
Here we will provide a brief summary of the main enhanced-sampling methods, to then move to a more detailed description of one of them, the Transition Path Sampling (TPS) algorithm.

Among the many algorithms designed to deal with the intrinsic separation of timescales of these problems, \emph{Replica Exchange Molecular Dynamics}~\cite{Sugita1999} (known also as \emph{parallel tempering}) has been extensively used to simulate various biological systems.
The idea behind the algorithm is to simulate several copies of the system at different temperatures, referred to as replica, and, at fixed time intervals, to swap the replica with temperatures close to each other, using a Metropolis-Monte-Carlo procedure.
This allows for an efficient simulation of the system without wasting computational resources in long trajectories locally exploring metastable states of the system.
However, this scheme has a disadvantage, in fact the stochastic changes in temperature do not allow for a rigorous study of the kinetics for the transition process.

\emph{Umbrella sampling}~\cite{Torrie1974} relies instead on studying the evolution of the system using some collective variables that capture the relevant features of the process.
The motion of the system in configuration space is then biased by adding some Gaussian potentials (\emph{umbrella potentials}) to the equations of motion of the variables, which will be located in the configuration space points already visited by the variables.
These history-dependent terms ensure an efficient exploration of the free energy landscape, in fact the addition of these Gaussian potentials in the most visited regions of the system (kinetic traps or metastable states) will cause an effective decrease in the free-energy barrier heights and a faster exploration of the landscape.
Subsequently, an unbiasing procedure is used to obtain the unbiased simulation data, the \emph{weighted histogram analysis method}~\cite{Kumar1992}.
This method has however some drawbacks, namely the fact that it can be difficult to obtain a uniform sampling of the free space by an \textit{a priori} definition of the Gaussian potential parameters~\cite{Bartels1997}, and for this reason some modifications of this algorithm were introduced, such as \emph{Adaptive umbrella sampling}~\cite{Mezei1987}.

In a similar fashion to Umbrella sampling, \emph{Metadynamics}~\cite{Laio2002} assumes that a finite number of relevant reaction coordinates for the process exist, and relies on studying the exploration of the free energy space along these coordinates.
This exploration is analyzed by simulating several independent replica of the system at the same temperature, in each of which the coordinates are constrained on having a specific set of values.
From the statistically independent replica an estimate of the thermodynamic observables can then be obtained.
The procedure that is employed to ensure an efficient exploration of the free energy landscape involves again the introduction of a history-dependent term to the equations of motion of the variables, which also in this case is a Gaussian potential located in the configurations visited by the system.
The major disadvantages of metadynamics are the fact that the Gaussian bias potentials do not converge to the free energy landscape for the process but oscillates around it, and also the fact that the collective variables to be used are not easily determined \textit{a priori}.
To solve these problems, modified versions of the algorithm, such as \emph{well-tempered metadynamics}~\cite{Barducci2008,Barducci2011}, were introduced, although they cannot be easily implemented for all systems.

\emph{Markov state models}~\cite{Husic2018} on the other hand, are based on a quite different approach, in fact they rely on providing a coarse-grained representation of the dynamics of the system.
Provided that the system is at thermal equilibrium and Markovian, the exploration mechanism of the different microstates of the system (which represent the metastable states) can be encoded in a transition probability matrix, which satisfies detailed balance and provides the kinetic information on the system.
In this case, since the system is Markovian, the transition probability of moving from a specific microstate to another is not influenced by the history of the system but only from the current state in which the system is found.
Using the transition probability matrix, a master equation for the evolution of the system can be built, and the populations of the states can yield estimates of the free energies associated with each state.

\subsection{Transition Path Sampling algorithm}
\label{sec:TPS_base}
The \emph{Transition Path Sampling} (TPS)~\cite{Dellago1998,Bolhuis2002,Dellago2002} algorithm characterizes the transition processes between metastable states by analyzing the \emph{transition path ensemble}, \textit{i.e.} the ensemble of reactive paths linking the initial state of the system (reactant) to a final state for the process (product), with a proper definition of the two states and an initial guess of a reactive trajectory as the only requirements.
The idea of the algorithm relies on generating reactive trajectories linking the reactant and the product states and then accepting them in the ensemble of reactive paths according to a Metropolis-Monte-Carlo criterion.

Let us start from a formal definition of the probability associated with a single reactive path in configuration space.
A trajectory in a configuration space $\Omega$, $\mathcal{W}(\mathcal{T})$ (where $\mathcal{T}$ is the time duration of the trajectory), will have an associated probability of being observed in the system, $\mathcal{P}[\mathcal{W}(\mathcal{T})]$.
To ensure that this trajectory is reactive, however, two conditions need to be imposed on the endpoints, and the probability of such a trajectory of being reactive therefore reads:
\begin{equation}
\mathcal{P}_{\text{RP}}[\mathcal{W}(\mathcal{T})] = \frac{1}{\mathcal{Z}_{\text{RP}}(\mathcal{T})}h_{\text{R}}[w_{0}]\mathcal{P}[\mathcal{W}(\mathcal{T})]h_{\text{P}}[w_{\mathcal{T}}] \; ,
\label{eq:react_traj}
\end{equation}
where $h_{\text{R}}$ and $h_{\text{P}}$ are characteristic functions for the reactant and the product state, which ensure that the initial point $w_{0}$ of the trajectory $\mathcal{W}(\mathcal{T})$ is found within R and the final one $w_{\mathcal{T}}$ is found within P.
Namely:
\begin{equation}
h_{\text{R}}[w_{0}] = \begin{cases}
1 & \text{if } w_{0} \in \text{R} \\
0 & \text{otherwise}
\end{cases} \; ,
\end{equation}
and:
\begin{equation}
h_{\text{P}}[w_{\mathcal{T}}] = \begin{cases}
1 & \text{if } w_{\mathcal{T}} \in \text{P} \\
0 & \text{otherwise}
\end{cases} \; .
\end{equation}
Finally, $\mathcal{Z}_{\text{RP}}(\mathcal{T})$ is a normalization factor for the distribution of the trajectories of length $\mathcal{T}$, which can be written as a path integral in trajectory space and reads:
\begin{equation}
\mathcal{Z}_{\text{RP}}(\mathcal{T}) = \int \mathcal{D} \mathcal{W}(\mathcal{T}) h_{\text{R}}[w_{0}]\mathcal{P}[\mathcal{W}(\mathcal{T})]h_{\text{P}}[w_{\mathcal{T}}] \; .
\end{equation}
The ensemble of all reactive paths described by Eq.~\ref{eq:react_traj} constitutes the transition path ensemble, which contains the relevant information on how the transitions between R and P occur.

Now, the trajectory $\mathcal{W}(\mathcal{T})$ can be discretized as a sequence of $N$ microstates $w_{i} \in \Omega$ (with $i = 0, ..., N$) visited by the system during its time evolution, such that $N \Delta t = \mathcal{T}$, where $\Delta t$ is the time interval separating the time steps.
If the system is Markovian (\textit{i.e.} it doesn't have memory and the new state which will be visited after a time $\Delta t$ will depend only on the previous state and not on the complete history of the system) the probability of a specific trajectory can be written as:
\begin{equation}
\mathcal{P}[\mathcal{W}(\mathcal{T})] = \rho(w_{0}) \prod_{i = 0}^{N-1} p(w_{i}\! \rightarrow\! w_{i\!+\!1}) \; ,
\end{equation}
where $\rho(w_{0})$ is the steady-state distribution of the initial states of the path in the system (which, in the case of an equilibrium system, is the equilibrium distribution), and $p(w_{i}\! \rightarrow\! w_{i\!+\!1})$ is the probability of transitioning from the state $w_{i}$ at time $i \Delta t$ to the state $w_{i\!+\!1}$ at time $(i\!+\!1) \Delta t$.
For a reactive trajectory, this yields:
\begin{equation}
\mathcal{P}_{\text{RP}}[\mathcal{W}(\mathcal{T})] \propto h_{\text{R}}[w_{0}] h_{\text{P}}[w_{N}] \rho(w_{0}) \prod_{i = 0}^{N-1} p(w_{i}\! \rightarrow\! w_{i\!+\!1}) \; ,
\end{equation}
where the proportionality relation takes into account the normalization factor.

Now that the probability for a reactive trajectory has been derived, we can delve into the details of the TPS algorithm.
As aforementioned, the idea of the algorithm is based on performing a Monte-Carlo search in trajectory space, gradually building up the transition path ensemble by generating and accepting new reactive trajectories.
If an initial guess of a reactive trajectory is provided to the system, the algorithm generates a Markov chain of reactive trajectories, which are then accepted into the transition path ensemble using a Metropolis criterion.

\begin{figure}[H]
\centering
\includegraphics[height=50mm]{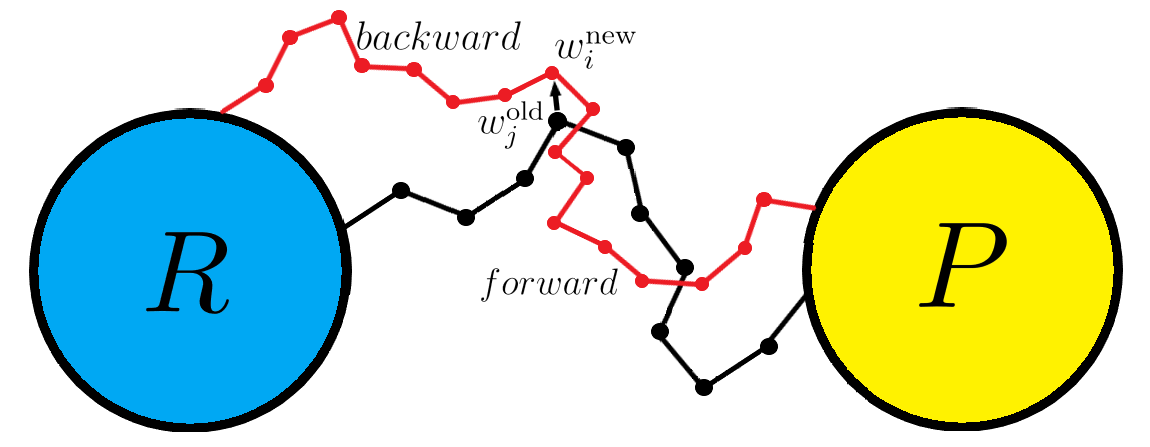}
\caption{Sketch of a shooting move performed by the TPS algorithm. A point $w_{j}^{\text{old}}$ is selected along an initial path (in black) and is perturbed into a new point $w_{i}^{\text{new}}$. From this new point two branches of a new trajectory (in red) are generated by evolving forward and backward in time the equations of motion of the system. The new trajectory is accepted with a probability $\mathcal{P}_{\text{acc}}[\mathcal{W}^{\text{old}}\! \rightarrow\! \mathcal{W}^{\text{new}}]$ in the transition path ensemble. If it is accepted, the new trajectory will become the starting point for the generation of the following one using an iterative procedure.}
\label{fig:TPS_shooting}
\end{figure}

In particular, if an initial reactive trajectory is provided, the algorithm starts by randomly selecting a point along this trajectory and perturbing it.
From that point, evolving the system forward and backward in time following its equations of motion, the algorithm generates two branches of a new trajectory (shooting procedure) (see Fig.~\ref{fig:TPS_shooting} for a sketch of this mechanism).
The new trajectory is then accepted into the transition path ensemble with a certain probability, that is obtained by imposing a detailed-balance condition in trajectory space.
This detailed-balance condition reads:
\begin{equation}
\mathcal{P}_{\text{RP}}[\mathcal{W}^{\text{old}}] \pi[\mathcal{W}^{\text{old}}\! \rightarrow\! \mathcal{W}^{\text{new}}] = \mathcal{P}_{\text{RP}}[\mathcal{W}^{\text{new}}] \pi[\mathcal{W}^{\text{new}}\! \rightarrow\! \mathcal{W}^{\text{old}}] \; ,
\end{equation}
where $\pi[\mathcal{W}^{\text{old}}\! \rightarrow\! \mathcal{W}^{\text{new}}]$ represents the probability of transitioning from the old trajectory to the new one in trajectory space (and vice versa for $\pi[\mathcal{W}^{\text{new}}\! \rightarrow\! \mathcal{W}^{\text{old}}]$).
This transition probability can be expressed as the product of the probability of generating the new trajectory starting from the old one and the probability of accepting the new trajectory in the ensemble of reactive trajectories:
\begin{equation}
\pi[\mathcal{W}^{\text{old}}\! \rightarrow\! \mathcal{W}^{\text{new}}] = \mathcal{P}_{\text{gen}}[\mathcal{W}^{\text{old}}\! \rightarrow\! \mathcal{W}^{\text{new}}] \times \mathcal{P}_{\text{acc}}[\mathcal{W}^{\text{old}}\! \rightarrow\! \mathcal{W}^{\text{new}}] \; .
\end{equation}
By combining this equation with the detailed balance condition in trajectory space, one obtains:
\begin{equation}
\frac{\mathcal{P}_{\text{acc}}[\mathcal{W}^{\text{old}}\! \rightarrow\! \mathcal{W}^{\text{new}}]}{\mathcal{P}_{\text{acc}}[\mathcal{W}^{\text{new}}\! \rightarrow\! \mathcal{W}^{\text{old}}]} = \frac{\mathcal{P}_{\text{RP}}[\mathcal{W}^{\text{new}}] \mathcal{P}_{\text{gen}}[\mathcal{W}^{\text{new}}\! \rightarrow\! \mathcal{W}^{\text{old}}]}{\mathcal{P}_{\text{RP}}[\mathcal{W}^{\text{old}}] \mathcal{P}_{\text{gen}}[\mathcal{W}^{\text{old}}\! \rightarrow\! \mathcal{W}^{\text{new}}]} \; ,
\end{equation}
from which the Metropolis rule for accepting the new trajectories can be obtained:
\begin{equation}
\mathcal{P}_{\text{acc}}[\mathcal{W}^{\text{old}}\! \rightarrow\! \mathcal{W}^{\text{new}}] = \text{min} \Bigg\{ 1, \frac{\mathcal{P}_{\text{RP}}[\mathcal{W}^{\text{new}}] \mathcal{P}_{\text{gen}}[\mathcal{W}^{\text{new}}\! \rightarrow\! \mathcal{W}^{\text{old}}]}{\mathcal{P}_{\text{RP}}[\mathcal{W}^{\text{old}}] \mathcal{P}_{\text{gen}}[\mathcal{W}^{\text{old}}\! \rightarrow\! \mathcal{W}^{\text{new}}]} \Bigg\} \; ,
\end{equation}
which can be further simplified considering that the old trajectory is always reactive by definition, becoming:
\begin{equation}
\label{eq:acc}
\mathcal{P}_{\text{acc}}[\mathcal{W}^{\text{old}}\! \rightarrow\! \mathcal{W}^{\text{new}}] = h[\mathcal{W}^{\text{new}}] \text{min} \Bigg\{ 1, \frac{\mathcal{P}[\mathcal{W}^{\text{new}}] \mathcal{P}_{\text{gen}}[\mathcal{W}^{\text{new}}\! \rightarrow\! \mathcal{W}^{\text{old}}]}{\mathcal{P}[\mathcal{W}^{\text{old}}] \mathcal{P}_{\text{gen}}[\mathcal{W}^{\text{old}}\! \rightarrow\! \mathcal{W}^{\text{new}}]} \Bigg\} \; ,
\end{equation}
where $h[\mathcal{W}^{\text{new}}]$ is a compact form for $h_{\text{R}}[w_{0}^{\text{new}}]h_{\text{P}}[w_{\mathcal{T}}^{\text{new}}]$.

This scheme allows for an efficient and rigorous sampling of the transition path ensemble without wasting computational resources on the simulation of the local relaxation in metastable states, provided that the acceptance probability does not become too small.

Now, since we already provided a definition for a general $\mathcal{P}[\mathcal{W}]$, we only need an expression for the probability of generating a path $\mathcal{P}_{\text{gen}}$.

The shooting moves, which generate new trajectories, are performed by first selecting a random point $w_{j}$ along the previous path with a selection probability $\mathcal{P}_{\text{sel}}(w_{j}^{\text{old}}|\mathcal{W}^{\text{old}})$, to then use it as a starting point for generating two branches of the new trajectory.
This shooting point might first be perturbed with a probability $\mathcal{P}_{\text{pert}}(w_{j}^{\text{old}}\! \rightarrow\! w_{i}^{\text{new}})$, to ensure a better sampling of the trajectory space.
If the dynamics of the system is deterministic, this perturbation is required to ensure a new trajectory different from the previous one, while if the dynamics is stochastic this perturbation might not be necessary.
The two new trajectory branches are then generated by evolving the system forward (for $N^{\text{new}}-i$ steps, where $N^{\text{new}}$ is the number of steps of the new path, which might in general be different from the length of the old path $N^{\text{old}}$) and backward in time (for $i$ steps) following its equations of motion.

This can be formalized by writing the generation probability as:
\begin{equation}
\begin{aligned}[t]
\mathcal{P}_{\text{gen}}[\mathcal{W}^{\text{old}}\!& \rightarrow\! \mathcal{W}^{\text{new}}] = \mathcal{P}_{\text{sel}}(w_{j}^{\text{old}}|\mathcal{W}^{\text{old}})\mathcal{P}_{\text{pert}}(w_{j}^{\text{old}}\! \rightarrow\! w_{i}^{\text{new}}) \times \\
& \times \prod_{k = i}^{N^{\text{new}}-1} p(w_{k}^{\text{new}}\! \rightarrow\! w_{k+1}^{\text{new}}) \prod_{k = 0}^{i-1} \bar{p}(w_{k+1}^{\text{new}}\! \rightarrow\! w_{k}^{\text{new}}) \; ,
\end{aligned}
\end{equation}
where $\bar{p}(w_{k+1}^{\text{new}}\! \rightarrow\! w_{k}^{\text{new}})$ represents the probability of transitioning from the microstate $w_{k+1}^{\text{new}}$ to the microstate $w_{k}^{\text{new}}$ following a dynamics backward in time.

Now, combining this equation with Eq.~\ref{eq:acc}, one finds:
\begin{equation}
\begin{aligned}[t]
&\frac{\mathcal{P}[\mathcal{W}^{\text{new}}] \mathcal{P}_{\text{gen}}[\mathcal{W}^{\text{new}}\! \rightarrow\! \mathcal{W}^{\text{old}}]}{\mathcal{P}[\mathcal{W}^{\text{old}}] \mathcal{P}_{\text{gen}}[\mathcal{W}^{\text{old}}\! \rightarrow\! \mathcal{W}^{\text{new}}]} = \frac{\rho(w_{0}^{\text{new}}) \prod_{k = 0}^{N^{\text{new}}-1} p(w_{k}^{\text{new}}\! \rightarrow\! w_{k\!+\!1}^{\text{new}})}{\rho(w_{0}^{\text{old}}) \prod_{k = 0}^{N^{\text{old}}-1} p(w_{k}^{\text{old}}\! \rightarrow\! w_{k\!+\!1}^{\text{old}})} \times \\
\times & \frac{\mathcal{P}_{\text{sel}}(w_{i}^{\text{new}}|\mathcal{W}^{\text{new}})\mathcal{P}_{\text{pert}}(w_{i}^{\text{new}}\! \rightarrow\! w_{j}^{\text{old}})\prod_{k = j}^{N^{\text{old}}-1} p(w_{k}^{\text{old}}\! \rightarrow\! w_{k+1}^{\text{old}}) \prod_{k = 0}^{j-1} \bar{p}(w_{k+1}^{\text{old}}\! \rightarrow\! w_{k}^{\text{old}})}{\mathcal{P}_{\text{sel}}(w_{j}^{\text{old}}|\mathcal{W}^{\text{old}})\mathcal{P}_{\text{pert}}(w_{j}^{\text{old}}\! \rightarrow\! w_{i}^{\text{new}})\prod_{k = i}^{N^{\text{new}}-1} p(w_{k}^{\text{new}}\! \rightarrow\! w_{k+1}^{\text{new}}) \prod_{k = 0}^{i-1} \bar{p}(w_{k+1}^{\text{new}}\! \rightarrow\! w_{k}^{\text{new}})} \; ,
\end{aligned}
\end{equation}
which after canceling some terms yields:
\begin{equation}
\label{eq:acc_ratio}
\begin{aligned}[t]
&\frac{\mathcal{P}[\mathcal{W}^{\text{new}}] \mathcal{P}_{\text{gen}}[\mathcal{W}^{\text{new}}\! \rightarrow\! \mathcal{W}^{\text{old}}]}{\mathcal{P}[\mathcal{W}^{\text{old}}] \mathcal{P}_{\text{gen}}[\mathcal{W}^{\text{old}}\! \rightarrow\! \mathcal{W}^{\text{new}}]} = \frac{\rho(w_{0}^{\text{new}}) \prod_{k = 0}^{i-1} p(w_{k}^{\text{new}}\! \rightarrow\! w_{k\!+\!1}^{\text{new}})}{\rho(w_{0}^{\text{old}}) \prod_{k = 0}^{j-1} p(w_{k}^{\text{old}}\! \rightarrow\! w_{k\!+\!1}^{\text{old}})} \times \\
\times & \frac{\mathcal{P}_{\text{sel}}(w_{i}^{\text{new}}|\mathcal{W}^{\text{new}})\mathcal{P}_{\text{pert}}(w_{i}^{\text{new}}\! \rightarrow\! w_{j}^{\text{old}})\prod_{k = 0}^{j-1} \bar{p}(w_{k+1}^{\text{old}}\! \rightarrow\! w_{k}^{\text{old}})}{\mathcal{P}_{\text{sel}}(w_{j}^{\text{old}}|\mathcal{W}^{\text{old}})\mathcal{P}_{\text{pert}}(w_{j}^{\text{old}}\! \rightarrow\! w_{i}^{\text{new}})\prod_{k = 0}^{i-1} \bar{p}(w_{k+1}^{\text{new}}\! \rightarrow\! w_{k}^{\text{new}})} \; .
\end{aligned}
\end{equation}

If the system obeys microscopic reversibility, then:
\begin{equation}
\label{eq:micr_rev}
\rho_{0}(w_{k}) p(w_{k}\! \rightarrow\! w_{k\!+\!1}) = \rho_{0}(w_{k\!+\!1}) \bar{p}(w_{k\!+\!1}\! \rightarrow\! w_{k}) 
\end{equation}
where $\rho_{0}(w_{k})$ is the equilibrium distribution in the state $w_{k}$ and analogously $\rho_{0}(w_{k\!+\!1})$ is the equilibrium distribution for $w_{k\!+\!1}$.
This condition can be expressed as:
\begin{equation}
\frac{\rho_{0}(w_{k\!+\!1})}{\rho_{0}(w_{k})} = \frac{p(w_{k}\! \rightarrow\! w_{k\!+\!1})}{\bar{p}(w_{k\!+\!1}\! \rightarrow\! w_{k})} \; ,
\end{equation}
which can in turn be inserted in Eq.~\ref{eq:acc_ratio}, yielding:
\begin{equation}
\begin{aligned}[t]
&\frac{\mathcal{P}[\mathcal{W}^{\text{new}}] \mathcal{P}_{\text{gen}}[\mathcal{W}^{\text{new}}\! \rightarrow\! \mathcal{W}^{\text{old}}]}{\mathcal{P}[\mathcal{W}^{\text{old}}] \mathcal{P}_{\text{gen}}[\mathcal{W}^{\text{old}}\! \rightarrow\! \mathcal{W}^{\text{new}}]} = \frac{\rho(w_{0}^{\text{new}})\mathcal{P}_{\text{sel}}(w_{i}^{\text{new}}|\mathcal{W}^{\text{new}})}{\rho(w_{0}^{\text{old}})\mathcal{P}_{\text{sel}}(w_{j}^{\text{old}}|\mathcal{W}^{\text{old}})} \times \\
& \times \frac{\mathcal{P}_{\text{pert}}(w_{i}^{\text{new}}\! \rightarrow\! w_{j}^{\text{old}})}{\mathcal{P}_{\text{pert}}(w_{j}^{\text{old}}\! \rightarrow\! w_{i}^{\text{new}})} \prod_{k = 0}^{i-1} \frac{\rho_{0}(w_{k\!+\!1}^{\text{new}})}{\rho_{0}(w_{k}^{\text{new}})} \prod_{k = 0}^{j-1} \frac{\rho_{0}(w_{k}^{\text{old}})}{\rho_{0}(w_{k\!+\!1}^{\text{old}})}\; .
\end{aligned}
\end{equation}

This result can be further simplified, obtaining:
\begin{equation}
\begin{aligned}[t]
&\frac{\mathcal{P}[\mathcal{W}^{\text{new}}] \mathcal{P}_{\text{gen}}[\mathcal{W}^{\text{new}}\! \rightarrow\! \mathcal{W}^{\text{old}}]}{\mathcal{P}[\mathcal{W}^{\text{old}}] \mathcal{P}_{\text{gen}}[\mathcal{W}^{\text{old}}\! \rightarrow\! \mathcal{W}^{\text{new}}]} = \frac{\rho(w_{0}^{\text{new}})\mathcal{P}_{\text{sel}}(w_{i}^{\text{new}}|\mathcal{W}^{\text{new}})}{\rho(w_{0}^{\text{old}})\mathcal{P}_{\text{sel}}(w_{j}^{\text{old}}|\mathcal{W}^{\text{old}})} \times \\
& \times \frac{\mathcal{P}_{\text{pert}}(w_{i}^{\text{new}}\! \rightarrow\! w_{j}^{\text{old}})}{\mathcal{P}_{\text{pert}}(w_{j}^{\text{old}}\! \rightarrow\! w_{i}^{\text{new}})} \frac{\rho_{0}(w_{i}^{\text{new}})}{\rho_{0}(w_{0}^{\text{new}})} \frac{\rho_{0}(w_{0}^{\text{old}})}{\rho_{0}(w_{j}^{\text{old}})} \; ,
\end{aligned}
\end{equation}
which, if the system is at equilibrium and consequently the initial distribution of the points is the equilibrium distribution (so $\rho(w_{0}) = \rho_{0}(w_{0})$), can be written as:
\begin{equation}
\begin{aligned}[t]
\frac{\mathcal{P}[\mathcal{W}^{\text{new}}] \mathcal{P}_{\text{gen}}[\mathcal{W}^{\text{new}}\! \rightarrow\! \mathcal{W}^{\text{old}}]}{\mathcal{P}[\mathcal{W}^{\text{old}}] \mathcal{P}_{\text{gen}}[\mathcal{W}^{\text{old}}\! \rightarrow\! \mathcal{W}^{\text{new}}]} = & \, \frac{\mathcal{P}_{\text{sel}}(w_{i}^{\text{new}}|\mathcal{W}^{\text{new}})}{\mathcal{P}_{\text{sel}}(w_{j}^{\text{old}}|\mathcal{W}^{\text{old}})} \times \\
& \times \frac{\mathcal{P}_{\text{pert}}(w_{i}^{\text{new}}\! \rightarrow\! w_{j}^{\text{old}})}{\mathcal{P}_{\text{pert}}(w_{j}^{\text{old}}\! \rightarrow\! w_{i}^{\text{new}})} \frac{\rho_{0}(w_{i}^{\text{new}})}{\rho_{0}(w_{j}^{\text{old}})} \; .
\end{aligned}
\end{equation}

From this expression, the acceptance probability can be finally obtained from Eq.~\ref{eq:acc} as:
\begin{equation}
\label{eq:acc_final}
\begin{aligned}[t]
\mathcal{P}_{\text{acc}}[\mathcal{W}^{\text{old}}\! \rightarrow\! \mathcal{W}^{\text{new}}] = & \, h[\mathcal{W}^{\text{new}}] \times \\
& \times \text{min} \Bigg\{ 1, \frac{\mathcal{P}_{\text{sel}}(w_{i}^{\text{new}}|\mathcal{W}^{\text{new}})\mathcal{P}_{\text{pert}}(w_{i}^{\text{new}}\! \rightarrow\! w_{j}^{\text{old}})\rho_{0}(w_{i}^{\text{new}})}{\mathcal{P}_{\text{sel}}(w_{j}^{\text{old}}|\mathcal{W}^{\text{old}})\mathcal{P}_{\text{pert}}(w_{j}^{\text{old}}\! \rightarrow\! w_{i}^{\text{new}})\rho_{0}(w_{j}^{\text{old}})} \Bigg\} \; ,
\end{aligned}
\end{equation}

Finally, in the specific case of a passive Brownian particle obeying the following stochastic equation of motion:
\begin{equation}
\label{eom_p}
\bm{r}_{i\!+\!1} = \bm{r}_{i} - \mu \bm{\nabla} U(\bm{r}_{i}) \Delta t + \sqrt{2D\Delta t} \, \bm{\xi}_i\;,
\end{equation}
where $\bm{r}_{i}$ is a vector identifying the microstate $w_{i}$ for the system, then the probability of transitioning from the microstate $w_{i}$ to $w_{i+1}$ reads (for more details see Appendix~\ref{app:PI}, where this equation is obtained in the case of a 2D system):
\begin{equation}
p(w_{i}\! \rightarrow\! w_{i\!+\!1}) = \mathcal{N} e^{- \frac{[\bm{r}_{i\!+\!1} - \bm{r}_{i} + \mu\! \Delta\! t \bm{\nabla}_{\bm{r}}\! U(\bm{r}_{i})]^{2}}{4\! D\! \Delta\! t}} \; ,
\end{equation}
where $\mathcal{N}$ is a constant. From this equation, one can steadily derive the corresponding expression for the probability of a reactive trajectory as:
\begin{equation}
\mathcal{P}_{\text{RP}}[\mathcal{W}(\mathcal{T})] \propto h_{\text{R}}[w_{0}] h_{\text{P}}[w_{N}] \rho(w_{0}) \prod_{i = 0}^{N-1} e^{- \frac{[\bm{r}_{i\!+\!1} - \bm{r}_{i} + \mu\! \Delta\! t \bm{\nabla}_{\bm{r}}\! U(\bm{r}_{i})]^{2}}{4\! D\! \Delta\! t}} \; ,
\end{equation}
where $\rho(w_{0})$ is the equilibrium distribution for the system (Boltzmann distribution).

In this case, the perturbation of the shooting point is not necessary to guarantee the generation of a new path and consequently the ratio of the steady state distributions $\rho_{st}$ in Eq.~\ref{eq:acc_final} simplifies, yielding an acceptance probability of:
\begin{equation}
\mathcal{P}_{\text{acc}}[\mathcal{W}^{\text{old}}\! \rightarrow\! \mathcal{W}^{\text{new}}] = h[\mathcal{W}^{\text{new}}] \text{min} \Bigg\{ 1, \frac{\mathcal{P}_{\text{sel}}(w_{i}^{\text{new}}|\mathcal{W}^{\text{new}})}{\mathcal{P}_{\text{sel}}(w_{j}^{\text{old}}|\mathcal{W}^{\text{old}})} \Bigg\} \; ,
\end{equation}
which in the case of a uniform probability of selecting the shooting points in the two paths of length $N^{\text{old}}$ for the old one and $N^{\text{new}}$ for the new one gives:
\begin{equation}
\label{eq:acc_final_passive}
\mathcal{P}_{\text{acc}}[\mathcal{W}^{\text{old}}\! \rightarrow\! \mathcal{W}^{\text{new}}] = h[\mathcal{W}^{\text{new}}] \text{min} \Bigg\{ 1, \frac{N^{\text{old}}}{N^{\text{new}}} \Bigg\} \; .
\end{equation}

We conclude by pointing out that if the system is at equilibrium and is governed by a stochastic dynamics such as the one provided by Eq.~\ref{eom_p}, the system will obey the detailed-balance condition:
\begin{equation}
\label{eq:det_bal}
\rho_{0}(w_{k}) p(w_{k}\! \rightarrow\! w_{k\!+\!1}) = \rho_{0}(w_{k\!+\!1}) p(w_{k\!+\!1}\! \rightarrow\! w_{k}) \; ,
\end{equation}
which, combined with microscopic reversibility provided by Eq.~\ref{eq:micr_rev}, has:
\begin{equation}
\label{eq:for_back_dyn}
p(w_{k\!+\!1}\! \rightarrow\! w_{k}) = \bar{p}(w_{k\!+\!1}\! \rightarrow\! w_{k})
\end{equation}
as a consequence. This has the important implication that the backward part of the new trajectory generated by TPS can be obtained from the same equations of motion used for the forward evolution of the system.

\subsection{TPS limitations and alternatives}
While the TPS algorithm offers many advantages in studying transition processes involving rare events, namely a way to rigorously and efficiently sample the transition path ensemble without any previous knowledge or bias on the optimal reaction coordinates of the process, it also has some drawbacks and limitations.
In particular, the calculation of rate constants from the TPS algorithm might require expensive calculations depending on the complexity of the problem~\cite{Buijsman2020}, and usually relies on a first simulation to obtain the transition path ensemble, and then on a second simulation that uses a combination of TPS and umbrella sampling to extract thermodynamic observables from the transition path ensemble~\cite{vanErp2003}.
Additionally, while TPS can be used to study equilibrium rare transitions in a smart way, it cannot be easily generalized to describe rare transitions in the many non-equilibrium systems that are found in nature.
In fact, TPS requires microscopic reversibility in order to perform the backward shooting move, which is lacking for non-equilibrium systems.

To achieve a more efficient estimation of the transition rates, a new technique was proposed, called \emph{Transition Interface Sampling} (TIS)~\cite{vanErp2003,vanErp2005}.
This method first introduced the possibility of varying the length of the reactive trajectories generated with TPS to reduce the length of the paths to the shortest possible amount, consequently increasing the efficiency of the algorithm.
Additionally, this algorithm provided the possibility of calculating the transition rates in a more efficient way than the one provided by TPS.
In TPS the transition rates can be obtained by measuring a correlation function $C(t)$, which describes the probability of observing the system in the product state P at time $t$ provided that it was in R at time $0$.
The calculation of $C(t)$ can be performed by writing it as a free energy difference and by using umbrella sampling to estimate it~\cite{Dellago1999}.
In TIS the rates are instead computed by measuring the effective positive fluxes through some surfaces, or interfaces, separating the initial and final states (so without counting multiple times the same transition moving forward and backward across a surface), in this way achieving a reduction of the computational time required to estimate the rates.

Due to the impossibility of applying TPS to out-of-equilibrium settings, several algorithms were later developed to tackle the problem of simulating rare events for non-equilibrium systems.

The first of these was based on the idea of producing the transition paths by generating the noise history associated with the trajectory rather than directly obtaining a new trajectory through the dynamics of the system with its problematic backward part~\cite{Crooks2001}.
However, this noise sampling algorithm can only be used to sample paths that do not change considerably with respect to changes in the noise terms~\cite{Buijsman2020}.

Subsequently, other algorithms such as \emph{Forward Flux Sampling} (FFS), \emph{Milestoning}, or \emph{Non-equilibrium Umbrella Sampling} (NEUS) were introduced to cope with the problem of sampling rare events in non-equilibrium systems.

FFS~\cite{Allen2005,Allen2006}, similarly to TIS, uses the concept of several interfaces located along a chosen reaction coordinate for the problem, in order to deal with the impossibility of generating a backward trajectory for the system~\cite{Buijsman2020,Dellago2009}.
Since only forward shooting is allowed, this algorithm relies on collecting short trajectories linking two subsequent interfaces along the reaction coordinate, which start from the first interface and reach the second by evolving forward in time the equations of motion of the system.
Then, the end points of these short trajectories on the second interface are collected and used as starting points for new trajectories moving forward in time.
Some (or many, depending on the rarity of the transitions for that system) of these trajectories will be unable to reach the subsequent interface, while others will manage to, and will therefore link the current interface to the following one.
This ratchet procedure is iterated until the final state for the system is reached.
This algorithm allows for an efficient generation of productive paths linking the initial to the final state of the system by just shooting forward in time, and, through the calculation of the transition probabilities between an interface and the following one, allows also for an efficient estimation of the transition rate for the reaction.
However, this algorithm relies on the knowledge (or on a good guess) of a reaction coordinate for the process, whose lack can otherwise backfire in generating a transition ensemble depending on a wrong choice of the reaction coordinate.

Similarly, Milestoning~\cite{Faradjian2004} requires an initial guess of a reaction coordinate whose associated timescale is slower than the one of the other degrees of freedom of the system.
Starting from this initial guess, a set of surfaces perpendicular to the reaction coordinate (milestones) are identified and for each of the surfaces a set of short trajectories are started, which terminate when reaching one of the neighboring milestones.
From these short trajectories, the probabilities of successfully reaching the two nearest milestones after a time $\tau$ are achieved, and are then used to obtain the kinetics of the system.

Finally, NEUS~\cite{Warmflash2007,Dickson2009} relies on the idea that if a system is simulated only within a specific region of its space and the flux from outside the region into it is used to weight the trajectory slices correctly, one could obtain the same results as if one was running a long ergodic trajectory and then checking the trajectory slices contained in that portion of space only.
The algorithm works by dividing the space of the system of interest into regions and on performing standard simulations within these regions.
If the system tries to exit from one of these boxes, it is brought back to a boundary configuration obtained from the information provided by the flux coming from the neighboring boxes.
These fluxes are adjusted iteratively with some constant weighting factors to account for the non-physical number of system simulations per box, although this increases the computational cost of the procedure compared to FFS~\cite{Dickson2010}.
Additionally, the computational cost of the algorithm scales with the number of boxes and the boxes have to be chosen small enough to visit all the states in every box, further increasing the computational cost~\cite{Warmflash2007}.

Recently, two new transition-path-sampling algorithms~\cite{Buijsman2020} have been proposed to study rare transitions in non-equilibrium processes, with an application on a specific active particle system, modeled through the Vicsek model.
These two algorithms are based on a mixed approach between TPS and FFS.
The main idea of these algorithms is the sampling of non-equilibrium transition paths using a Monte-Carlo search in trajectory space without the need of a predefined reaction coordinate.
In the first of the two, the initial path is used to compute a biasing field that favors the system transition towards the final state.
In the second, a path variable is used to define interfaces that guide the system towards the product state.
However, both of these algorithms circumvent the problem of irreversible dynamics in non-equilibrium systems by shooting only forward in time, without the possibility of generating backward branches, and for this reason may be inefficient in some scenarios.

A generalization of TPS to non-equilibrium systems characterized by rare events is therefore still missing.
The community struggled in the past few years to find algorithms to simulate rare events in non-equilibrium settings that feature a conceptual simplicity and efficiency as TPS does and that do not rely on prior knowledge on the system's reaction coordinates.
For this reason, a generalization of this algorithm to out-of-equilibrium systems is highly desirable.
In the next sections we will describe how such a generalization can be achieved in the case of an active Brownian particle, opening the road to the use of this algorithm to the more general class of non-equilibrium transitions.

\section{Generalizing TPS to active Brownian particles}

As pointed out in section \ref{sec:TPS_base}, while in equilibrium systems the backward probability is related to the forward probability thanks to microscopic reversibility, this property is lacking for systems out of equilibrium.
Therefore, when the TPS algorithm is applied to non-equilibrium settings, the main obstacle resides in the generation of a backward dynamics for the system and on its related shooting move.
In particular, the active Brownian particle, thanks to its self-propulsion velocity, has a microscopically irreversible dynamics with entropy production along trajectories~\cite{Dabelow2019}.
Here, we will show how the problem of not knowing the exact microscopic backward dynamics for the system can be overcome and how backward shooting can be performed in the case of an active Brownian particle performing target search in a double-well potential.

We start by considering an active Brownian particle in 2D with equations of motion defined by a set of Langevin equations in It\^{o} discretized form:
\begin{subequations}
\begin{eqnarray}\label{eom1}
\bm{r}_{i\!+\!1} &=& \bm{r}_{i} + v\, \bm{u}_{i} \, \Delta t - \mu \bm{\nabla} U(\bm{r}_{i}) \Delta t + \sqrt{2D\Delta t} \, \bm{\xi}_i\;,\\ \label{eom21}
\vartheta_{i\!+\!1} &=& \vartheta_{i} + \sqrt{2D_{\vartheta}\Delta t} \, \eta_i\;.
\end{eqnarray}
\end{subequations}
Note that here we adopt a notation for the noise, positions and indexing different from the one reported in Eqs.~(\ref{eom:ppi1},\ref{eom:ppi2}) as we will discuss below.

First, we will proceed to derive the path probabilities for trajectories performed by ABPs, which will lead to the formulation of a stochastic path integral for ABPs.
Then, we will show how a backward dynamics for this system can be defined to perform the backward shooting and we will use the explicit expression of the path probabilities to derive the acceptance rule for the Metropolis criterion.
Finally, we will use this generalized TPS algorithm to study the problem of an ABP trying to reach a target located on the opposite side of an high energy barrier in a paradigmatic double-well potential and we will validate the algorithm with brute-force molecular-dynamics simulations.

\subsection{Path-integral formulation for ABPs}
\label{sec:PI_ABP}
Similarly to the case of passive particles discussed in Appendix~\ref{app:PI}, the ABP system is Markovian, therefore its trajectory in configuration space is uniquely defined once the sequence of noise terms is determined.
Therefore, the process for obtaining the path probability for a trajectory of an ABP works as follows:
\begin{itemize}
\item first, the probabilities for the discrete series of the random variables $\xi_{x}$, $\xi_{y}$ and $\eta$ are derived separately;
\item these probabilities are then used with a change of variables to obtain the probabilities for a series of positions $x$, $y$, and angles $\vartheta$;
\item the probabilities for the series of positions and angles are then combined to obtain the discretized probability of observing a specific path;
\item finally, introducing a continuous notation, we obtain a stochastic path-integral description for the trajectories of an ABP.
\end{itemize}

\subsubsection{Probability of a series of angles}
The probability distribution associated with a discrete time series of $N$ independent Gaussian random variables is trivially the product of the probabilities of extracting the random variable at each discretized time step, therefore:
\begin{equation}
P\big[\eta(t_{0})\! \rightarrow\! \eta(t_{1}) \rightarrow ... \rightarrow \eta(t_{N\!-\!1})\big] = \Bigg( \frac{1}{\sqrt{2 \pi}} \Bigg)^{N} \prod_{k = 0}^{N\!-\!1} e^{-\frac{\eta_{k}^{2}}{2}} \; .
\end{equation}

Now, if we perform a change of variables from $\eta_{i}$ to $\vartheta_{i\!+\!1}$ according to $\eta_{i} = (\vartheta_{i\!+\!1} - \vartheta_{i})/\sqrt{2 D_{\vartheta} \Delta t}$, we can transform the probability for the series of noise variables to a probability of a series of angles\footnote{Note that while the angular variable $\vartheta_{i}$ has $N+1$ discrete values, the noise term has only $N$ values, because the noise term $\eta_{i}$ is used to obtain the angle $\vartheta_{i\!+\!1}$ and the angle $\vartheta_{0}$ is fixed by initial conditions. Therefore, the time indexes for the angle series changed from $0, ..., N-1$ of the noise series to $1, ..., N$ for the angle. This also implies that since the initial angle $\vartheta_{0}$ is fixed, to obtain the probability for the series of angles we need to perform the change of variable $\eta_{i}\! \rightarrow\! \vartheta_{i\!+\!1}$ to be consistent with our choice of indexing angles and noises. Our choice is analogous to what is reported in~\cite{Adib2008}, while a different indexing choice for the noise is adopted for instance in~\cite{Elber2000} with a different change of variables to obtain the same probability for the series of positions.}:
\begin{equation}
P\big[\vartheta(t_{1})\! \rightarrow\! \vartheta(t_{2})\! \rightarrow\! ...\! \rightarrow\! \vartheta(t_{N})\big] = \Bigg( \frac{1}{\sqrt{2 \pi}} \Bigg)^{N} \prod_{k = 0}^{N\!-\!1} e^{-\frac{(\vartheta_{k\!+\!1}-\vartheta_{k})^{2}}{4D_{\vartheta} \Delta t}} J \Bigg[ \frac{\partial \eta}{\partial \vartheta}\Bigg] \; ,
\end{equation}
where $J \Big[ \frac{\partial \eta}{\partial \vartheta}\Big]$ is the determinant of the Jacobian for the chosen change in coordinates. It reads:
\begin{center}
$J \left[ \frac{\partial \eta}{\partial \vartheta} \right] = det$ \(
\begin{bmatrix}
\frac{\partial \eta_{0}}{\partial \vartheta_{1}} & \frac{\partial \eta_{0}}{\partial \vartheta_{2}} & \dots & \frac{\partial \eta_{0}}{\partial \vartheta_{N}} \\
\frac{\partial \eta_{1}}{\partial \vartheta_{1}} & \frac{\partial \eta_{1}}{\partial \vartheta_{2}} & \dots & \frac{\partial \eta_{1}}{\partial \vartheta_{N}} \\
\vdots & \vdots & \ddots & \vdots \\
\frac{\partial \eta_{N\!-\!1}}{\partial \vartheta_{1}} & \frac{\partial \eta_{N\!-\!1}}{\partial \vartheta_{2}} & \dots & \frac{\partial \eta_{N\!-\!1}}{\partial \vartheta_{N}} \\
\end{bmatrix}
\) ,
\end{center}
and therefore:
\begin{center}
$J \left[ \frac{\partial \eta}{\partial \vartheta} \right] = det$
\(
\begin{bmatrix}
\frac{1}{\sqrt{2 D_{\vartheta} \Delta t}} & 0 & \dots & 0 \\
-\frac{1}{\sqrt{2 D_{\vartheta} \Delta t}} & \frac{1}{\sqrt{2 D_{\vartheta} \Delta t}} & \dots & 0 \\
\vdots & \ddots & \ddots & \vdots \\
0 & \dots & \frac{1}{\sqrt{2 D_{\vartheta} \Delta t}} & 0 \\
0 & \dots & -\frac{1}{\sqrt{2 D_{\vartheta} \Delta t}} & \frac{1}{\sqrt{2 D_{\vartheta} \Delta t}} \\
\end{bmatrix}
\)
$= \Big(\frac{1}{\sqrt{2 D_{\vartheta} \Delta t}}\Big)^{N}$ .
\end{center}

The probability distribution for the angles therefore reads:
\begin{equation}
P\big[\vartheta(t_{1})\! \rightarrow\! \vartheta(t_{2})\! \rightarrow\! ...\! \rightarrow\! \vartheta(t_{N})\big] = \mathcal{N}_{\vartheta} \prod_{k = 0}^{N\!-\!1} e^{-\frac{(\vartheta_{k\!+\!1}-\vartheta_{k})^{2}}{4D_{\vartheta} \Delta t}} \; ,
\end{equation}
where $\mathcal{N}_{\vartheta}$ is a normalizing constant.

\subsubsection{Probability of a series of positions}

Similarly, the probability for time series of independent random variables for the spatial components can be obtained as:
\begin{equation}
P\big[\xi_{\alpha}(t_{0})\! \rightarrow\! \xi_{\alpha}(t_{1})\! \rightarrow\! ...\! \rightarrow\! \xi_{\alpha}(t_{N\!-\!1})\big] = \Bigg( \frac{1}{\sqrt{2 \pi}} \Bigg)^{N} \prod_{k = 0}^{N\!-\!1} e^{-\frac{\xi_{\alpha,k}^{2}}{2}} \; ,
\end{equation}
where $\alpha = (x,y)$ in the case of our 2D system.
Let's focus on the probability for the series $\xi_{x,i}$ for simplicity (the procedure can be analogously extended to $\xi_{y,i}$).
If the change of variables $\xi_{x,i} \rightarrow x_{i\!+\!1}$ is adopted according to $\xi_{x,i} = \big(x_{i\!+\!1}-x_{i}-v\cos(\vartheta_{i}) \Delta t + \mu \nabla_{x} U (x_{i},y_{i}) \Delta t\big)/\sqrt{2D\Delta t}$ (where $\nabla_{x} = \partial/\partial x$), the probability for the series of positions can be obtained as:
\begin{equation}
P\big[x(t_{1})\! \rightarrow\! x(t_{2})\! \rightarrow\! ...\! \rightarrow\! x(t_{N})\big] = \Bigg( \frac{1}{\sqrt{2 \pi}} \Bigg)^{N} \prod_{k = 0}^{N\!-\!1} e^{-\frac{(x_{k\!+\!1}-x_{k}-v\cos(\vartheta_{k}) \Delta t + \mu \nabla_{x} U (x_{k},y_{k})\Delta t)^{2}}{4D \Delta t}} J \Bigg[ \frac{\partial \xi_{x}}{\partial x}\Bigg] \; ,
\end{equation}
where this time the Jacobian matrix is built as:
\begin{center}
\(
\begin{bmatrix}
\frac{\partial \xi_{x,0}}{\partial x_{1}} & \frac{\partial \xi_{x,0}}{\partial x_{2}} & \dots & \frac{\partial \xi_{x,0}}{\partial x_{N}} \\
\frac{\partial \xi_{x,1}}{\partial x_{1}} & \frac{\partial \xi_{x,1}}{\partial x_{2}} & \dots & \frac{\partial \xi_{x,1}}{\partial x_{N}} \\
\vdots & \vdots & \ddots & \vdots \\
\frac{\partial \xi_{x,N-1}}{\partial x_{1}} & \frac{\partial \xi_{x,N-1}}{\partial x_{2}} & \dots & \frac{\partial \xi_{x,N-1}}{\partial x_{N}} \\
\end{bmatrix}
\) ,
\end{center}
and its determinant is obtained as:
\begin{center}
$J \left[ \frac{\partial \eta}{\partial \vartheta} \right] = det$
\(
\begin{bmatrix}
\frac{1}{\sqrt{2 D \Delta t}} & 0 & \dots & 0 \\
-\frac{(1-\mu \nabla_{x}^{2}U(x_{1},y_{1}))}{\sqrt{2 D \Delta t}} & \frac{1}{\sqrt{2 D \Delta t}} & \dots & 0 \\
\vdots & \ddots & \ddots & \vdots \\
0 & \dots & \frac{1}{\sqrt{2 D \Delta t}} & 0 \\
0 & \dots & -\frac{(1-\mu \nabla_{x}^{2}U(x_{N\!-\!1},y_{N\!-\!1}))}{\sqrt{2 D \Delta t}} & \frac{1}{\sqrt{2 D \Delta t}} \\
\end{bmatrix} =
\)
$= \Big(\frac{1}{\sqrt{2 D \Delta t}}\Big)^{N}$ .
\end{center}

The probability distribution for the series of $x$ positions therefore reads:
\begin{equation}
P\big[x(t_{1})\! \rightarrow\! x(t_{2})\! \rightarrow\! ...\! \rightarrow\! x(t_{N})\big] = \mathcal{N}_{\alpha} \prod_{k = 0}^{N\!-\!1} e^{-\frac{(x_{k\!+\!1}-x_{k}-v\cos(\vartheta_{k}) \Delta t + \mu \nabla_{x} U (x_{k},y_{k})\Delta t)^{2}}{4D \Delta t}} \; ,
\end{equation}
and analogously for $y$:
\begin{equation}
P\big[y(t_{1})\! \rightarrow\! y(t_{2})\! \rightarrow\! ...\! \rightarrow\! y(t_{N})\big] = \mathcal{N}_{\alpha} \prod_{k = 0}^{N\!-\!1} e^{-\frac{(y_{k\!+\!1}-y_{k}-v\sin(\vartheta_{k}) \Delta t + \mu \nabla_{y} U (x_{k},y_{k})\Delta t)^{2}}{4D \Delta t}} \; ,
\end{equation}
where $\mathcal{N}_{\alpha}$ is a constant with the same value for $x$ and $y$.

\subsubsection{Complete path probability and stochastic path integral for ABPs}
\label{sec:stoc_PI_abp}

Now, to obtain the total probability for a complete path, we can trivially combine the expressions for the series of positions and angles:
\begin{equation}
\begin{aligned}[t]
& P[x(t_{1})\! \rightarrow\! ...\! \rightarrow\! x(t_{N}),y(t_{1})\! \rightarrow\! ...\! \rightarrow\! y(t_{N}),\vartheta(t_{1})\! \rightarrow\! ...\! \rightarrow\! \vartheta(t_{N})] = \\
&= \mathcal{N}_{\alpha}^{2} \mathcal{N}_{\vartheta} \prod_{k=0}^{N\!-\!1} e^{-\frac{(\vartheta_{k\!+\!1}-\vartheta_{k})^2}{4D_{\vartheta}\Delta t}} e^{-\frac{\left(x_{k\!+\!1}-x_{k}-v \cos(\vartheta_{k}) \Delta t +\mu \nabla_{x} U(x_{k},y_{k}) \Delta t \right)^2}{4D\Delta t}} \cdot \\
&\cdot e^{-\frac{\left(y_{k\!+\!1}-y_{k}- v \sin(\vartheta_{k} \Delta t)+ \mu \nabla_{y} U(x_{k},y_{k}) \Delta t \right)^2}{4D\Delta t}} \; ,
\end{aligned}
\end{equation}
which can then be rewritten as:
\begin{equation}
\begin{aligned}[t]
& P[x(t_{1})\! \rightarrow\! ...\! \rightarrow\! x(t_{N}),y(t_{1})\! \rightarrow\! ...\! \rightarrow\! y(t_{N}),\vartheta(t_{1})\! \rightarrow\! ...\! \rightarrow\! \vartheta(t_{N})] = \\
&= \mathcal{N} e^{ \sum_{k = 0}^{N\!-\!1} -\frac{(\vartheta_{k\!+\!1}-\vartheta_{k})^2}{4D_{\vartheta}\Delta t} -\frac{\left(x_{k\!+\!1}-x_{k}-v \cos(\vartheta_{k}) \Delta t +\mu \nabla_{x} U(x_{k},y_{k}) \Delta t \right)^2+\left(y_{k\!+\!1}-y_{k}- v \sin(\vartheta_{k} \Delta t)+ \mu \nabla_{y} U(x_{k},y_{k}) \Delta t \right)^2}{4D\Delta t}} \; ,
\end{aligned}
\end{equation}
where $\mathcal{N} = \mathcal{N}^{2}_{\alpha} \mathcal{N}_{\vartheta}$

Finally, the propagator for the system, which represents the probability of observing the system in a state $w_{\text{f}} = (x_{\text{f}},y_{\text{f}},\vartheta_{\text{f}})$ at time $t$ provided that at time $0$ it was found in the state $w_{0} = (x_{0},y_{0},\vartheta_{0})$, can be obtained performing a path integration along all possible paths linking the initial state $w_{0}$ to the final state $w_{\text{f}}$ and lasting exactly a time $t$.
Turning from the discrete to the continuous notation and using the path-integral formalism, the stochastic path integral can be expressed symbolically as:
\begin{equation}
\begin{aligned}[t]
& p(w_{\text{f}},t|w_{0},0) = \mathcal{Z}^{-1} \int \mathcal{D} x \int \mathcal{D} y \int \mathcal{D} \vartheta e^{-\frac{1}{4D_{\theta}} \int_{0}^{t} d\tau \dot{\vartheta}^{2}(\tau)} \cdot \\
\cdot & e^{-\frac{1}{4D}\int_{0}^{t}d\tau [(\dot{x}(\tau) -v \cos\vartheta(\tau) +\mu \nabla_{x} U(x(\tau),y(\tau)))^{2}+(\dot{y}(\tau) -v\sin\vartheta(\tau) +\mu \nabla_{y} U(x(\tau),y(\tau)))^{2}]} \; ,
\end{aligned}
\end{equation}
where $\mathcal{Z}$ is a normalization constant. This can be finally rewritten as:
\begin{equation}
\label{eq:PI_ABP}
p(w_{\text{f}},t|w_{0},0) = \mathcal{Z}^{-1} \int \mathcal{D} \bm{r} \int \mathcal{D} \vartheta e^{-\frac{1}{4D_{\theta}} S_{\text{rot}}[\vartheta]} e^{-\frac{1}{4D} S_{\text{trans}}[\bm{r},\vartheta]} \; ,
\end{equation}
where $S_{\text{rot}}$ and $S_{\text{trans}}$ are functionals encoding the rotational and translational noise:
\begin{equation}
S_{\text{rot}}[\vartheta] = \int_{0}^{t} d\tau \big[ \dot{\vartheta}(\tau) \big]^{2} \; ,
\end{equation}
\begin{equation}
S_{\text{trans}}[\bm{r},\vartheta] = \int_{0}^{t}d\tau \big[ \dot{\bm{r}}(\tau) -v \bm{u}(\tau) +\mu \bm{\nabla}U(\bm{r}(\tau)) \big]^{2} \; .
\end{equation}
We conclude by pointing out that $S_{\text{trans}}$ is the active version of the Onsager-Machlup functional of a passive particle~\cite{Onsager1953,Machlup1953}.

\subsection{Backward shooting and modified acceptance probability}

\subsubsection{Backward shooting}

In the previous section we derived an expression for the path integral probability of the propagator $p(w_{\text{f}},t|w_{0},0)$ (Eq.~\ref{eq:PI_ABP}).
The probability associated with a single step forward in time along a transition path is then found as:
\begin{equation}
\begin{aligned}[t]
& p(w_{i}\! \rightarrow\! w_{i\!+\!1}) = \Big(\frac{1}{\sqrt{2 D_{\vartheta} \Delta t}}\Big) \Big(\frac{1}{\sqrt{2 D \Delta t}}\Big)^{2} e^{-\frac{(\vartheta_{i\!+\!1}-\vartheta_{i})^2}{4D_{\vartheta}\Delta t}} \cdot \\
&\cdot e^{-\frac{\left(x_{i\!+\!1}-x_{i}-v \cos(\vartheta_{i}) \Delta t +\mu \nabla_{x} U(x_{i},y_{i}) \Delta t \right)^2}{4D\Delta t}} e^{-\frac{\left(y_{i\!+\!1}-y_{i}- v \sin(\vartheta_{i} \Delta t)+ \mu \nabla_{y} U(x_{i},y_{i}) \Delta t \right)^2}{4D\Delta t}} \; ,
\end{aligned}
\end{equation}
or equivalently in vector notation as:
\begin{equation}
\label{eq:sing_step_ABP_for}
p(w_{i}\! \rightarrow\! w_{i\!+\!1}) = \Big(\frac{1}{\sqrt{2 D_{\vartheta} \Delta t}}\Big) \Big(\frac{1}{\sqrt{2 D \Delta t}}\Big)^{2} e^{-\frac{(\vartheta_{i\!+\!1}-\vartheta_{i})^2}{4D_{\vartheta}\Delta t}} e^{-\frac{\left(\bm{r}_{i\!+\!1}-\bm{r}_{i}-v \bm{u}_{i} \Delta t +\mu \bm{\nabla}_{\bm{r}} U(\bm{r}_{i}) \Delta t \right)^2}{4D\Delta t}} \; .
\end{equation}

This equation can also be rewritten in a factorized form that we then use to compute the acceptance probability:
\begin{equation}
\label{eq:sing_step_ABP_for_fact}
\begin{aligned}[t]
&p(w_{i}\! \rightarrow\! w_{i\!+\!1}) = \Big(\frac{1}{\sqrt{2 D_{\vartheta} \Delta t}}\Big) \Big(\frac{1}{\sqrt{2 D \Delta t}}\Big)^{2} e^{-\frac{(\vartheta_{i\!+\!1}-\vartheta_{i})^2}{4D_{\vartheta}\Delta t}} e^{-\frac{v^{2} \Delta t}{4D}}\cdot \\
& \cdot e^{-\frac{v \bm{u}_{i}}{2D} \cdot (\bm{r}_{i}-\bm{r}_{i\!+\!1}-\mu \bm{\nabla}_{\bm{r}} U(\bm{r}_{i}) \Delta t)} \pi(\bm{r}_{i}\!\rightarrow\!\bm{r}_{i\!+\!1}) \; ,
\end{aligned}
\end{equation}
where we used the fact that $\bm{u}_{i}$ is a versor and therefore $\bm{u}^{2}_{i} = 1$ and we indicated with $\pi(\bm{r}_{i}\!\rightarrow\!\bm{r}_{i\!+\!1})$ the probability for a passive particle of transitioning from $\bm{r}_{i}$ to $\bm{r}_{i\!+\!1}$.

To perform the backward shooting in TPS we then need to find an expression for the probability of transitioning backward in time.
Since the system is out of equilibrium, the microscopic reversibility condition expressed in Eq.~\ref{eq:micr_rev} does not hold anymore.
Additionally, also the detailed-balance condition between microstates (Eq.~\ref{eq:det_bal}) is broken for an active-particle system~\cite{bech2016,Cates2012,Dal_Cengio2021}.

Consequently, the advantage of using the same laws of motion for deriving the forward and backward trajectories, as it is done in the case of a passive particle, cannot be applied to the out-of-equilibrium system of the ABP.

However, the impossibility of shooting backward in time can be circumvented.
We will show that one can in principle generate the backward branch of the new trajectories by choosing whichever set of equations of motion for the system evolution backward in time, as long as the choice is accounted for when the acceptance probability of the new trajectories is computed.
The acceptance probability will therefore include a correction term that will depend on the choice of the backward dynamics and will ensure the correct sampling of the transition path ensemble for the system.

We will now proceed by choosing a possible set of equations for the evolution of an ABP backward in time and we will then derive the associated acceptance probability for the new path, which is found to reduce to the usual TPS acceptance probability in the case of vanishing activity.

Let's consider the following choice for the backward equations of motion for an ABP, discretized in It\^{o} form:
\begin{subequations}
\begin{eqnarray}\label{eq:eom_back1}
\bm{r}_{i} &=& \bm{r}_{i\!+\!1} - v\, \bm{u}_{i\!+\!1} \, \Delta t - \mu \bm{\nabla} U(\bm{r}_{i\!+\!1}) \Delta t + \sqrt{2D\Delta t} \, \bm{\xi}_{i\!+\!1}\;,\\ \label{eq:eom_back2}
\vartheta_{i} &=& \vartheta_{i\!+\!1} + \sqrt{2D_{\vartheta}\Delta t} \, \eta_{i\!+\!1}\;.
\end{eqnarray}
\end{subequations}
We point out that Eq.~\ref{eq:eom_back1} is equivalent to Eq.~\ref{eom1} but with an opposite sign on the self-propulsion term (for more details on this choice and the discussion of a different choice of a backward dynamics see section \ref{sec:backward_choice}).
Additionally, the minus sign for the potential ensures that the dynamics reduces to the passive case in the limit of vanishing activity, even though one in principle could have chosen a positive sign for the potential (see again section \ref{sec:backward_choice}).

Now, we can derive the backward probability for a single step starting from Eqs.~(\ref{eq:eom_back1},\ref{eq:eom_back2}) and following an analogous procedure to the one presented in section \ref{sec:PI_ABP}, leading to an expression similar to Eq.~\ref{eq:sing_step_ABP_for}:
\begin{equation}
\label{eq:sing_step_ABP_back}
\bar{p}(w_{i\!+\!1}\! \rightarrow\! w_{i}) = \Big(\frac{1}{\sqrt{2 D_{\vartheta} \Delta t}}\Big) \Big(\frac{1}{\sqrt{2 D \Delta t}}\Big)^{2} e^{-\frac{(\vartheta_{i}-\vartheta_{i\!+\!1})^2}{4D_{\vartheta}\Delta t}} e^{-\frac{\left(\bm{r}_{i}-\bm{r}_{i\!+\!1}+v \bm{u}_{i\!+\!1} \Delta t +\mu \bm{\nabla}_{\bm{r}} U(\bm{r}_{i\!+\!1}) \Delta t \right)^2}{4D\Delta t}} \; .
\end{equation}

This equation can then be factorized (analogously to Eq.~ \ref{eq:sing_step_ABP_for_fact}) as:
\begin{equation}
\label{eq:sing_step_ABP_back_fact}
\begin{aligned}[t]
&\bar{p}(w_{i\!+\!1}\! \rightarrow\! w_{i}) = \Big(\frac{1}{\sqrt{2 D_{\vartheta} \Delta t}}\Big) \Big(\frac{1}{\sqrt{2 D \Delta t}}\Big)^{2} e^{-\frac{(\vartheta_{i}-\vartheta_{i\!+\!1})^2}{4D_{\vartheta}\Delta t}} e^{-\frac{v^{2} \Delta t}{4D}}\cdot \\
& \cdot e^{-\frac{v \bm{u}_{i\!+\!1}}{2D} \cdot (\bm{r}_{i}-\bm{r}_{i\!+\!1}+\mu \bm{\nabla}_{\bm{r}} U(\bm{r}_{i\!+\!1}) \Delta t)} \pi(\bm{r}_{i\!+\!1}\!\rightarrow\!\bm{r}_{i}) \; .
\end{aligned}
\end{equation}

\subsubsection{New acceptance probability}
To compute the acceptance probability for the new trajectories we can then start from Eq.~\ref{eq:acc_ratio} and use the two expressions for the forward and backward single-step probabilities of Eq.~\ref{eq:sing_step_ABP_for_fact} and Eq.~\ref{eq:sing_step_ABP_back_fact}, which, canceling the constant factors, yields:
\begin{equation}
\begin{aligned}[t]
&\frac{\mathcal{P}[\mathcal{W}^{\text{new}}] \mathcal{P}_{\text{gen}}[\mathcal{W}^{\text{new}}\! \rightarrow\! \mathcal{W}^{\text{old}}]}{\mathcal{P}[\mathcal{W}^{\text{old}}] \mathcal{P}_{\text{gen}}[\mathcal{W}^{\text{old}}\! \rightarrow\! \mathcal{W}^{\text{new}}]} = \\
& = \frac{\rho(w_{0}^{\text{new}}) \mathcal{P}_{\text{sel}}(w_{i}^{\text{new}}|\mathcal{W}^{\text{new}})\mathcal{P}_{\text{pert}}(w_{i}^{\text{new}}\! \rightarrow\! w_{j}^{\text{old}})}{\rho(w_{0}^{\text{old}})\mathcal{P}_{\text{sel}}(w_{j}^{\text{old}}|\mathcal{W}^{\text{old}})\mathcal{P}_{\text{pert}}(w_{j}^{\text{old}}\! \rightarrow\! w_{i}^{\text{new}})} \times \\
& \times \frac{\prod_{k = 0}^{i-1} e^{-\frac{(\vartheta^{\text{new}}_{k\!+\!1}-\vartheta^{\text{new}}_{k})^2}{4D_{\vartheta}\Delta t}} e^{-\frac{v^{2} \Delta t}{4D}} e^{-\frac{v \bm{u}^{\text{new}}_{k}}{2D} \cdot (\bm{r}^{\text{new}}_{k}-\bm{r}^{\text{new}}_{k\!+\!1}-\mu \bm{\nabla}_{\bm{r}} U(\bm{r}^{\text{new}}_{k}) \Delta t)} \pi(\bm{r}^{\text{new}}_{k}\!\rightarrow\!\bm{r}^{\text{new}}_{k\!+\!1})}{\prod_{k = 0}^{j-1} e^{-\frac{(\vartheta^{\text{old}}_{k\!+\!1}-\vartheta^{\text{old}}_{k})^2}{4D_{\vartheta}\Delta t}} e^{-\frac{v^{2} \Delta t}{4D}} e^{-\frac{v \bm{u}^{\text{old}}_{k}}{2D} \cdot (\bm{r}^{\text{old}}_{k}-\bm{r}^{\text{old}}_{k\!+\!1}-\mu \bm{\nabla}_{\bm{r}} U(\bm{r}^{\text{old}}_{k}) \Delta t)} \pi(\bm{r}^{\text{old}}_{k}\!\rightarrow\!\bm{r}^{\text{old}}_{k\!+\!1})} \times \\
& \times \frac{\prod_{k = 0}^{j-1} e^{-\frac{(\vartheta^{\text{old}}_{k}-\vartheta^{\text{old}}_{k\!+\!1})^2}{4D_{\vartheta}\Delta t}} e^{-\frac{v^{2} \Delta t}{4D}} e^{-\frac{v \bm{u}^{\text{old}}_{k\!+\!1}}{2D} \cdot (\bm{r}^{\text{old}}_{k}-\bm{r}^{\text{old}}_{k\!+\!1}+\mu \bm{\nabla}_{\bm{r}} U(\bm{r}^{\text{old}}_{k\!+\!1}) \Delta t)} \pi(\bm{r}^{\text{old}}_{k\!+\!1}\!\rightarrow\!\bm{r}^{\text{old}}_{k})}{\prod_{k = 0}^{i-1} e^{-\frac{(\vartheta^{\text{new}}_{k}-\vartheta^{\text{new}}_{k\!+\!1})^2}{4D_{\vartheta}\Delta t}} e^{-\frac{v^{2} \Delta t}{4D}} e^{-\frac{v \bm{u}^{\text{new}}_{k\!+\!1}}{2D} \cdot (\bm{r}^{\text{new}}_{k}-\bm{r}^{\text{new}}_{k\!+\!1}+\mu \bm{\nabla}_{\bm{r}} U(\bm{r}^{\text{new}}_{k\!+\!1}) \Delta t)} \pi(\bm{r}^{\text{new}}_{k\!+\!1}\!\rightarrow\!\bm{r}^{\text{new}}_{k})} \; ,
\end{aligned}
\end{equation}
where $\rho(w)$ is the steady state distribution for an active particle in the state $w$ that is in general different from the equilibrium distribution for a passive particle in the same state, $\rho_{0}(w) = \pi(w)$ (Boltzmann distribution).

This expression can be considerably simplified by canceling some terms and with the following assumptions:
\begin{itemize}
\item if the old path has a length of $N^{\text{old}}$ states and the probability of selecting each state along the path is the same for every state, the probability of selecting the state $w_{j}^{\text{old}}$ is obtained as $1/N^{\text{old}}$ (and similarly if the new path has a length of $N^{\text{new}}$ the probability of selecting the state $w_{i}^{\text{new}}$ will be $1/N^{\text{new}}$)
\item if we chose not to perturb the shooting point the probability of perturbing the state $w_{j}^{\text{old}}$ into the state $w_{i}^{\text{new}}$ will be one if $w_{j}^{\text{old}} = w_{i}^{\text{new}}$ and similarly for the vice versa perturbing probability.
\end{itemize}
In this case we obtain:
\begin{equation}
\begin{aligned}[t]
&\frac{\mathcal{P}[\mathcal{W}^{\text{new}}] \mathcal{P}_{\text{gen}}[\mathcal{W}^{\text{new}}\! \rightarrow\! \mathcal{W}^{\text{old}}]}{\mathcal{P}[\mathcal{W}^{\text{old}}] \mathcal{P}_{\text{gen}}[\mathcal{W}^{\text{old}}\! \rightarrow\! \mathcal{W}^{\text{new}}]} = \\
& = \frac{\rho(w_{0}^{\text{new}}) N^{\text{old}} \prod_{k = 0}^{i-1} e^{-\frac{v \bm{u}^{\text{new}}_{k}}{2D} \cdot (\bm{r}^{\text{new}}_{k}-\bm{r}^{\text{new}}_{k\!+\!1}-\mu \bm{\nabla}_{\bm{r}} U(\bm{r}^{\text{new}}_{k}) \Delta t)} \pi(\bm{r}^{\text{new}}_{k}\!\rightarrow\!\bm{r}^{\text{new}}_{k\!+\!1})}{\rho(w_{0}^{\text{old}}) N^{\text{new}} \prod_{k = 0}^{j-1} e^{-\frac{v \bm{u}^{\text{old}}_{k}}{2D} \cdot (\bm{r}^{\text{old}}_{k}-\bm{r}^{\text{old}}_{k\!+\!1}-\mu \bm{\nabla}_{\bm{r}} U(\bm{r}^{\text{old}}_{k}) \Delta t)} \pi(\bm{r}^{\text{old}}_{k}\!\rightarrow\!\bm{r}^{\text{old}}_{k\!+\!1})} \times \\
& \times \frac{\prod_{k = 0}^{j-1} e^{-\frac{v \bm{u}^{\text{old}}_{k\!+\!1}}{4D} \cdot (\bm{r}^{\text{old}}_{k}-\bm{r}^{\text{old}}_{k\!+\!1}+\mu \bm{\nabla}_{\bm{r}} U(\bm{r}^{\text{old}}_{k\!+\!1}) \Delta t)} \pi(\bm{r}^{\text{old}}_{k\!+\!1}\!\rightarrow\!\bm{r}^{\text{old}}_{k})}{\prod_{k = 0}^{i-1} e^{-\frac{v \bm{u}^{\text{new}}_{k\!+\!1}}{2D} \cdot (\bm{r}^{\text{new}}_{k}-\bm{r}^{\text{new}}_{k\!+\!1}+\mu \bm{\nabla}_{\bm{r}} U(\bm{r}^{\text{new}}_{k\!+\!1}) \Delta t)} \pi(\bm{r}^{\text{new}}_{k\!+\!1}\!\rightarrow\!\bm{r}^{\text{new}}_{k})} \; .
\end{aligned}
\end{equation}

Now we can apply the detailed-balance condition for the passive probability:
\begin{equation}
\pi(\bm{r}^{\text{new}}_{k\!+\!1}\!\rightarrow\!\bm{r}^{\text{new}}_{k}) = \frac{\pi(\bm{r}^{\text{new}}_{k})}{\pi(\bm{r}_{k\!+\!1}^{\text{new}})} \pi(\bm{r}^{\text{new}}_{k}\!\rightarrow\!\bm{r}^{\text{new}}_{k\!+\!1}) \; ,
\end{equation}
(and analogously for the old path), which leads to:
\begin{equation}
\begin{aligned}[t]
&\frac{\mathcal{P}[\mathcal{W}^{\text{new}}] \mathcal{P}_{\text{gen}}[\mathcal{W}^{\text{new}}\! \rightarrow\! \mathcal{W}^{\text{old}}]}{\mathcal{P}[\mathcal{W}^{\text{old}}] \mathcal{P}_{\text{gen}}[\mathcal{W}^{\text{old}}\! \rightarrow\! \mathcal{W}^{\text{new}}]} = \\
& = \frac{\rho(w_{0}^{\text{new}}) N^{\text{old}} \prod_{k = 0}^{i-1} e^{-\frac{v \bm{u}^{\text{new}}_{k}}{2D} \cdot (\bm{r}^{\text{new}}_{k}-\bm{r}^{\text{new}}_{k\!+\!1}-\mu \bm{\nabla}_{\bm{r}} U(\bm{r}^{\text{new}}_{k}) \Delta t)} \pi(\bm{r}^{\text{new}}_{k}\!\rightarrow\!\bm{r}^{\text{new}}_{k\!+\!1})}{\rho(w_{0}^{\text{old}}) N^{\text{new}} \prod_{k = 0}^{j-1} e^{-\frac{v \bm{u}^{\text{old}}_{k}}{2D} \cdot (\bm{r}^{\text{old}}_{k}-\bm{r}^{\text{old}}_{k\!+\!1}-\mu \bm{\nabla}_{\bm{r}} U(\bm{r}^{\text{old}}_{k}) \Delta t)} \pi(\bm{r}^{\text{old}}_{k}\!\rightarrow\!\bm{r}^{\text{old}}_{k\!+\!1})} \times \\
& \times \frac{\prod_{k = 0}^{j-1} e^{-\frac{v \bm{u}^{\text{old}}_{k\!+\!1}}{4D} \cdot (\bm{r}^{\text{old}}_{k}-\bm{r}^{\text{old}}_{k\!+\!1}+\mu \bm{\nabla}_{\bm{r}} U(\bm{r}^{\text{old}}_{k\!+\!1}) \Delta t)} \frac{\pi(\bm{r}^{\text{old}}_{k})}{\pi(\bm{r}^{\text{old}}_{k\!+\!1})} \pi(\bm{r}^{\text{old}}_{k}\!\rightarrow\!\bm{r}^{\text{old}}_{k\!+\!1})}{\prod_{k = 0}^{i-1} e^{-\frac{v \bm{u}^{\text{new}}_{k\!+\!1}}{2D} \cdot (\bm{r}^{\text{new}}_{k}-\bm{r}^{\text{new}}_{k\!+\!1}+\mu \bm{\nabla}_{\bm{r}} U(\bm{r}^{\text{new}}_{k\!+\!1}) \Delta t)} \frac{\pi(\bm{r}^{\text{new}}_{k})}{\pi(\bm{r}^{\text{new}}_{k\!+\!1})} \pi(\bm{r}^{\text{new}}_{k}\!\rightarrow\!\bm{r}^{\text{new}}_{k\!+\!1})} \; .
\end{aligned}
\end{equation}

Finally, after canceling some terms, the ratio can be written as:
\begin{equation}
\label{eq:final_gen_acc}
\begin{aligned}[t]
&\frac{\mathcal{P}[\mathcal{W}^{\text{new}}] \mathcal{P}_{\text{gen}}[\mathcal{W}^{\text{new}}\! \rightarrow\! \mathcal{W}^{\text{old}}]}{\mathcal{P}[\mathcal{W}^{\text{old}}] \mathcal{P}_{\text{gen}}[\mathcal{W}^{\text{old}}\! \rightarrow\! \mathcal{W}^{\text{new}}]} = \frac{N^{\text{old}} \rho(w_{0}^{\text{new}}) \pi(\bm{r}^{\text{old}}_{0}) \pi(\bm{r}^{\text{new}}_{i})}{N^{\text{new}} \rho(w_{0}^{\text{old}}) \pi(\bm{r}^{\text{old}}_{j}) \pi(\bm{r}^{\text{new}}_{0})} \times \\
& \times \frac{\prod_{k = 0}^{i-1} e^{-\frac{v \bm{u}^{\text{new}}_{k}}{2D} \cdot (\bm{r}^{\text{new}}_{k}-\bm{r}^{\text{new}}_{k\!+\!1}-\mu \bm{\nabla}_{\bm{r}} U(\bm{r}^{\text{new}}_{k}) \Delta t)} \prod_{k = 0}^{j-1} e^{-\frac{v \bm{u}^{\text{old}}_{k\!+\!1}}{4D} \cdot (\bm{r}^{\text{old}}_{k}-\bm{r}^{\text{old}}_{k\!+\!1}+\mu \bm{\nabla}_{\bm{r}} U(\bm{r}^{\text{old}}_{k\!+\!1}) \Delta t)}}{\prod_{k = 0}^{j-1} e^{-\frac{v \bm{u}^{\text{old}}_{k}}{2D} \cdot (\bm{r}^{\text{old}}_{k}-\bm{r}^{\text{old}}_{k\!+\!1}-\mu \bm{\nabla}_{\bm{r}} U(\bm{r}^{\text{old}}_{k}) \Delta t)} \prod_{k = 0}^{i-1} e^{-\frac{v \bm{u}^{\text{new}}_{k\!+\!1}}{2D} \cdot (\bm{r}^{\text{new}}_{k}-\bm{r}^{\text{new}}_{k\!+\!1}+\mu \bm{\nabla}_{\bm{r}} U(\bm{r}^{\text{new}}_{k\!+\!1}) \Delta t)}} \; .
\end{aligned}
\end{equation}

Therefore, the acceptance probability of our generalized TPS algorithm is provided by Eq.~\ref{eq:acc} where the value of the ratio of probabilities is computed according to Eq.~\ref{eq:final_gen_acc}.
Note that the value of the ratio in Eq.~\ref{eq:final_gen_acc} for the limit of vanishing activity ($v\!\rightarrow\!0$) reduces to the ratio of a passive particle (Eq.~\ref{eq:acc_final_passive}), provided that the same assumptions on the probabilities of selecting and perturbing points hold.

We conclude this section by providing an alternative form to Eq.~\ref{eq:sing_step_ABP_back_fact} , which has some relevant implications.
If we use the detailed-balance condition for the passive probability (as we did when calculating the new acceptance probability) we obtain:
\begin{equation}
\begin{aligned}[t]
&\bar{p}(w_{i\!+\!1}\! \rightarrow\! w_{i}) = \Big(\frac{1}{\sqrt{2 D_{\vartheta} \Delta t}}\Big) \Big(\frac{1}{\sqrt{2 D \Delta t}}\Big)^{2} e^{-\frac{(\vartheta_{i}-\vartheta_{i\!+\!1})^2}{4D_{\vartheta}\Delta t}} e^{-\frac{v^{2} \Delta t}{4D}} \times \\
& \times e^{-\frac{v \bm{u}_{i\!+\!1}}{2D} \cdot (\bm{r}_{i}-\bm{r}_{i\!+\!1}+\mu \bm{\nabla}_{\bm{r}} U(\bm{r}_{i\!+\!1}) \Delta t)} \pi(\bm{r}_{i}\!\rightarrow\!\bm{r}_{i\!+\!1}) \frac{\pi(\bm{r}_{i})}{\pi(\bm{r}_{i\!+\!1})} \; ,
\end{aligned}
\end{equation}
which can be expressed as:
\begin{equation}
\label{eq:back_act_pass}
\bar{p}(w_{i\!+\!1}\! \rightarrow\! w_{i}) = p(w_{i}\! \rightarrow\! w_{i\!+\!1}) \frac{\pi(\bm{r}_{i})}{\pi(\bm{r}_{i\!+\!1})} \frac{e^{-\frac{v \bm{u}_{i\!+\!1}}{2D} \cdot (\bm{r}_{i}-\bm{r}_{i\!+\!1}+\mu \bm{\nabla}_{\bm{r}} U(\bm{r}_{i\!+\!1}) \Delta t)}}{e^{-\frac{v \bm{u}_{i}}{2D} \cdot (\bm{r}_{i}-\bm{r}_{i\!+\!1}-\mu \bm{\nabla}_{\bm{r}} U(\bm{r}_{i}) \Delta t)}} \; .
\end{equation}
Here, if the last ratio of the second member of the equation is equal to one, we recover the backward probability for a passive particle, which for our choice of the backward dynamics is in fact verified for $v\!\rightarrow\!0$.
If instead the velocity of the particle is different from $0$, the last ratio of Eq.~\ref{eq:back_act_pass} assumes the form of a correction term accounting for the microscopic irreversibility of the process, which resembles a Crooks-like relation~\cite{Crooks2011} for the entropy production along these trajectories.

\section{Validation of the generalized TPS in a double-well system}

\subsection{Target-search transitions in a double-well potential}

\subsubsection{Problem characterization}
\label{sec:pr_car}

Now that we provided a novel generalization of TPS to the out-of-equilibrium setting of an ABP, we will validate it against brute-force simulations in the case of an ABP searching for a target located on the opposite side of an high energy barrier.

\begin{figure}[H]
\centering
\subfigure{\label{fig:side}\includegraphics[height=55.mm]{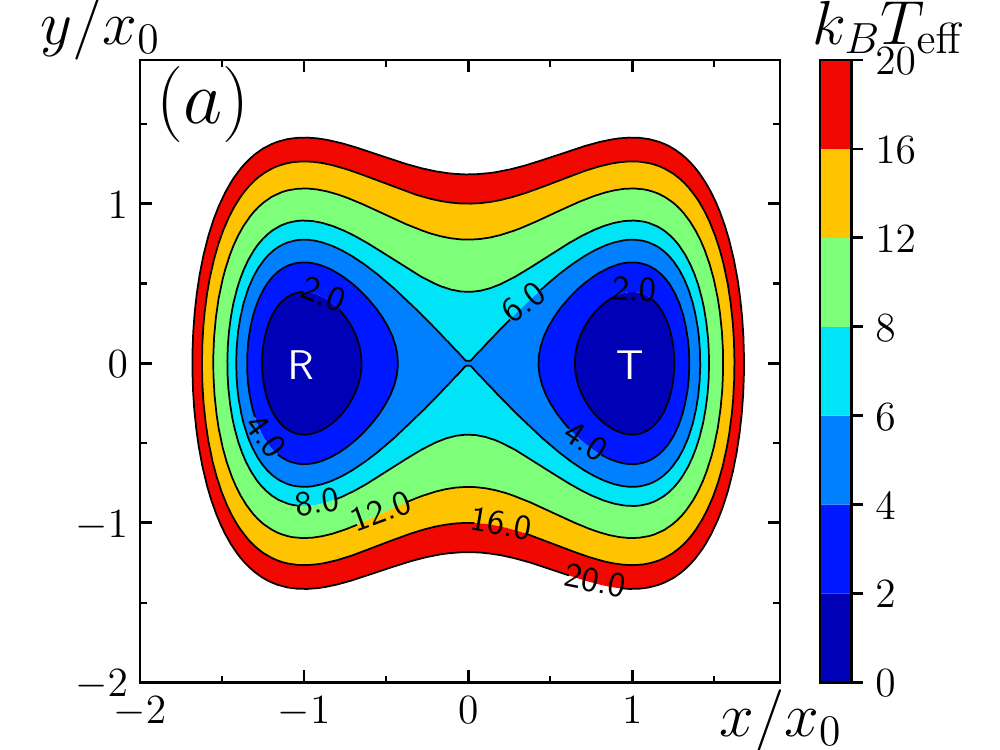}}
\subfigure{\label{fig:top}\includegraphics[height=55.mm]{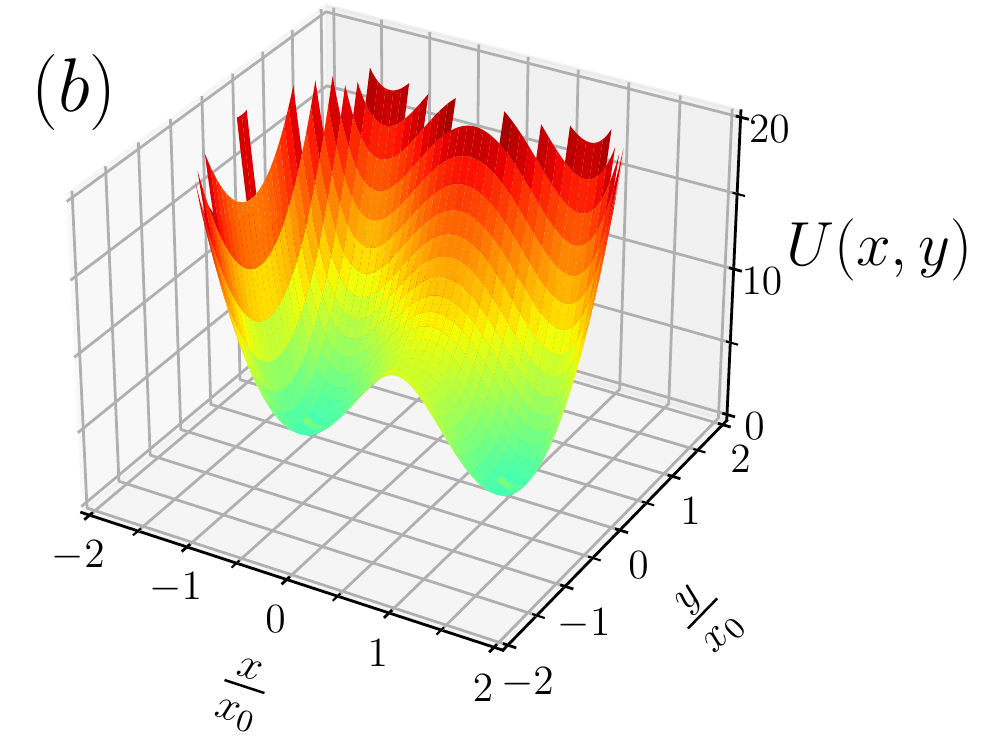}}
\caption{(a) Projection on the $xy$ plane of the double-well potential. The energy value in units of $k_{B}T_{\text{eff}}$ is reported along the potential contour lines. R identifies the reactant basin and T the target basin. (b) 3D representation of the double-well potential, where $U(x,y)$ is reported in units of $k_{B}T_{\text{eff}}$. The values of the potential are represented to an upper limit of $20 k_{B}T_{\text{eff}}$ similarly to panel (a).}
\label{fig:dw_tps}
\end{figure}

The ABP will navigate in a paradigmatic double-well potential, starting from one of the two wells (R state) and trying to reach a target state T located at the bottom of the other well (which is equivalent to the state P in the notation of transition processes).
The energy barrier separating the two states is chosen to be relatively high, such that a separation of timescales emerges in the system, with a fast scale associated with the local sampling of the metastable states for the system (the R and T basins) and a slow one associated with the transitions over the barrier.
At the same time, the barrier is not chosen to be too high, so that a direct comparison of our generalized TPS algorithm with brute-force molecular-dynamics simulations is still possible in order to validate our algorithm.

The double-well potential is defined as:
\begin{equation}
\label{eq:pot_dw_ch1}
U(x,y) = k_{x} (x^{2}-x_{0}^{2})^{2} + \frac{k_{y}}{2} y^{2} \; ,
\end{equation}
where we set $x_{0} = 1$, $k_{x} = 6$, $k_{y} = 20$ and we measure the energies in units of an effective thermal energy $k_{B}T_{\text{eff}} = D/\mu$ where $T_{\text{eff}}$ is the temperature of the external bath in the passive limit, as we introduced in section \ref{sec:ABP_model}.
The R state will be selected as the set of positions $(x,y)$ for which $U(x,y) \leq 2 k_{B}T_{\text{eff}}$ and $x < 0$, while T will be the region of the energy landscape for which $U(x,y) \leq 2 k_{B}T_{\text{eff}}$ and $x > 0$.
A representation of the double-well potential is displayed in Fig.~\ref{fig:dw_tps}.

We proceed by analyzing the ABP behavior for three different regimes, identified by three values of the P{\'e}clet number.
We define the dimensionless P{\'e}clet number as:
\begin{equation}
\text{Pe} := \frac{v\sqrt{3}}{2\sqrt{DD_{\vartheta}}} \; ,
\end{equation}
and the three regimes will be identified by $\text{Pe} = 0$, which is the limit case of the passive particle, and by two gradually increasing values of the activity of the particle, representing an intermediate level of activity ($\text{Pe} = 5$) and a high value of activity ($\text{Pe} = 10$).
These values will be achieved by modifying $v$ while keeping fixed $D = 0.1$, $D_{\vartheta} = 1$ and $\mu = 0.1$ (these parameters are considered fixed up to section \ref{sec:rot_diff_coeff}, where they are varied in the subsequent analysis).
Therefore the values that we employed for the three chosen $\text{Pe}$ numbers are $v = 0$, $v = 1.83$, and $v = 3.65$ for a passive particle, a particle with intermediate activity and a particle with high activity respectively.

We start by generating a set of reactive trajectories departing from R and reaching T, and from these trajectories we then extract the TPTs and compute the transition probability densities and currents for the process.

\subsubsection{Steady-state distribution in the R basin}

To generate these reactive trajectories, we first need to obtain the steady-state distribution in the R state that will be then used to compute the acceptance probability of each path.
To do so, we simulate the system by brute force (\textit{i.e.} by numerically solving Eqs.~(\ref{eom1},\ref{eom21})) until a proper sampling of the distribution is achieved.
While in the passive particle case the steady state distribution in the R state is the Boltzmann distribution in that region, in the case of an active particle this distribution will not be an equilibrium distribution and will depend on the particle parameters, therefore we will refer to it as ‘‘steady-state distribution''.

\begin{figure}[H]
\centering
\includegraphics[height=50mm]{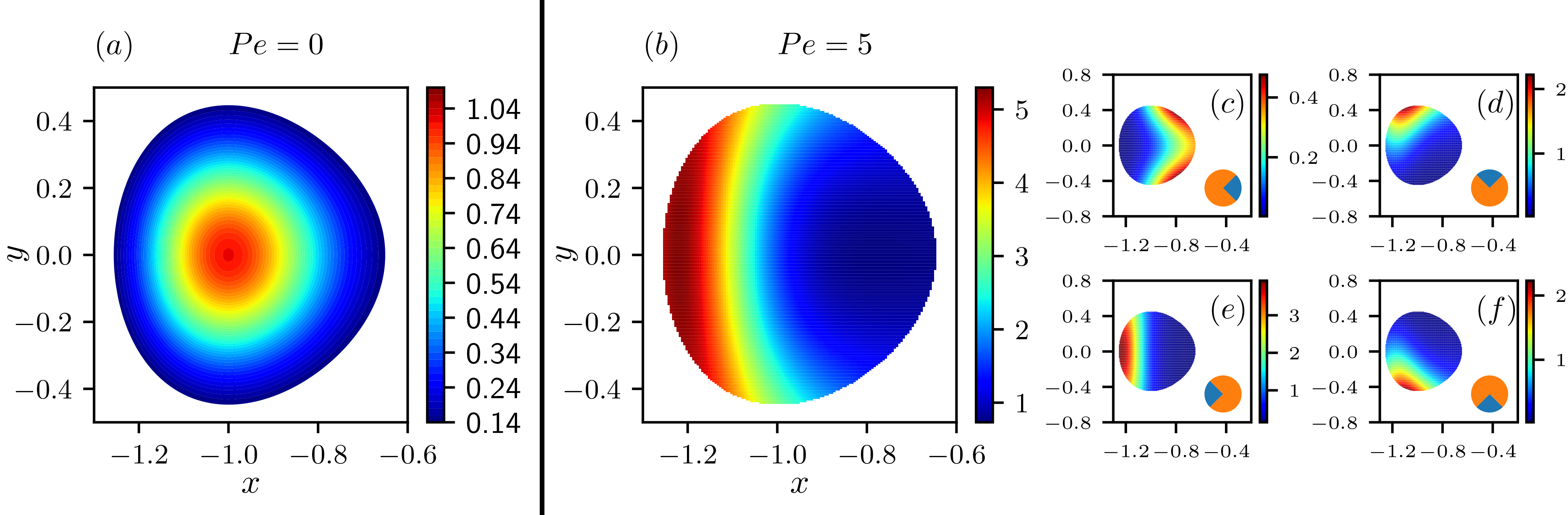}
\caption{Steady-state distributions within the R basin. (a) equilibrium distribution in the case of a passive particle. (b) steady-state distribution of an active particle in the R basin, in the case of $\text{Pe} = 5$. For each point $(x,y)$ of the grid, the value of the probability distribution is obtained as the cumulative distribution over the $64$ angular slices of $\vartheta$ used to histogram the self-propulsion direction during the simulation. (c-f) steady-state distribution within the R basin in the case of $\text{Pe} = 5$ marginalized over the angular component $\vartheta$. Each plot represents the marginalization performed over $16$ angular slices. These slices $\Delta \vartheta$ are located between $-\pi/4$ and $\pi/4$ for panel (c), $\pi/4$ and $3\pi/4$ for panel (d), $3\pi/4$ and $5\pi/4$ for panel (e), and between $5\pi/4$ and $7\pi/4$ for panel (f) (the blue quadrants in the orange pie plots). The figure has been reproduced from the Supplemental Material of our publication Ref.~\cite{Zanovello2021}.}
\label{fig:rho}
\end{figure}

The equilibrium distribution of a passive particle in the R state, being a Boltzmann distribution, is peaked with a maximum at the center of the well (see Fig.~\ref{fig:rho}(a)).
Instead, the active particle in the R basin will have higher chances of being found on the left side than on the right side (see Fig.~\ref{fig:rho}(b)), a behavior emerging due to the effects of the persistence of the particle swimming direction combined with the more steeper potential in that direction.
In fact, if the particle is pointing in the other directions it will be easier to exit from the well (especially towards the right), while if the particle is pointing towards the left, it is more likely to remain inside the R state.
This will happen for at least a time of the order $\tau_{\vartheta} := 1/D_{\vartheta}$, which is the typical time required by the particle to reorient its self-propulsion away from the steep potential in more favorable directions.
Additionally, the probability distribution of the ABP inside the R basin displays a strong dependence on the angle $\vartheta$ which represents the particle orientation (see Fig.~\ref{fig:rho}(c-f)).
From panels (c-f) one infers that the particle is more frequently found with an orientation pointing away from the center of the well and towards the outside.
This happens because the particle crosses quickly the center of the well and spends longer times at the boundaries.

\subsubsection{Validation of the generalized TPS algorithm and TPT distributions}
\label{sec:tpt_dw_validation}

Now that the steady-state distribution in the R basin is known, we can proceed to generate the ensemble of reactive paths using our generalized TPS algorithm and we can validate it by comparing the statistics to the one obtained from brute-force integration of Eqs.~(\ref{eom1},\ref{eom21}).

After generating the ensemble of reactive trajectories (which we obtained by generating $4 \times 10^{8}$ trajectories and then performing an undersampling by retaining only a trajectory every $400$, to reduce the correlation between paths and still achieving a statistics of $10^{6}$ reactive paths), we compute the TPT associated with each path and we construct a TPT distribution for each of the three considered values of the activity. 
Then, we compare these distributions with the ones obtained from $10^{6}$ reactive paths generated through brute-force simulations with the same parameters.

\begin{figure}[H]
\centering
\includegraphics[height=90mm]{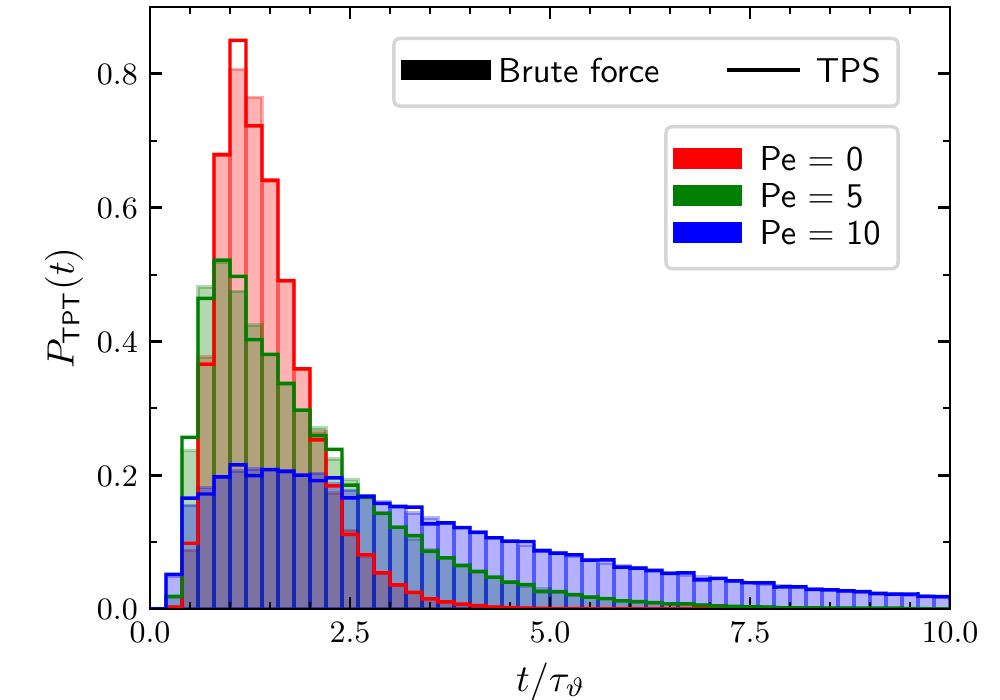}
\caption{TPT distributions for different values of the activity of the particle, represented by the P{\'e}clet number. Histograms with filled bars represent the distributions obtained from brute-force integration of the equations of motion, histograms with empty bars represent the distributions obtained with our generalized TPS. The TPTs are reported in units of $\tau_{\vartheta}$. The figure has been reproduced from our publication Ref.~\cite{Zanovello2021}.}
\label{fig:tpt_tps}
\end{figure}

The results show how the brute-force TPT distributions are correctly reproduced by our generalized TPS algorithm for all the values of the activity considered (see Fig.~\ref{fig:tpt_tps}).
Additionally, from Fig.~\ref{fig:tpt_tps} one infers that the average TPTs increase with increasing activity and concomitantly the TPT distributions broaden with thicker tails at long times.

\subsubsection{Typical target-finding paths}

\begin{figure}[H]
\centering
\includegraphics[height=100mm]{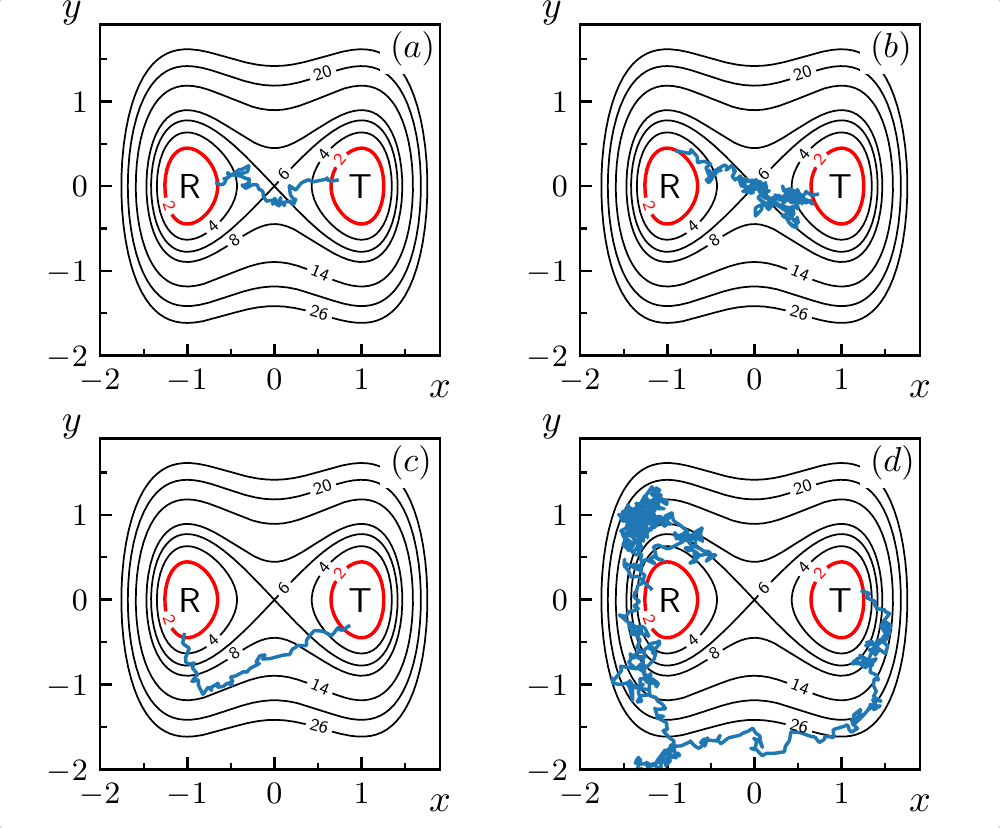}
\caption{Typical target-finding paths for different values of the activity. For $\text{Pe} = 0$ (passive particle) panel (a) shows a typical fast transition path with a TPT such that $t_{\text{TPT}}/\tau_{\vartheta} \simeq 0.65$, while panel (b) represents a slow transition path with $t_{\text{TPT}}/\tau_{\vartheta} \simeq 2.37$. Similarly, panel (c) shows a typical fast target-finding path in the case of $\text{Pe} = 10$, with $t_{\text{TPT}}/\tau_{\vartheta} \simeq 0.53$, while panel (d) represents a slow reactive path for $\text{Pe} = 10$ with $t_{\text{TPT}}/\tau_{\vartheta} \simeq 8.36$. The energy contour lines for the double-well potential are depicted in black, while the red contour lines indicate the boundary of the R and T state. The figure has been reproduced from our publication Ref.~\cite{Zanovello2021}.}
\label{fig:typical_paths}
\end{figure}

To gather a better understanding on the underlying process leading to these results, it is instructive to analyze the typical target-finding paths\footnote{From here on we will use target-finding paths with an equivalent meaning of reactive paths or transition paths.} observed at different values of the activity and reported in Fig.~\ref{fig:typical_paths}).

As is apparent from the figure, for the passive particle the fast and slow reactive paths are quite similar (panels (a),(b)): both the trajectories cross the barrier at the saddle point, the only difference being that the slow path is showing a longer time spent to reach the top of the barrier.
Unsurprisingly, passive particles follow the minimum energy path leading from R to T (which is the straight line with $y = 0$ going from R to T and passing at the saddle point of the barrier, located at $(x,y) = (0,0)$).
Instead, the picture drastically changes in the case of an ABP (panels (c),(d)).
While the fast trajectories are similar to the passive ones, crossing the barrier at the saddle point faster than a passive particle thanks to the active contribution to the motion of the agent, slow trajectories exhibit a strikingly different behavior.
These trajectories take long detours in higher regions of the energy landscape and cross the barrier avoiding the minimum energy path.
The contribution of the TPTs obtained from these paths causes the emergence of the thicker tails at long times in the TPT distributions.

\subsubsection{Transition probability density and transition current}
\label{sec:tpd_tc_dw_prl}

A statistically more sound analysis of the process can be achieved by computing the transition probability densities and the transition currents.
From the paths harvested using our modified TPS a normalized histogram $m(\bm{r})$ of the visited configurations can be built, accounting for the probability that a reactive path visits a position $\bm{r}$ in the transition region $\Omega_{\text{T}} = \Omega \setminus (\text{R} \cup \text{T})$.
This distribution will provide us the information of which regions of the landscape are more frequently visited during successful target-finding events.
Complementary information can then be obtained by computing the transition current $\bm{J}(\bm{r})$ associated with the reactive paths: this is achieved by computing the instantaneous velocities along the two components $x$ and $y$ for all the steps of a reactive path, averaging all the instantaneous velocities for each bin of the histogram and multiplying the result for the transition probability density value found in that bin.

\begin{figure}[H]
\centering
\includegraphics[height=40mm]{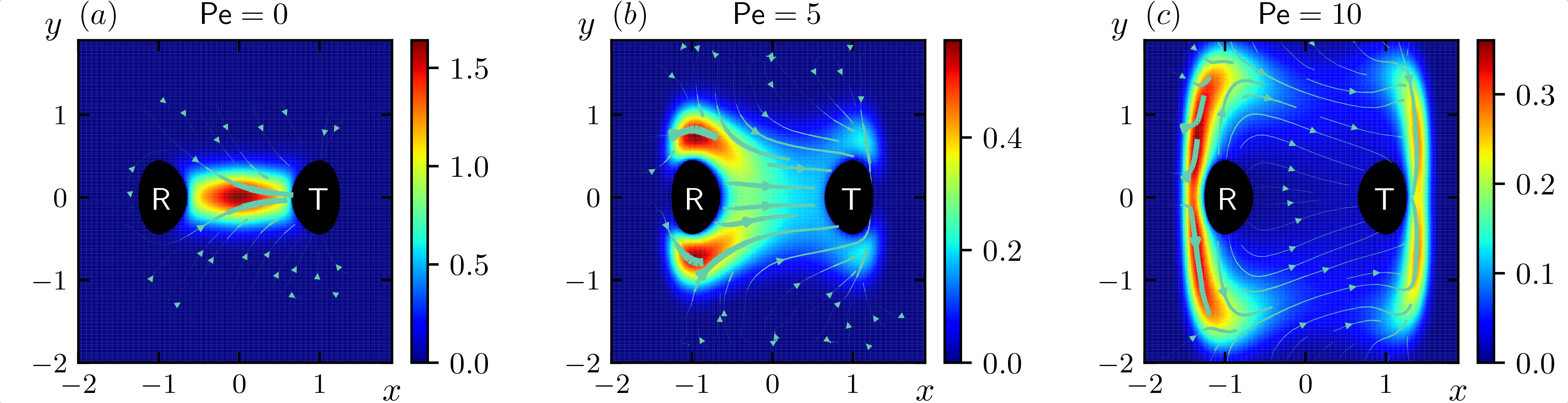}
\caption{Transition probability densities $m(\bm{r})$ (color maps) and field lines of the transition currents $\bm{J}(\bm{r})$ (cyan arrows) for the three different values of the P{\'e}clet number, obtained from a statistics of $4 \cdot 10^{8}$ reactive paths. The figure has been reproduced from our publication Ref.~\cite{Zanovello2021}.}
\label{fig:trprd_trc}
\end{figure}

The results of the transition probability densities and currents for the three values of the activity are reported in Fig.~\ref{fig:trprd_trc}.
In the case of a passive particle with $\text{Pe} = 0$ (panel (a)) the transition probability density is mostly focused around the minimum energy path, as expected.
Additionally, the field lines of the transition current show how the highest portion of the flux in the target-finding process is located in the configurations around the minimum energy path.
The picture changes considerably when activity is introduced in the problem.
In the case of an intermediate level of activity ($\text{Pe} = 5$, panel (b)) the transition probability density becomes considerably different from the passive one.
In particular, it becomes evident how the regions where the particle spends most of its transition time shift from the ones around the minimum energy path to those surrounding the R basin.
This fact is corroborated also from the field lines of the transition current, which are thicker in those regions indicating a larger probability flux across those configurations.
This behavior is enhanced when the activity increases to $\text{Pe} = 10$ (panel (c)), where the most frequently visited regions become the ones behind the R basin, in the opposite direction with respect to the T basin.
In this case the field lines of the transition current are mostly found in those regions and lead the transition through high-energy regions of the landscape far from the minimum energy path.
As we will see in section \ref{sec:trans_rates}, this exploration of high-energy regions is associated with increased target-finding rates of the active particles compared to the passive ones.

This behavior suggests that the target-finding dynamics of active particles is counterintuitive and drastically different from the one observed in passive systems.
While passive particles cross more frequently the barrier at the saddle point where the barrier is smaller, compatibly with Kramers theory, active particles cross the barrier more often in unexpected high-energy regions that are virtually inaccessible to passive particles.
Their self-propulsion and the persistence of their motion allows them to climb energy barriers more effectively than passive particles, allowing for an increased exploration capability of energy landscapes.
Additionally, when the activity is high enough (for example in the case with $\text{Pe} = 10$), the typical target-search trajectories are more likely to behave in an unexpected and counterintuitive fashion: they exit from the back of the R region, then cross the barrier in high-energy regions, and finally they reach the back of the T basin, falling into it thanks to thermal fluctuations or after the particle had enough time to reorient and move towards T.
These long-lasting paths are associated with longer TPTs compared to the paths that cross the barrier along the minimum energy path and are observed more frequently the more the activity of the particle increases, as can be inferred from the TPT distributions (Fig.~\ref{fig:tpt_tps}).
Interestingly, this increase of the TPTs associated with an enhanced activity is the opposite trend to the one observed in 1D~\cite{Carlon2018}, however in 1D systems the target-finding paths cannot even occur with the mechanism of exiting from the back of R and reaching the back of T, suggesting a nontrivial interplay between the activity of the particle, the environment topology and the system dimensionality.

\subsubsection{Additional analysis for Pe = 5 and Pe = 10}
\label{sec:add_analysis_Pe5_10_hist}

\begin{figure}[H]
\centering
\includegraphics[height=80mm]{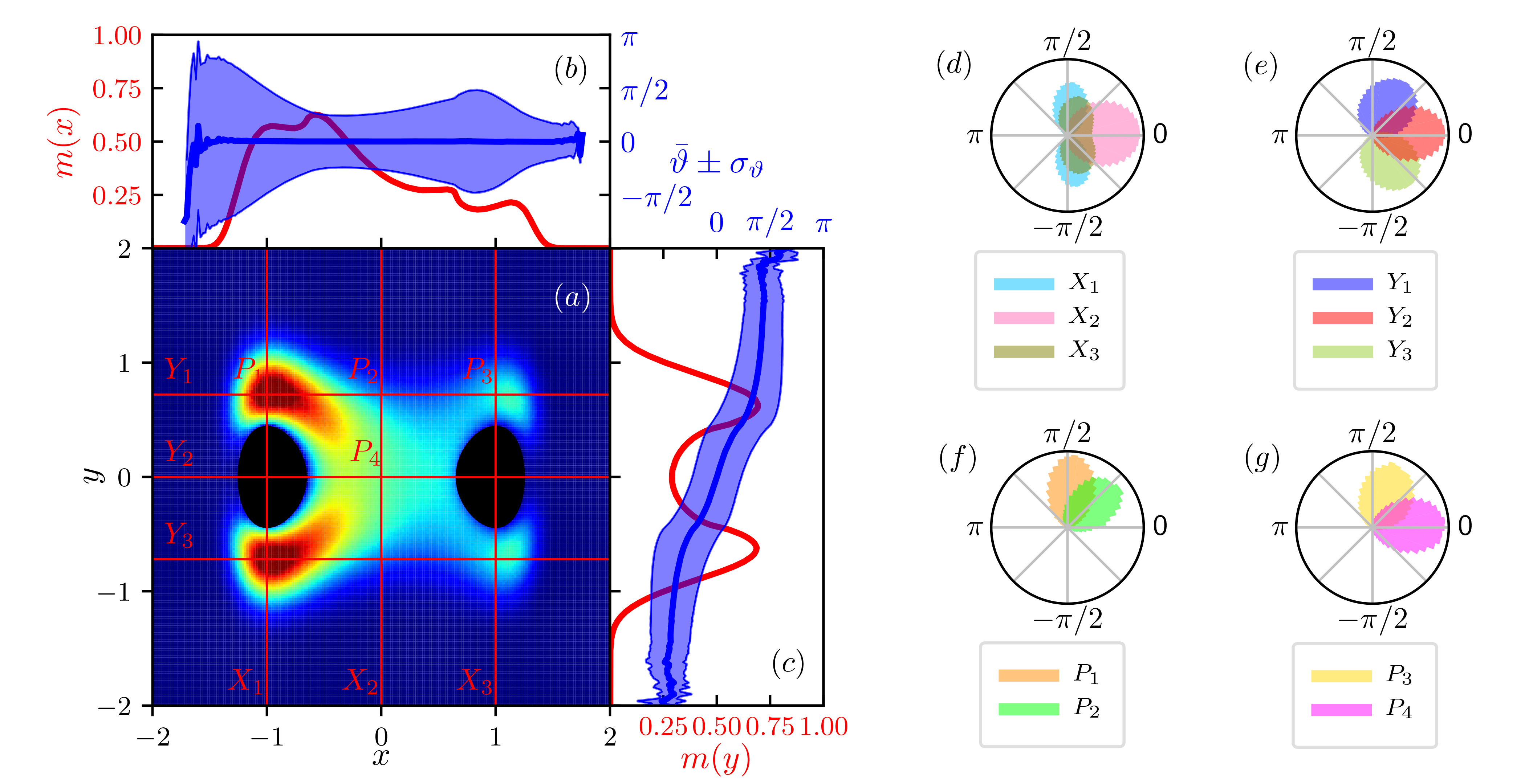}
\caption{(a) Transition probability density $m(\bm{r})$ for an ABP with $\text{Pe} = 5$, obtained from $4 \cdot 10^{8}$ reactive paths. (b-c) Marginalized transition probability density $m(x)$ (b) and $m(y)$ (c) (red lines), and average angle $\bar{\vartheta}$ (thick blue lines) as a function of $x$ (b) or $y$ (c) with its standard deviation $\sigma_{\vartheta}$ (thin blue lines). (d-e) Normalized distribution for the angle $\vartheta$ for the states within the intervals $X_{1}$, $X_{2}$, $X_{3}$ (d) and $Y_{1}$, $Y_{2}$, $Y_{3}$ (e), where $X_{1} : [-1 -\Delta, -1 +\Delta]$, $X_{2} : [-\Delta,+\Delta]$, $X_{3} : [1 -\Delta, 1 +\Delta]$, $Y_{1} : [-0.72 -\Delta, -0.72 +\Delta]$, $Y_{2} : [-\Delta, +\Delta]$, $Y_{3} : [0.72 -\Delta, 0.72 +\Delta]$, and $\Delta = 0.04$. (f-g) Normalized angular distribution for the states $P_{1},...,P_{4}$, obtained as intersections of the previous intervals as shown in panel (a). The figure has been reproduced from the Supplemental Material of our publication Ref.~\cite{Zanovello2021}.}
\label{fig:Pe5_add}
\end{figure}

\begin{figure}[H]
\centering
\includegraphics[height=80mm]{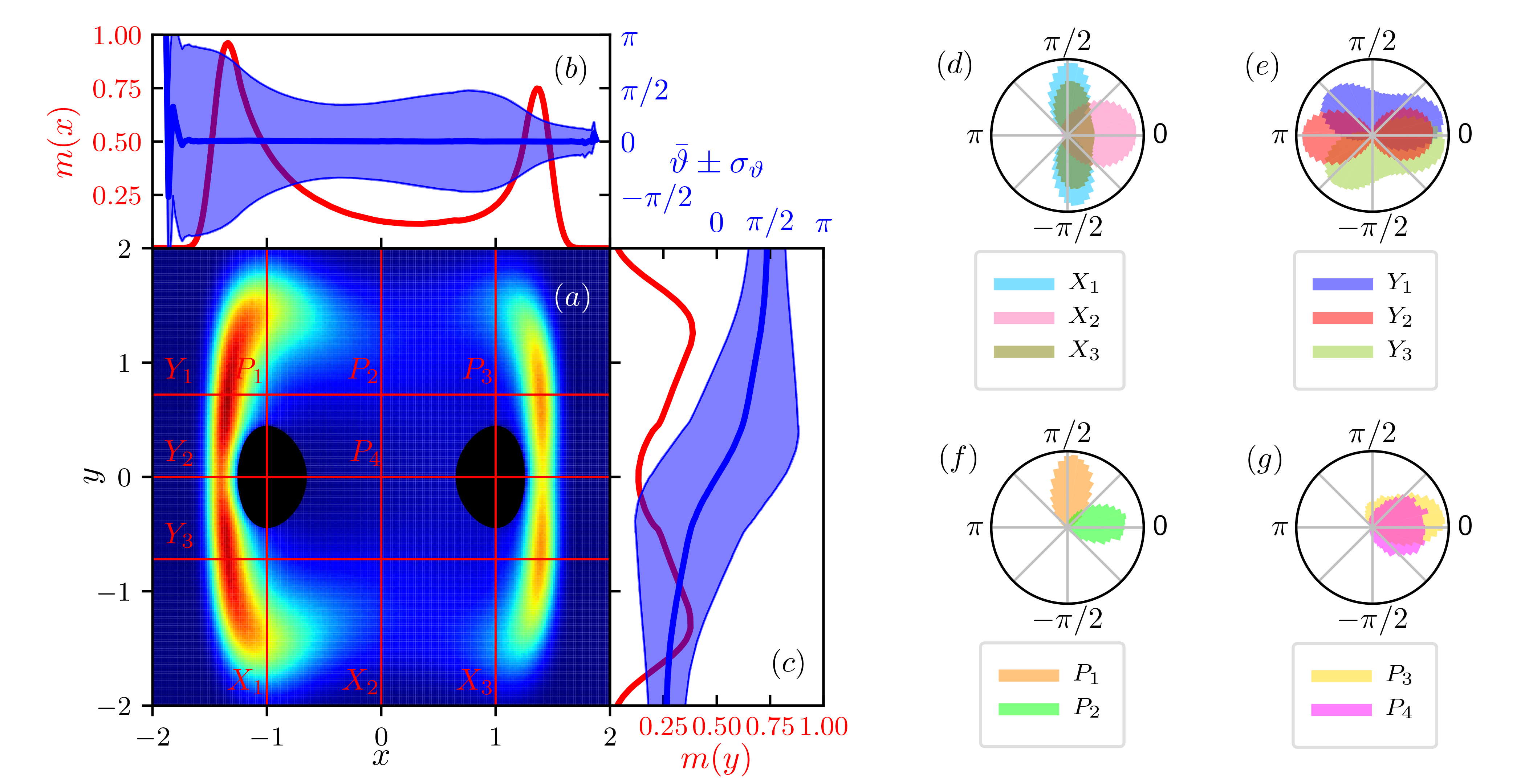}
\caption{(a) Transition probability density $m(\bm{r})$ for an ABP with $\text{Pe} = 10$, obtained from $4 \cdot 10^{8}$ reactive paths. (b-c) Marginalized transition probability density $m(x)$ (b) and $m(y)$ (c) (red lines), and average angle $\bar{\vartheta}$ (thick blue lines) as a function of $x$ (b) or $y$ (c) with its standard deviation $\sigma_{\vartheta}$ (thin blue lines). (d-e) Normalized distribution for the angle $\vartheta$ for the states within the intervals $X_{1}$, $X_{2}$, $X_{3}$ (d) and $Y_{1}$, $Y_{2}$, $Y_{3}$ (e), where $X_{1} : [-1 -\Delta, -1 +\Delta]$, $X_{2} : [-\Delta,+\Delta]$, $X_{3} : [1 -\Delta, 1 +\Delta]$, $Y_{1} : [-0.72 -\Delta, -0.72 +\Delta]$, $Y_{2} : [-\Delta, +\Delta]$, $Y_{3} : [0.72 -\Delta, 0.72 +\Delta]$, and $\Delta = 0.04$. (f-g) Normalized angular distribution for the states $P_{1},...,P_{4}$, obtained as intersections of the previous intervals as shown in panel (a). The figure has been reproduced from the Supplemental Material of our publication Ref.~\cite{Zanovello2021}.}
\label{fig:Pe10_add}
\end{figure}

Here we provide an additional detailed analysis on the transition path ensemble by studying the behavior of the angle $\vartheta$ for points belonging to the reactive paths (see Fig.~\ref{fig:Pe5_add} and \ref{fig:Pe10_add}).

As found from the previous analysis of the transition probability densities and currents, in the case of $\text{Pe} = 5$ (Fig.~\ref{fig:Pe5_add}) also the marginalized distributions $m(x) = \int dy \, m(x,y)$ (panel (b)) and $m(y) = \int dx \, m(x,y)$ (panel (c)) show how the particles during successful target-finding events are more likely to be found close to the R basin, and in particular $m(y)$ displays a bimodal distribution with maximal values shifted away from the minimum energy path.

The average angle $\bar{\vartheta}$ as a function of $x$ is zero due to the symmetry of the problem along the $y$ axis (panel (b)).
However, its standard deviation $\sigma_{\vartheta}$ is not constant and shows a larger magnitude associated with the $X_{1}$ and $X_{3}$ position compared to the one observed at position $X_{2}$.
This is concomitant with the emergence of a bimodal distribution for the angles $\vartheta$ for those slices compared to the distribution for the slice $X_{2}$, which is instead focused around $\vartheta = 0$ (panel (d)).
The average angle $\bar{\vartheta}$ as a function of $y$ is instead showing a different behavior, with values larger than $0$ for $y>0$ and values smaller than $0$ for $y<0$, which is confirmed also from the distributions for the angles constrained to some constant values of $y$ (panel (e)).
This shows once again how active particles tend to be found with a self-propulsion directed on average towards the higher-energy regions due to the persistence of their motion.
Finally, the standard deviation $\sigma_{\vartheta}$ for $\bar{\vartheta}(y)$ is found approximately constant, in contrast to the standard deviation for $\bar{\vartheta}(x)$.

The distributions for the angles constrained to a specific position show how for the points $P_{1}$, $P_{2}$, and $P_{4}$ the particles are usually found directed in the opposite direction with respect to the R basin during successful target-finding events (panels (f-g)).
Additionally, the distribution for $P_{3}$ shows that the particles that successfully reach T do not directly point in the direction of the target when they are close to the T basin, but once again they are usually found directed towards the higher-energy regions, surfing on the back of T.

Fig.~\ref{fig:Pe10_add} shows the equivalent analysis in the case of $\text{Pe}=10$.
This time, also in the marginalized transition probability density $m(x)$ (panel (b)) a bimodal distribution emerges, indicating how, with an increase in the value of the self-propulsion, the ABPs are even more likely to be found in the region of the landscape behind the two basins, where the potential is steeper.
Additionally, the standard deviation for $\bar{\vartheta}(y)$ (panel (c)) is not constant anymore, with larger values for points along $Y_{2}$, related to a bimodal distribution with the two modes pointing in the opposite directions, with an angle of $\vartheta$ close to $0$ if the particle is behind the T basin and around $\pi$ if the particle is behind the R basin (panel (e)).
Interestingly, the distributions for the angles found at point $P_{2}$ and $P_{3}$ display a maximum around $0$ differently from what is observed in the case of $\text{Pe} = 5$.

These findings together with the behavior of the transition probability density suggest that the more the activity of the particle increases, the more often the ABP performs target-searches by surfing along the high-energy regions of the landscape up to falling into the T basin, with a decrease in the trajectories that travel close to the minimum energy path linking R and T.

\begin{figure}[H]
\centering
\includegraphics[height=75mm]{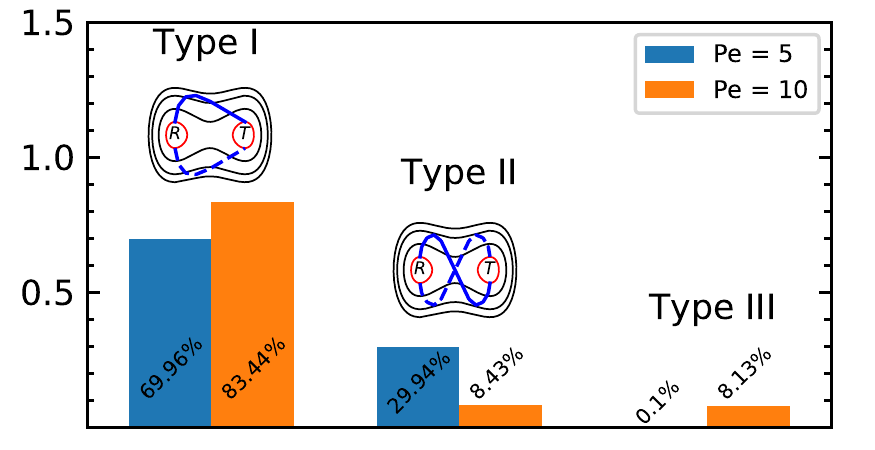}
\caption{Frequency histogram of the different trajectory types for two values of the $\text{Pe}$ number, $\text{Pe} = 5$ and $\text{Pe} = 10$, obtained from a statistics of $4 \cdot 10^{8}$ reactive paths for each value of $\text{Pe}$. A sketch of the different trajectory types is reported on top of the histograms. The figure has been reproduced from the Supplemental Material of our publication Ref.~\cite{Zanovello2021}.}
\label{fig:traj_hist}
\end{figure}

We conclude this analysis by providing a characterization of the different target-finding-paths frequencies classified according to their shape, for different values of the activity of the particle.
In particular, we divide the target-finding paths in three categories depending on their behavior in the region $-1<x<1$.
We categorize as Type I all target-finding paths exhibiting only positive or only negative $y$ values in traveling from R to T within this region.
Type II will refer to all target-finding paths that have positive $y$ when close to the R region and negative $y$ when close to the T region or vice versa.
Finally, Type III paths will be all other types of trajectories.
Fig.~\ref{fig:traj_hist} displays that at both levels of activity ($\text{Pe} = 5$ and $\text{Pe} = 10$) the trajectory type observed more frequently is Type I.
However, while at $\text{Pe} = 5$ Type II trajectories show a much larger frequency compared to Type III and are the second most represented type, at $\text{Pe} = 10$ their number decreases considerably, increasing both the number of Type I trajectories (because of the increased tendency to surf in the high-energy regions of the landscape as $\text{Pe}$ increases) and the number of Type III, which becomes as frequent as Type II.

\subsection{Additional analysis and dependence on the system parameters}

\subsubsection{Target search in different energy landscapes}

We now want to obtain a first insight into the effect of the details of the energy landscape on the dynamics of the particle.

\begin{figure}[H]
\centering
\includegraphics[height=50mm]{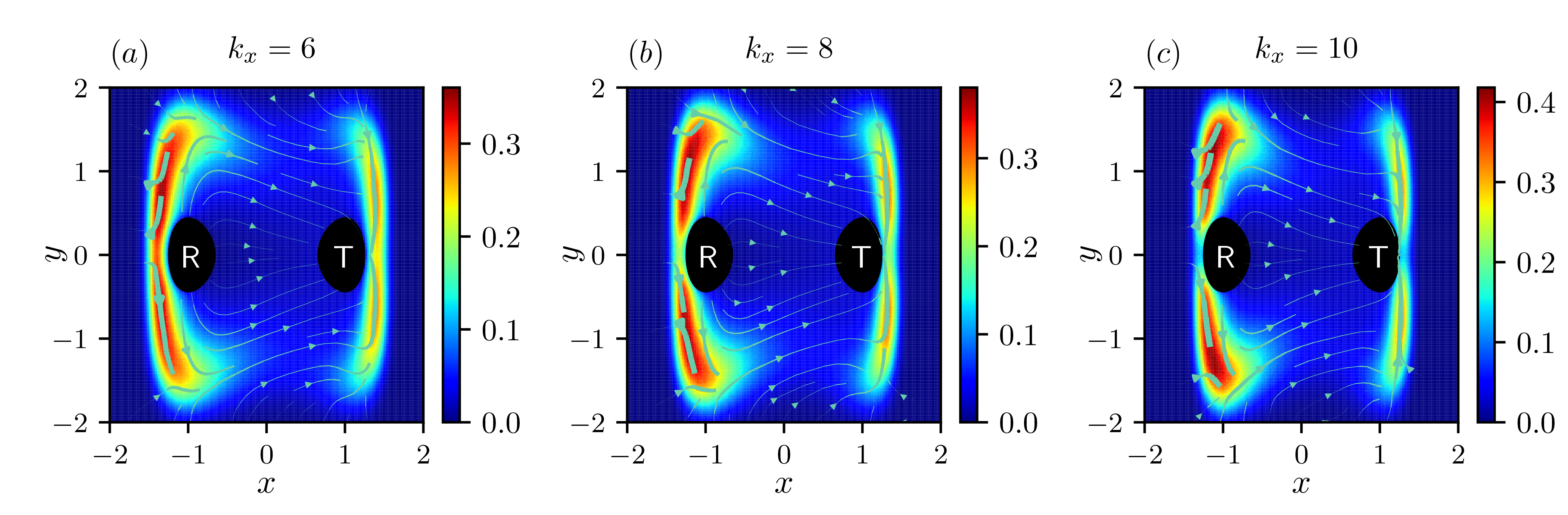}
\caption{Transition probability density $m(\bm{r})$ (color maps) and field lines of the transition currents $\bm{J}(\bm{r})$ (cyan arrows) for three different values of the stiffness parameter along the $x$ direction, $k_{x}$. The particles have a value of $\text{Pe}=10$ and the other parameter of the potential is fixed at $k_{y} = 20$. The pictures are obtained from a statistics of $4 \cdot 10^{8}$ reactive paths. The figure has been reproduced from the Supplemental Material of our publication Ref.~\cite{Zanovello2021}.}
\label{fig:dw_stiffness}
\end{figure}

Let's start by considering the effects of an increased stiffness of the potential along the $x$ direction in the double well.
Let's consider two double wells that differ from the one defined in the previous sections by the parameter $k_{x}$, which in one case will be equal to $8$ and in the other to $10$.
The behavior of an active particle in these landscapes will not differ too much from the one in the original landscape.
The transition probability densities and the transition currents will be quite similar across the three considered cases (see Fig.~\ref{fig:dw_stiffness} for the case of $\text{Pe} = 10$), with the only difference being the fact that the increased stiffness along the $x$ direction causes the particle to be closer to the R and T basins as $k_{x}$ increases.
This will also prompt a splitting in the most frequented regions of $m(\bm{r})$ on the back of the basins, turning from a single compact region to two smaller regions (one at positive $y$ and the other at negative $y$), characterized by higher values of $m(\bm{r})$.

\begin{figure}[H]
\centering
\includegraphics[height=57mm]{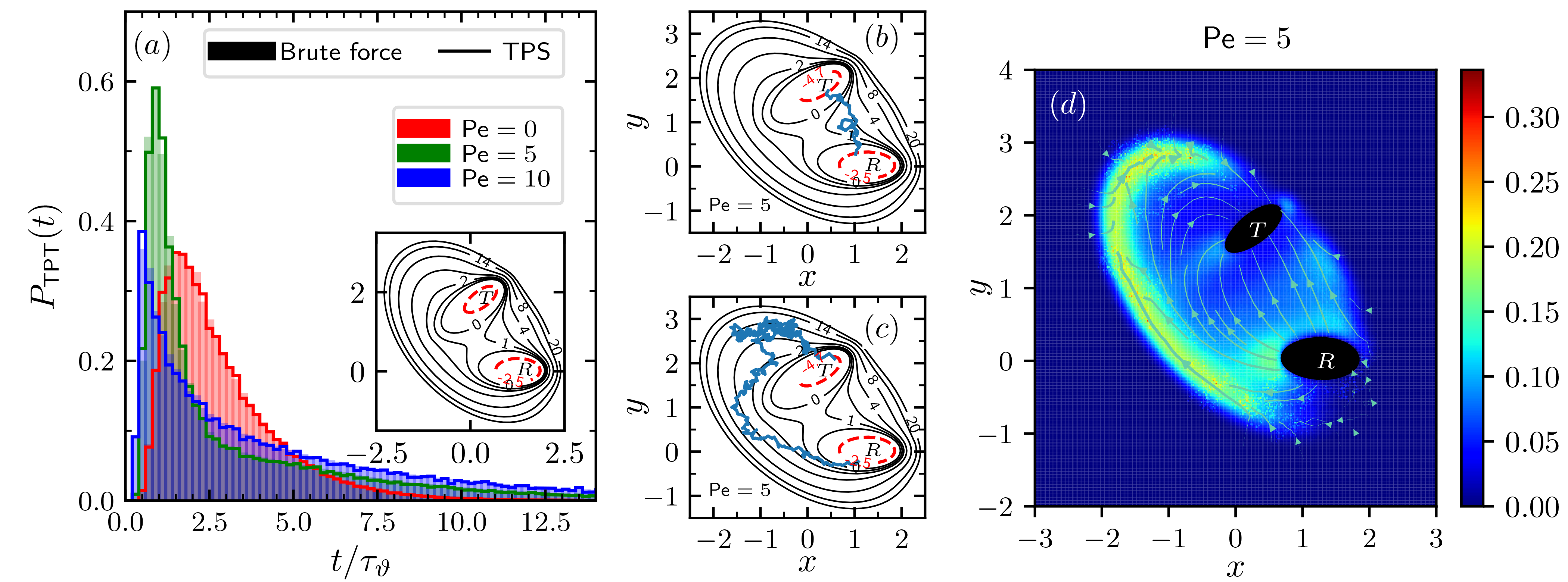}
\caption{Analysis of the target-search paths in the M\"uller-like potential. (a) TPT distributions at different values of the particles activity. The distributions are obtained from a statistics of $10^{8}$ target-finding paths, undersampled to $10^{6}$ paths. Inset: contour lines of the energy landscape, with red dashed lines indicating the boundaries of the R and T region. (b) typical short target-finding path in the case with $\text{Pe} = 5$. (c) typical long target-finding path for $\text{Pe} = 5$. (d) transition probability density (color map) and field lines of the transition current (cyan arrows) of the target-finding paths for $\text{Pe} = 5$. The figure has been reproduced from the Supplemental Material of our publication Ref.~\cite{Zanovello2021}.}
\label{fig:muller_mod}
\end{figure}

We now change the underlying energy landscape to something qualitatively different from the double-well potential that we used throughout the previous discussion.
Explicitly, we will consider a M\"uller-like potential (see inset in Fig.~\ref{fig:muller_mod}(a)) defined by:
\begin{equation}
U(x,y) = \sum_{i=1}^{3} K_{i} e^{[a_{i}(x-x_{0,i})^{2}+b_{i}(x-x_{0,i})(y-y_{0,i})+c_{i}(y-y_{0,i})^{2}]} \; ,
\end{equation}
where the parameters are:
\begin{center}
\begin{tabular}{|c|c|c|c|c|c|c|}
\hline
$i$ & $K_{i}$ & $a_{i}$ & $b_{i}$ & $c_{i}$ & $x_{0,i}$ & $y_{0,i}$ \\
\hline
$1$ & $-10$ & $-0.8$ & $0$ & $-5$ & $1.7$ & $0$ \\
$2$ & $-10$ & $-3$ & $6$ & $-5$ & $1.7$ & $0$ \\
$3$ & $1$ & $0.7$ & $0.6$ & $0.7$ & $0$ & $1$ \\
\hline
\end{tabular}
\end{center}

For this form of the energy landscape, we define as R basin the region with energy lower than $-2.5 \, k_{B}T_{\text{eff}}$ around the local minimum near $(x_{0,1},y_{0,1})$, which has a depth of about $3.4 \, k_{B}T_{\text{eff}}$.
The T basin will instead be chosen as the region with energy lower than $-4.7 \, k_{B}T_{\text{eff}}$ around the local minimum near $(x_{0,2},y_{0,2})$, with a depth of about $2.7 \, k_{B}T_{\text{eff}}$.
The saddle point of the barrier separating R and T has an energy value of about $0.9 \, k_{B}T_{\text{eff}}$.
This landscape is markedly asymmetric, it has a curved minimum energy path and the basins R and T exhibit different depth and shape compared to the double-well case.

Once more, the TPT distributions obtained with our generalized TPS match the ones observed by performing brute-force simulations (see Fig.~\ref{fig:muller_mod}(a)).
However, this time the increase in activity is producing different results in the shape of the TPT distributions compared to the double-well case.
In particular, even though similarly to the double well the tails at long times of the distributions are increasing with activity, a shift of the most probable TPT length towards lower times is observed as activity increases, which is not as pronounced in the double well.
This indicates how particles with higher activities are displaying short target-finding paths (similar to the one reported in Fig.~\ref{fig:muller_mod}(b)) faster than the passive case, but at the same time the particle is also more likely to take long-lasting target-finding paths by performing the typical ``surfing'' in high-energy regions (as the one reported in Fig.~\ref{fig:muller_mod}(c)).
This is supported also by the transition probability density and current (Fig.~\ref{fig:muller_mod}(d)), which show most frequently visited configurations in the regions along the steep energy walls provided by the confining part of the potential, with a direction of the current following these boundaries.

We conclude this paragraph by pointing out that this analysis shows how our generalized TPS algorithm can be used for active particles exploring virtually any energy landscape characterized by high-energy barriers, as long as a rigorous definition of the R and T states for the process is provided.
The case of landscapes with many metastable states, however, might introduce difficulties for the TPS algorithm by decreasing its efficiency.
In the past these issues were dealt with by using the classic TPS combined with a quenching technique~\cite{Dellago1998a}, and we believe that this approach could be applied also to the case of our generalized TPS.

\subsubsection{Influence of the backward dynamics choice}
\label{sec:backward_choice}

We now proceed to evaluating the effects of choosing different backward dynamics from the one presented in Eqs.~(\ref{eq:eom_back1},\ref{eq:eom_back2}), which we recall here for convenience:

\begin{subequations}
\begin{eqnarray}\label{eq:eom_back11}
\bm{r}_{i} &=& \bm{r}_{i\!+\!1} - v\, \bm{u}_{i\!+\!1} \, \Delta t - \mu \bm{\nabla} U(\bm{r}_{i\!+\!1}) \Delta t + \sqrt{2D\Delta t} \, \bm{\xi}_{i\!+\!1}\;,\\ \label{eq:eom_back12}
\vartheta_{i} &=& \vartheta_{i\!+\!1} + \sqrt{2D_{\vartheta}\Delta t} \, \eta_{i\!+\!1}\;.
\end{eqnarray}
\end{subequations}

\begin{figure}[H]
\centering
\includegraphics[height=60mm]{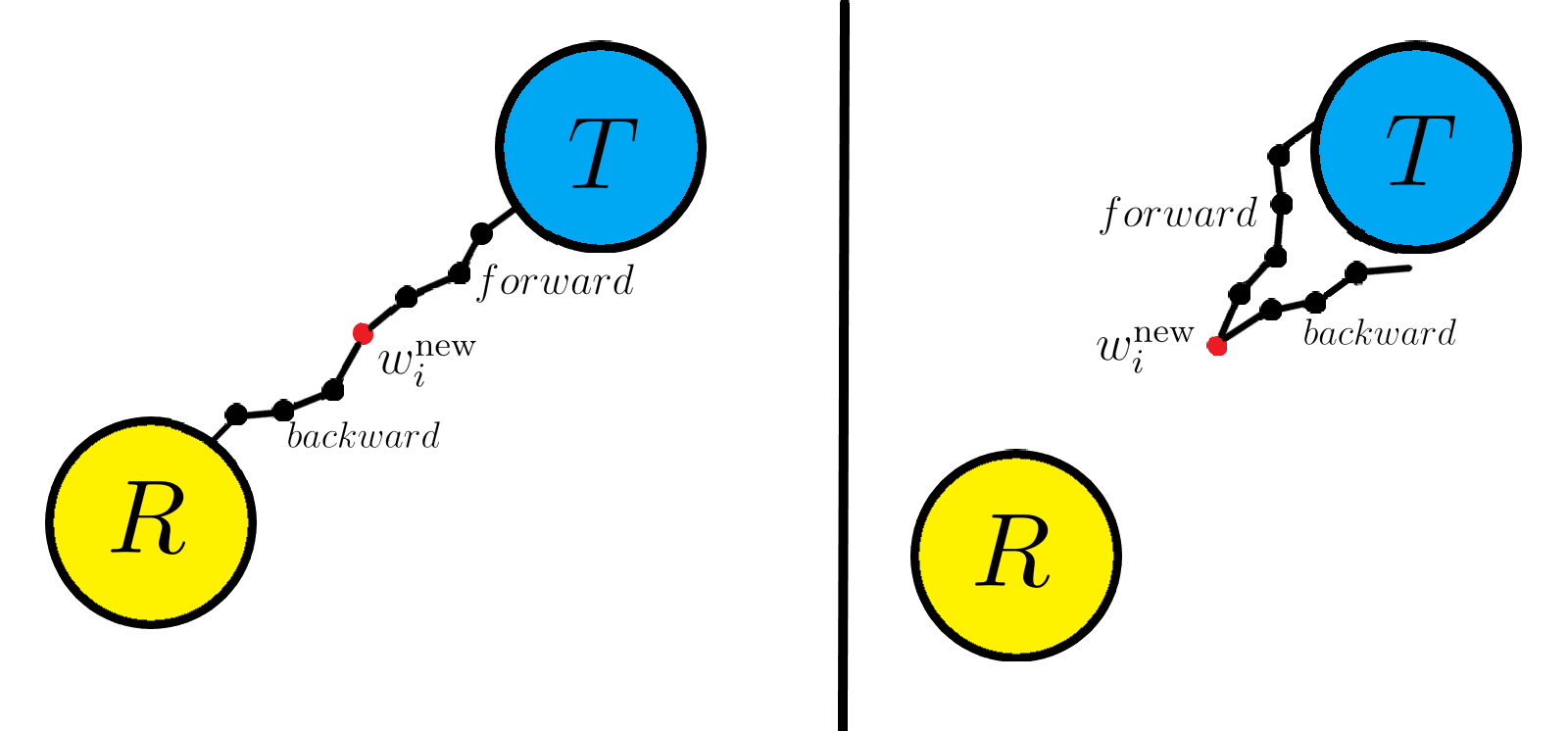}
\caption{Sketch of trajectories generated with our modified TPS in the case where activity dominates over the translational diffusion. \textit{Left panel}: trajectory obtained in the case of an odd choice for the self-propulsion term under time reversal. The backward branch of the new trajectory generated from the shooting point $w_{i}^{\text{new}}$ can reach the R state. \textit{Right panel}: trajectory obtained in the case of an even choice for the self-propulsion term under time reversal. The backward branch of the new trajectory generated from the shooting point $w_{i}^{\text{new}}$ rarely reaches the R state.}
\label{fig:back_choice}
\end{figure}

\begin{figure}[H]
\centering
\includegraphics[height=90mm]{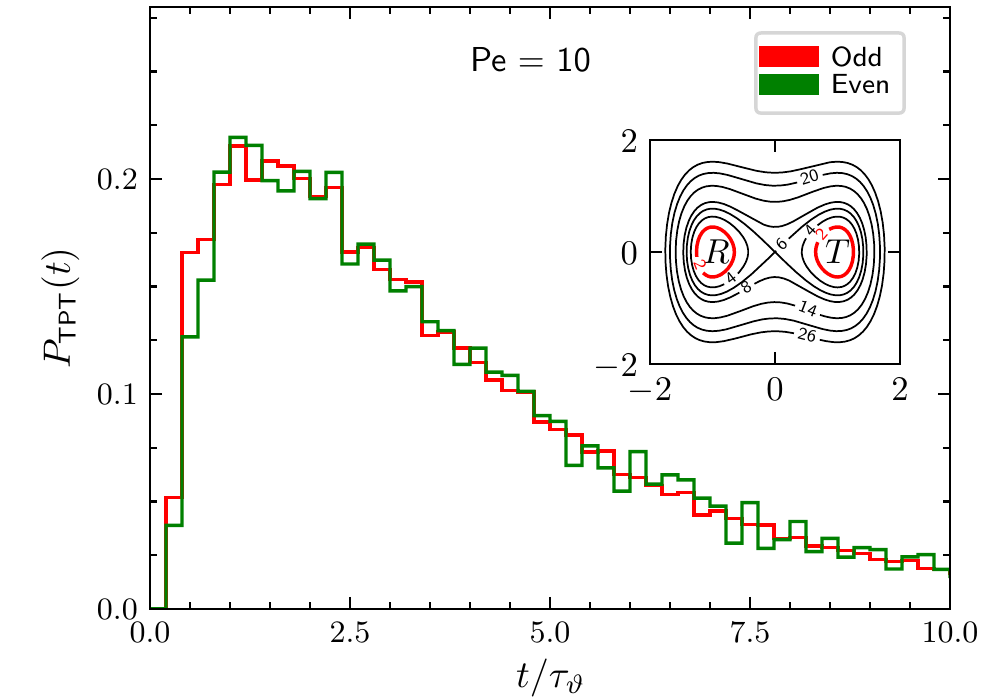}
\caption{TPT distributions for an active particle with $\text{Pe} = 10$ for an odd (red) or even (green) choice of the self-propulsion for the backward dynamics. The ABP parameters are: $D = 0.1$, $v = 3.65$, $\mu = 0.1$, $D_{\vartheta} = 1$, and the potential parameters are the one used in section \ref{sec:pr_car}. The odd curve is obtained from a distribution of $10^{6}$ TPTs after undersampling a set of $4 \times 10^{8}$ TPTs, while the curve for the even distribution is obtained from $10^{6}$ TPTs with an undersampling of a starting distribution of $10^{8}$ TPTs. The figure has been reproduced from the Supplemental Material of our publication Ref.~\cite{Zanovello2021}.}
\label{fig:even_odd}
\end{figure}

First of all, we point out that the choice for the minus sign in the self-propulsion term of the backward dynamics has been chosen so that our backward dynamics resembles as much as possible the forward dynamics reverted in time.
For this reason, the term is treated as an odd term under the effect of time reversal symmetry, so it changes sign with time reversal.
However, in principle one could adopt also an even convention for the particle self-propulsion.
This would be equivalent to treating the self-propulsion as a non-conservative force, which does not change sign under time reversal (for more on this topic see~\cite{Shankar2018}).
In this case, one would need to modify accordingly the backward probability and the acceptance probability of the new paths.
Once the new rules of the acceptance probability are obtained, one could sample the transition path ensemble with this new choice of the backward dynamics.
However, one can do so only in the case of non-negligible translational diffusion, because if the activity dominates over the translational motion the backward branch of a new trajectory with this choice of the self-propulsion term will look similar to a forward branch and would rarely lead to R (see Fig.~\ref{fig:back_choice}).
Since the translational diffusion process is non-negligible for the parameters we used in our work, we implemented this backward dynamics, which proved to sample the same distributions with a 10-fold decrease in the efficiency of the algorithm (\textit{i.e.} the acceptance probability of the new paths).
The comparison of the TPT distributions for the choice of an odd and even signature for the self-propulsion is illustrated in Fig.~\ref{fig:even_odd} in the case of $\text{Pe} = 10$.

Similarly to the choice of the sign of the self-propulsion speed, one could decide to use a different sign for the conservative force related to the external potential.
We decided to use a minus sign for two reasons:
\begin{itemize}
\item similarly to the case of the even signature for the self-propulsion speed, the algorithm becomes more inefficient if a plus sign is adopted for the conservative forces. In this case, the acceptance probability of the new trajectories drops by a factor of $4-6$.
\item our choice of the backward dynamics with its acceptance probability reduces to the passive rule in the limit of vanishing P{\'e}clet number, which has the advantage of using the same dynamics to generate the forward and backward branches of the new trajectories thanks to microscopic reversibility and detailed balance (see Eq.~\ref{eq:for_back_dyn}).
\end{itemize}

\subsubsection{Target search in strictly confining potentials at D = 0}
The complete analysis for the target-search process performed by an ABP navigating in an energy landscape that we provided above was always performed in a regime where the self-propulsion (in the cases with $\text{Pe} \neq 0$) is large enough to ensure that the barriers separating R from T can be always crossed even in the absence of thermal fluctuations (in the limit $D\!\rightarrow\!0$).
We now want to analyze the behavior of the system when the external potential becomes strictly confining in the limit $D\!\rightarrow\!0$, so that the particle cannot cross the barrier without the contributions of the thermal fluctuations, similarly to what has been done elsewhere~\cite{Woillez2019}.

\begin{figure}[H]
\centering
\includegraphics[height=60mm]{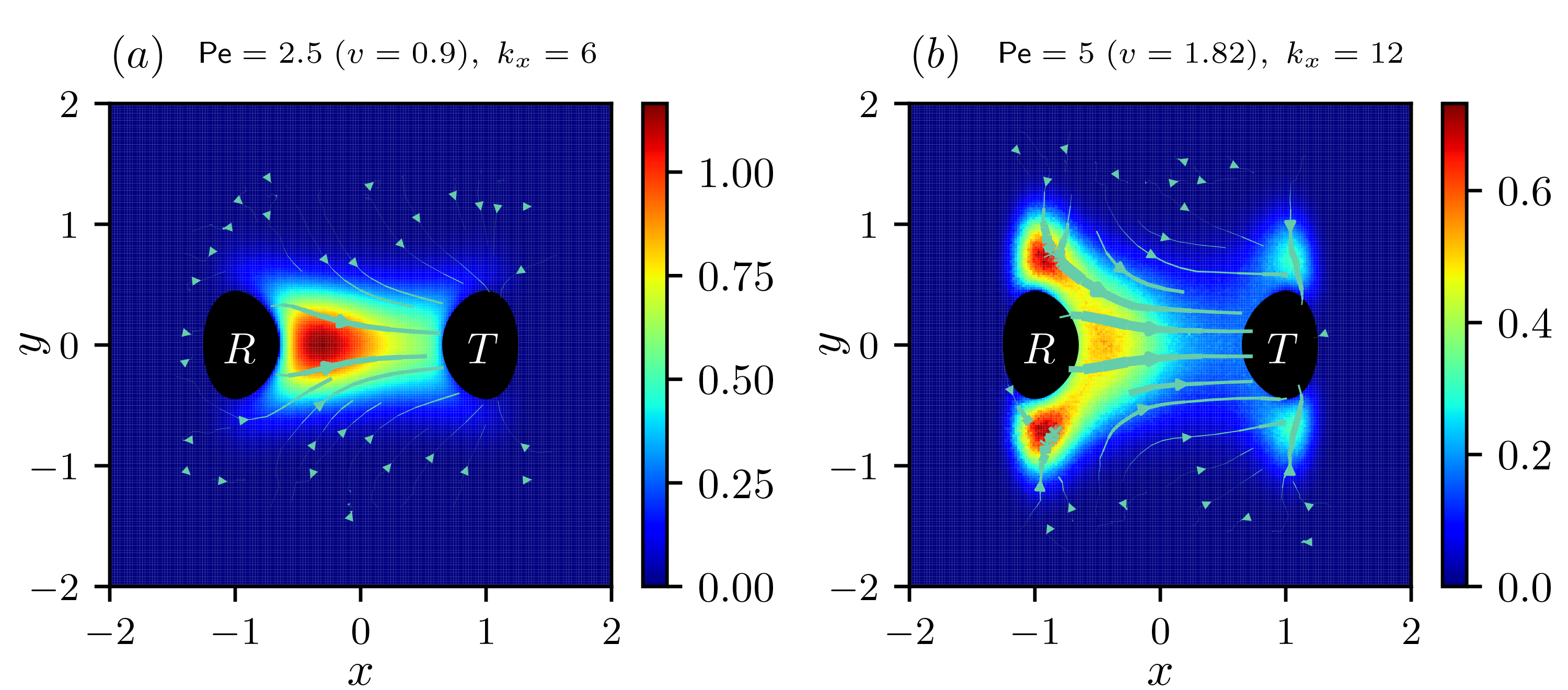}
\caption{Transition probability densities (color maps) and field lines of the transition currents (cyan arrows) in the two limiting cases of a strictly confining potential with parameters reported on top of the figures, obtained from a statistics of $10^{8}$ reactive paths. The other parameters of the problem are kept fixed with $k_{y} = 20$, $D = 0.1$, $D_{\vartheta} = 1$ and $\mu = 0.1$. The figure has been reproduced from the Supplemental Material of our publication Ref.~\cite{Zanovello2021}.}
\label{fig:conf_pot}
\end{figure}

For given shape of the external potential (the double well used throughout the previous analysis, with parameters $x_{0} = 1$, $k_{x} = 6$ and $k_{y} = 20$) we can compute the limiting value of $v$ for which the system is strictly confining in the case of $D = 0$, which yields a maximum value of $v = 0.9$ ($\text{Pe} = 2.46$) for the particle to be confined.
In this case of low activity, the behavior of the transition probability density and current (where we set $D = 0.1$ to make the transitions possible) shows features similar to the case of a passive particle, with most frequently visited regions lying along the minimum energy path and a current funneled along this channel (see Fig.~\ref{fig:conf_pot}(a)).

As an alternative scenario, one can consider a more active particle exploring a double well with increased stiffness along the $x$ direction to ensure a confinement of the agent on one side of the barrier.
We do so by employing a value for $k_{x}$ of $12$, with a corresponding limiting velocity for the confinement of $v = 1.82$ ($\text{Pe} = 5$).
In this case, the particle's target-finding paths are more focused around the minimum energy path compared to the case with $Pe = 5$ and $k_{x} = 6$ (see Fig.~\ref{fig:conf_pot}(b) and Fig.~\ref{fig:trprd_trc}(b) for a comparison).
At the same time, however, the target-finding paths are qualitatively different from those of a passive particle, displaying most frequently visited regions lying above and below the R basin.
This suggests that these paths usually leave R by exiting from the top or the bottom of the basin, rather than from the right side as those of passive particles.
Since their distribution in the R basin will not be substantially different from to the one reported in the case of $k_{x} = 6$ (Fig.~\ref{fig:rho}(b)), they will usually accumulate in the back of the basin.
However, the steep potential along the $x$ direction will not allow them to exit from the back of the basin and start productive target searches from there.
At the same time, the probability of finding the particle on the right side of the R basin will be rather low, so a low number of target-finding paths will start from there.
On the top and bottom regions of R, instead, the probability to find the particle there will assume larger values, and the potential will not be too steep for the particle to exit from the basin and start reactive paths.

\subsubsection{Effects of the rotational diffusion coefficient}
\label{sec:rot_diff_coeff}

Finally, we want to investigate the behavior of the target-finding paths in the case where the P{\'e}clet number is varied by changing the rotational diffusion coefficient $D_{\vartheta}$ while the speed $v$ is kept fixed.
Here, we characterize the behavior for the ABP again for three different values of $\text{Pe}$ in the usual double-well potential (with system parameters $k_{x} = 6$, $k_{y} = 20$, $D = 0.1$, and $\mu = 0.1$).
These values are obtained by setting $v = 1.83$, and by changing $D_{\vartheta}$ to $2$ (with corresponding $\text{Pe} = 3.5$), $0.5$ ($\text{Pe} = 7$), and $0.25$ ($\text{Pe} = 10$).

\begin{figure}[H]
\centering
\includegraphics[height=55mm]{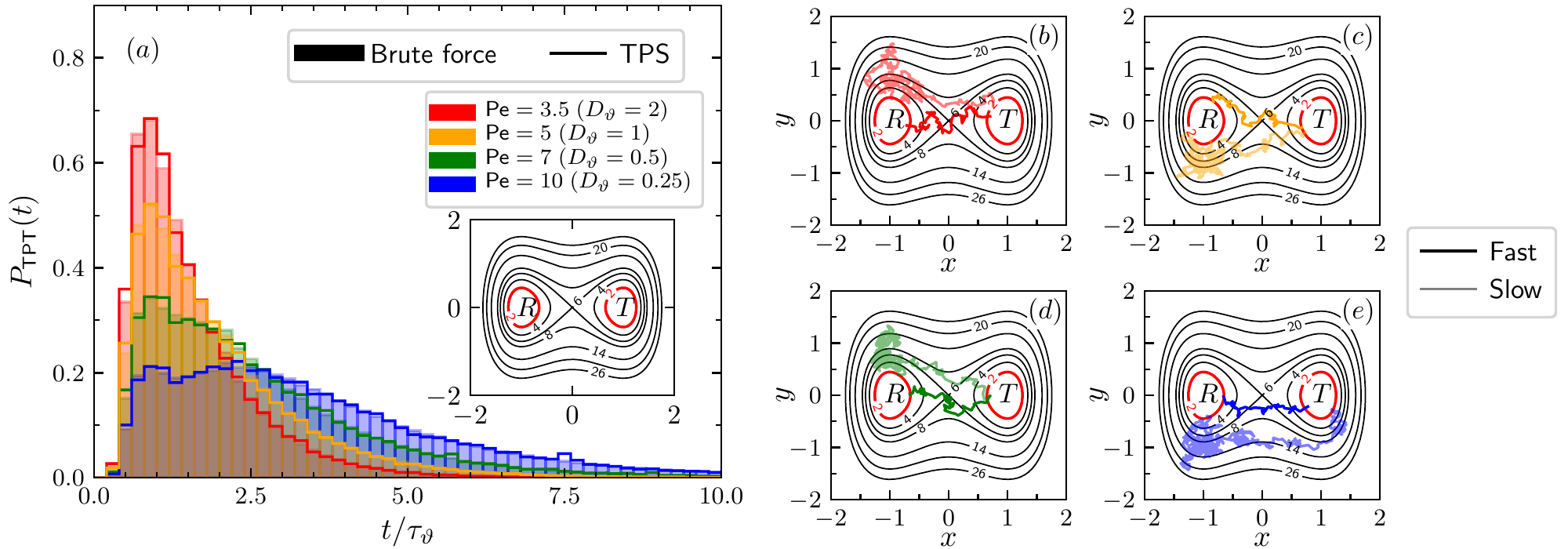}
\caption{Analysis of the target-search process for ABPs at different $\text{Pe}$ numbers, obtained for different values of $D_{\vartheta}$. (a) TPT distributions for the three new values of $\text{Pe}$ ($3.5$, $7$, and $10$) and for the old value of $\text{Pe} = 5$ for comparison, obtained from a statistics of $10^{8}$ paths with an undersampling to $10^{6}$ TPTs. (b-e) a fast and a slow target-finding path for $\text{Pe} = 3.5$ (panel (b)), $\text{Pe} = 5$ (panel (c)), $\text{Pe} = 7$ (panel (d)), and $\text{Pe} = 10$ (panel (e)). The figure has been reproduced from the Supplemental Material of our publication Ref.~\cite{Zanovello2021}.}
\label{fig:dr_tpt}
\end{figure}

\begin{figure}[H]
\centering
\includegraphics[height=50mm]{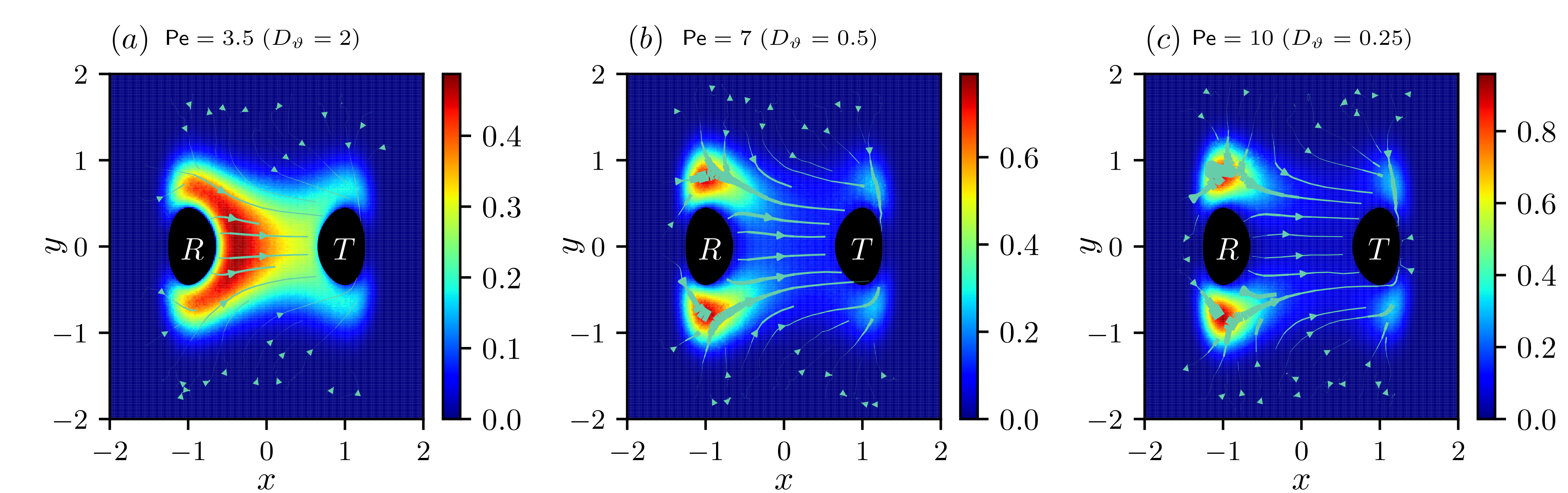}
\caption{Transition probability density (color maps) and field lines of the transition current (cyan arrows) in the three different cases of $\text{Pe} = 3.5$ (panel (a)), $\text{Pe} = 7$ (panel (b)), and $\text{Pe} = 10$ (panel (c)), obtained from $10^{8}$ target-finding paths. The figure has been reproduced from the Supplemental Material of our publication Ref.~\cite{Zanovello2021}.}
\label{fig:dr_rp_rc}
\end{figure}

Once more, the TPT distributions show that our generalized TPS samples the same TPT distributions obtained from direct integration of the equations of motion (Fig.~\ref{fig:dr_tpt}(a)).
The TPT distributions display a behavior similar to the one observed in the previous TPT analysis in the double well (Fig.~\ref{fig:tpt_tps}).
In particular, the average TPT increases with $\text{Pe}$, concomitantly with the tails of the distributions becoming thicker, with an increasing number of long TPTs.
Differently from the case where the P{\'e}clet number is increased by changing $v$, however, in this case the distribution at large $\text{Pe}$ displays the emergence of a bimodal feature, with a more distinct separation between the TPTs associated with the short target-finding paths traveling close to the minimum energy path and the long ones where the particle surfs on the high-energy walls (see blue distribution in Fig.~\ref{fig:dr_tpt}(a)).

The differences between the cases where $\text{Pe}$ is modified by changing $v$ and by modifying $D_{\vartheta}$ are further exemplified by the transition probability densities and transition currents analysis (Fig.~\ref{fig:dr_rp_rc}).
In particular, while an increase in $\text{Pe}$ by changing $v$ results in distributions with frequently-visited regions shifting behind the R basin (see Fig.~\ref{fig:trprd_trc} (b-c)), the case where this is achieved by modifying $D_{\vartheta}$ results instead in the high-probability parts of the distributions remaining concentrated in the region above and below the R basin (Fig.~\ref{fig:dr_rp_rc}(b-c)).

\begin{figure}[H]
\centering
\includegraphics[height=75mm]{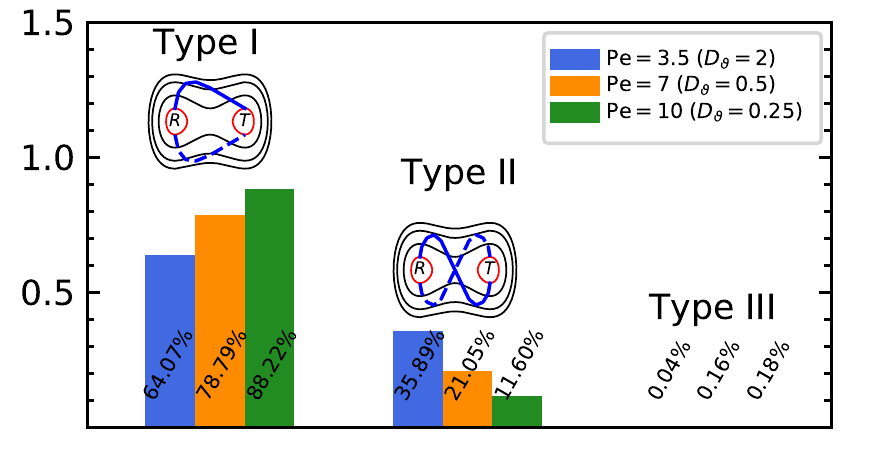}
\caption{Frequency histogram for the different types of trajectories, reported for different values of $\text{Pe}$, $3.5$ (blue), $7$ (orange), and $10$ (green), and obtained from $10^{8}$ reactive paths. A sketch of the different trajectory types is reported on top of the histogram bars. The figure has been reproduced from the Supplemental Material of our publication Ref.~\cite{Zanovello2021}.}
\label{fig:dr_hist}
\end{figure}

We conclude this study by classifying the different target-finding paths according to their shape, performing this analysis for changes in $D_{\vartheta}$ as previously done for changes in the self-propulsion speed (for a discussion on the trajectory types see section \ref{sec:add_analysis_Pe5_10_hist}).
From this classification of the reactive paths (see Fig.~\ref{fig:dr_hist}) one infers that for all values of the activity the most represented group of trajectories is again Type I, with lower percentages for Type II trajectories.
Interestingly, this time small percentages of Type III trajectories are observed regardless of the value of $\text{Pe}$, in contrast to the case where the activity is varied by changes in $v$ (Fig.~\ref{fig:traj_hist}).
Finally, Fig.~\ref{fig:dr_hist} highlights that the percentage of Type II trajectories decreases for increasing $\text{Pe}$, resulting in an increased percentage of Type I trajectories and also on a smaller increase in Type III paths, following the same trend discussed in section \ref{sec:add_analysis_Pe5_10_hist}.

\subsubsection{Transition rates from R to T}
\label{sec:trans_rates}

\begin{table}[H]
\begin{center}
\resizebox{\columnwidth}{!}{
\begin{tabular}{|c|c|c|c|}
\hline
 & $\text{Pe} = 0$ & $\text{Pe} = 3.5$ & $\text{Pe} = 5$ \\
 & $(D_{\vartheta} = 1, v = 0)$ & $(D_{\vartheta} = 2, v = 1.83)$ & $(D_{\vartheta} = 1, v = 1.83)$ \\
\hline
double well $(k_{x} = 6, k_{y} = 20)$ & $6.32 \cdot 10^{-4}$ & $5.61 \cdot 10^{-2}$ & $5.63 \cdot 10^{-2}$ \\
\hline
double well $(k_{x} = 8, k_{y} = 20)$ & $1.15 \cdot 10^{-4}$ & $4.15 \cdot 10^{-2}$ & $4.6 \cdot 10^{-2}$ \\
\hline
double well $(k_{x} = 10, k_{y} = 20)$ & $1.93 \cdot 10^{-5}$ & $2.59 \cdot 10^{-2}$ & $3.25 \cdot 10^{-2}$ \\
\hline
M\"uller-like potential & $2.61 \cdot 10^{-4}$ & $3.99 \cdot 10^{-2}$ & $3.94 \cdot 10^{-2}$ \\
\hline
 & & & \\
\hline
 & $\text{Pe} = 7$ & $\text{Pe} = 10$ & $\text{Pe} = 10$ \\
 & $(D_{\vartheta} = 0.5, v = 1.83)$ & $(D_{\vartheta} = 0.25, v = 1.83)$ & $(D_{\vartheta} = 1, v = 3.65)$ \\
 \hline
double well $(k_{x} = 6, k_{y} = 20)$ & $4.46 \cdot 10^{-2}$ & $2.93 \cdot 10^{-2}$ & $5.62 \cdot 10^{-2}$ \\
\hline
double well $(k_{x} = 8, k_{y} = 20)$ & $3.92 \cdot 10^{-2}$ & $2.7 \cdot 10^{-2}$ & $6.07 \cdot 10^{-2}$ \\
\hline
double well $(k_{x} = 10, k_{y} = 20)$ & $3.1 \cdot 10^{-2}$ & $2.31 \cdot 10^{-2}$ & $6.25 \cdot 10^{-2}$ \\
\hline
M\"uller-like potential & $3.05 \cdot 10^{-2}$ & $2.03 \cdot 10^{-2}$ & $4.85 \cdot 10^{-2}$ \\
\hline
\end{tabular}
}
\caption{Target-finding rates in units of the inverse computational time step. In all cases considered $D = 0.1$ and $\mu = 0.1$.}
\label{table:rates_tps}
\end{center}
\end{table}

We now proceed to discuss the target-finding rates\footnote{Note that we call these target-finding rates also as ‘‘transition rates'' because for the purpose of this study the two concepts are equivalent.} for all energy landscapes and parameters considered in the previous analysis.
Even though as discussed previously one could obtain the transition rates from TPS~\cite{vanErp2003,vanErp2005}, for the purpose of this analysis we obtained the transition rates directly from the brute-force simulations we have run for validating our TPS.
The rates are obtained by evolving the ABP position in the energy landscape with Eqs.~(\ref{eom1},\ref{eom21}) and by measuring how much time $t$ it is required to observe $10^{4}$ target-finding events and then computing the ratio $10^{4}/t$.
Note that these are only approximations of the real rates that one would obtain by restarting the position of the particle at the bottom of R every time a productive event is observed, in fact they account not only for the time required to observe a productive event but also for the time lost in unproductive paths going from T to R and in local fluctuations within T.
These rates are therefore used only as a qualitative measure for the process, and not as a quantitative one, differently to the rates that are computed in Chapter \ref{ch:nav_strat}.
However, they can still serve as a comparative measure for the success odds between the different parameter sets. 
The rates are reported in Table \ref{table:rates_tps}.

The first characteristic that can be extracted is the fact that active particles show, for all the considered parameter sets, rates that are at least two order of magnitude larger than the ones of a passive particle.
However, if one considers only the rates of the ABPs it is less clear how the parameter sets influence them.
In particular, the transition rates increase with $\text{Pe}$ if this is done by increasing the self-propulsion speed, while the behavior is different if $\text{Pe}$ is increased by decreasing $D_{\vartheta}$.
This suggests that the relationship between the parameters and the transition rates is non trivial, and a systematic study on the parameter space of the model is needed to find out if an optimal set of parameters exists and what this implies for the shape of the target-finding paths and the dynamics of the system.
This topic will be the focus of chapter \ref{ch:nav_strat} of this thesis.

\subsubsection{Algorithm efficiency}

We conclude this study by providing some numbers on the efficiency of the generalized TPS compared to brute-force simulations.
In the double-well landscape with parameters $k_{x} = 6$ and $k_{y} = 20$, the ratio of the time needed to gather a statistics of $10^{6}$ reactive paths with our TPS to the time needed to gather the same statistics using brute-force simulations is about $0.001$ in the case of a passive particle with $\text{Pe} = 0$.
Instead, when our TPS is used to compute the same statistics in the case of an ABP, this ratio becomes about $0.12$ for $\text{Pe} = 5$ and $0.27$ in the case of $\text{Pe} = 10$.

This decrease in efficiency goes along with the increase in the self-propulsion of the particle, and we believe it to be caused by two distinct mechanisms.
First, the increase in activity has the effect of increasing considerably the transition rates for the process, as discussed in the previous section.
Therefore, the particles will be able to move much more frequently from one side of the barrier to the other, decreasing the need of enhanced-sampling techniques and probably accounting for most of the decrease in the algorithm efficiency.
Of course, increasing the barrier height, we expect a decrease in the transition rates with a consequent increase in the efficiency of the algorithm compared to brute-force simulations.
Second, the increase in activity is associated with the emergence of longer transition paths, which spend a lot of their TPT in regions virtually inaccessible to passive particles.
Consequently, due to the increased length of the reactive paths, a longer computational time is required to generate the Monte-Carlo trial moves and compute their acceptance probabilities, resulting in a further decrease in efficiency.

\section{Chapter conclusions}
In this chapter we have introduced the concept of enhanced sampling algorithms to study rare-event problems, with a particular focus on Transition Path Sampling and its mechanism.
We have then provided for the first time a generalization of this algorithm to non-equilibrium systems, disproving the common belief that it is not possible to shoot backward in time.

We have shown that the microscopic reversibility is not a necessary condition to study a rare-event process with TPS, and we have provided a protocol to perform the backward shooting using any sort of backward laws of motion, provided that the acceptance probability for the new paths is computed accordingly to ensure a correct sampling of the underlying distribution for the process.

Subsequently, we have used this generalized TPS to study the rare paths undertaken by an ABP to find a target in an environment characterized by a high energy barrier.
After providing the rules for the backward shooting for this particular system and a path-integral formulation for the propagator of an ABP, we have consistently derived the new acceptance rules for the trajectories generated with our generalized TPS.
We found the usual acceptance probability with a corrective term explicitly depending on the particle self-propulsion, which reduces to the usual acceptance rule in the case of vanishing activity.

After validating this extended TPS with brute-force simulations, we have used it to characterize the target-finding paths of an ABP in a double-well potential, discovering how these active agents perform rare transitions in a counterintuitive and totally different fashion compared to passive particles.
In fact, while passive particles perform their transition through a bundle of trajectories focused around the minimum energy path linking the initial state R to the target state T, active particles are more likely to be found in higher-energy regions, adopting target-finding paths with an unexpected shape that reach the target from the back and ‘‘surf'' along the confining energy walls of the landscape.

Finally, an analysis of the rates for this target-finding process indicated that the relationship between the success rate in the target-search and the system parameters is non-trivial, and that an optimal set of parameters for the problem may exist, requiring a more systematic analysis to verify it.

\afterpage{\null\newpage}

\pagestyle{plain}

\chapter{Navigation strategies of ABPs in target-search problems}
\label{ch:nav_strat}
\pagestyle{fancy}

In the past few years, several studies investigated how active particle can achieve optimal navigation strategies.
In particular, special emphasis has been devoted to the analysis of optimal bacterial transport and similar properties near walls and surfaces~\cite{PerezIpiña2019,Schaar2015}, bacterial navigation in rugged environments characterized by the presence of chemoattractant gradients~\cite{Gosztolai2020}, optimal strategies in the case of heterogeneous environments~\cite{Yang2018,Volpe2017,Yang2020,Daddi-Moussa-Ider2021}, in scarce-resources environments~\cite{Haeufle2016}, and in shear-flow processes~\cite{Liebchen2019}.
Once more, however, an analysis of optimal strategies for ABPs in target-search problems taking place in energy landscapes with high energy barriers is still lacking.

Our previous analysis on how active particles perform target search in environments with high energy barriers revealed that these agents perform this task in a counterintuitive and strikingly different fashion compared to passive particles.
Furthermore, our analysis underlined how the model parameters influence the target-finding rates in a non-trivial way, which requires an in-depth analysis of the parameter space to be elucidated.
In this section we want to extend our previous study, where we characterized only one-dimensional parameter slices, by providing a complete analysis on the full parameter space of the model.
This will allow us to learn if a set of parameters exists in order to achieve an optimal navigation strategy during a target-search problem with a fixed barrier height separating R and T.
Additionally, we will use these results to analyze the interplay between the particle motility and the shape of the underlying energy landscape.

Here we start by presenting the model and discussing the parameter space for the ABP, to then perform a detailed and quantitative analysis on the target-finding rates, on the average TPTs and on the typical target-finding paths of these active agents exploring a double-well potential.
Furthermore, we extend this analysis to the case of a Brown-M\"uller potential~\cite{Müller1979} and we study the effects of the external potential shape on the target-search problem.

\section{Target-search problem and model parameters}

\subsection{Problem definition and relevant observables}

Here we study an active Brownian particle exploring an energy landscape by starting from an initial region R and trying to reach a target region T, which is separated from R by a high energy barrier.

We recall that the equations of motion for the ABP employed to study the problem are given by Eqs.~(\ref{eom},\ref{eom2}):
\begin{subequations}
\begin{eqnarray}\label{eom2a}
\bm{r}_{i\!+\!1} &=& \bm{r}_{i} + v\, \bm{u}_{i} \, \Delta t - \mu \bm{\nabla} U(\bm{r}_{i}) \Delta t + \sqrt{2D\Delta t} \, \bm{\xi}_i\;,\\ \label{eom22b}
\vartheta_{i\!+\!1} &=& \vartheta_{i} + \sqrt{2D_{\vartheta}\Delta t} \, \eta_i\;,
\end{eqnarray}
\end{subequations}
where, as usual, $v$ is the self-propulsion speed of the particle with instantaneous orientation $\bm{u}_{i} = \big(\cos(\vartheta_{i}),\sin(\vartheta_{i})\big)$, $\mu$ is an effective mobility for the particle and $U(\bm{r_{i}})$ is the external potential evaluated in position $\bm{r}_{i}$.
Finally $D$ indicates the translational diffusion coefficient, $D_{\vartheta}$ the rotational diffusion coefficient, and the components of $\bm{\xi}_{i}$ are independent Gaussian centered random variables with unit variance, as well as $\eta_{i}$.

Once more, we will analyze the transition path ensemble, \textit{i.e.} the ensemble of all reactive paths, which are defined as those trajectory slices entering the transition region by last exiting R and leaving it by first entering T.

As aforementioned, we will characterize the transition rates  $r$ for the process, which provide a measure of how frequently the ABP manages to find the target.
Since we don't have yet any way to extract the transition rates from our generalized version of TPS, and ABPs display in general much larger transition rates compared to passive particles, for the purpose of this study we decided to rely exclusively on the numeric ‘‘brute-force'' integration of Eqs.~(\ref{eom2a},\ref{eom22b}).
In particular, we simulate extremely long trajectories that include several target-finding events.
The rates are then computed as the inverse of the average time $t_{r}$ required for a particle reentering region R (after performing a transition from T to R) to find the target T.
In this way, we account for the real time spent by the particle in unsuccessful attempts and the time required for a successful target-finding event.

Subsequently we analyze the average Transition Path Time $t_{\text{TPT}}$ as another observable for the process, which is obtained as an average of the times needed to move along the observed reactive paths.
We underline how the $t_{\text{TPT}}$ differs from $t_{r}$ for the time related to unsuccessful attempts that is considered in $t_{r}$ and not in $t_{\text{TPT}}$ and also for the time spent in thermal fluctuations within the R state.

Finally, we rely once more on the concept of the transition probability density $m(\bm{r})$ for the process to be able to investigate the regions where the ABP is most frequently found during successful target searches.

\subsection{Parameter space of the model}
\label{sec:nav_strat_model_par}

Since we want to study how the target-search dynamics is influenced by the parameters of the system, we start with an analysis of the parameter space for the model.

In the case of a passive particle, where the equations of motion of the position (Eq.~\ref{eom_p}) and of the orientation are decoupled due to the absence of the self-propulsion term, the model requires only the external potential $U(\bm{r})$, $\mu$, and $D$ as input parameters.
We can now reframe the problem through the introduction of some dimensionless parameters for the model.
In the case of a passive searcher at equilibrium, we can express the energy scale through the Einstein-Smoluchowski relation $k_{B}T = D/\mu$.
However, as we have seen in the previous chapter, also in the case of an out-of-equilibrium system we can define an energy scale through the use of an effective temperature for the thermal bath in the passive limit, using the same relation.
Subsequently, we can make the potential dimensionless as  $U(\bm{r})/k_{B}T$, and use it to derive a characteristic length scale $L$ for the problem.
In this case we define this characteristic length scale as $L := k_{B}T/F_{\text{max}}$, where $F_{\text{max}}$ is the maximum value of the force along the minimum energy path linking R and T.
Finally, the time scale is defined as $\tau := L^{2}/D$.

For a passive particle, once the external potential is specified, there are no further dimensionless parameters.
If the particle is active, instead, two additional parameters are introduced in the model, namely the self-propulsion speed $v$ and the rotational diffusion coefficient $D_{\vartheta}$.
We can then proceed to define two dimensionless parameters to account for these new model parameters of the problem.

To quantify the self-propulsion speed we introduce once again the P{\'e}clet number.
This time, however, we define this number differently from our previous definition in section \ref{sec:pr_car}, relying instead on the choice $\text{Pe} := vL/D$ that differs from the previous one by a factor of $2L\sqrt{D_{\vartheta}/3D}$.
This new definition allows for a direct comparison of the self-propulsion of the particle with respect to the translational diffusion process.
Additionally, we can now define a maximal drift speed as $v_{\text{max}} := DF_{\text{max}}/k_{B}T$, which resembles the maximal value of the drift velocity generated by the forces of the external landscape when the particle is traveling along the minimum energy path.
With this definition we can now express P{\'e}clet also as $\text{Pe} = v/v_{\text{max}}$, which implies that a particle with $\text{Pe}>1$ that is following the minimum energy path (neglecting the effects of the rotational diffusion process) can overcome the energy barrier without the need of thermal fluctuations.
The second dimensionless parameter that we introduce to account for $D_{\vartheta}$ is a dimensionless persistence $\ell^{*}$, which serves as a measure to compare the self-propulsion speed of the particle with the typical time required for the particle to reorient.
To do so, we use the persistence length $\ell$ defined in section \ref{sec:ABP_model} and we divide it by the length scale $L$ to make it dimensionless, $\ell^{*} := \ell/L$, and therefore we obtain $\ell^{*} := v/D_{\vartheta}L$.

When the dimensionless potential $U(\bm{r})/k_{B}T$ is specified, the two-dimensional parameter space for the ABP is spanned by $(\text{Pe},\ell^{*})$.
Note that on this plane the case of a passive particle is identified by the parameter set $(\text{Pe} = 0,\ell^{*} = 0)$.

We conclude this discussion on the parameters by showing that the points analyzed in the previous chapter identify one-dimensional slices in the new parameter space.
Since in the double-well potential the maximal force along the minimum energy path is obtained as $F_{\text{max}} = 8k_{x}x_{0}^{3}/3\sqrt{3}$, we can convert the coordinates of the parameter space points that we analyzed in the previous chapter to coordinates in our new parameter space.

The set of points in the old space with constant $D_{\vartheta} = 1$ and a variable $\text{Pe}$ (tuned through $v$) correspond to the following points in the new parameter space: $(\text{Pe} = 0,\ell^{*} = 0)$ (for our passive particle), $(\text{Pe} \simeq 2,\ell^{*} \simeq 17)$ for the old parameter $\text{Pe} = 5$ obtained with $v = 1.83$, and finally $(\text{Pe} \simeq 4,\ell^{*} \simeq 34)$ for the old point with $\text{Pe} = 10$ and $v = 3.65$.
These three points therefore lie along a one-dimensional slice in our new space.

The points where $v = 1.83$ was constant and $\text{Pe}$ was varied by changing $D_{\vartheta}$ instead correspond to: $(\text{Pe} \simeq 2,\ell^{*} \simeq 8.5)$ for the old point with $\text{Pe} = 3.5$ and $D_{\vartheta} = 2$, $(\text{Pe} \simeq 2,\ell^{*} \simeq 34)$ for the point with the old values of $\text{Pe} = 7$ and $D_{\vartheta} = 0.5$, and $(\text{Pe} \simeq 2,\ell^{*} \simeq 68)$ for the point with old $\text{Pe} = 10$ and $D_{\vartheta} = 0.25$.
Also these three points lie along a one-dimensional slice in the new parameter space.

\section{Target-search strategies in a double well}
\label{sec:search_strat_dw}

\begin{figure}[H]
\centering
\includegraphics[height=65mm]{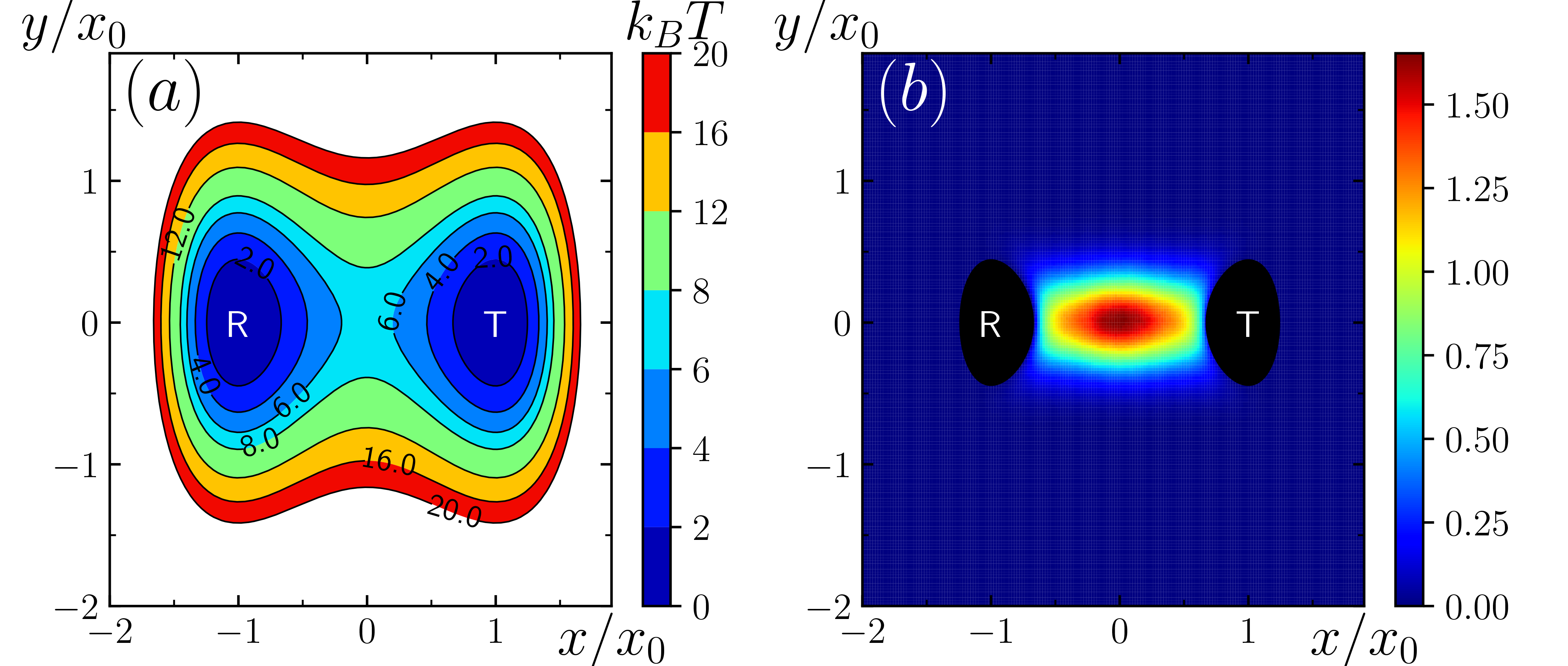}
\caption{(a) Representation of the double-well potential with $k_{x} = 6.5 k_{B}T/x_{0}^{4}$ and $k_{y} = 20k_{B}T/x_{0}^{2}$. For each contour line the energy value is reported in units of $k_{B}T$. R indicates the reactant basin and T the target basin, both delimited by their respective contour lines corresponding to $U(\bm{r}) = 2k_{B}T$. (b) Transition probability density for a passive particle during successful target searches in the considered landscape, obtained from $10^{5}$ reactive paths. The figure has been reproduced from our publication Ref.~\cite{Zanovello2021a}.}
\label{fig:dw_passive_tpd}
\end{figure}

We start by analyzing the behavior of the system with this new parameter space in the case of a double-well potential.
The double well is once again defined by:
\begin{equation}
U(\bm{r}) = k_{x}(x^{2}-x_{0}^{2})^{2} + \frac{k_{y}}{2} y^{2} \; .
\end{equation}

\begin{figure}[H]
\centering
\includegraphics[height=100mm]{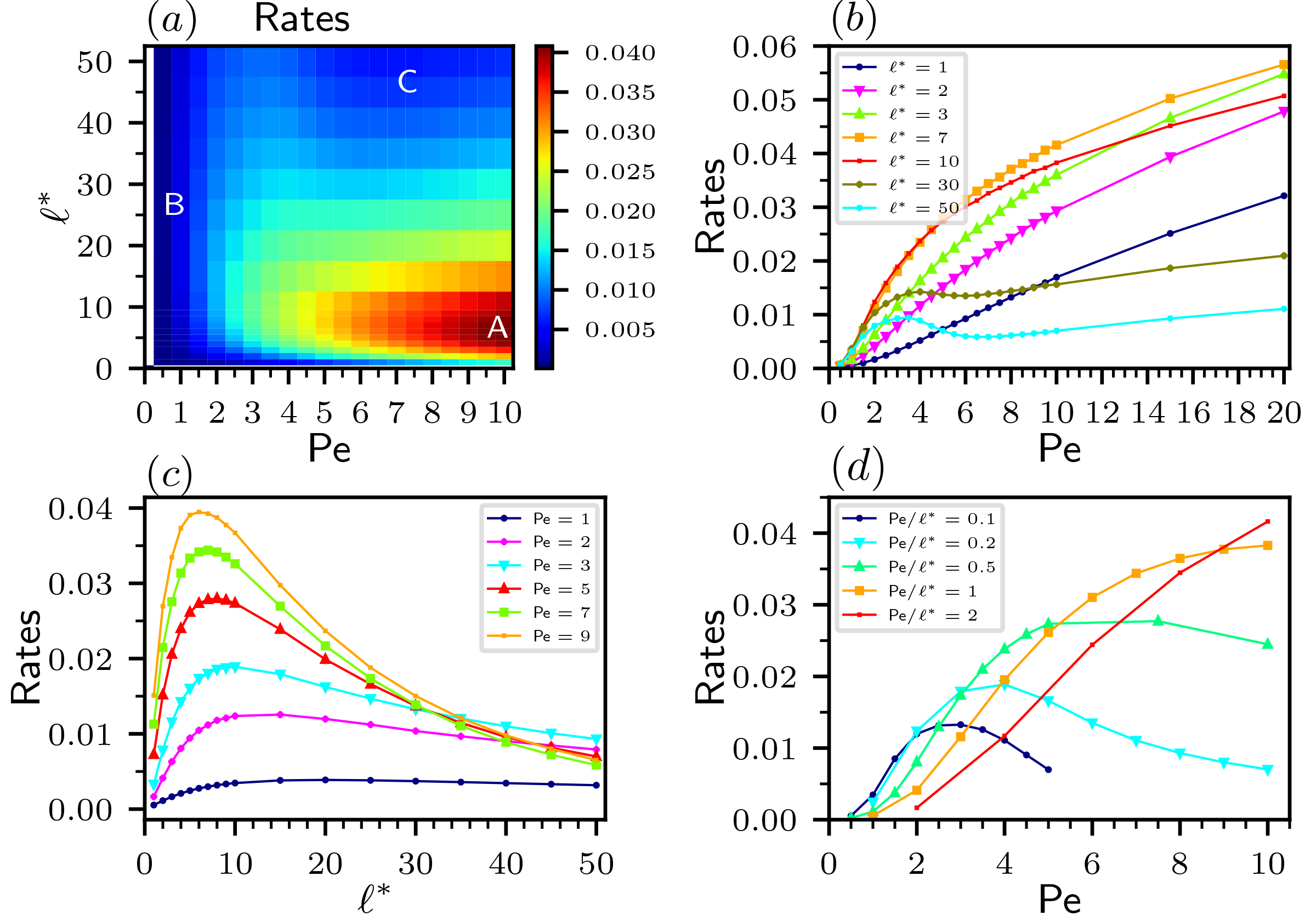}
\caption{(a) Heatmap of the target-finding rates represented as a function of the model parameters, $\text{Pe}$ and $\ell^{*}$, obtained from a statistics of $10^{5}$ successful target searches for each parameter set. Region A shows the highest transition rates, while regions B and C display much smaller rates. (b) Target-finding rates as a function of $\text{Pe}$, for different values of $\ell^{*}$. In the current landscape, the highest rates are found at high values of the P{\'e}clet number for persistence values of $3 \lesssim \ell^{*} \lesssim 10$. (c) Target-finding rates reported as a function of $\ell^{*}$ for different P{\'e}clet numbers. For each considered value of $\text{Pe}$, an $\ell^{*}$ that maximizes the rates emerges, which becomes smaller as $\text{Pe}$ increases. (d) Target-finding rates as a function of $\text{Pe}$ for different ratios of $\text{Pe}/\ell^{*}$. For each ratio, a $\text{Pe}$ that maximizes the rates is found, which increases together with the value of the ratio. The figure has been reproduced from our publication Ref.~\cite{Zanovello2021a}.}
\label{fig:dw_rates}
\end{figure}

We recall that this potential is characterized by two minima, located at $(\pm x_{0},0)$, and the minimum energy path linking the two lies along the axis $y = 0$, with a saddle point at $(0,0)$.
The barrier height is given by $k_{x}x_{0}^{4}$ while its shape is determined by $k_{y}/k_{x}x_{0}^{2}$, where $x_{0}$ is a typical geometric length of the landscape.
We select the barrier height by imposing $k_{x}x_{0}^{4} = 6.5 k_{B}T$, so a value slightly higher than the one employed in the previous section, which ensures that the transitions in the case of a passive particle are dominated by rare events.
To fix the shape of the energy landscape, we finally impose $k_{y}x_{0}^{2} = 20k_{B}T$.
As we mentioned above, the maximal force along this minimum energy path is $F_{\text{max}} = 8k_{x}x_{0}^{3}/3\sqrt{3}$, which in the case of our barrier height becomes $F_{\text{max}} \approx 10k_{B}T/x_{0}$.
Consequently, our length scale $L$ will read $L = 3\sqrt{3}k_{B}T/8k_{x}x_{0}^{3}$, so in our case $L \approx x_{0}/10$.
Finally, as in the previous chapter, we define R as the region where $U(x,y) \leq 2k_{B}T$ and $x<0$, and T as the region with $U(x,y) \leq 2k_{B}T$ and $x>0$.
A representation of the energy landscape is reported in Fig.~\ref{fig:dw_passive_tpd}(a).

A passive particle performing a target-finding process in such a landscape shows, unsurprisingly, a small target-finding rate due to the energy barrier separating R from T.
In particular, the rate $r$ is about $8.5 \cdot 10^{-5} \tau^{-1}$ (see the corresponding point in Fig.~\ref{fig:dw_rates}), and the passive particle, as we showed in the previous chapter, always crosses the barrier at the saddle point, following the minimum energy path from R to T (Fig.~\ref{fig:dw_passive_tpd}(b)).

\begin{figure}[H]
\centering
\includegraphics[height=70mm]{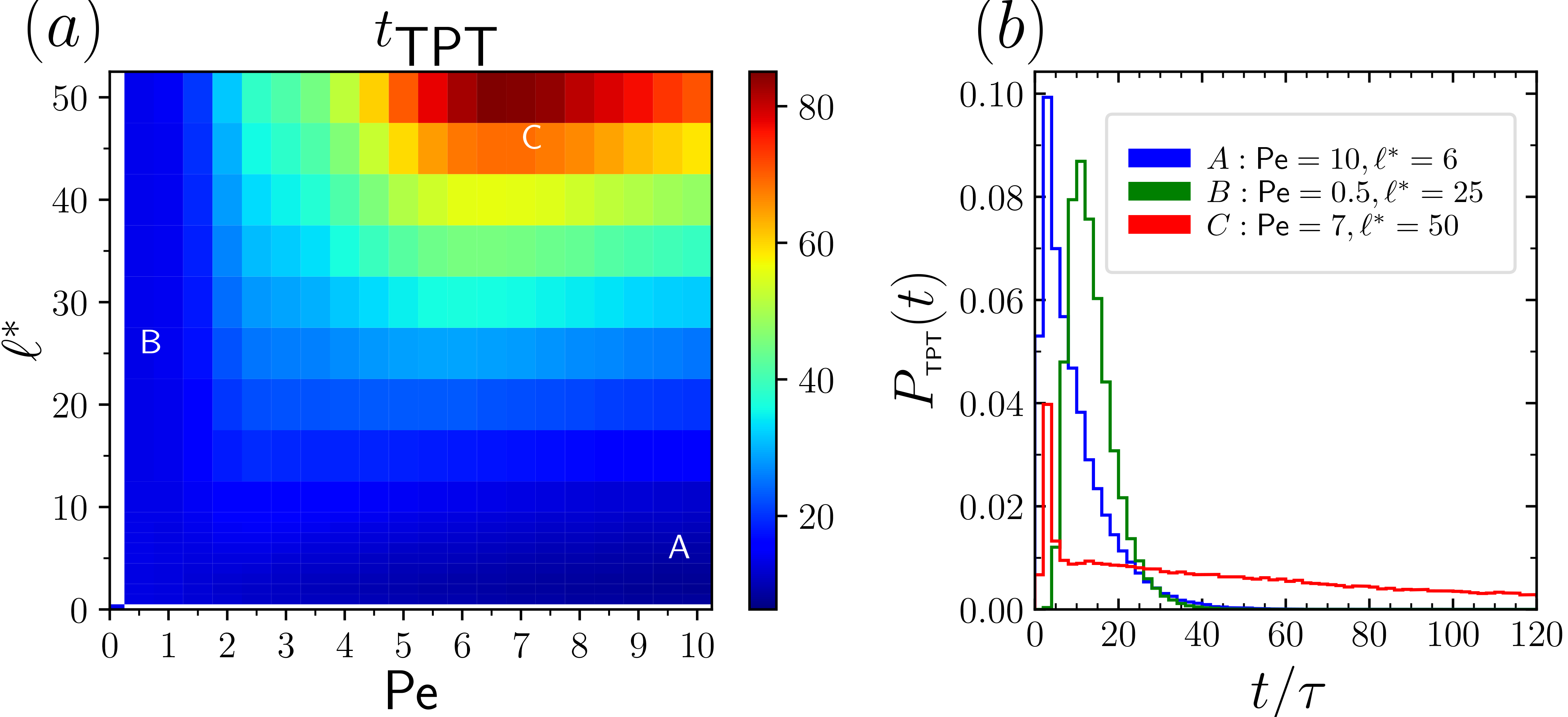}
\caption{Transition path time analysis for the double-well potential, obtained from a statistics of $10^{5}$ successful target searches for each point in parameter space and each distribution. (a) Heatmap of the average TPTs as a function of the model parameters, $\text{Pe}$ and $\ell^{*}$. Both region A and B display short average TPTs, while region C is characterized by longer times. (b) TPT distributions for different values of $\text{Pe}$ and $\ell^{*}$. The figure has been reproduced from our publication Ref.~\cite{Zanovello2021a}.}
\label{fig:dw_tpts}
\end{figure}

\begin{figure}[H]
\centering
\includegraphics[height=110mm]{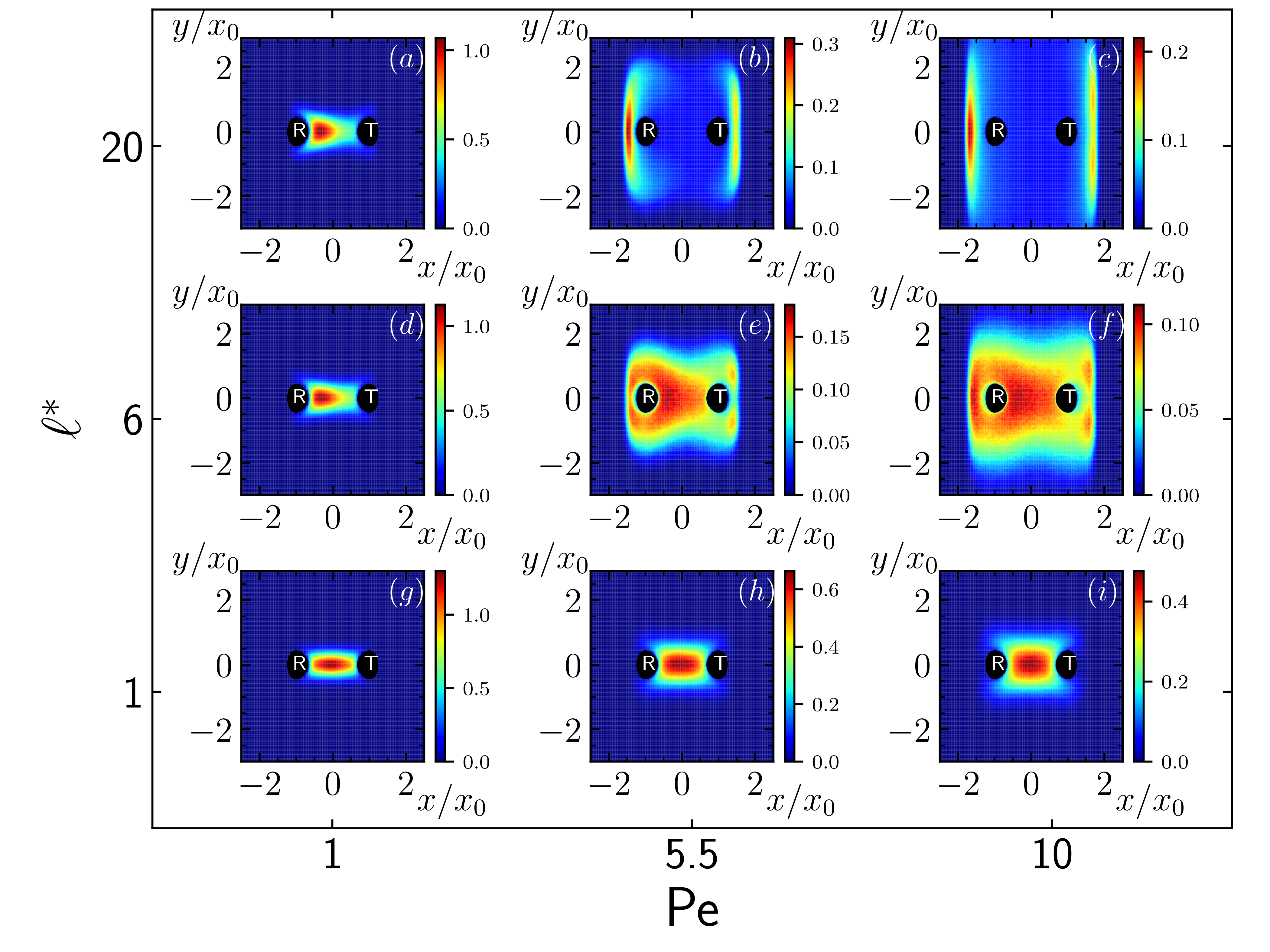}
\caption{Transition probability density $m(\bm{r})$ for different values of the P{\'e}clet number and persistence in the double-well potential, obtained from a statistics of $10^{5}$ target-finding events. For small values of $\text{Pe}$ (panels (a), (d), and (g)) the behavior of the particle resembles the one of a passive particle, with the transition probability density showing most frequented regions along the minimum energy path. This similarity decreases with an increase in $\ell^{*}$. An increase in $\text{Pe}$ allows the particle to explore larger regions in the surrounding landscape. For small values of the persistence (panels (h) and (i)), $m(\bm{r})$ resembles again the distribution for a passive particle with a translational diffusion coefficient increasing with $\text{Pe}$. At intermediate values of $\ell^{*}$ (panels (e) and (f)), the particle can efficiently navigate the surrounding regions in the transition space. A further increase in the persistence (panels (b) and (c)) causes the emergence of long-lasting target-finding paths, which spend long times at the boundaries of the system. The figure has been reproduced from our publication Ref.~\cite{Zanovello2021a}.}
\label{fig:dw_tpd}
\end{figure}

Instead, active particles show a behavior that is strongly influenced by the system parameters and that is reflected in the emergence of different transition mechanisms.
The most favorable target-finding rates are observed for intermediate values of the persistence and large values of $\text{Pe}$ (region A in Fig.~\ref{fig:dw_rates}(a)), where the rates have a magnitude of about $4 \cdot 10^{-2} \tau^{-1}$.
For particles belonging to region A, a value of $\text{Pe}\gtrsim 5$ allows them to efficiently climb over the energy barrier, and an intermediate level of the persistence $2 \lesssim \ell^{*} \lesssim 10$ ensures an efficient exploration of the landscape.
This behavior is confirmed by a short average TPT (see Fig.~\ref{fig:dw_tpts}(a)), by a TPT distribution characterized by short times (Fig.~\ref{fig:dw_tpts}(b), blue curve), and by the transition probability density (Fig.~\ref{fig:dw_tpd}(e), (f)) that shows how the particle can properly explore the most relevant part of the transition region.
A decrease in the persistence causes instead the particles to lose quickly their orientation and behave similarly to passive particles with an increased effective translational diffusion coefficient.
This has the effect of decreasing the rates (see Fig.~\ref{fig:dw_rates}(a)) and of reshaping the transition probability density to one more similar to the the passive case (Fig.~\ref{fig:dw_tpd}(h), (i)).

Outside of region A, two unfavorable regions (B and C) are found, which display a significant decrease in the target-finding rates.
Region B is characterized by a small value of the P{\'e}clet number, which is not enough for the particle to easily climb over the energy barrier.
This has the immediate consequence of lowering the transition rate due to a behavior of the particle resembling the one of a passive agent (see Fig.~\ref{fig:dw_rates} (a)).
Particles in region B show in fact short average TPTs, and the distribution of these times (Fig.~\ref{fig:dw_tpts}(b), green curve) resembles the one of a passive particle, similarly to the transition probability density, which is concentrated around the minimum energy path (Fig.~\ref{fig:dw_tpd}(a), (d), and (g) and Fig.~\ref{fig:dw_passive_tpd}(b) as a comparison for the passive behavior).
The similarity of the transition probability density to the one of a passive particle decreases with an increasing persistence, which allows the small active contributions to the motion of the particle to be added in the same direction for longer times.

Region C instead shows a decrease in the target-finding rates (Fig.~\ref{fig:dw_rates} (a)) for a different reason.
Particles in region C are characterized by large values of $\text{Pe}$ and $\ell^{*}$: their large P{\'e}clet number allows them to easily climb over the barrier similarly to particles in region A, but at the same time, their increased persistence compared to region A induces them to take long transitions that surf along the high-energy boundaries of the landscape, without being able to successfully reach the target region T that is characterized by a smaller energy value (see Fig.~\ref{fig:dw_tpd}(b) and (c)).
This is reflected also in an increase on the average TPTs (Fig.~\ref{fig:dw_tpts}(a)), which displays values consistently larger than in the remaining of the parameter space and which is the main contributor for the increase in $t_{r}$ in this region.
Note that this increase is determined by the long tail of the TPT distribution at long times (Fig.~\ref{fig:dw_tpts}(b), red curve), which also shows a peak at short times for all the trajectories that begin their transition with a favorable orientation of the self-propulsion term.
We also point out that the decrease of the average TPT found when increasing the self-propulsion of the particle, which is observed at low values of the persistence in the transition between region A and B, resembles the behavior in the case of 1D system~\cite{Carlon2018}, while a different behavior is observed at large persistence (region C) and also along different slices of the parameter space (see section \ref{sec:tpt_dw_validation}).

Note that region B and C are separated by a section of parameter space which shows more favorable target-finding rates, which is observed also in the curves with fixed and large persistence $\ell^{*}$ in Fig.~\ref{fig:dw_rates} (b).
This region, which has no counterpart on the behavior of the average TPTs (Fig.~\ref{fig:dw_tpts}(b)), corresponds to those particles that possess a sufficiently high P{\'e}clet number to climb efficiently the energy barriers, and at the same time not too large to force them to long detours far from T due to the high persistence.

Furthermore, Fig.~\ref{fig:dw_rates} suggests how the success rate for the target-search process can be optimized.
If the particle possesses a fixed value of the persistence $\ell^{*}$, the best option to increase the success rate relies on increasing $\text{Pe}$ as much as possible (Fig.~\ref{fig:dw_rates}(b)).
While this can be easily understood for low levels of activity, where an increase in $\text{Pe}$ leads to an easier process in climbing the barrier, one would na\"ively expect that at high values of the activity an additional increase in $\text{Pe}$ would lead to even longer detours along the confining boundaries.
However, if $\ell^{*}$ is kept fixed, $\text{Pe}$ becomes directly proportional to the rotational diffusion coefficient $D_{\vartheta}$, thus an increase in $\text{Pe}$ leads also to an increase in $D_{\vartheta}$ with a consequent reduction of the time spent surfing along the energy contour lines.
This is reflected in the non-monotonic behavior of the average TPTs in Fig.~\ref{fig:dw_tpts}(a) for fixed and large values of $\ell^{*}$, which first increase with $\text{Pe}$ and then start decreasing, with a consequent effect on the target-finding rates.

If instead the particle is constrained to having a fixed value of $\text{Pe}$, the rates can still be optimized by tuning the persistence of the particle.
In fact, for each fixed value of $\text{Pe}$ a value of $\ell^{*}$ exists that maximizes the target-finding rates, which becomes smaller as $\text{Pe}$ increases (see Fig.~\ref{fig:dw_rates}(c)).
This phenomenon can be explained considering that at smaller $\text{Pe}$ the particle will more easily climb the energy barrier if the self-propulsion efforts are pushing in the same direction for longer times.

Finally, if the particle can adjust its self-propulsion speed while its rotational diffusion coefficient is fixed (so the particle has a constant ratio $\text{Pe}/\ell^{*}$), the rates can be optimized by selecting a specific value of $\text{Pe}$ (Fig.~\ref{fig:dw_rates}(d)).
In the case of such a particle with fixed ratio $\text{Pe}/\ell^{*}$, the best target-finding rates are observed at large values of the ratio and the value of $\text{Pe}$ that optimizes them increases with the ratio.

\section{Target-search strategies in the Brown-M\"uller potential}

Now, the question arises of whether some of these observations are influenced by the choice of such a simple potential, which is symmetric, has a single energy barrier located between R and T, and displays a straight minimum energy path linking the two states.
In particular, the effects of the shape of the landscape on the optimal strategy that the ABPs use to find their targets are still unexplored.
To answer these queries, we perform an equivalent analysis with a different energy landscape, the Brown-M\"uller potential, which is defined as:
\begin{equation}
U(x,y) = \sum_{i=1}^{4} K_{i} e^{[a_{i}(x-x_{0,i})^{2}+b_{i}(x-x_{0,i})(y-y_{0,i})+c_{i}(y-y_{0,i})^{2}]} \; ,
\end{equation}
where the parameters are:
\begin{center}
\begin{tabular}{|c|c|c|c|c|c|c|}
\hline
$i$ & $K_{i}/k_{B}T$ & $a_{i}/x_{0}^{2}$ & $b_{i}/x_{0}^{2}$ & $c_{i}/x_{0}^{2}$ & $x_{0,i}/x_{0}$ & $y_{0,i}/x_{0}$ \\
\hline
$1$ & $-9.69$ & $-1.2$ & $0$ & $-6$ & $1.7$ & $0$ \\
$2$ & $-4.41$ & $-1.2$ & $0$ & $-8.5$ & $-0.7$ & $0.3$ \\
$3$ & $-8.81$ & $-3$ & $6$ & $-5$ & $0.5$ & $2$ \\
$4$ & $0.88$ & $0.7$ & $0.6$ & $0.7$ & $0$ & $1$ \\
\hline
\end{tabular}
\end{center}

\begin{figure}[H]
\centering
\includegraphics[height=65mm]{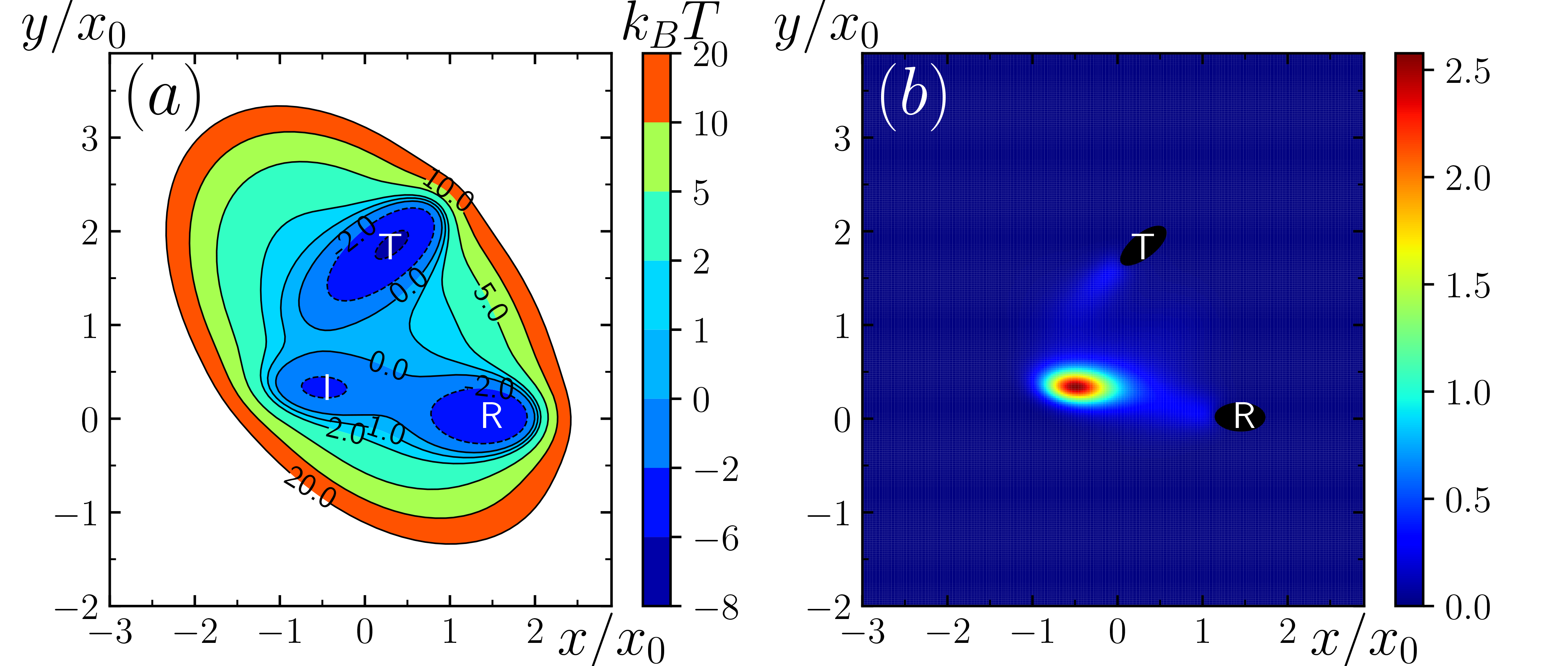}
\caption{(a) Representation of the Brown-M\"uller potential. For each contour line the energy value is reported in units of $k_{B}T$. R indicates the reactant basin, I is an intermediate state, and T the target basin. (b) Transition probability density for a passive particle during successful target searches in the considered landscape, obtained from $10^{5}$ reactive paths. The figure has been reproduced from our publication Ref.~\cite{Zanovello2021a}.}
\label{fig:muller_passive_tpd}
\end{figure}

We define the R state as the region where $U(x,y) \leq -4.5 k_{B}T$ and $y<0.7$ and the T state as the region where $U(x,y) \leq -5.5 k_{B}T$ and $y>0.7$.
This new landscape is asymmetric and displays a curved minimum energy path (see Fig.~\ref{fig:muller_passive_tpd}(a)).
This minimum energy path will visit a third metastable state in the system, the intermediate state I, which causes the energy barrier to be split into two smaller ones along the path, one located between R and I and the other one between I and T.
This time the energy barrier separating R from T has a maximum height of about $6 k_{B}T$ measured from the bottom of the R basin and the two basins R and T have a different shape and depth (about $1.3 k_{B}T$ for R and $1.1 k_{B}T$ for T).
The maximal force along the minimum energy path is by construction $F_{\text{max}} \simeq 10 k_{B}T/x_{0}$, to ensure a typical length scale $L$ equivalent to the one used in the double well.

\begin{figure}[H]
\centering
\includegraphics[height=100mm]{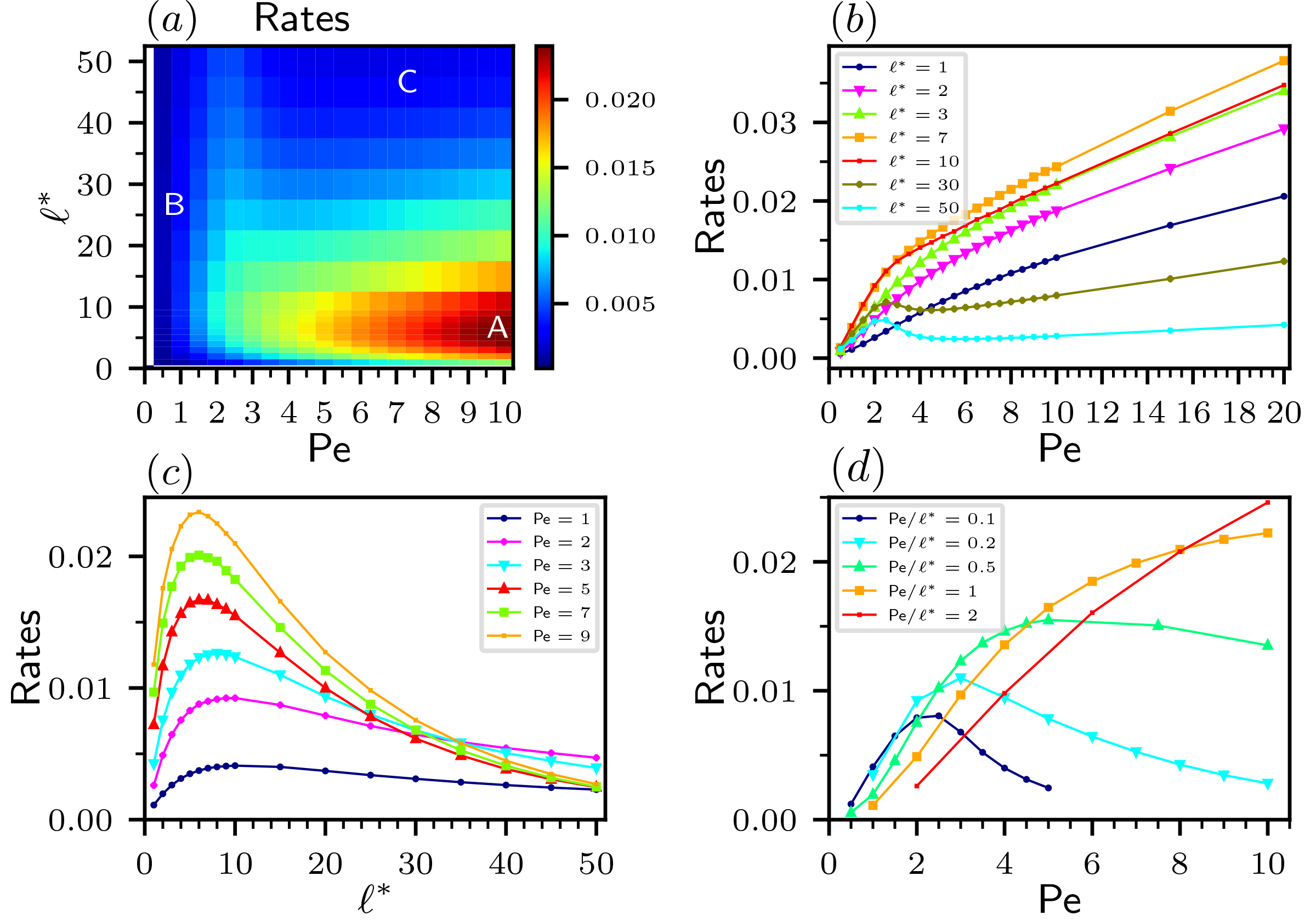}
\caption{(a) Heatmap of the target-finding rates represented as a function of the model parameters, $\text{Pe}$ and $\ell^{*}$, obtained from a statistics of $10^{5}$ successful target searches for each parameter set. Region A shows the highest transition rates, while regions B and C display much smaller rates. (b) Target-finding rates as a function of $\text{Pe}$, for different values of $\ell^{*}$. In the current landscape, the highest rates are found at high values of the P{\'e}clet number for persistence values of $3 \lesssim \ell^{*} \lesssim 10$. (c) Target-finding rates reported as a function of $\ell^{*}$ for different P{\'e}clet numbers. For each considered value of $\text{Pe}$, an $\ell^{*}$ that maximizes the rates emerges, which becomes smaller as $\text{Pe}$ increases. (d) Target-finding rates as a function of $\text{Pe}$ for different ratios of $\text{Pe}/\ell^{*}$. For each ratio, a $\text{Pe}$ that maximizes the rates is found, which increases together with the value of the ratio. The figure has been reproduced from our publication Ref.~\cite{Zanovello2021a}.}
\label{fig:muller_rates}
\end{figure}

The behavior of the rates in the Brown-M\"uller potential is found qualitatively similar to the one observed in the double-well case, regardless of the differences introduced by the energy landscape (compare the heatmaps in Fig.~\ref{fig:dw_rates}(a) and Fig.~\ref{fig:muller_rates}(a)).

Once again, the passive particle performs successful target searches by following the minimum energy path, which this time causes the particle to spend a long time in the intermediate state I (see Fig.~\ref{fig:muller_passive_tpd}(b)).
In the case of the Brown-M\"uller potential, the rates observed for a passive particle ($\text{Pe} = 0$, $\ell^{*} = 0$) are around $2.4 \cdot 10^{-4} \tau^{-1}$.
This rate is about three times the value found in the case of the double well due to the decrease of the energy barrier height in the Brown-M\"uller potential.

Also in this case, active particles show target-finding rates larger than the passive ones, with an increase of about two orders of magnitude.
In particular, the most favorable rates show values around $2 \cdot 10^{-2} \tau^{-1}$, which are approximately half of those observed in the double-well case.
Such a decrease in the target-finding rates would appear counterintuitive given the reduced barrier height in the case of the Brown-M\"uller potential.
However, active particles are not substantially influenced by such a small difference in the barrier height, especially in the cases with high values of the self-propulsion, but they rather show decreased rates due to the reduced geometrical dimension of basin T.

\begin{figure}[H]
\centering
\includegraphics[height=70mm]{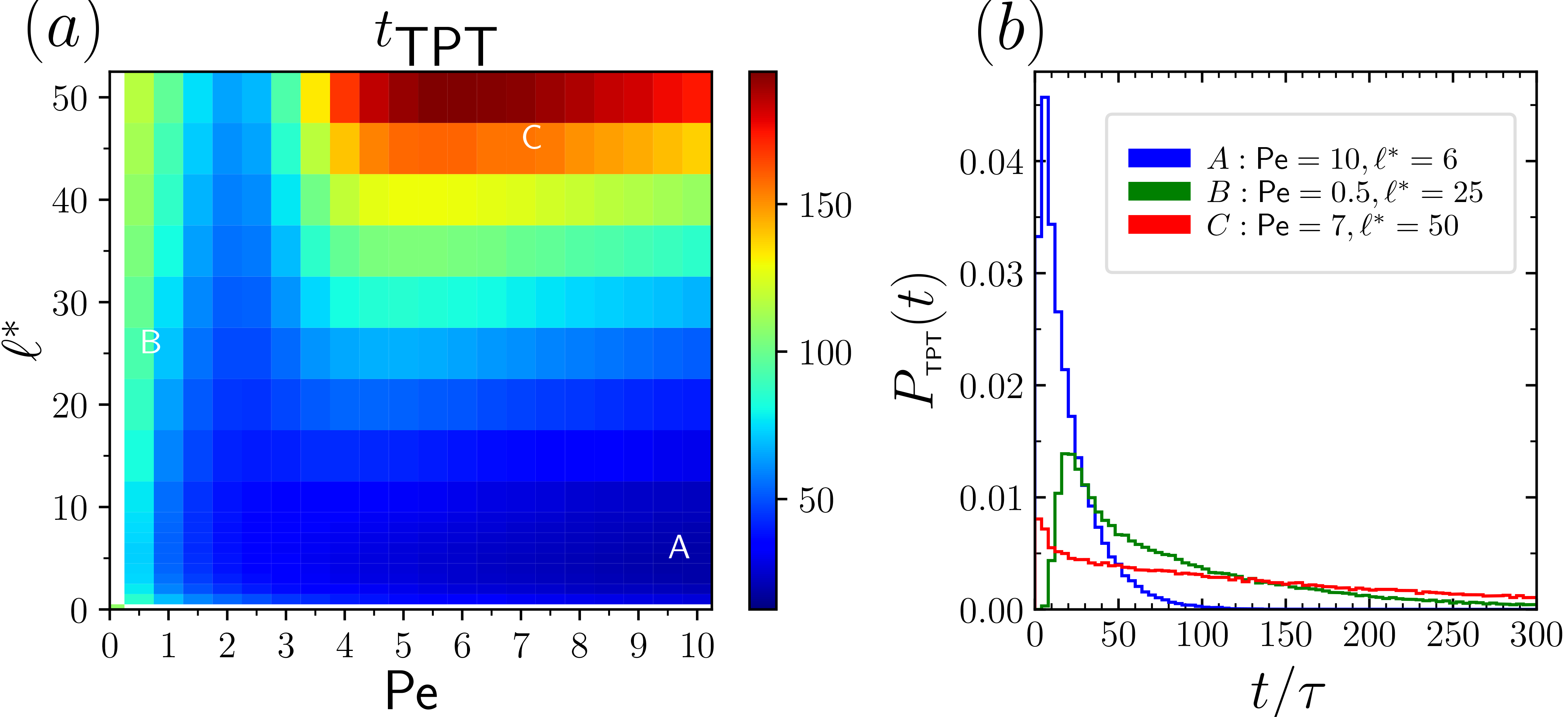}
\caption{Transition path time analysis for the Brown-M\"uller potential, obtained from a statistics of $10^{5}$ successful target searches for each point in parameter space and each distribution. (a) Heatmap of the average TPTs as a function of the model parameters, $\text{Pe}$ and $\ell^{*}$. Region A displays short average TPTs, while both region B and C are characterized by long average TPTs. (b) TPT distributions for different values of $\text{Pe}$ and $\ell^{*}$. The figure has been reproduced from our publication Ref.~\cite{Zanovello2021a}.}
\label{fig:muller_tpts}
\end{figure}

\begin{figure}[H]
\centering
\includegraphics[height=110mm]{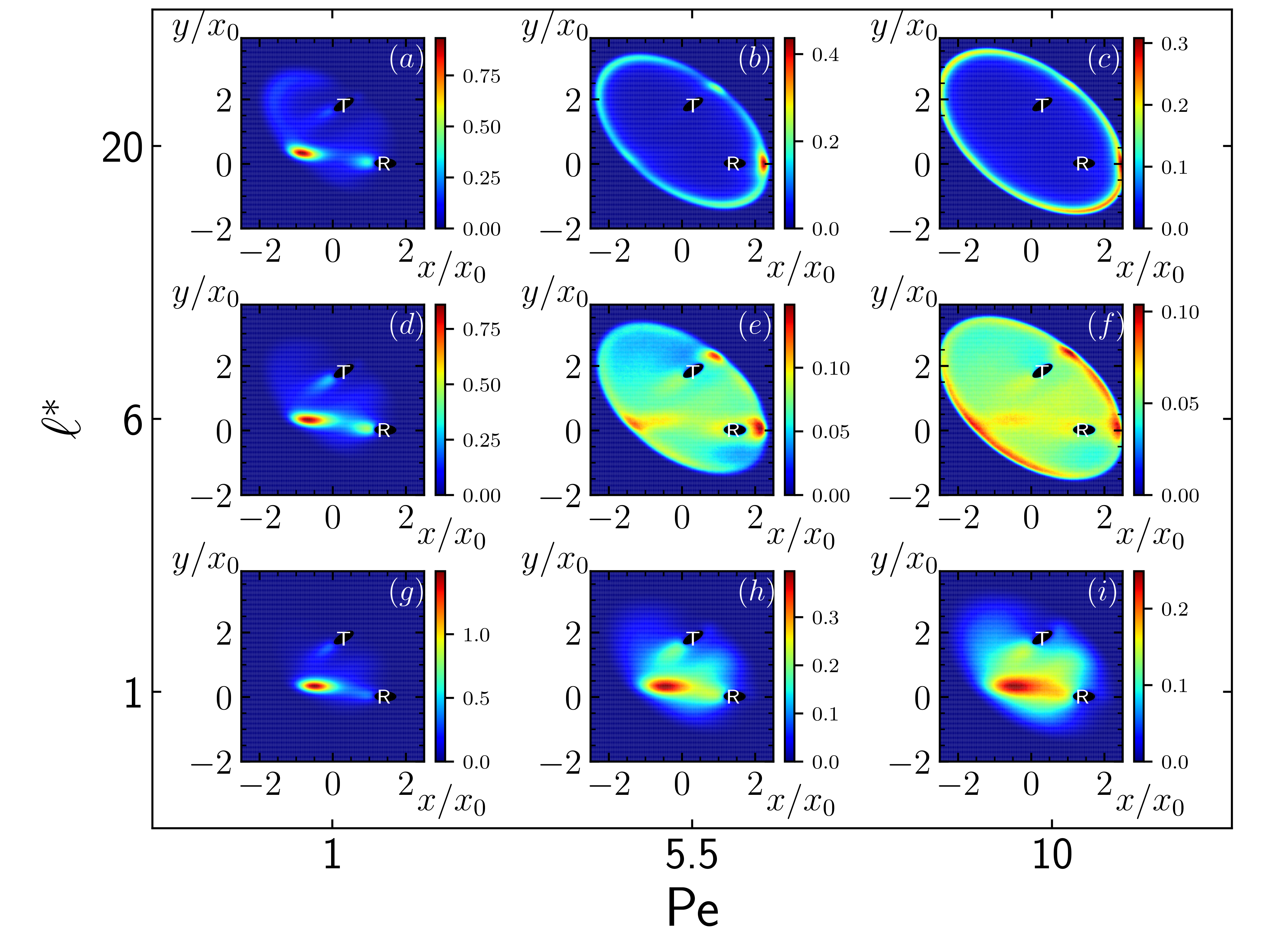}
\caption{Transition probability density $m(\bm{r})$ for different values of the P{\'e}clet number and persistence in the Brown-M\"uller potential, obtained from a statistics of $10^{5}$ target-finding events. For small values of $\text{Pe}$ (panels (a), (d), and (g)) the behavior of the agent resembles the one of a passive particle, with the transition probability density showing most frequented regions along the minimum energy path and in the intermediate state I. This similarity decreases with an increase in $\ell^{*}$. An increase in $\text{Pe}$ allows the particle to explore larger regions in the surrounding landscape. For small values of the persistence (panels (h) and (i)), $m(\bm{r})$ resembles again the distribution for a passive particle with a translational diffusion coefficient increasing with $\text{Pe}$. At intermediate values of $\ell^{*}$ (panels (e) and (f)), the particle can efficiently navigate the surrounding regions in the transition space. A further increase in the persistence (panels (b) and (c)) causes the emergence of long-lasting target-finding paths, which spend a long time at the boundaries. The figure has been reproduced from our publication Ref.~\cite{Zanovello2021a}.}
\label{fig:muller_tpd}
\end{figure}

Once more, the region showing the most favorable target-finding rates is identified as region A of the parameter space (see Fig.~\ref{fig:muller_rates}(a)), which is once again characterized by intermediate values of the persistence and large values of $\text{Pe}$.
The best target-finding mechanism is equivalent to the case discussed in the double well: it relies on finding a value of the persistence large enough to preserve the motion directionality and easily cross the barriers, but at the same time not too large to avoid long detours by surfing along the energy walls of the landscape.
This analysis is once more confirmed by short average TPTs (Fig.~\ref{fig:muller_tpts}(a))and a TPT distribution peaked at short times (Fig.~\ref{fig:muller_tpts}(b), blue curve).
The transition probability density in region A (Fig.~\ref{fig:muller_tpd}(e) and (f)) displays how these values of the parameters allow a complete and efficient exploration of the most relevant part of the landscape.
If the persistence of the agent is decreased, however, the particle will change its self-propulsion direction faster, with a consequent decrease in the rates.
The particle will then behave again similarly to a passive particle with increased translational diffusion coefficient, spending longer times in the intermediate state I (see Fig.~\ref{fig:muller_tpd}(h) and (i)).

The two unfavorable regions B and C are located in the same positions of the parameter space for the two energy landscapes, so at small values of $\text{Pe}$ and large values of $\ell^{*}$ respectively.

Again, particles in region B have a too small self-propulsion contribution to their motion and behave similarly to a passive particle, with this similarity decreasing as $\ell^{*}$ increases.
They follow mostly the minimum energy path and show decreased target-finding rates due to the difficulty in crossing the barriers.
Differently from the double well, however, due to the presence of an additional state, these particles will spend a long time in the intermediate state I, with a consequent increase in the average TPTs that is not observed in the double-well case (see Fig.~\ref{fig:muller_tpts}(a)), and a mode of the TPT distribution shifted towards slightly larger times compared to the one for region A (see Fig.~\ref{fig:muller_tpts}(b), green curve).
The long permanence in the metastable state I is highlighted also by the transition probability densities, see Fig.~\ref{fig:muller_tpd}(a), (d), and (g).

Similarly to the double-well case, particles in region C have a large value of the activity, which allows them to climb easily the energy barriers, but at the same time a large persistence, which forces them to spend a long time orbiting at the outskirts of the system (Fig.~\ref{fig:muller_tpd}(b) and (c)).
This generates difficulties in reaching specific lower energy regions such as the T basin (similarly to what happens in the case of an ABP looking for a target located far from the rigid boundaries of a confining environment~\cite{Wang2016}).
This phenomenon has the effect of increasing the average TPTs (see Fig.~\ref{fig:muller_tpts}(a)) and consequently decreasing the target-finding rates (see Fig.~\ref{fig:muller_rates}(a)).
Additionally, the TPT distribution for particles in this region is characterized by a much more pronounced tail at long times (Fig.~\ref{fig:muller_tpts}(b), red curve).
Once more, the average TPTs at fixed and large value of $\ell^{*}$ decrease with increasing $\text{Pe}$ after overcoming a local maximum due to the fact that an increase in $\text{Pe}$ at fixed $\ell^{*}$ has an increase in $D_{\vartheta}$ as consequence.
In contrast to the double-well case, however, regions B and C are well-separated not only in the rates, by the presence of a local maximum, but also in the average TPTs, where a local minimum divides the region where the long TPTs are generated by a long permanence in the metastable state I from the region where the TPTs are increased due to the long detours at the boundaries of the system.

Fig.~\ref{fig:muller_rates} suggests criteria to optimize the target-finding rates, which are equivalent to those observed in the double well.
Fig.~\ref{fig:muller_rates}(b) displays how the rates can be optimized by selecting a value of the persistence such that $3 \lesssim \ell^{*} \lesssim 10$ and then by achieving a self-propulsion as large as possible.
In general, the target-finding rates are found to increase with the $\text{Pe}$ number.
Fig.~\ref{fig:muller_rates}(c) shows once again how for a particle with fixed $\text{Pe}$ there exists a value of the persistence that optimizes the rates, and this value decreases with increases in $\text{Pe}$.
Finally, Fig.~\ref{fig:muller_rates}(d) shows that if the particle is constrained to having a fixed value of the ratio $\text{Pe}/\ell^{*}$, the rates can be optimized for a specific value of $\text{Pe}$, which increases with the value of the ratio.

These findings suggest the generality of this optimal target-finding strategy, which can be employed for any similar landscape, as long as the height of the energy barrier remains fixed as well as the distance between the R and T basin.
If the height of the energy barrier is modified or the distance between the basins is changed, we still expect to find some values of $\text{Pe}$ and $\ell^{*}$ that optimize the target-finding rates, even though they might be found at different values of the parameters.
The addition of other metastable states in the process does not seem to change the qualitative picture of the transition process, as long as their depth remain comparable to the one of the R and T basins.
Finally, as shown from the transition probability densities, the location of the target plays a relevant role in determining the success odds of the search: if the target is located in some region with high energy closer to the confining boundaries rather than in a basin, the particle is expected to show much larger target-finding rates at higher values of the persistence and activity.

\section{Chapter conclusions}
In this chapter we performed a systematic analysis on the parameter space of the ABP model for target-search processes in energy landscapes.

After showing that the parameter space for the model is two-dimensional and spanned by the parameters $\text{Pe}$ and $\ell^{*}$, we characterized the target-search dynamics in a double-well potential and in a more complex landscape, the Brown-M\"uller potential, through relevant observables for the process: the target-finding rates, the Transition Path Times, and the transition probability densities.
This analysis revealed how this process can be optimized depending on some tuning of the model parameters, resulting in higher rates.
We then analyzed the microscopic mechanisms behind this behavior, discovering how the particle can optimize its success rate by using an intermediate value of the persistence, so to avoid a passive-like behavior or long detours at the system boundaries, and using a large self-propulsion, to travel quickly along the resulting reactive paths.
We also described in detail the dynamics in the case of lower target-finding rates, showing how this decrease in the rates can be observed for two different reasons depending on the model parameters, namely the passive-like behavior or the long surfs at the system boundaries in the case of high activity and persistence.
Additionally, we found how the target-finding rates can be optimized also in the case that the particle is unable to tune one of the control parameters by acting on the remaining one.

The comparison between the two different landscapes highlighted how the qualitative behavior of the system is reproduced regardless of the details of the energy landscape at hand, and additionally how the favorable region in parameter space is substantially not influenced by the choice of the energy landscape.
In fact, in both cases the rates can be optimized by following the same protocol, and although differences emerge in the details of some observables (\textit{e.g.} the average TPTs) a different shape and characteristics of the underlying energy landscape does not modify the microscopic mechanism used by the particle to efficiently find its target.

Finally, the addition of the metastable state I in the Brown-M\"uller potential provided us with a better insight on how active particles navigate in rugged landscapes in order to reach a specific target location.
In particular, we found that while the presence of this state is reflected in the typical trajectories inferred from the transition probability densities, these kinetic traps have little effect on the target-finding rates, proving once more that activity is an advantageous feature to possess for target-finding processes.

\afterpage{\null\newpage}

\pagestyle{plain}

\chapter{Committor function application to active Brownian Particles}
\label{ch:comm_function}
\pagestyle{fancy}

\section{Generalizing the committor function to ABP in target-search problems}

In the previous chapters we provided an efficient way to simulate ABPs target-searches characterized by rare events and we analyzed the optimal strategy for ABPs in this class of processes.

Here, we want to extend our set of tools to study these processes by generalizing the concept of the committor function to target-search processes.
We will therefore develop an expression for the committor function for our system of an active particle navigating in an energy landscape and searching for a target region.
This new tool can then be used to obtain additional relevant information on how the transition occurs by quantifying the probability of successfully finding the target given a specific initial condition (see sketch in Fig.~\ref{fig:comm_sketch}).

\begin{figure}[H]
\centering
\includegraphics[height=50mm]{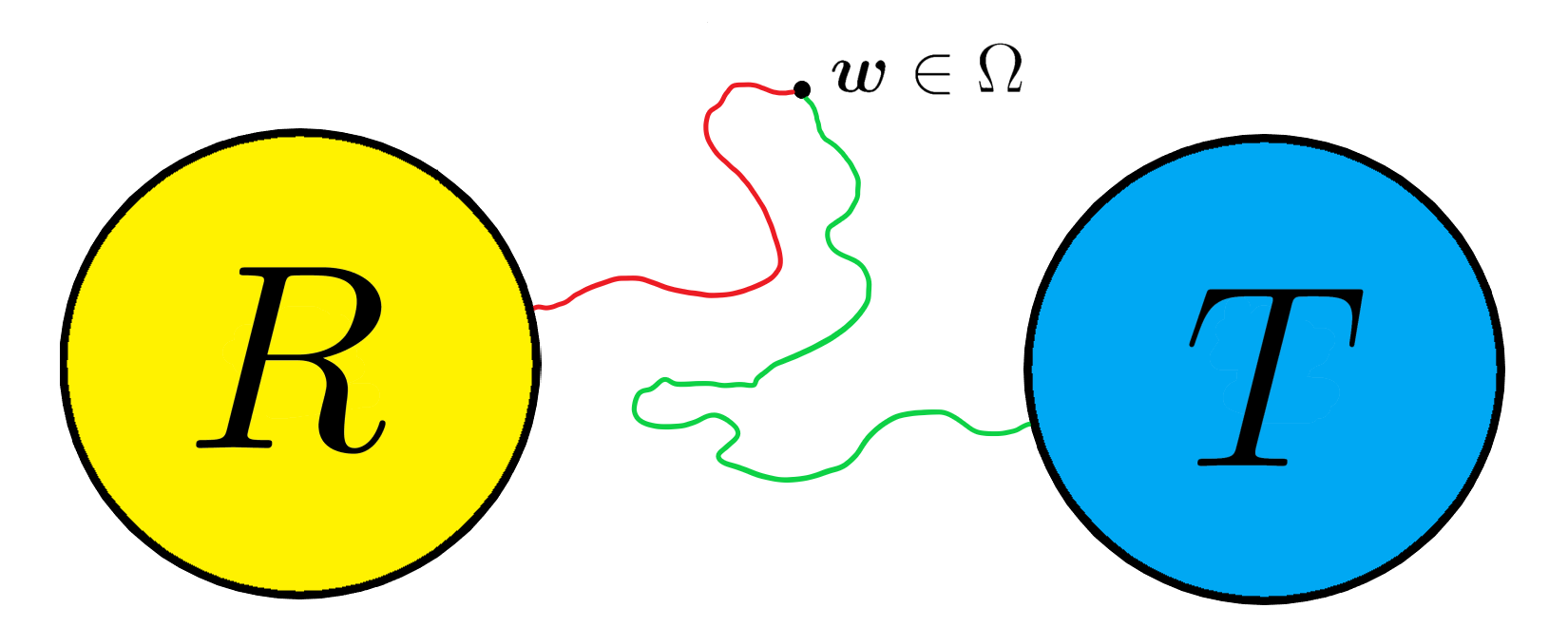}
\caption{Sketch of a committor function. The committor is evaluated as the number of trajectories starting from a given initial configuration $w \in \Omega$ and reaching T without touching R first (green line) divided by the total number of trajectories starting from $w$, which includes also the paths that arrive in R first (red line).}
\label{fig:comm_sketch}
\end{figure}

Here we start by deriving a Fokker-Planck (FP) equation for an active Brownian particle and a modified propagator for the system, which are then used to obtain an analytic expression for the committor function of an ABP.
We then prove that, similarly to the passive case, the active version of the committor satisfies the backward Kolmogorov equation for the system.
Subsequently, we discuss how the committor can be numerically obtained by solving this equation using a finite-difference algorithm.
This leads to an efficient computation of the committor for the system without the need of performing long brute-force simulations.
We then use this method to characterize the committor for an ABP with three different sets of $\text{Pe}$ and $\ell^{*}$, highlighted in the discussion in chapter \ref{ch:nav_strat}, during a target-search process in the paradigmatic double-well potential.
Finally, we validate the finite-difference solution against brute-force simulations, which serve as a control to ensure the correct analytic derivation of the committor for an ABP.

\subsection{Fokker-Planck equation for ABPs}

Given a stochastic variable $\bm{w}$ (which in the case of the ABP can be the position of the particle in configuration space), the Fokker-Planck equation for the system describes how the probability distribution of this variable changes in time (for more on this see Appendix \ref{app:FP}).

The ABP equations of motion can be described by a stochastic equation in the form:
\begin{equation}
\label{eq:stoc_abp_fp}
d\bm{w} = \bm{A}(\bm{w},t)dt + \bm{B}(\bm{w},t)d\bm{W}(t) \; ,
\end{equation}
where $\bm{A}(\bm{w},t)$ is the drift vector for the system, representing the deterministic contributions to the motion of the variable, while the product of the matrix $\bm{B}(\bm{w},t)$ with the Wiener process $d\bm{W}(t)$ provides the stochastic contribution to the motion.
This equation, in the case of the ABP, has the equivalent representation in 2D given by the usual set of overdamped discretized Langevin equations:
\begin{subequations}
\begin{eqnarray}\label{eom3}
x_{i\!+\!1} &=& x_{i} + v\, \cos \vartheta_{i} \, \Delta t - D \beta \nabla_{x} U(x_{i},y_{i}) \Delta t + \sqrt{2D\Delta t} \, \bm{\xi}_{x,i}\;,\\ \label{eom23}
y_{i\!+\!1} &=& y_{i} + v\, \sin \vartheta_{i} \, \Delta t - D \beta \nabla_{y} U(x_{i},y_{i}) \Delta t + \sqrt{2D\Delta t} \, \bm{\xi}_{y,i}\;,\\ \label{eom33}
\vartheta_{i\!+\!1} &=& \vartheta_{i} + \sqrt{2D_{\vartheta}\Delta t} \, \eta_i\;,
\end{eqnarray}
\end{subequations}
where this time we are using Cartesian coordinates and we substituted the effective mobility of the particle with the translational diffusion coefficient and the effective temperature that sets the energy scale $k_{B}T$, which is encoded in $\beta = 1/k_{B}T$ (see discussion in section \ref{sec:ABP_model}).
This implies:
\begin{equation}
d\bm{w} = \begin{pmatrix} dx\\dy\\d\vartheta \end{pmatrix} \; ,
\end{equation}
\begin{equation}
    \bm{A}(\bm{w},t) = \begin{pmatrix} v\, \cos\vartheta \, - D\, \beta\, \nabla_{x}U(x,y) \\ v\, \sin\vartheta \, - D\, \beta\, \nabla_{y}U(x,y) \\ 0 \end{pmatrix}\;,
\end{equation}
\begin{equation}
    \bm{B}(\bm{w},t) = \begin{pmatrix} \sqrt{2D} & 0 & 0 \\ 0 & \sqrt{2D} & 0 \\ 0 & 0 & \sqrt{2D_{\vartheta}} \end{pmatrix}\;,
\end{equation}
\begin{equation}
    d\bm{W}(t) = \begin{pmatrix} \xi_{x}\, \sqrt{dt} \\ \xi_{y}\, \sqrt{dt} \\ \eta\, \sqrt{dt} \end{pmatrix}\;.
\end{equation}
Given an equation in the form of Eq.~\ref{eq:stoc_abp_fp}, the probability $p(\bm{w})$ for the stochastic variable will satisfy a Fokker-Planck equation in the form of Eq.~\ref{eq:SDEND}.
In particular, we recall that:
\begin{equation}
D_{ij}(\bm{w},t) = \frac{1}{2} \sum_{k=1}^{3}B_{ik}(\bm{w},t)B_{jk}(\bm{w},t)
\end{equation}
and by noting that $D_{ij}(\bm{w},t) \neq 0 \iff i = j$, $D_{11} = D_{22} = D$, and $D_{33} = D_{\vartheta}$, one can derive the Fokker-Planck equation for the system as:
\begin{equation}
\begin{aligned}[t]
   \frac{\partial p}{\partial t} = -\frac{\partial}{\partial x} \bigg[ \big( v\, \cos{\vartheta}\, - D\, \beta\, & \nabla_{x}U(x,y) \big) p \bigg]\, -\frac{\partial}{\partial y} \bigg[ \big( v\, \sin{\vartheta}\, - D\, \beta\, \nabla_{y}U(x,y) \big) p \bigg]\, + \\ & + D \frac{\partial^{2}}{\partial x^{2}} p\, + D \frac{\partial^{2}}{\partial y^{2}} p\, + D_{\vartheta} \frac{\partial^{2}}{\partial \vartheta^{2}} p \; ,
\end{aligned}
\end{equation}
where we omitted the dependence of $p$ from $\bm{w}$ for simplicity.
Finally, in a more compact form:
\begin{equation}
    \frac{\partial p}{\partial t} = - \bm{\nabla}_{\bm{r}} \bigg[ \big( v\, \bm{u}\, - D\, \beta\, \bm{\nabla}_{\bm{r}}U(\bm{r}) \big) p \bigg]\, + D\, \nabla_{\bm{r}}^{2} p\, + D_{\vartheta}\, \partial_{\vartheta}^{2} p\;,
    \label{eq:fpabp}
\end{equation}
where:
\begin{equation}
\bm{\nabla}_{\bm{r}} = \begin{pmatrix} \partial_{x} \\ \partial_{y} \end{pmatrix}\;.
\end{equation}

From the Fokker-Planck equation an expression of the Fokker-Planck operator for an ABP can trivially be derived as:
\begin{equation}
\begin{aligned}[t]
    \frac{\partial p}{\partial t} & = D \bm{\nabla}_{\bm{r}} \bigg[ \bm{\nabla}_{\bm{r}} p\, - \bigg( \frac{v\bm{u}}{D} - \beta \bm{\nabla}_{\bm{r}}U(\bm{r}) \bigg) p \bigg] + D_{\vartheta}\, \partial_{\vartheta}^{2} p\\
    & =  D \bm{\nabla}_{\bm{r}} \Bigg[ \bigg[ \bm{\nabla}_{\bm{r}}\, - \bigg( \frac{v\bm{u}}{D} - \beta \bm{\nabla}_{\bm{r}}U(\bm{r}) \bigg) \bigg] p \Bigg] + D_{\vartheta}\, \partial_{\vartheta}^{2} p\\
    & = -\hat{H}_{\text{FP}} \; p \; ,
\end{aligned}
\end{equation}
where $\hat{H}_{\text{FP}}$ is the Fokker-Planck operator for an active Brownian particle:
\begin{equation}
    \hat{H}_{\text{FP}} = -D \bm{\nabla}_{\bm{r}} \bigg[ \bm{\nabla}_{\bm{r}}\, - \bigg( \frac{v\bm{u}}{D} - \beta \bm{\nabla}_{\bm{r}}U(\bm{r}) \bigg) \bigg] - D_{\vartheta}\, \partial_{\vartheta}^{2} \; .
    \label{eq:fpoabp}
\end{equation}

Finally, eq.~\ref{eq:fpabp} can be rewritten as a continuity equation:
\begin{equation}
    \frac{\partial p}{\partial t} = - \bm{\nabla}_{\bm{r}} \cdot \bm{J}_{\bm{r}} - \partial_{\vartheta} J_{\vartheta}
    \label{eq:fpcabp}
\end{equation}
where:
\begin{equation}
    \bm{J}_{\bm{r}} = \big[v \bm{u} - D \bm{\nabla}_{\bm{r}} - D \beta \bm{\nabla}_{\bm{r}} U(\bm{r})\big]p\; ,
\end{equation}
\begin{equation}
    J_{\vartheta} = - D_{\vartheta} \partial_{\vartheta}p\; .
\end{equation}

Eq.~\ref{eq:fpcabp} can be expressed in a more compact form, as:
\begin{equation}
    \frac{\partial p}{\partial t} = - \sum_{\nu=1}^{2} \partial_{\nu} J_{\nu} \; ,
\end{equation}
where:
\begin{equation}
    \partial_{\nu} = \begin{pmatrix} \bm{\nabla}_{\bm{r}} \\ \partial_{\vartheta} \end{pmatrix}
\end{equation}
and
\begin{equation}
    J_{\nu} = \begin{pmatrix} \bm{J}_{\bm{r}} \\ J_{\vartheta} \end{pmatrix} \; .
\end{equation}
We conclude by pointing out that the propagator for the ABP $p(\bm{w},t|\bm{w}_{0},0)$ is a solution to the Fokker-Planck equation for the system.

\subsection{Modified propagator for an ABP}

The next step in the definition of a committor function relies on the knowledge of the propagator for the system, and in particular on the definition of a modified propagator for the problem.

As we derived in section \ref{sec:stoc_PI_abp} of this thesis, the propagator for an active Brownian particle can be written in the form of a stochastic path integral over all possible trajectories linking the initial state $\bm{w}_{0}$ to the final one $\bm{w}_{\text{f}}$ and lasting a time $t$ as:
\begin{equation}
\label{eq:PI_ABP3}
p(\bm{w}_{\text{f}},t|\bm{w}_{0},0) = \mathcal{Z}^{-1} \int \mathcal{D} \bm{r} \int \mathcal{D} \vartheta e^{-\frac{1}{4D_{\theta}} S_{\text{rot}}[\vartheta]} e^{-\frac{1}{4D} S_{\text{trans}}[r,\vartheta]} \; ,
\end{equation}
where $\bm{w} = (\bm{r},\vartheta)$ and:
\begin{equation}
S_{\text{rot}}[\vartheta] = \int_{0}^{t} d\tau \big[ \dot{\vartheta}(\tau) \big]^{2} \; ,
\end{equation}
and:
\begin{equation}
S_{\text{trans}}[\bm{r},\vartheta] = \int_{0}^{t}d\tau \big[ \dot{\bm{r}}(t) -v \bm{u}(t) +\mu \bm{\nabla}U(\bm{r}(t)) \big]^{2} \; .
\end{equation}

Now, similarly to what we presented in section \ref{sec:comm_func_pass}, one needs the definition of a modified propagator for the system to obtain the flux of the probability current across the boundary $\partial \text{T}$ of the target region.
Once more, this can be easily obtained starting from Eq.~\ref{eq:PI_ABP3} through the introduction of a characteristic function $\Omega_{W}(\bm{r},\vartheta)$.
This function is again 0 if the particle is outside $\text{W} = \text{R} \cup \text{T}$ and very large inside W, thus suppressing the contributions to the path integral coming from trajectories that reach the boundaries of R or T in going from $\bm{w}_{0}$ to $\bm{w}_{\text{f}}$ in a time $t$.
Using this characteristic function, the probability of going from $\bm{w}_{0}$ to $\bm{w}_{\text{f}}$ in a time $t$ through any path that doesn't reach the boundary of W can be obtained as:
\begin{equation}
    p_{\partial \text{W}}^{*}(\bm{w}_{\text{f}},t|\bm{w}_{0},0) = \mathcal{Z}^{-1} \int \mathcal{D}\bm{r} \int\mathcal{D}\vartheta\, e^{-\frac{1}{4D_{\vartheta}}S_{rot}[\vartheta] - \frac{1}{4D} S_{trans}[\bm{r},\vartheta] - \int_{0}^{t}d\tau \Omega_{\text{W}}[\bm{r}(\tau),\vartheta(\tau)]} \; .
\end{equation}
Subsequently, again by imposing $\Omega_{\text{W}}(\bm{r}_{\text{f}},\vartheta_{\text{f}})p_{\partial \text{W}}^{*}(\bm{w}_{\text{f}},t|\bm{w}_{0},0) = 0$ and since $p(\bm{w}_{\text{f}},t|\bm{w}_{0},0)$ and $p_{\partial \text{W}}^{*}(\bm{w}_{\text{f}},t|\bm{w}_{0},0)$ differ only from the boundary conditions imposed by $\Omega_{\text{W}}(\bm{r},\vartheta)$, which does not explicitly depend on time, they are solution of the same FP equation, thus:
\begin{equation}
    \frac{\partial p_{\partial \text{W}}^{*}}{\partial t} = - \bm{\nabla}_{\bm{r}} \bigg[ \big( v\, \bm{u}\, - D\, \beta\, \bm{\nabla}_{\bm{r}}U(\bm{r}) \big) p_{\partial \text{W}}^{*} \bigg]\, + D\, \nabla_{\bm{r}}^{2} p_{\partial \text{W}}^{*}\, + D_{\vartheta}\, \partial_{\vartheta}^{2} p_{\partial \text{W}}^{*}\;.
\end{equation}
Once again, this equation can be rewritten as a continuity equation, that reads:
\begin{equation}
    \frac{\partial p_{\partial \text{W}}^{*}}{\partial t} = - \bm{\nabla}_{\bm{r}} \cdot \bm{J}_{\partial \text{W},\bm{r}}^{*} - \partial_{\vartheta} J_{\partial \text{W},\vartheta}^{*} \; ,
\end{equation}
and consequently:
\begin{equation}
    \frac{\partial p_{\partial \text{W}}^{*}}{\partial t} = - \sum_{\nu=1}^{2} \partial_{\nu} J_{\partial \text{W},\nu}^{*} \; ,
\end{equation}
where:
\begin{equation}
    J_{\partial \text{W},\nu}^{*} = \begin{pmatrix} \bm{J}_{\partial \text{W},\bm{r}}^{*} \\ J_{\partial \text{W},\vartheta}^{*} \end{pmatrix} \; ,
\end{equation}
\begin{equation}
    \bm{J}_{\partial \text{W},\bm{r}}^{*} = \big[v \bm{u} - D \bm{\nabla}_{\bm{r}} - D \beta \bm{\nabla}_{\bm{r}} U(\bm{r})\big]p_{\partial \text{W}}^{*}\; ,
\end{equation}
and:
\begin{equation}
    J_{\partial \text{W},\vartheta}^{*} = - D_{\vartheta} \partial_{\vartheta}p_{\partial \text{W}}^{*}\; .
\end{equation}

\subsection{Committor function for ABPs}
Finally, in an analogous fashion to the passive case, the committor is obtained by integrating the modified probability current $J_{\partial \text{W},\nu}^{*}$ over the boundary of region T, so to retain the contribution of the current associated with the trajectories reaching $\partial \text{T}$ while still suppressing the contribution of those reaching $\partial \text{R}$, and over time, so to account for the probability of reaching region T at any time.
Therefore, the committor for an ABP reads:
\begin{equation}
    q(\bm{r},\vartheta) = - \int_{0}^{\infty} dt \int_{\partial \text{T}} d\Sigma'_{\nu} \cdot J^{*}_{\partial \text{W},\nu} (\bm{r}'\!,\!\vartheta'\!,\!t|\bm{r}\!,\!\vartheta\!,\!0) \; ,
\end{equation}
where:
\begin{equation}
\label{eq:boundary_int}
    \int_{\partial \text{T}} d\Sigma'_{\nu} = \int_{\partial \text{T}_{\bm{r}}} d\bm{\sigma}' \int_{-\pi}^{\pi} d\vartheta' \; .
\end{equation}
Note that this time, differently from the passive case where the orientation of the particle doesn't play any role in determining the outcome of the target-search, in the case of the ABP the angular variable has to be taken into account, so in general the committor will be a function of the position of the particle and of the self-propulsion orientation.
This implies that the integral of the current over the boundary hyper-surface of the target region has to be carried out in a space with higher dimensionality than the passive one, as becomes clear from Eq.~\ref{eq:boundary_int}.
In all the cases we considered in this study, however, the characterization of the target region relies on a definition of T in Cartesian space (so only for the radial coordinate of the problem), with no constraint imposed on the angle.
Therefore the angular integration is performed in the interval $[-\pi,\pi]$, while the integration of the infinitesimal surface element $d\bm{\sigma}'$ on the radial hyper-surface $\partial \text{T}_{\bm{r}}$ is performed by imposing the spatial definition of region T.

\subsection{Backward Kolmogorov equation for ABPs}
In the previous section, we derived an analytic expression for the committor function.
The problem now becomes how to practically compute the values for this function.
In principle, one could trivially obtain the committor for a point $\bm{w} = (\bm{r},\vartheta)$ in configuration space by simulating $N$ trajectories starting from $\bm{w}$ with direct integration of the equations of motion, and by then checking how many $N_{\text{T}}$ of these trajectories reach T before visiting R and computing the committor for the point as the ratio $N_{\text{T}}/N$.
However, this method rapidly becomes unpractical as the number of configuration-space points for which a committor is required increases.

One can instead use the information from the transition path ensemble to derive the values of the committor only in the relevant fraction of the transition region in a more efficient way: one can use the procedure outlined above of running $N$ trajectories for each point, but, this time, starting only from the points visited by reactive paths, thus reducing the computational cost of the procedure~\cite{Dellago2002}.

However, in the case of low-dimensional systems, such as the ABP model, one can use a different procedure~\cite{Metzner2006} to obtain the committor without the need of simulating trial trajectories.
To this end, we exploit one of the properties of the committor function, namely the fact that it satisfies the Backward Kolmogorov equation, to compute it more efficiently in a larger set of configuration-space points that constitutes the relevant part of the transition region.

Here we limit ourselves to discussing how the Backward Kolmogorov equation can be derived for the ABP model, while the numeric implementation will be presented in the following sections.

Similarly to the case of a passive particle (see Appendix \ref{app:BK}), also in the case of an ABP the committor satisfies the Backward Kolmogorov equation.
The only requirements for this statement to hold are the Markovianity of the system and the fact that the system must obey the Forward Kolmogorov equation (Fokker-Planck equation), which are both verified in the case of an ABP.

To prove this statement in the active case, we start by deriving the adjoint of the Fokker-Planck operator of an ABP (Eq.~\ref{eq:fpoabp}), which is also called the Backward Kolmogorov operator.
The Backward Kolmogorov operator can be obtained through the definition of adjoint operator using the standard scalar product:
\begin{equation}
    \langle \hat{H}_{\text{FP}} f(\bm{w}),\, g(\bm{w}) \rangle = \langle f(\bm{w}),\, \hat{H}^{\dag}_{\text{FP}} g(\bm{w}) \rangle \, ,
\end{equation}
where, once again, $\bm{w} = (\bm{r},\vartheta)$.

From this definition then follows:
\begin{align*}
    & \langle \hat{H}_{\text{FP}} f(\bm{w}),\, g(\bm{w}) \rangle = - \int d\bm{w} g(\bm{w}) \cdot \Bigg( D \bm{\nabla}_{\bm{r}} \cdot \Bigg[ \bm{\nabla}_{\bm{r}} - \frac{v \bm{u}}{D} + \beta \bm{\nabla}_{\bm{r}} U (\bm{r}) \Bigg] + D_{\vartheta} \partial_{\vartheta}^{2} \Bigg) f(\bm{w}) = \\
    & = - \int d\bm{w} g(\bm{w}) \cdot D \bm{\nabla}_{\bm{r}} \Bigg[ \bm{\nabla}_{\bm{r}} f(\bm{w}) - \bigg( \frac{v \bm{u}}{D} - \beta \bm{\nabla}_{\bm{r}} U (\bm{r}) \bigg) f(\bm{w}) \Bigg] - \int d\bm{w} g(\bm{w}) D_{\vartheta} \partial_{\vartheta}^{2} f(\bm{w}) = \\
    & = - \int d\vartheta \int d\bm{r} g(\bm{r},\vartheta) \cdot D \bm{\nabla}_{\bm{r}} \Bigg[ \bm{\nabla}_{\bm{r}} f(\bm{r},\vartheta) - \bigg( \frac{v \bm{u}}{D} - \beta \bm{\nabla}_{\bm{r}} U (\bm{r}) \bigg) f(\bm{r},\vartheta) \Bigg] + \\
    & - \int d\vartheta \int d\bm{r} g(\bm{r},\vartheta) D_{\vartheta} \partial_{\vartheta}^{2} f(\bm{r},\vartheta)
\end{align*}
Then, by assuming that $\bm{r}$ and $\vartheta$ are independent variables, one can integrate by parts getting:
\begin{align*}
& - D \int d\vartheta \int d\bm{r} g(\bm{r},\vartheta)\nabla_{\bm{r}}^{2} f(\bm{r},\vartheta) + D \int d\vartheta \int d\bm{r} g(\bm{r},\vartheta) \bm{\nabla}_{\bm{r}} \Bigg[ \bigg( \frac{v \bm{u}}{D} - \beta \bm{\nabla}_{\bm{r}} U (\bm{r}) \bigg) f(\bm{r},\vartheta) \Bigg] + \\
& - D_{\vartheta} \int d\bm{r} \int d\vartheta g(\bm{r},\vartheta) \partial_{\vartheta}^{2} f(\bm{r},\vartheta) = \\
& = - D \int d\vartheta \Bigg( \cancel{\bigg[ g(\bm{r},\vartheta) \bm{\nabla}_{\bm{r}} f(\bm{r},\vartheta) \bigg]} - \int d\bm{r} \bm{\nabla}_{\bm{r}}g(\bm{r},\vartheta) \bm{\nabla}_{\bm{r}} f(\bm{r},\vartheta)\Bigg) + \\
& + D \int d\vartheta \Bigg( \cancel{\bigg[ g(\bm{r},\vartheta) \bigg( \frac{v \bm{u}}{D} - \beta \bm{\nabla}_{\bm{r}} U (\bm{r}) \bigg) f(\bm{r},\vartheta) \bigg]} - \int d\bm{r} f(\bm{r},\vartheta) \bigg( \frac{v \bm{u}}{D} - \beta \bm{\nabla}_{\bm{r}} U (\bm{r}) \bigg) \bm{\nabla}_{\bm{r}} g(\bm{r},\vartheta) \Bigg) + \\
& - D_{\vartheta} \int d\bm{r} \Bigg( \cancel{\bigg[ g(\bm{r},\vartheta) \partial_{\vartheta} f(\bm{r},\vartheta) \bigg]} - \int d\vartheta \partial_{\vartheta}g(\bm{r},\vartheta) \partial_{\vartheta} f(\bm{r},\vartheta)\Bigg) = \\
& = + D \int d\vartheta \Bigg( \cancel{\bigg[ \bm{\nabla}_{\bm{r}} g(\bm{r},\vartheta) f(\bm{r},\vartheta) \bigg]} - \int d\bm{r} f(\bm{r},\vartheta) \nabla_{\bm{r}}^{2}g(\bm{r},\vartheta) \Bigg) + \\
& - D \int d\vartheta \int d\bm{r} f(\bm{r},\vartheta) \bigg( \frac{v \bm{u}}{D} - \beta \bm{\nabla}_{\bm{r}} U (\bm{r}) \bigg) \bm{\nabla}_{\bm{r}} g(\bm{r},\vartheta) + \\
& + D_{\vartheta} \int d\bm{r} \Bigg( \cancel{\bigg[ \partial_{\vartheta} g(\bm{r},\vartheta) f(\bm{r},\vartheta) \bigg]} - \int d\vartheta f(\bm{r},\vartheta) \partial_{\vartheta}^{2}g(\bm{r},\vartheta) \Bigg) = \\
& = - D \int d\vartheta \int d\bm{r} f(\bm{r},\vartheta) \nabla_{\bm{r}}^{2}g(\bm{r},\vartheta) - D \int d\vartheta \int d\bm{r} f(\bm{r},\vartheta) \bigg( \frac{v \bm{u}}{D} - \beta \bm{\nabla}_{\bm{r}} U (\bm{r}) \bigg) \bm{\nabla}_{\bm{r}} g(\bm{r},\vartheta) + \\
& - D_{\vartheta} \int d\bm{r} \int d\vartheta f(\bm{r},\vartheta) \partial_{\vartheta}^{2}g(\bm{r},\vartheta) = \\
& = - \int d\vartheta \int d\bm{r} f(\bm{r},\vartheta) \bigg( v \bm{u} - D \beta \bm{\nabla}_{\bm{r}} U (\bm{r}) \bigg) \bm{\nabla}_{\bm{r}} g(\bm{r},\vartheta) + f(\bm{r},\vartheta) D \nabla_{\bm{r}}^{2}g(\bm{r},\vartheta) + f(\bm{r},\vartheta) D_{\vartheta} \partial_{\vartheta}^{2}g(\bm{r},\vartheta) = \\
& = - \int d\vartheta \int d\bm{r} f(\bm{r},\vartheta) \cdot \Bigg[ \bigg( v \bm{u} - D \beta \bm{\nabla}_{\bm{r}} U (\bm{r}) \bigg) \bm{\nabla}_{\bm{r}} g(\bm{r},\vartheta) + D \nabla_{\bm{r}}^{2}g(\bm{r},\vartheta) + D_{\vartheta} \partial_{\vartheta}^{2}g(\bm{r},\vartheta) \Bigg] = \\
& = - \int d\bm{w} f(\bm{w}) \cdot \Bigg[ \bigg( v \bm{u} - D \beta \bm{\nabla}_{\bm{r}} U (\bm{r}) \bigg) \bm{\nabla}_{\bm{r}} g(\bm{w}) + D \nabla_{\bm{r}}^{2}g(\bm{w}) + D_{\vartheta} \partial_{\vartheta}^{2}g(\bm{w}) \Bigg] = \\
& = - \int d\bm{w} f(\bm{w}) \cdot \Bigg[ D \nabla_{\bm{r}}^{2} + \bigg( v \bm{u} - D \beta \bm{\nabla}_{\bm{r}} U (\bm{r}) \bigg) \bm{\nabla}_{\bm{r}} + D_{\vartheta} \partial_{\vartheta}^{2} \Bigg] g(\bm{w}) = \langle f(\bm{w}) ,\, \hat{H}^{\dag}_{\text{FP}} g(\bm{w}) \rangle \numberthis \label{eq:end_of_block}
\end{align*}
where the surface terms cancel assuming that $f(\bm{r},\vartheta)$ and $g(\bm{r},\vartheta)$ tend to $0$ sufficiently fast for $r \rightarrow \infty$ and they are periodic with period $2\pi$.
The active Backward Kolmogorov operator then reads:
\begin{equation}
\hat{H}^{\dag}_{\text{FP}} = - D \nabla_{\bm{r}}^{2} - \bigg( v \bm{u} - D \beta \bm{\nabla}_{\bm{r}} U (\bm{r}) \bigg) \bm{\nabla}_{\bm{r}} - D_{\vartheta} \partial_{\vartheta}^{2} \; .
\end{equation}

The active Backward Kolmogorov operator can then be used to obtain the Backward Kolmogorov equation for the active system, which is satisfied by the active committor provided that $q(\bm{x}_{i})|_{\partial \text{R}}\!=\!0$ and $q(\bm{x}_{i})|_{\partial \text{T}}\!=\!1$:
\begin{equation}
\label{eq:bk_eq_abp}
\hat{H}^{\dag}_{\text{FP}} \, q(\bm{r},\vartheta) = 0
\end{equation}

The proof of this is analogous to the passive case (see Appendix \ref{app:BK}):
\begin{equation}
\begin{aligned}
    & \hat{H}^{\dag}_{\text{FP}} \, q(\bm{r},\vartheta) = - \hat{H}^{\dag}_{\text{FP}} \int_{0}^{\infty} dt \int_{\partial \text{T}} d\Sigma'_{\nu} \cdot J^{*}_{\partial \text{W},\nu}(\bm{r}'\!,\!\vartheta'\!,\!t|\bm{r}\!,\!\vartheta\!,\!0) = \\
    & = \hat{H}^{\dag}_{\text{FP}} \int_{0}^{\infty} dt \int_{\partial \text{T}} d\Sigma'_{\nu} \cdot \bigg[ D \bm{\nabla}_{\bm{r}'} -  v \bm{u}' + D \beta \bm{\nabla}_{\bm{r}'} U (\bm{r}') + D_{\vartheta'} \partial_{\vartheta'} \bigg] p_{\partial \text{W}}^{*}(\bm{r}'\!,\!\vartheta'\!,\!t|\bm{r}\!,\!\vartheta\!,\!0) = \\
    & = \int_{0}^{\infty} dt \int_{\partial \text{T}} d\Sigma'_{\nu} \cdot \bigg[ D \bm{\nabla}_{\bm{r}'} -  v \bm{u}' + D \beta \bm{\nabla}_{\bm{r}'} U (\bm{r}') + D_{\vartheta'} \partial_{\vartheta'} \bigg] \hat{H}^{\dag}_{\text{FP}} p_{\partial \text{W}}^{*}(\bm{r}'\!,\!\vartheta'\!,\!t|\bm{r}\!,\!\vartheta\!,\!0)
    \label{eq:cabpint}
\end{aligned}
\end{equation}

Now, since our system is Markovian, it satisfies the forward Kolmogorov equation, and the probability in the system is conserved, the distribution $p_{\partial \text{W}}^{*}$ satisfies also the backward Kolmogorov equation, namely:
\begin{equation}
    -\partial_{t_{0}} p_{\partial \text{W}}^{*} (\bm{w}_{\text{f}},t|\bm{w}_{0},t_{0}) = - \hat{H}^{\dag}_{\text{FP}} p_{\partial \text{W}}^{*} (\bm{w}_{\text{f}},t|\bm{w}_{0},t_{0}) \; ,
\end{equation}
which, assuming time homogeneity, can be rewritten as:
\begin{equation}
    \partial_{t} p_{\partial \text{W}}^{*} (\bm{w}_{\text{f}},t|\bm{w}_{0},t_{0}) = - \hat{H}^{\dag}_{\text{FP}} p_{\partial \text{W}}^{*} (\bm{w}_{\text{f}},t|\bm{w}_{0},t_{0}) \; .
\end{equation}
Therefore, from Eq.~\ref{eq:cabpint} one gets:
\begin{equation}
\begin{aligned}
    \hat{H}^{\dag}_{\text{FP}} \, q(\bm{r},\vartheta) & = - \int_{0}^{\infty} dt \frac{\partial}{\partial t} \int_{\partial \text{T}} d\Sigma'_{\nu} \cdot \bigg[ D \bm{\nabla}_{\bm{r}'} -  v \bm{u}' + D \beta \bm{\nabla}_{\bm{r}'} U (\bm{r}') + D_{\vartheta'} \partial_{\vartheta'} \bigg] p_{\partial \text{W}}^{*}(\bm{r}'\!,\!\vartheta'\!,\!t|\bm{r}\!,\!\vartheta\!,\!0) = \\
    & = \int_{0}^{\infty} dt \frac{\partial}{\partial t} \int_{\partial \text{T}} d\Sigma'_{\nu} \cdot J^{*}_{\partial \text{W},\nu}(\bm{r}'\!,\!\vartheta'\!,\!t|\bm{r}\!,\!\vartheta\!,\!0) = \\
    & = F(t = \infty) - F(t = 0) = 0 \; ,
\end{aligned}
\end{equation}
where we used the definition of the flux of the probability current through the hyper-surface $\partial \text{T}$ associated with the trajectories starting in $\bm{w}$:
\begin{equation}
    F(\partial \text{T}, t ; \bm{w}) = \int_{\partial \text{T}} d\Sigma'_{\nu} \cdot J^{*}_{\partial \text{W},\nu}(\bm{r}',\vartheta',t|\bm{r},\vartheta,0) \; ,
\end{equation}
and the fact that for an ergodic system (such as an ABP) $F(t = \infty) = 0$ (for $t \rightarrow \infty$ the system will have visited all configurations) and $F(t = 0) = 0$ if the initial configuration is at a finite distance from T.

\section{Calculating the committor function for ABPs through simulations}

\subsection{Finite-difference method}
Now that we have an expression for the Backward Kolmogorov equation in the case of active Brownian particles, we can numerically solve it on a grid using a finite-difference method for elliptic partial differential equations (PDEs)~\cite{Press2007}.
We point out that this method becomes inefficient as the dimensionality of the system increases~\cite{Metzner2006}, but for our ABP moving in two dimensions it is computationally much more efficient than computing the committor from brute-force integration of the equations of motion for all the grid points in our system.

\begin{figure}[H]
\centering
\includegraphics[height=80mm]{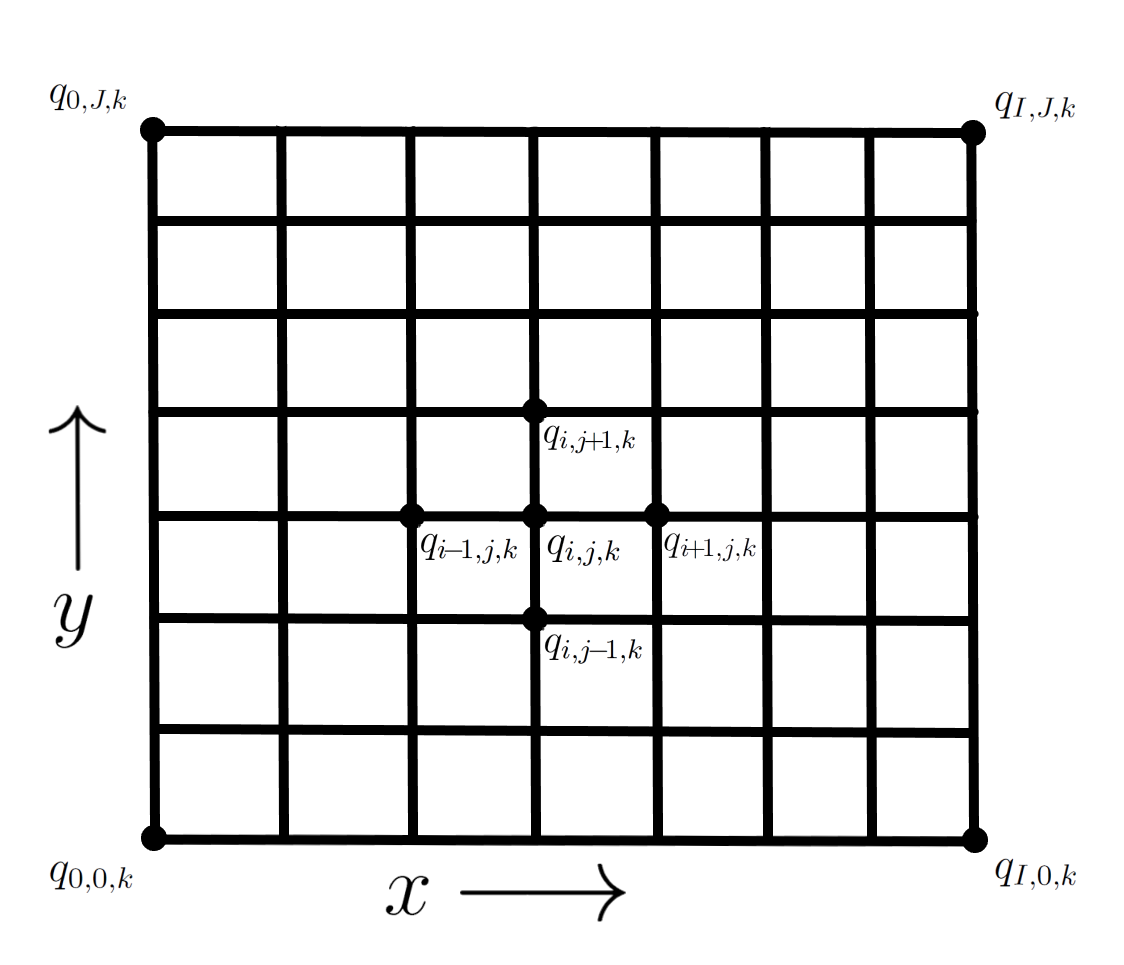}
\caption{Sketch of a two-dimensional slice of the 3D grid obtained by discretizing the system along $x$, $y$, and $\vartheta$, for a fixed value $k$ of the angle $\vartheta$, in the ABP system.}
\label{fig:finitediff_grid}
\end{figure}

The first step in implementing this algorithm relies on discretizing the space where $q$ is evaluated.
In the passive-particle case in 2D this is equivalent to applying a spatial grid along the $x$ and $y$ components of the system, resulting in a two-dimensional grid, while in the ABP case, since the committor is a function of the position and the angle, this results in a three-dimensional grid with $x$, $y$ and $\vartheta$ components.
In the ABP case, the committor values on the grid will be denoted as $q_{i,j,k}$, where $i = 0,...,I$ represents the index of the discretized position along the $x$ component, $j = 0,...,J$ the index along $y$, and $k = 0,...,K$ the index along $\vartheta$.
Then the partial derivatives can be approximated using the neighboring values of the committor on the grid in the finite-difference approach, which in the case of $x$ gives:
\begin{equation}
\frac{\partial^{2} q}{\partial x^{2}} = \frac{q_{i\!+\!1,j,k}-2q_{i,j,k}+q_{i\!-\!1,j,k}}{\Delta x^{2}} \; , \; \frac{\partial q}{\partial x} = \frac{q_{i\!+\!1,j,k}-q_{i\!-\!1,j,k}}{2\Delta x} \; ,
\end{equation}
where $\Delta x$ is the grid spacing along the $x$ component.
Analogous approximations can be obtained for the $y$ and $\vartheta$ derivatives.
A sketch of the grid is reported in Fig.~\ref{fig:finitediff_grid}.

The Backward Kolmogorov equation (Eq.~\ref{eq:bk_eq_abp}) can then be rewritten in Cartesian form as:
\begin{equation}
D \frac{\partial^{2} q}{\partial x^{2}} + D \frac{\partial^{2} q}{\partial y^{2}} + D_{\vartheta} \frac{\partial^{2} q}{\partial \vartheta^{2}} + v \cos \vartheta \frac{\partial q}{\partial x} + v \sin \vartheta \frac{\partial q}{\partial y} - D \beta \nabla_{x} U(x,y) \frac{\partial q}{\partial x} - D \beta \nabla_{y} U(x,y) \frac{\partial q}{\partial y} = 0 \; ,
\end{equation}
where we omitted the dependence of $q$ from $x$, $y$, and $\vartheta$ for simplicity.
This equation can be represented in a more compact form as:
\begin{equation}
D \frac{\partial^{2} q}{\partial x^{2}} + D \frac{\partial^{2} q}{\partial y^{2}} + D_{\vartheta} \frac{\partial^{2} q}{\partial \vartheta^{2}} + a(x,y,\vartheta) \frac{\partial q}{\partial x} + b(x,y,\vartheta) \frac{\partial q}{\partial y} = 0 \; ,
\end{equation}
with $a(x,y,\vartheta) = (v \cos \vartheta - D \beta \nabla_{x} U(x,y))$ and $b(x,y,\vartheta) = (v \sin \vartheta - D \beta \nabla_{y} U(x,y))$.
By using the discretized expressions for the partial derivatives, one then obtains:
\begin{equation}
\begin{aligned}[t]
& D \frac{q_{i\!+\!1,j,k}-2q_{i,j,k}+q_{i\!-\!1,j,k}}{\Delta x^{2}} + D \frac{q_{i,j\!+\!1,k}-2q_{i,j,k}+q_{i,j\!-\!1,k}}{\Delta y^{2}} + D_{\vartheta} \frac{q_{i,j,k\!+\!1}-2q_{i,j,k}+q_{i,j,k\!-\!1}}{\Delta \vartheta^{2}} + \\
& + a(x,y,\vartheta) \frac{q_{i\!+\!1,j,k}-q_{i\!-\!1,j,k}}{2\Delta x} + b(x,y,\vartheta) \frac{q_{i,j\!+\!1,k}-q_{i,j\!-\!1,k}}{2\Delta y} = 0 \; ,
\end{aligned}
\end{equation}
which can be finally rewritten as:
\begin{equation}
\begin{aligned}[t]
\label{eq:final_discrete_PDE}
& q_{i\!+\!1,j,k} \Big( \frac{D}{\Delta x^{2}} + \frac{a(x,y,\vartheta)}{2\Delta x} \Big) + q_{i,j\!+\!1,k} \Big( \frac{D}{\Delta y^{2}} + \frac{b(x,y,\vartheta)}{2\Delta y} \Big) + \\
& + q_{i,j,k\!+\!1} \frac{D_{\vartheta}}{\Delta \vartheta^{2}} - q_{i,j,k} \Big( \frac{2D}{\Delta x^{2}} + \frac{2D}{\Delta y^{2}} + \frac{2D_{\vartheta}}{\Delta \vartheta^{2}} \Big) + q_{i,j,k\!-\!1} \frac{D_{\vartheta}}{\Delta \vartheta^{2}} + \\
& + q_{i,j\!-\!1,k} \Big( \frac{D}{\Delta y^{2}} - \frac{b(x,y,\vartheta)}{2\Delta y} \Big) + q_{i\!-\!1,j,k} \Big( \frac{D}{\Delta x^{2}} - \frac{a(x,y,\vartheta)}{2\Delta x} \Big) = 0 \; .
\end{aligned}
\end{equation}

To solve this discretized PDE on a grid one must provide the required boundary conditions for the problem.
In particular, we impose $q_{i,j,k} = 0$ if $(x_{i},y_{j},\vartheta_{k}) \in R$ and $q_{i,j,k} = 1$ if $(x_{i},y_{j},\vartheta_{k}) \in T$.
Additionally, one needs to impose the correct boundary conditions at the edges of the grid.
This can be done straightforwardly for the angular component by imposing periodic boundary conditions, but the same cannot be done for the spatial components.
The committor, in fact, is a continuous function that assumes discrete values between $0$ and $1$ on the grid depending on how easy it is to reach T before R from that position, so on one side of the grid it might show values close to $0$ and on the opposite side values close to $1$.
In this case, a periodic boundary condition would introduce discontinuities on the boundaries of the grid, leading to an incorrect evaluation of the committor within the grid.
Additionally, the procedure of using Neumann boundary conditions at the edges of the grid that was used elsewhere~\cite{Metzner2006} might prove difficult to implement in the cases of active systems, where the space spanned by an active agent (and therefore the dimension of the grid required to impose a no-flux condition on the boundaries) can become considerably larger than the one of a passive searcher.
Therefore, we decided to obtain the boundary values at the edges of the grid along the $x$ and $y$ components by employing brute-force simulations starting from each of the edges grid points, and consequently estimating the committor as the number of successful trajectories reaching T before R over the total number of trials for that point.

The discretized partial differential equation can then be solved by finding the solution to the linear system:
\begin{equation}
A q = B \; ,
\end{equation}
where $q$ is a one-dimensionalized vector for $q_{i,j,k}$, with length $N = (I+1)(J+1)(K+1)$, $A$ is a matrix with dimension $N \times N$ and B is a one-dimensional vector with length $N$.
The rows of the $A$ matrix and the elements of the $B$ vector are then filled according to the discretized version of the PDE.
In particular, the rows of the matrix $A$ that multiply a value $q_{i,j,k}$ that lies in the inner part of the grid will be made of the coefficients obtained from Eq.~\ref{eq:final_discrete_PDE}, which will be located with a position $n$ in the row so that they multiply the corresponding $q_{n}$ value in the $q$ vector.
At the same time, the element of $B$ corresponding to $q_{i,j,k}$ will be zero.

For example, in the general case of a point $q_{i,j,k}$, the $n^{\text{th}}$ row of A (where $n = i(J+1)(K+1)+j(K+1)+k$) will have:
\begin{itemize}
\item $\Big( \frac{D}{\Delta x^{2}} - \frac{a(x,y,\vartheta)}{2\Delta x} \Big)$ at position $(i-1)(J+1)(K+1)+j(K+1)+k$,
\item $\Big( \frac{D}{\Delta y^{2}} - \frac{b(x,y,\vartheta)}{2\Delta y} \Big)$ at position $i(J+1)(K+1)+(j-1)(K+1)+k$,
\item $\frac{D_{\vartheta}}{\Delta \vartheta^{2}}$ at position $i(J+1)(K+1)+j(K+1)+k-1 = n-1$,
\item $-\Big( \frac{2D}{\Delta x^{2}} + \frac{2D}{\Delta y^{2}} + \frac{2D_{\vartheta}}{\Delta \vartheta^{2}} \Big)$ at position $n$,
\item $\frac{D_{\vartheta}}{\Delta \vartheta^{2}}$ at position $i(J+1)(K+1)+j(K+1)+k+1 = n+1$,
\item $\Big( \frac{D}{\Delta y^{2}} + \frac{b(x,y,\vartheta)}{2\Delta y} \Big)$ at position $i(J+1)(K+1)+(j+1)(K+1)+k$,
\item $\Big( \frac{D}{\Delta x^{2}} + \frac{a(x,y,\vartheta)}{2\Delta x} \Big)$ at position $(i+1)(J+1)(K+1)+j(K+1)+k$,
\item and finally $0$ in element $n$ of $B$.
\end{itemize}

The periodic boundary conditions along the angular component can be implemented in a similar fashion, by replacing the points with indexes $(i,j,K+1)$ with those with indexes $(i,j,0)$ (where $i \neq 0,I$ and $j \neq 0,J$), and vice versa the points with indexes $(i,j,-1)$ will be replaced by $(i,j,K)$.

Finally, the boundary conditions for R, T, and on the edges of the grid along the $x$ and $y$ dimensions can be implemented by filling the $\text{n}^{\text{th}}$ row of $A$ with zeros, except for the $\text{n}^{\text{th}}$ position, which will be $1$.
Correspondingly, the $\text{n}^{\text{th}}$ value of $B$ will be $0$ if the point is in R, $1$ if the point is in T, and $N_{\text{T}}/N$ if the point lies along the $x$ or $y$ edges, where $N$ is the total number of brute force trajectories started from there and $N_{\text{T}}$ is the number of trajectories that reach T before R.

After constructing $A$ and $B$, we solved the linear system using the function \emph{scipy.linalg.solve()} from the \emph{Python 3} library \emph{SciPy}.


\subsection{Committor on a grid for a passive particle}
We proceed now with discussing the committor function for a 2D passive searcher (with equation of motion given by Eq.~\ref{eom_p}) in a double-well potential (see Eq.~\ref{eq:pot_dw_ch1}), which will serve as a comparison with the ABP committor in the same landscape.

The particle parameters are given by $\mu = 0.1$, $D = 0.1$, and the energy scale is provided once again as $k_{B}T = D/\mu$.
The parameters employed for the double-well are $k_{x} x_{0}^{4} = 6.5 k_{B}T$ and $k_{y}x_{0}^{2} = 20 k_{B}T$, so, analogously to the case discussed in section \ref{sec:search_strat_dw}, the typical length scale is found as $L = k_{B}T/F_{\text{max}} \simeq x_{0}/10$.
Finally, the time scale is again given by $\tau = L^{2}/D$.
We defined once again the region R for the process as the set of points with $U(x,y) \leq 2 k_{B}T$ and $x < 0$, and T as the set of points with $U(x,y) \leq 2 k_{B}T$ and $x > 0$.
The square grid used to compute the committor with the finite difference algorithm was imposed from $-30 L$ and $30 L$ along both $x$ and $y$, with $60$ grid points along both dimensions, resulting in a width of the grid points of $\delta = L$ in both directions.
The boundary values of the committor on the grid were estimated from $10^{3}$ trajectories starting from each grid point, simulated using an integration time step of $0.1 \tau$.

\begin{figure}[H]
\centering
\includegraphics[height=100mm]{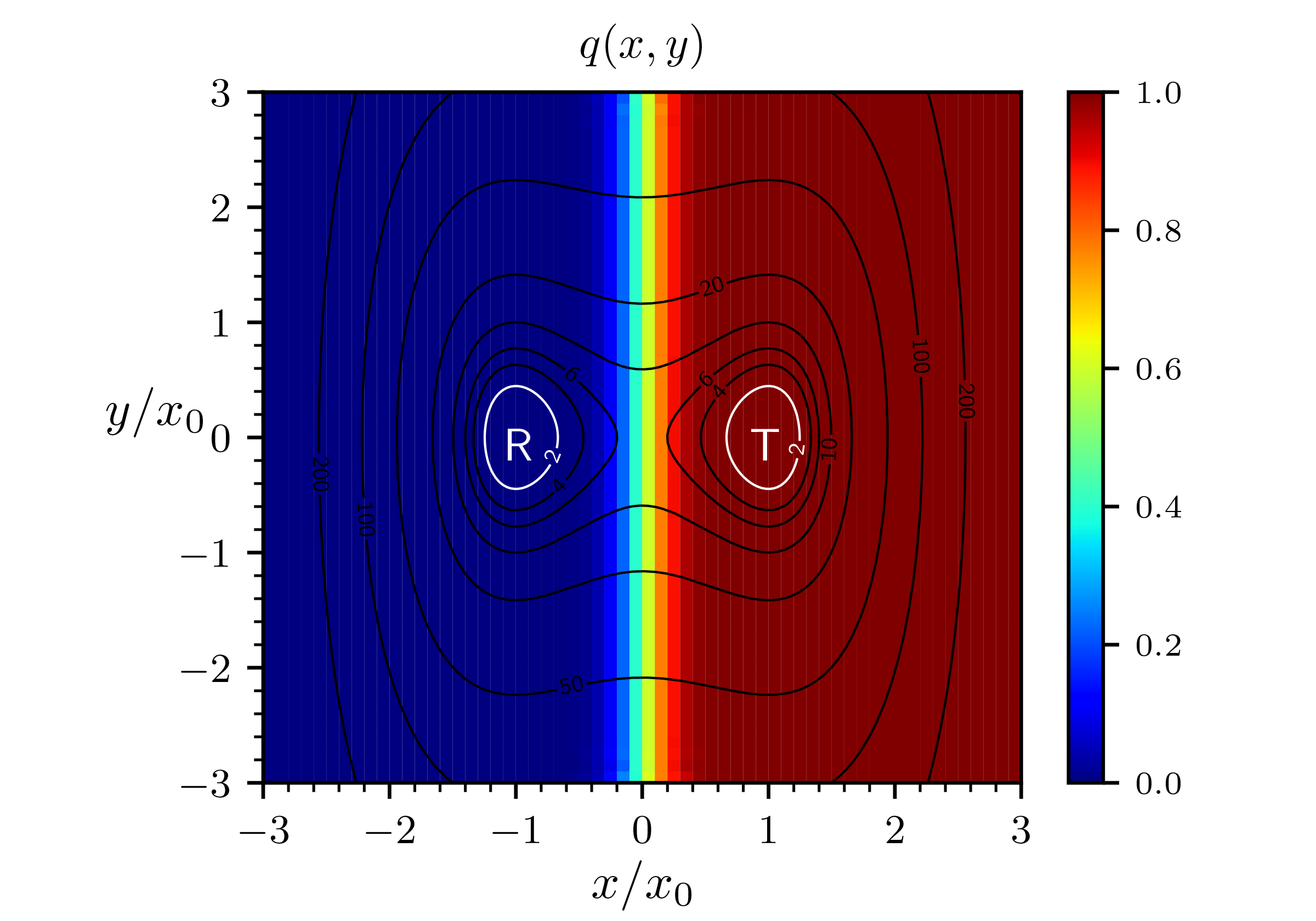}
\caption{Committor for a 2D passive particle in the double-well potential, obtained from the finite difference algorithm with brute-force simulations on the grid boundary. The contour lines represent the underlying energy landscape, with the energies in units of $k_{B}T$ reported for each contour line. The white lines with $U(x,y) = 2 k_{B}T$ represent the boundaries of the R and T regions.}
\label{fig:finitediff_comm_passive}
\end{figure}

The committor for a passive particle in the double-well potential is symmetric with respect to the $x$ axis (see Fig.~\ref{fig:finitediff_comm_passive}), due to the shape of the potential, which, since the translational diffusion doesn't possess any preferential direction, determines the success odds in the case of a passive particle.
The iso-committor surfaces (\textit{i.e.} the hypersurfaces along which the committor is constant) are therefore straight lines parallel to the $y$ axis due to the decoupling of the motion along $x$ and $y$ and the symmetry of the potential along the $y$ direction.
The value of the committor in the passive case is largely determined from which side of the barrier is currently being visited by the passive searcher: if the searcher is on the side closer to the R state, the committor will show values near $0$ due to the external force pushing the particle towards R, while if the searcher is on the side closer to T the committor will be close to $1$.
Trivially, the regions where the committor shows intermediate values are those found on top of the energy barrier, along the line parallel to $y$ and passing through the saddle point located in $(0,0)$.
This line will constitute the \emph{separatrix} of the system (\textit{i.e.} an hypersurface dividing the two states R and T, constituted from all the transition states where the committor has a value of $0.5$~\cite{Dellago2002}).
If the searcher is located on top of the barrier, in fact, the force exerted by the energy landscape on the particle will be zero, and the outcome of the process will be determined exclusively by the subsequent noise terms which will push the searcher downhill either towards R or towards T.

\subsection{Committor on a grid for an ABP}
We now discuss the committor for an active Brownian particle in the same landscape, where we use the same values of the parameters $D$ and $\mu$ employed in the passive particle case.
We will analyze the committor for three different sets of $\text{Pe}$ and $\ell^{*}$, which are defined as in section \ref{sec:nav_strat_model_par}.
The three points discussed in this section represent the behavior of the searcher in the three parameter-space regions A, B, and C, discussed in chapter \ref{ch:nav_strat}.
These points are defined as A: ($\text{Pe} \simeq 9$, $\ell^{*} \simeq 7$), which represent an optimal set of parameters for the target-finding process, B: ($\text{Pe} \simeq 1$, $\ell^{*} \simeq 25$), which has a behavior similar to the passive particle, and C: ($\text{Pe} \simeq 7$, $\ell^{*} \simeq 45$), which shows decreased target-finding rates due to the large value of the persistence\footnote{Note that from here on in the following discussion we use equivalently ‘‘region A'' and ‘‘point A'' to refer to a particle with parameter specified here, and similarly for B and C.}.
To compute the committor in the case of an ABP we use the same grid we defined in the passive particle case, with the addition of a third dimension to account for the angular component.
The grid along the angular component is imposed from $0$ to $2\pi$ with $8$ grid points, with a dimension of each grid slice of $\pi/4$.
The committor at the boundaries of the grid is again obtained from $10^{3}$ brute-force trajectories for each point, with an integration time step of $0.01\tau$, where we decided to reduce the integration time step compared to the passive case to account for the large self-propulsion speed in points A and C.
To have a direct comparison with the committor obtained in the passive case, we first average the grid along the angular component, and then we discuss the dependence of the committor from the angular components separately.

\subsubsection{Point A}

\begin{figure}[H]
\centering
\includegraphics[height=100mm]{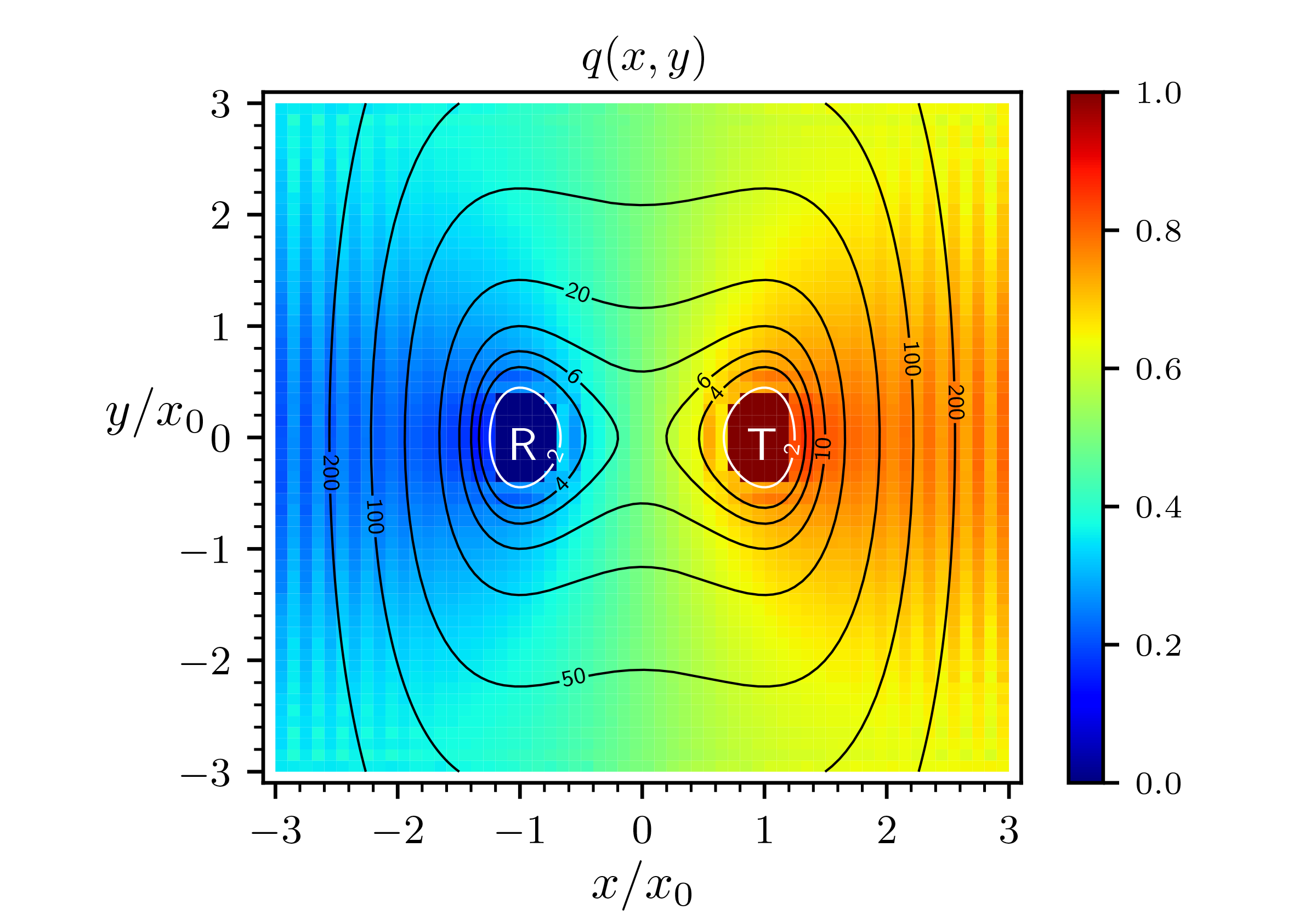}
\caption{Committor for a 2D active Brownian particle with parameters $(\text{Pe},\ell^{*})$ from region A of parameter space, in the double-well potential. The plot is obtained from the finite difference algorithm with brute-force simulations on the grid boundary, averaged along the angular component. The contour lines represent the underlying energy landscape, with the energies in units of $k_{B}T$ reported for each contour line. The white lines with $U(x,y) = 2 k_{B}T$ represent the boundaries of the R and T regions.}
\label{fig:finitediff_comm_point_A}
\end{figure}

\begin{figure}[H]
\centering
\includegraphics[height=100mm]{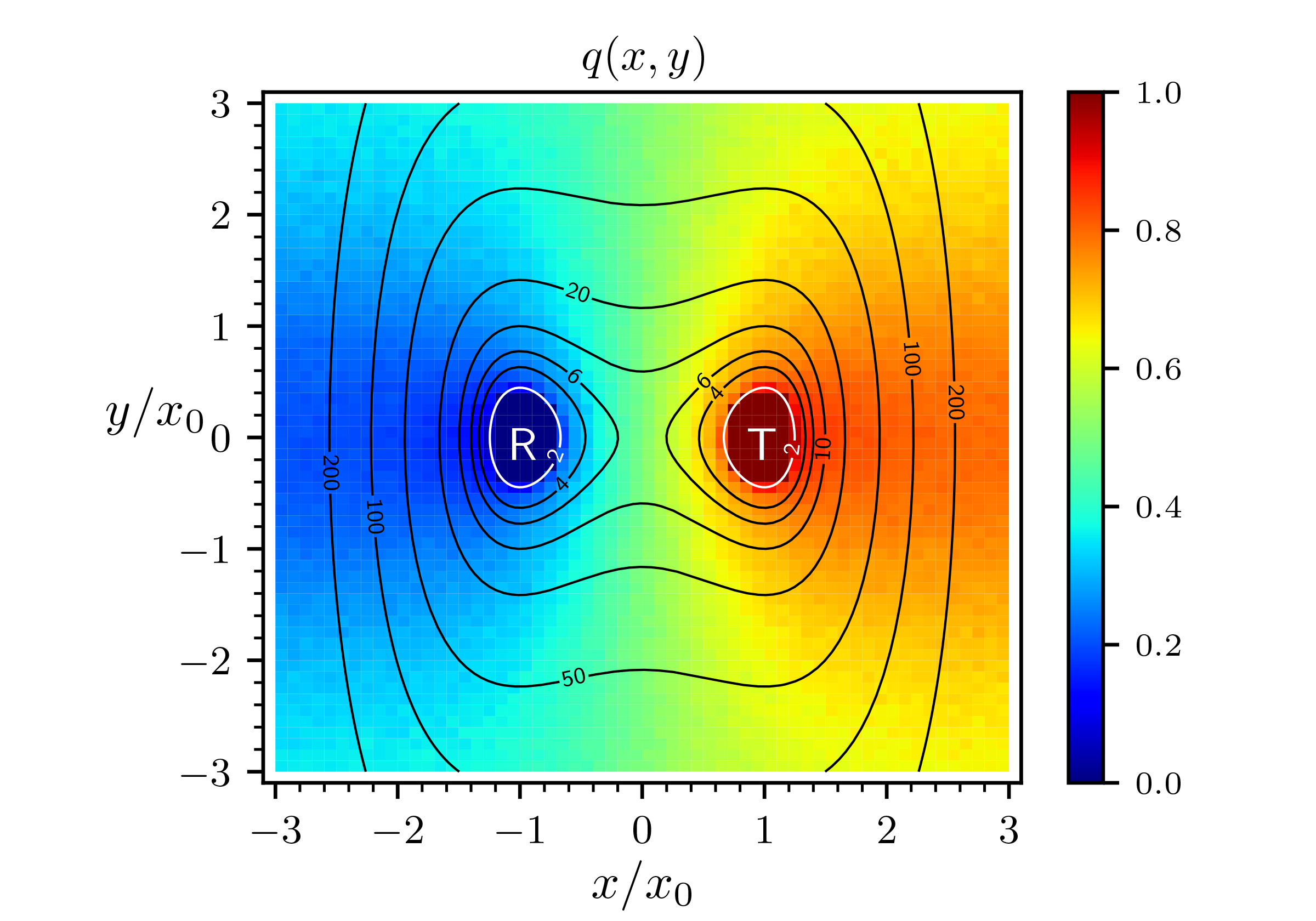}
\caption{Committor as in Fig.~\ref{fig:finitediff_comm_point_A} but with data obtained from $10^{3}$ trajectories for each grid point. The trajectories were generated with a direct integration of the equations of motion.}
\label{fig:bf_comm_point_A}
\end{figure}

The committor for an ABP averaged along the angular component for a point in region A (see Fig.~\ref{fig:finitediff_comm_point_A} for the solution with the finite difference algorithm and Fig.~\ref{fig:bf_comm_point_A} for a validation against brute-force simulations) displays striking differences compared to the committor of a passive particle in the same landscape (Fig.~\ref{fig:finitediff_comm_passive}).
In particular, the iso-committor surfaces projected on the 2D plane are not anymore straight lines but they rather become curved lines.
Additionally, the regions which yield a value of the committor close to $0$ or $1$ become much narrower than in the passive case and are concentrated on the back of the R and T basins.
In contrast, the regions displaying intermediate values of the committor become much broader, occupying most of the transition region $\Omega_{T}$ for the process.
Therefore, regardless of the self-propulsion orientation (which is averaged out), active particles located on the same side of the barrier compared to R have larger odds of successfully finding the target compared to the passive case, thanks to the self-propulsion term which allows an easier climb of the energy barrier.
In turn, an active particle found on the same side of the barrier as the T basin has smaller odds of successfully finding the target compared to a passive particle.
Notwithstanding this differences, the symmetry of the committor with respect to the $x$ axis is preserved and the separatrix of the system is still a straight line parallel to $y$ passing through the saddle point in the origin.

\begin{figure}[H]
\centering
\includegraphics[height=110mm]{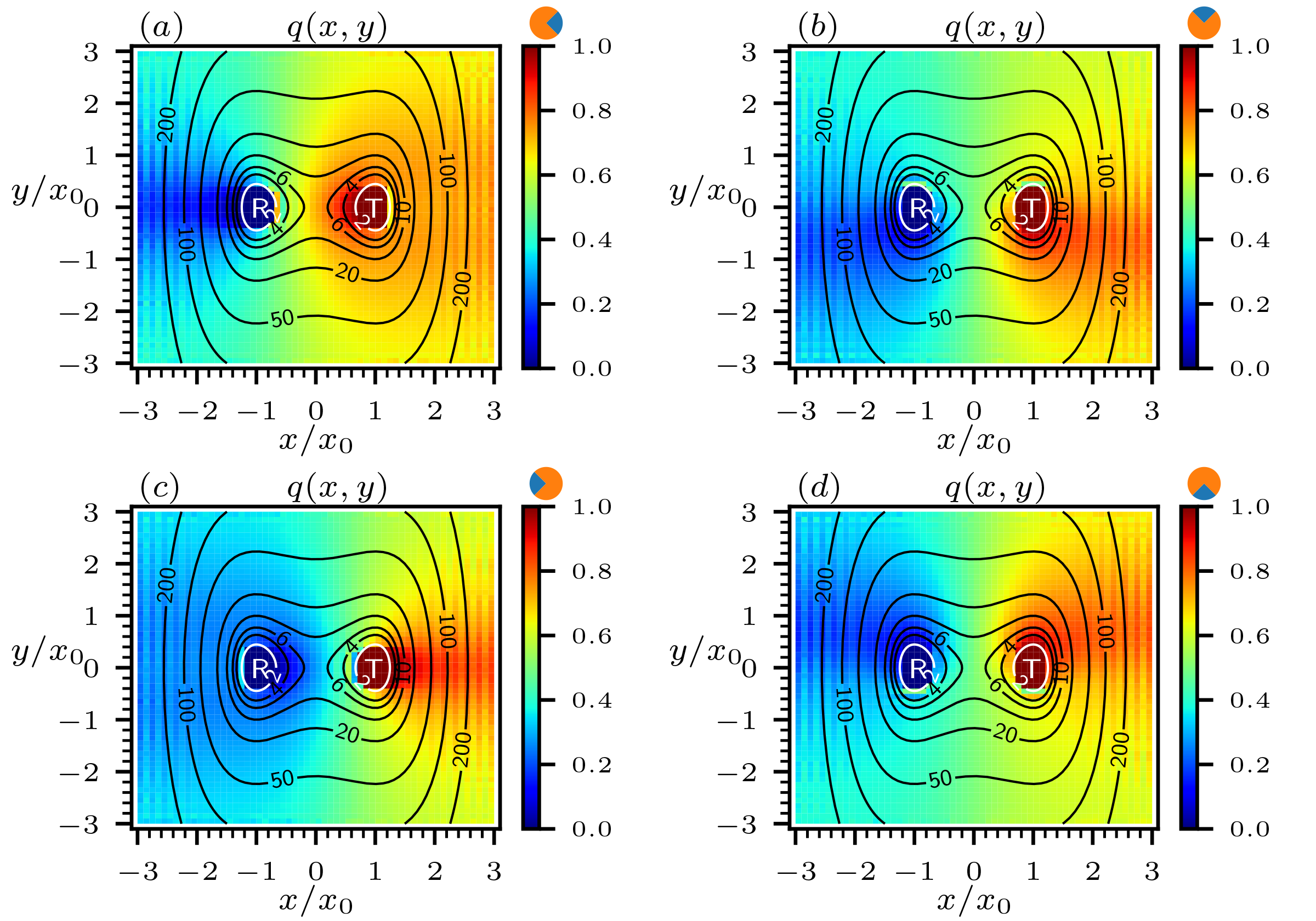}
\caption{Committor for a 2D active Brownian particle with parameters $(\text{Pe},\ell^{*})$ from region A of parameter space, in the double-well potential. The plot is obtained from the finite difference algorithm with brute-force simulations on the grid boundary. The contour lines represent the underlying energy landscape, with the energies in units of $k_{B}T$ reported for each contour line. The white lines with $U(x,y) = 2 k_{B}T$ represent the boundaries of the R and T regions. (a) Committor obtained from an average of the two angular slices located between $-\pi/4\!\leq\!\vartheta\!\leq\!\pi/4$. (b) Committor obtained from an average of the two angular slices located between $\pi/4\!\leq\!\vartheta\!\leq\!3\pi/4$. (c) Committor obtained from an average of the two angular slices located between $3\pi/4\!\leq\!\vartheta\!\leq\!5\pi/4$. (d) Committor obtained from an average of the two angular slices located between $5\pi/4\!\leq\!\vartheta\!\leq\!7\pi/4$. The angular slices over which the average is computed are represented by the blue quadrant in the orange pie plot on top of the bars.}
\label{fig:finitediff_comm_point_A_quarter}
\end{figure}

\begin{figure}[h]
\centering
\includegraphics[height=110mm]{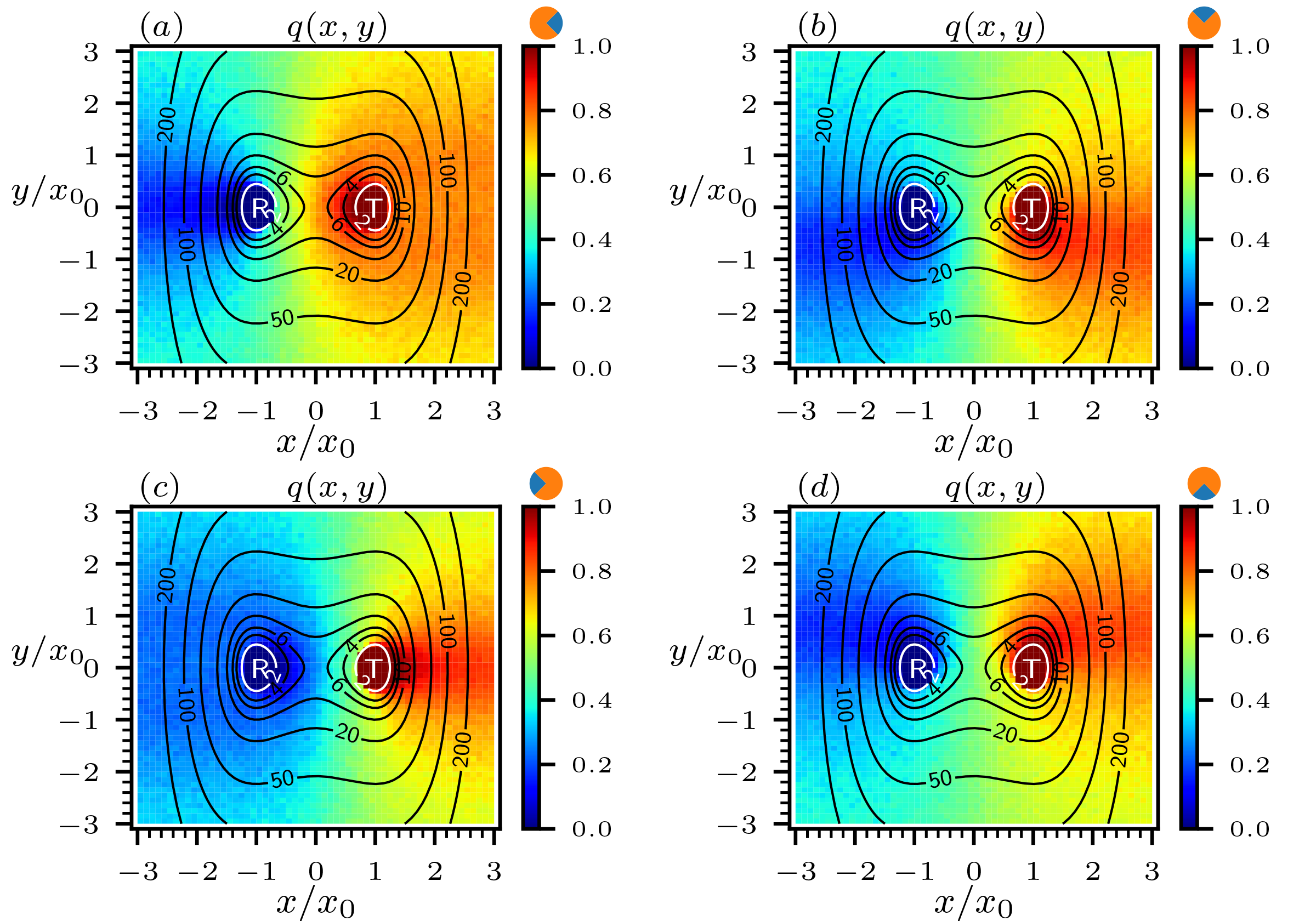}
\caption{Committor as in Fig.~\ref{fig:finitediff_comm_point_A_quarter} but with data obtained from $10^{3}$ trajectories for each grid point. The trajectories were generated with a direct integration of the equations of motion.}
\label{fig:bf_comm_point_A_quarter}
\end{figure}

If the angular information is explicitly taken into account, instead, the committor displays prominent differences depending on the orientation of the self-propulsion speed.
In particular, the largest fraction of transition region displaying high odds of successfully finding the target ($q(x,y) \geq 0.5$) is found for particles with an initial direction along the positive $x$ axis (see Fig.~\ref{fig:finitediff_comm_point_A_quarter}(a) and Fig.~\ref{fig:bf_comm_point_A_quarter}(a) for the brute-force comparison).
In this case, starting from the iso-committor surface passing at the saddle point $(0,0)$ yields high chances of finding the target region T, and, differently to the case with averaged angular component reported in Fig.~\ref{fig:finitediff_comm_point_A}, this iso-committor surface is curved.
We point out that, compared to the figure with averaged angular component and to the passive case, also particles located in the regions above or below the R basin have values of the committor around $0.5$, shifting the separatrix much closer to the R basin than in the other two cases.
Additionally, the committor shows values close to $0$ only if the particle is starting from the left of the R basin, because an initial orientation pointing towards the positive $x$ axis will likely lead the particle to R, and a committor close to $1$ is instead found only if the particle is close to T on the left side.
Finally, note that a particle on the same side of the barrier with respect to the T basin shows lower chances of finding the target T than a passive particle in the same configuration, similarly to the averaged case.
This occurs due to the self-propulsion mechanism, which, thanks due to the rotational diffusion process, may also lead the particle away from T and on the opposite side of the barrier, reaching R more frequently.

In contrast, particles with an initial self-propulsion directed along the negative $x$ axis (see Fig.~\ref{fig:finitediff_comm_point_A_quarter}(c) for the finite-difference solution and Fig.~\ref{fig:bf_comm_point_A_quarter}(c) for brute-force simulations) show the opposite behavior.
This time, the separatrix of the system is shifted much closer to the T region and the particle has overall lower point-wise chances of finding the target T compared to the case presented in Fig.~\ref{fig:finitediff_comm_point_A_quarter}(a).
However, this tendency shows some exceptions, in particular in the regions close to the R and T basins.
For points on the right side of T and on the left side of R, a direction pointing towards negative $x$ values will be more favorable than a direction towards positive $x$, and, in contrast, for the region located between the two basins and including the saddle point of the barrier an orientation pointing towards positive $x$ values will be more favorable to find the target.
Finally, note that even if the separatrix of the system is closer to T in this case compared to the passive case (and consequently also compared to Fig.~\ref{fig:finitediff_comm_point_A_quarter}(a)), the success odds for particles close to R are still larger than in the passive case, and they are again smaller if the particle is closer to T.

We conclude this analysis by inspecting the committor for particles initially directed along the positive $y$ axis (Fig.~\ref{fig:finitediff_comm_point_A_quarter}(b) for the finite-difference solution and Fig.~\ref{fig:bf_comm_point_A_quarter}(b) for the brute-force comparison) and along the negative $y$ axis (see Fig.~\ref{fig:finitediff_comm_point_A_quarter}(d) and Fig.~\ref{fig:bf_comm_point_A_quarter}(d)).
These two cases are symmetric with respect to the $x$ axis due to the symmetry of the landscape and the opposite self-propulsion directions, oriented perpendicularly to the symmetry axis.
In this case, for a direction pointing towards positive (negative) $y$ values, the transition region located above (below) the R and T states shows a much larger fraction of space with intermediate values of the committor compared to the region below (above) the R and T basins.
Additionally, the values of the committor closer to $0$ are found in the left-lower (left-upper) corner of the transition region with respect to R due to the presence of the potential pushing the particle towards lower-energy regions and a self-propulsion speed directed along positive (negative) $y$ values, while for similar reasons the committor values closer to $1$ are found in the lower-right (upper-right) corner with respect to the T state.
Finally, also in these two cases, the iso-committor surfaces projected onto the 2D plane are curved lines.
However, the separatrix for the system is a straight line, which is found again passing at the saddle point.

By combining these information together, we conclude that while the iso-committor hypersurfaces in the passive case are just straight lines perpendicular to the $x$ axis in the 2D plane, in the active case they become general hypersurfaces in 3D space provided by the spatial and angular components, and they acquire concave or convex shapes depending on the angular orientation.

\subsubsection{Point B}

\begin{figure}[H]
\centering
\includegraphics[height=100mm]{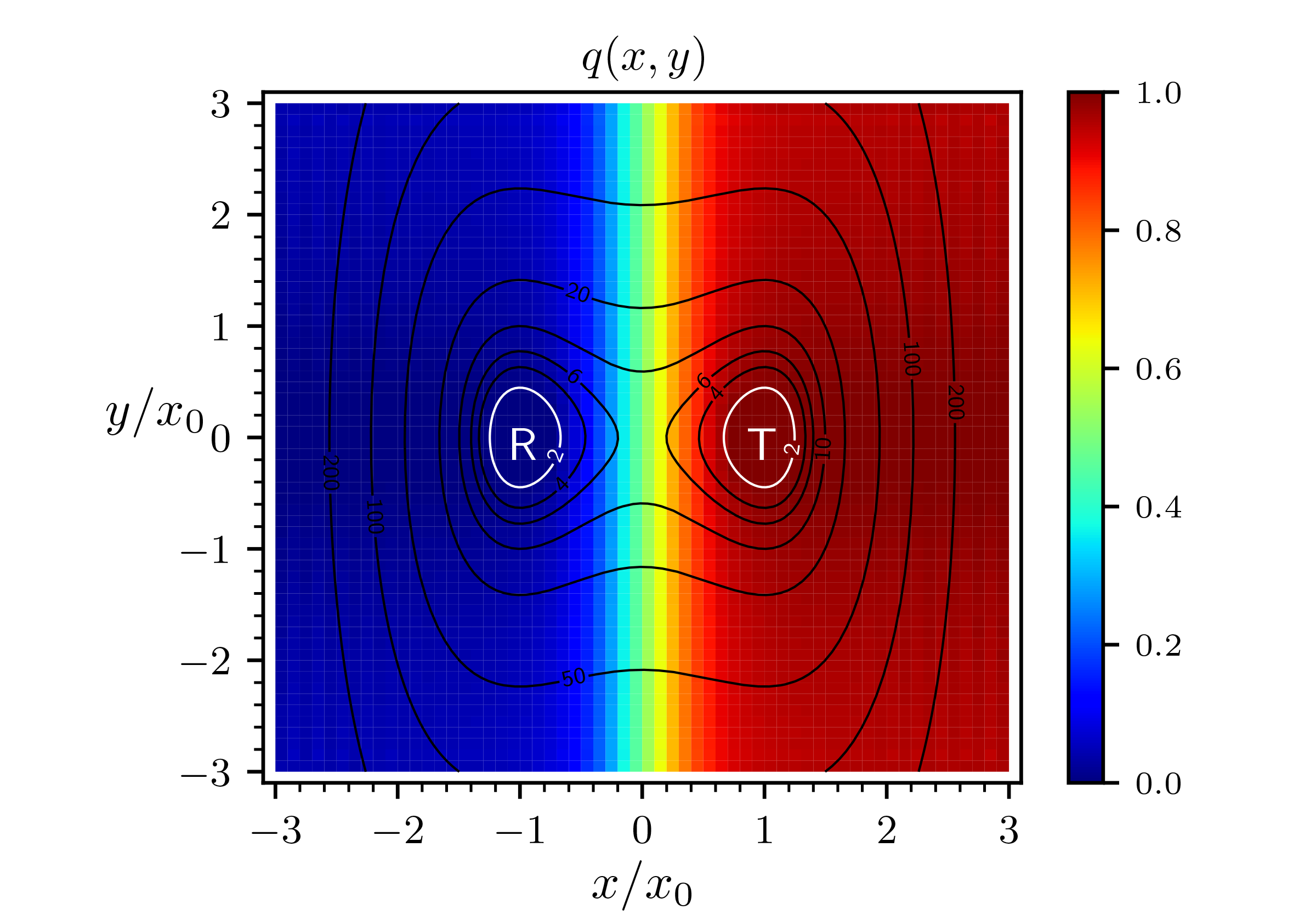}
\caption{Committor for a 2D active Brownian particle with parameters $(\text{Pe},\ell^{*})$ from region B of parameter space, in the double-well potential. The plot is obtained from the finite difference algorithm with brute-force simulations on the grid boundary, averaged along the angular component. The contour lines represent the underlying energy landscape, with the energies in units of $k_{B}T$ reported for each contour line. The white lines with $U(x,y) = 2 k_{B}T$ represent the boundaries of the R and T regions.}
\label{fig:finitediff_comm_point_B}
\end{figure}

\begin{figure}[H]
\centering
\includegraphics[height=100mm]{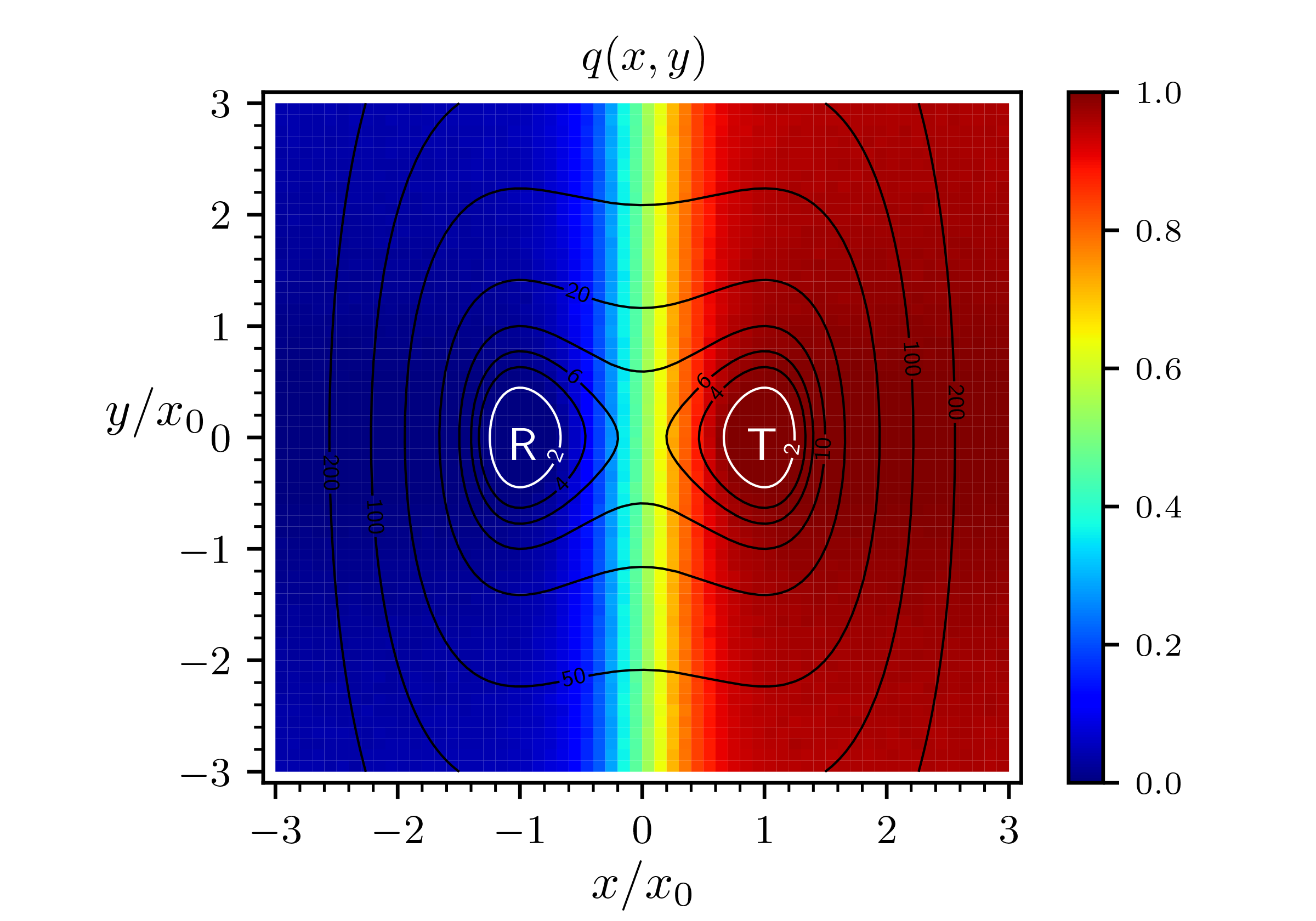}
\caption{Committor as in Fig.~\ref{fig:finitediff_comm_point_B} but with data obtained from $10^{3}$ trajectories for each grid point. The trajectories were generated with a direct integration of the equations of motion.}
\label{fig:bf_comm_point_B}
\end{figure}

Compared to point A (Fig.~\ref{fig:finitediff_comm_point_A}), the committor averaged along the angular component for a point in region B (see Fig.~\ref{fig:finitediff_comm_point_B} and Fig.~\ref{fig:bf_comm_point_B} for the brute-force comparison) is more similar to the committor for a passive particle (Fig.~\ref{fig:finitediff_comm_passive}).
The iso-committor lines, differently from the point in region A, are almost straight lines, and, similarly to the passive case, the regions where the committor is close to $0$ and $1$ are much broader in parameter space than in point A.
The values of the committor about $0.5$ are narrower compared to point A and once again are found around the $x = 0$ axis, which also in this case is the separatrix of the transition.

\begin{figure}[H]
\centering
\includegraphics[height=110mm]{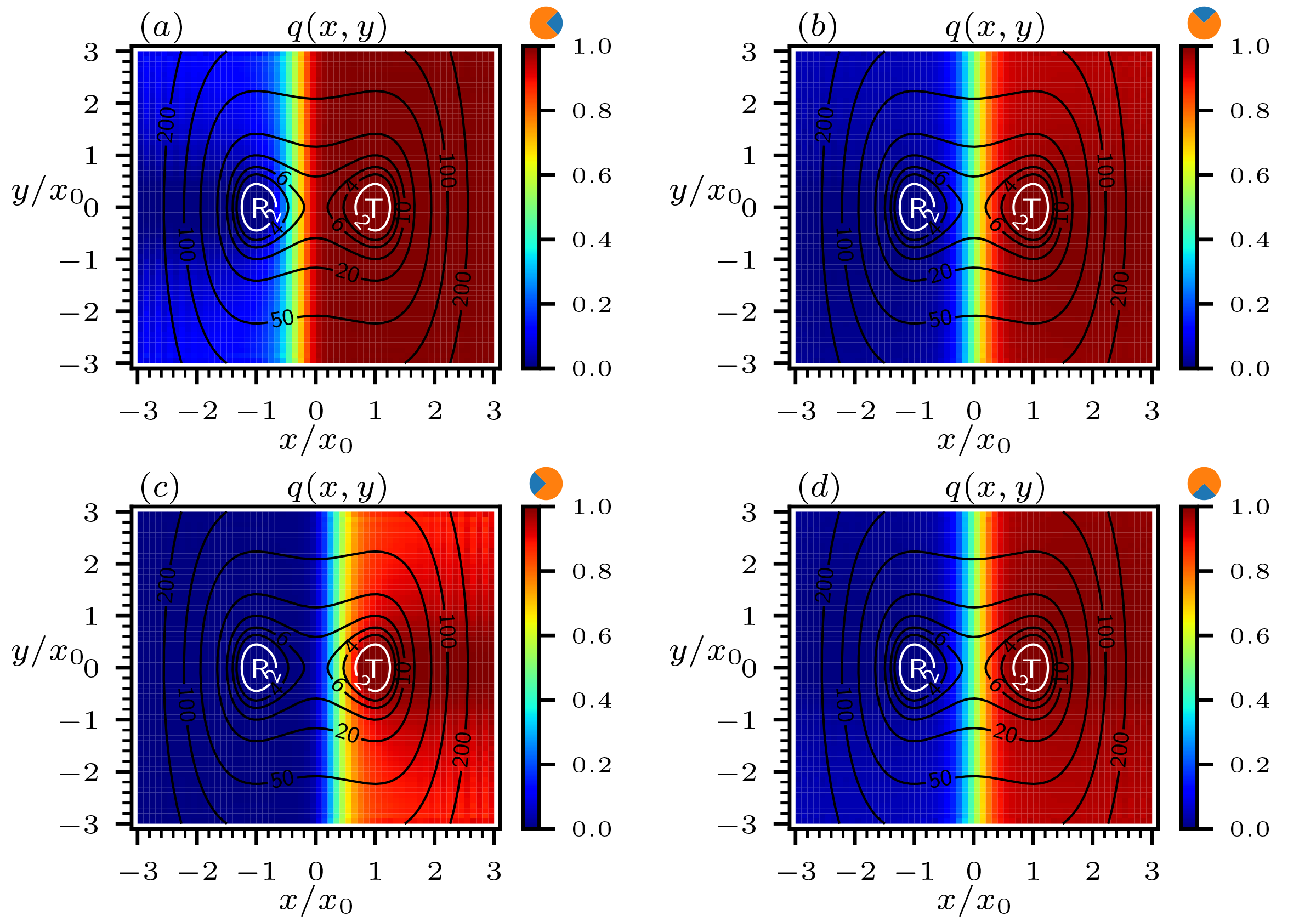}
\caption{Committor for a 2D active Brownian particle with parameters $(\text{Pe},\ell^{*})$ from region B of parameter space, in the double-well potential. The plot is obtained from the finite difference algorithm with brute-force simulations on the grid boundary. The contour lines represent the underlying energy landscape, with the energies in units of $k_{B}T$ reported for each contour line. The white lines with $U(x,y) = 2 k_{B}T$ represent the boundaries of the R and T regions. (a) Committor obtained from an average of the two angular slices located between $-\pi/4\!\leq\!\vartheta\!\leq\!\pi/4$. (b) Committor obtained from an average of the two angular slices located between $\pi/4\!\leq\!\vartheta\!\leq\!3\pi/4$. (c) Committor obtained from an average of the two angular slices located between $3\pi/4\!\leq\!\vartheta\!\leq\!5\pi/4$. (d) Committor obtained from an average of the two angular slices located between $5\pi/4\!\leq\!\vartheta\!\leq\!7\pi/4$. The angular slices over which the average is computed are represented by the blue quadrant in the orange pie plot on top of the bars.}
\label{fig:finitediff_comm_point_B_quarter}
\end{figure}

\begin{figure}[H]
\centering
\includegraphics[height=110mm]{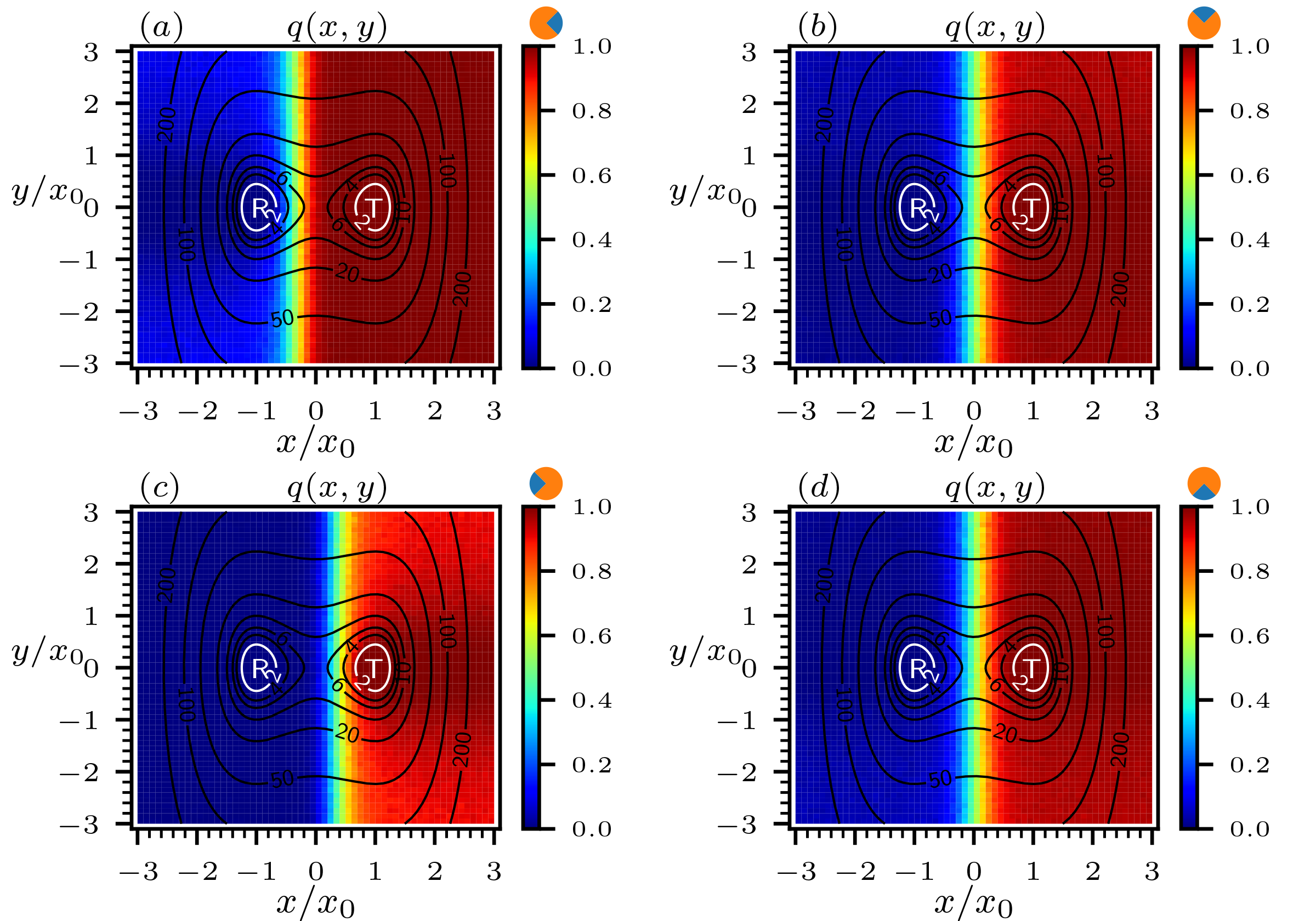}
\caption{Committor as in Fig.~\ref{fig:finitediff_comm_point_B_quarter} but with data obtained from $10^{3}$ trajectories for each grid point. The trajectories were generated with a direct integration of the equations of motion.}
\label{fig:bf_comm_point_B_quarter}
\end{figure}

The small value of the self propulsion found in point B makes the particle behavior quite similar to the one of a passive particle.
However, differences emerge if the contribution of the angular component to the committor are taken into account.
In particular, the committor constrained on some values of the angular orientation for point B (see Fig.~\ref{fig:finitediff_comm_point_B_quarter} and Fig.~\ref{fig:bf_comm_point_B_quarter} for the brute force comparison) resembles a version of point A (Fig.~\ref{fig:finitediff_comm_point_A_quarter}) where the orientation differences are smeared out.

A particle with parameters from region B and self-propulsion orientation directed along the positive $x$ axis (Fig.~\ref{fig:finitediff_comm_point_B_quarter}(a)) shows that, compared to the passive case (Fig.~\ref{fig:finitediff_comm_passive}), the transition is overall more advantageous regardless of the position of the particle.
If the agent is found on top of the barrier or on the same side of the barrier compared to T, an initial orientation pointing towards positive $x$ values always yields a committor close to $1$, similarly to the passive case.
The separatrix of the transition, however, is found closer to R than to T, increasing the success odds also for points on the side of the barrier closer to R.
This occurs thanks to the considered orientation, and the behavior observed is qualitatively similar to the results found for point A.
Additionally, particles located on the same side of the barrier as R but not on the left of it, show committor values larger than those of a passive particle, while only the particles starting in the region on the back of R display committor values similar to those of the passive particle. 

In contrast, if the self-propulsion orientation of the particle is initially directed along the negative $x$ direction, the target finding odds are overall less advantageous compared to a passive particle.
In this case (see Fig.~\ref{fig:finitediff_comm_point_B_quarter}(c)), the active particle shows a separatrix closer to T than to R, again similarly to point A.
Additionally, in this case the values of the committor are overall smaller than those of a passive particle, except on the back of the T basin and in the regions surrounding R where they are comparable.

Finally, the cases in which the self-propulsion orientation is directed along positive (negative) $y$ values, see Fig.~\ref{fig:finitediff_comm_point_B_quarter}(b) (Fig.~\ref{fig:finitediff_comm_point_B_quarter}(d)), the behavior of the committor resembles the one found in region A (Fig.~\ref{fig:finitediff_comm_point_A_quarter}).
This time, however, the regions with committor values close to $0$ and $1$ are larger than those found for point A, with a consequent reduction of the regions with intermediate values of the committor.
Additionally, the iso-committor lines look more straight than those observed in region A.
Consequently, from the comparison with point A, we infer that the self-propulsion speed and rotational diffusion process of the ABP determine the degree of convexity of the iso-committor hypersurfaces in the 3D space spanned by $x$, $y$, and $\vartheta$.
This has as a result that the self-propulsion parameters of the ABP determine the relevance of the angular component of the committor with respect to the spatial ones.

\subsubsection{Point C}

An active particle with large self-propulsion and persistence, as one found in point C in parameter space, shows instead a behavior totally different from the one observed in the other points and in the case of a passive particle.

\begin{figure}[H]
\centering
\includegraphics[height=100mm]{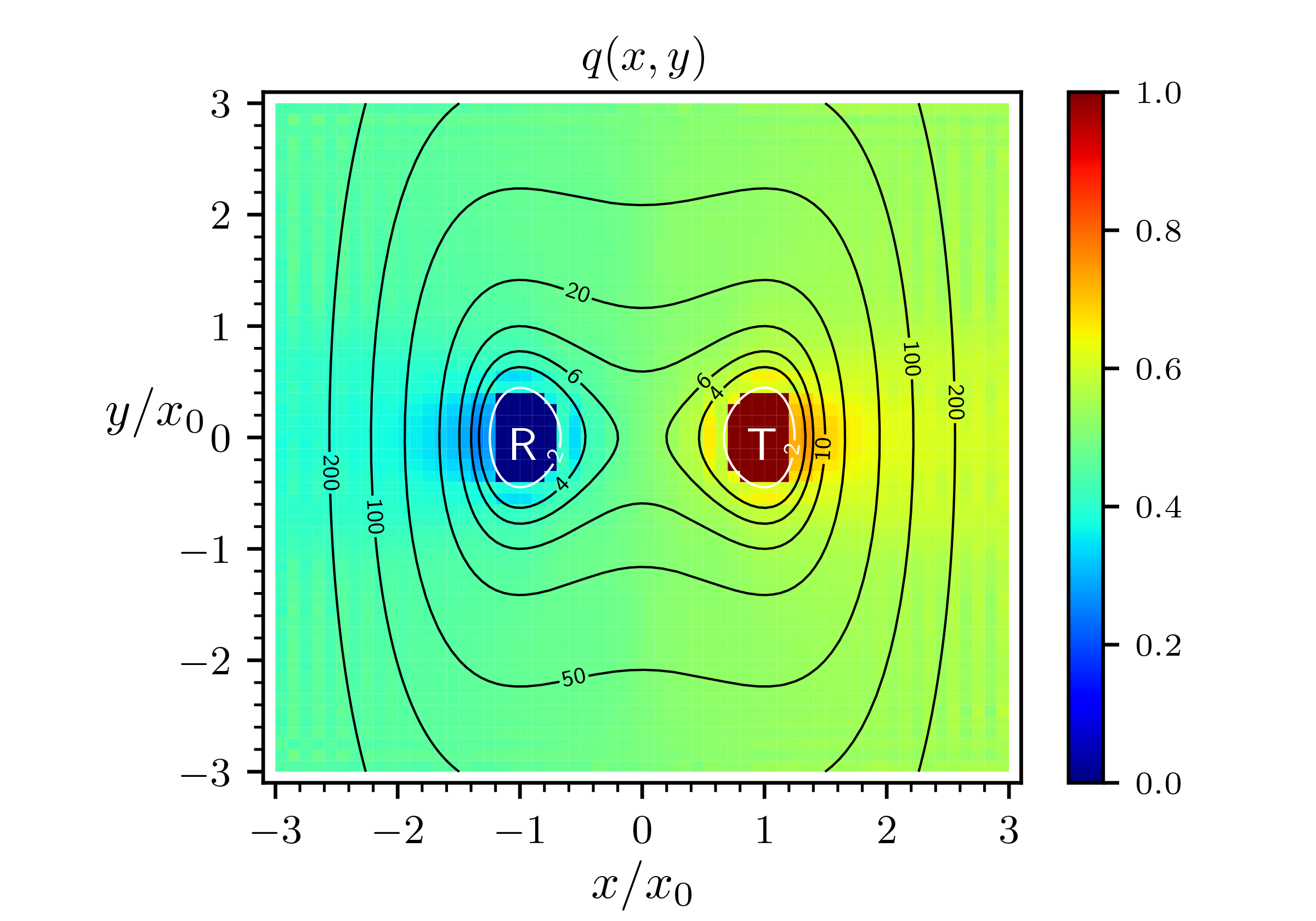}
\caption{Committor for a 2D active Brownian particle with parameters from region C of parameter space, in the double-well potential. The plot is obtained from the finite difference algorithm with brute-force simulations on the grid boundary, averaged along the angular component. The contour lines represent the underlying energy landscape, with the energies in units of $k_{B}T$ reported for each contour line. The white lines with $U(x,y) = 2 k_{B}T$ represent the boundaries of the R and T regions.}
\label{fig:finitediff_comm_point_C}
\end{figure}

\begin{figure}[H]
\centering
\includegraphics[height=100mm]{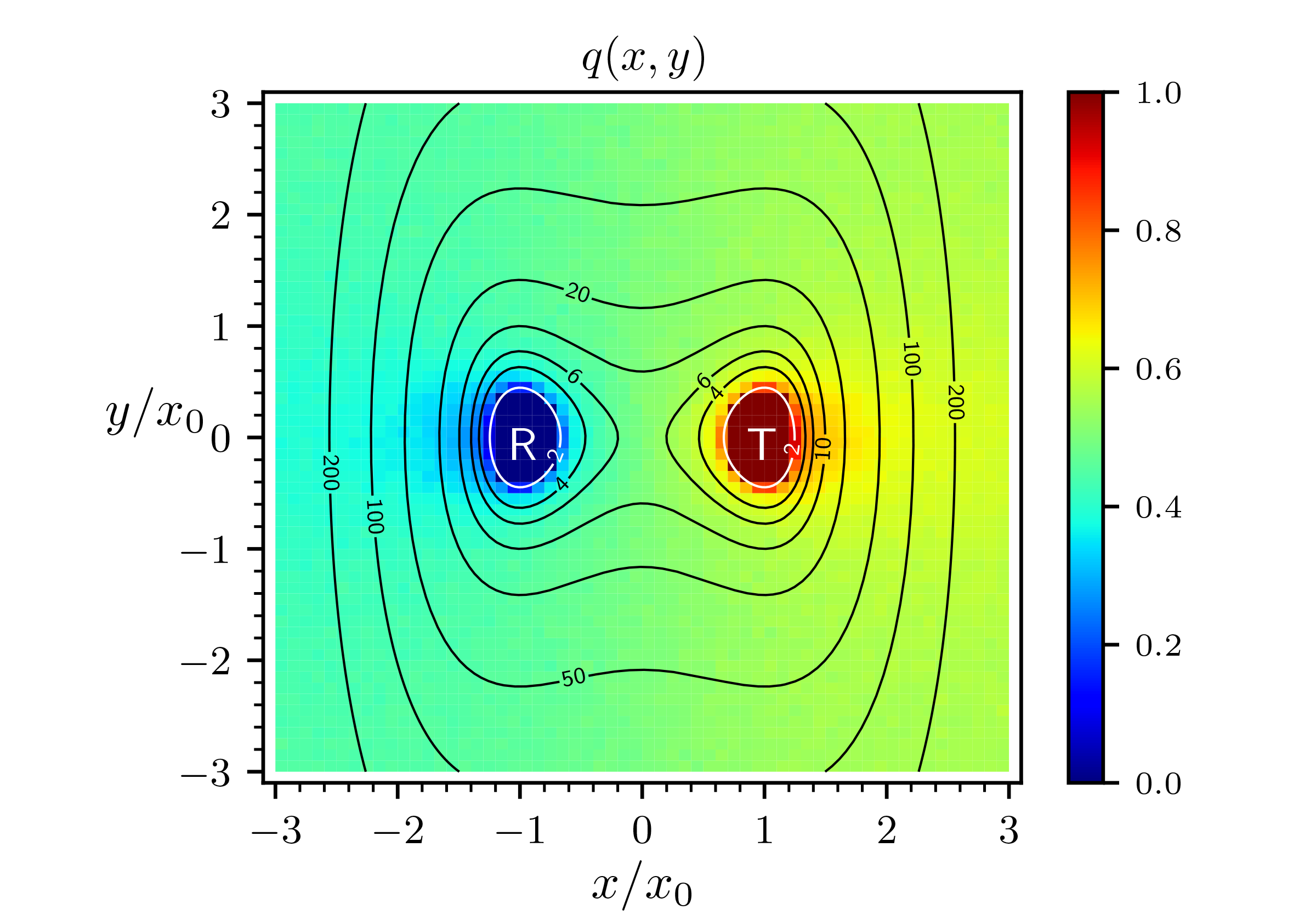}
\caption{Committor as in Fig.~\ref{fig:finitediff_comm_point_C} but with data obtained from $10^{3}$ trajectories for each grid point. The trajectories were generated with a direct integration of the equations of motion.}
\label{fig:bf_comm_point_C}
\end{figure}

If the committor is averaged over the angular component (see Fig.~\ref{fig:finitediff_comm_point_C} and Fig.~\ref{fig:bf_comm_point_C} for the brute-force validation), the regions where the committor shows intermediate values occupy most of the spatial configurations, while only the regions close to the back of the R and T basins display values close to $0$ and $1$ respectively.
In this case, it becomes difficult to identify the separatrix of the system, even though, thanks to the symmetry introduced by the averaging procedure, its position can be inferred once again as the straight iso-committor line passing at the saddle point of the barrier.

\begin{figure}[ht]
\centering
\includegraphics[height=110mm]{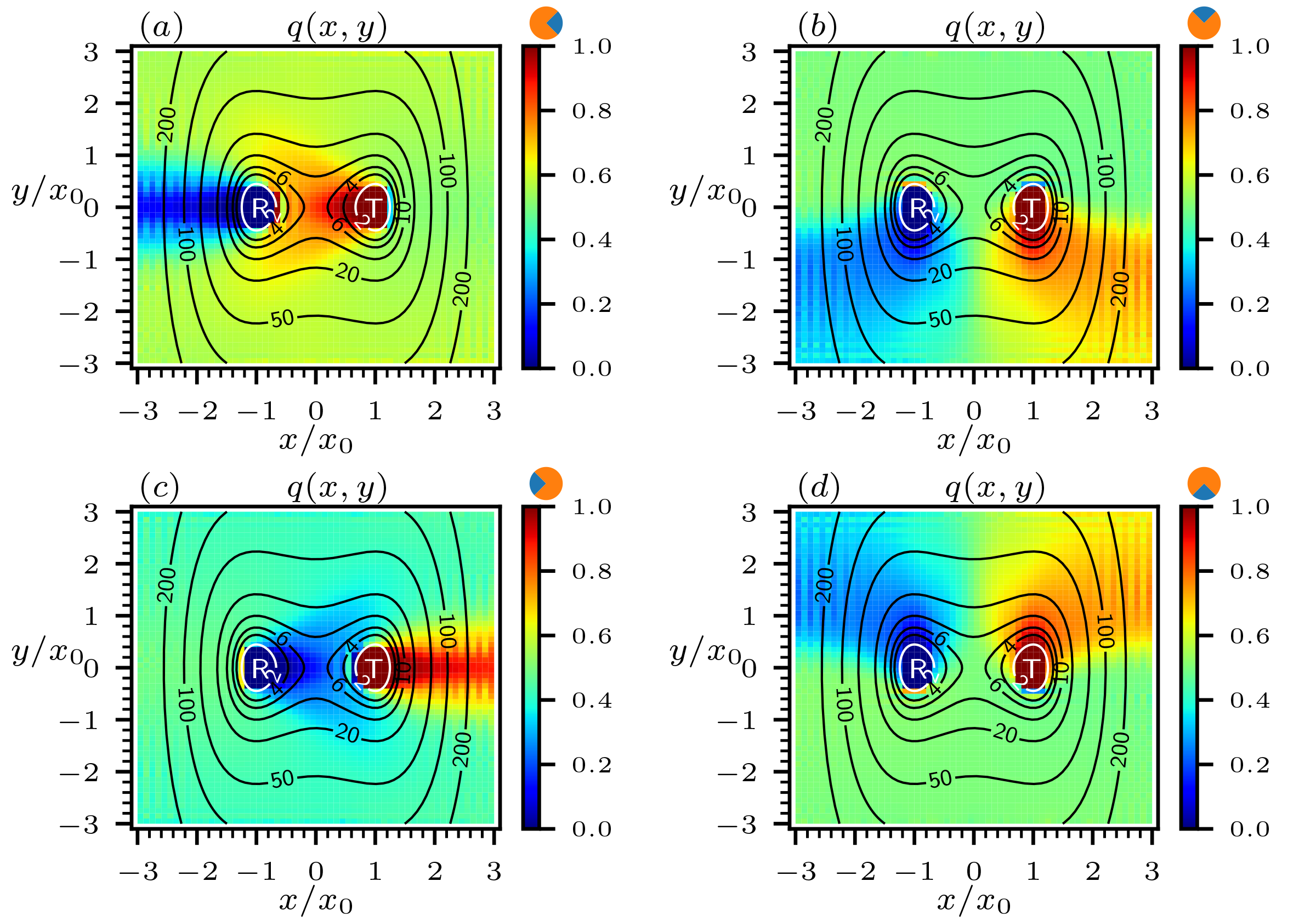}
\caption{Committor for a 2D active Brownian particle with parameters from region C of parameter space, in the double-well potential. The plot is obtained from the finite difference algorithm with brute-force simulations on the grid boundary. The contour lines represent the underlying energy landscape, with the energies in units of $k_{B}T$ reported for each contour line. The white lines with $U(x,y) = 2 k_{B}T$ represent the boundaries of the R and T regions. (a) Committor obtained from an average of the two angular slices located between $-\pi/4\!\leq\!\vartheta\!\leq\!\pi/4$. (b) Committor obtained from an average of the two angular slices located between $\pi/4\!\leq\!\vartheta\!\leq\!3\pi/4$. (c) Committor obtained from an average of the two angular slices located between $3\pi/4\!\leq\!\vartheta\!\leq\!5\pi/4$. (d) Committor obtained from an average of the two angular slices located between $5\pi/4\!\leq\!\vartheta\!\leq\!7\pi/4$. The angular slices over which the average is computed are represented by the blue quadrant in the orange pie plot on top of the bars.}
\label{fig:finitediff_comm_point_C_quarter}
\end{figure}

\begin{figure}[H]
\centering
\includegraphics[height=110mm]{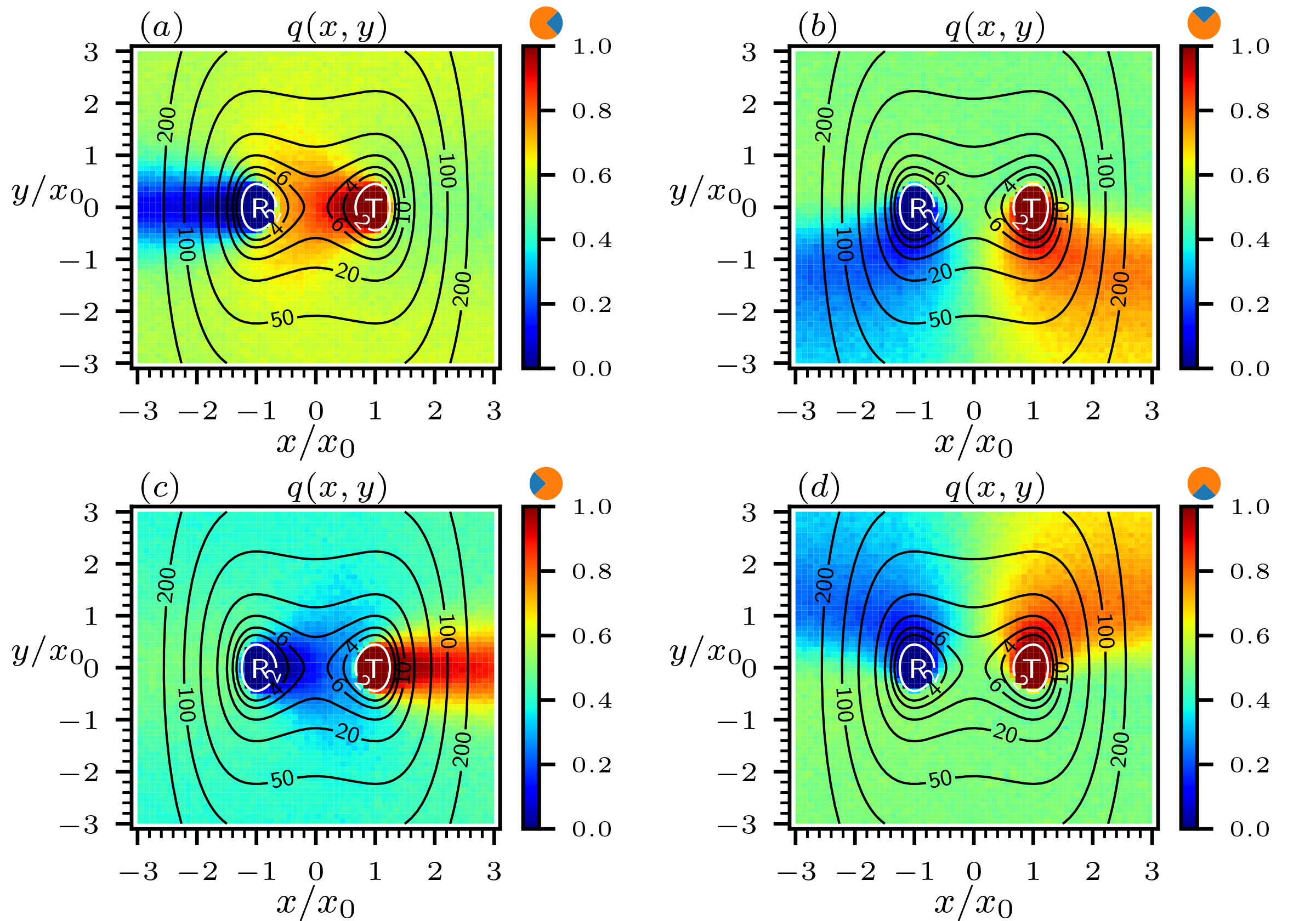}
\caption{Committor as in Fig.~\ref{fig:finitediff_comm_point_C_quarter} but with data obtained from $10^{3}$ trajectories for each grid point. The trajectories were generated with a direct integration of the equations of motion.}
\label{fig:bf_comm_point_C_quarter}
\end{figure}

If the angular information is retained (see Fig.~\ref{fig:finitediff_comm_point_C_quarter} and Fig.~\ref{fig:bf_comm_point_C_quarter} for the brute-force comparison), a strikingly different behavior emerges if compared to point A, point B, and the passive case.

For a particle with an initial orientation along the positive $x$ axis (Fig.~\ref{fig:finitediff_comm_point_C_quarter}(a)), once again the regions located on the back of the R state show values of the committor close to $0$, but at the same time, the regions displaying a committor close to $1$ are located on the opposite side of T compared to point A (Fig.~\ref{fig:finitediff_comm_point_A_quarter}(a)).
In particular, regions located between R and the barrier and close to R still display favorable target-finding odds as in point A, but the separatrix is shifted much closer to R and the regions on the back of T show instead much smaller values of successfully hitting the target.
Additionally, while in point A and point B the separatrix can be inferred as a straight line due to the opposite convexity of the iso-committor lines close to R compared to those close to T, in point C the separatrix of the system is split in two parts: one located between region R and the barrier, facing the rightmost boundary of R, and one located on the right of T for $x$ values larger than those lying in the T state.
This second separatrix highlights that the particles starting from the right of it (so with $x$ values larger than those found in the T basin, $y$ values around $0$, and a self-propulsion directed along positive $x$) move in the opposite direction compared to T, and due to the ‘‘surfing'' mechanism of active Brownian particles have higher chances to just slide along the confining boundaries and hit the R basin instead.

Similarly, particles with the opposite initial orientation (Fig.~\ref{fig:finitediff_comm_point_C_quarter}(c)), show large chances of hitting T if they start from the back of the T basin compared to the position of the barrier.
At the same time, the committor is larger than $0.5$ if they are located on the left of the R basin because of the surfing mechanism which makes them surf on the opposite side of the barrier by following the confining boundaries.
By the time employed to get to the opposite side of the barrier, the particle will have had enough time to reorient and forget its initial orientation, thus leading to better chances of finding T than R.
Also in this case the separatrix of the system is composed of two surfaces, which, projected on the 2D plane, become two curves: one located between T and the barrier, and one between R and the region located at smaller $x$ values than R.

Particles with an initial self-propulsion directed along the positive and negative $y$ axis (Fig.~\ref{fig:finitediff_comm_point_C_quarter}(b) and (d) respectively), show once again a symmetric behavior with respect to the $x$ axis.
Here we focus on the analysis for the former case, since the latter can be analogously derived.
In the case of a self propulsion directed along the positive $y$ axis, the region yielding a high value of the committor is located beneath region T.
However, note that, for a fixed value of $x$ lying below T, the more negative the value of $y$ becomes and the more the committor decreases, similarly to the behavior observed for point A (Fig.~\ref{fig:finitediff_comm_point_A_quarter}(b)).
This is due to the increased distance from T, which has as consequence a larger travel time before meeting T that allows the particle to forget its initial orientation more easily than a particle closer to T, leading those trajectories to R more frequently.
Finally, differently from point A, in this case a large value of the persistence allows a particle starting from the upper-left corner compared to R to have a committor larger than $0.5$, again due to the surfing mechanism of active particles.
This leads to a separatrix made of two lines, one enclosing the region with both $x$ and $y$ smaller than the values in R, and one enclosing the region with $x$ and $y$ values larger than those found in T.

\subsubsection{Final remarks on the committor for ABPs}

We conclude this analysis with some closing remarks.
It is interesting to carry out a comparison between the information provided by the committor function and the one obtained from the transition probability densities and currents, especially in the case of the counterintuitive target-search paths discussed in section \ref{sec:tpd_tc_dw_prl}.
In Fig.~\ref{fig:trprd_trc}(c) are reported the transition probability density and current for a particle with parameters ($\text{Pe}=4$,$\ell^{*}=34$) in the new parameter space.
Such a particle belongs to the edges of region C (see Fig.~\ref{fig:dw_rates}(a)), and therefore it has a behavior comparable to the point C discussed in this chapter.
By analyzing the behavior of the committor, one could assume that such a particle would find the target more easily by exiting the R region with an initial orientation along positive $x$.
Instead, from the transition probability densities and currents emerges how the most frequent paths leading from R to T are those exiting R from the back and reaching T by surfing along the confining boundaries.
The reason for this behavior is two-fold.
First, the reactive probability density represents the most frequently visited regions during the successful events, and it is obtained from the transition path ensemble.
Therefore, longer paths provide a larger contribution to the probability density, and consequently the trajectories ‘‘surfing'' the landscape are more visible than those going directly from R to T through the saddle point with a short travel time.
Second, the number of trajectories exiting R from the right side with a self-propulsion directed along positive $x$ values is much smaller than those exiting from the left side with a direction pointing towards negative $x$ (see the steady-state distribution in Fig.~\ref{fig:rho}(c) and (e) as an example).
Therefore, even though the committor function for particles exiting R on the left with a direction along negative $x$ is much smaller than the one for particles exiting from the opposite side with an opposite self-propulsion orientation (compare Fig.~\ref{fig:finitediff_comm_point_C_quarter}(c) and Fig.~\ref{fig:finitediff_comm_point_C_quarter}(a)), the number of attempts in the first case is considerably larger than in the second, and this fact, together with the increased length of the paths, generates the features displayed by the transition probability density.

Finally, the committor also suggests an efficient and intuitive way to find targets for particles capable of modifying their navigation parameters on the fly.
If the two basins R and T are located in metastable states separated by an energy barrier and a particle exits R without choosing a specific initial orientation, the picture offered by the committor would suggest that it is better, on average, to start the navigation with a persistence and self-propulsion speed typical of points in region C (Fig.~\ref{fig:finitediff_comm_point_C}), to then decrease the self-propulsion speed and increase the rotational diffusion coefficient as soon as the barrier is overcome (to end up with a behavior similar to the one in Fig.~\ref{fig:finitediff_comm_point_B}).
In this way, the particle would shift from a regime of high activity to a passive-like behavior, reaching T through the effects of the external potential without wasting additional resources in the self-propulsion mechanism.

\subsection{Finite-difference advantages and drawbacks compared to brute-force simulations}

The comparison of the committor obtained through the finite difference algorithm and through brute force in the cases discussed in the previous section highlights that the finite-difference calculation based on the Backward Kolmogorov equation can reproduce the features of the brute-force committor, validating our analytic generalization.
However, some differences emerge between the two, in particular at high values of the self-propulsion speed (point A and C).
From the finite-difference solution, in fact, a ‘‘stripe'' pattern on the grid cells emerges, which has no counterpart in the brute-force solution.
\begin{figure}[ht]
\centering
\includegraphics[height=100mm]{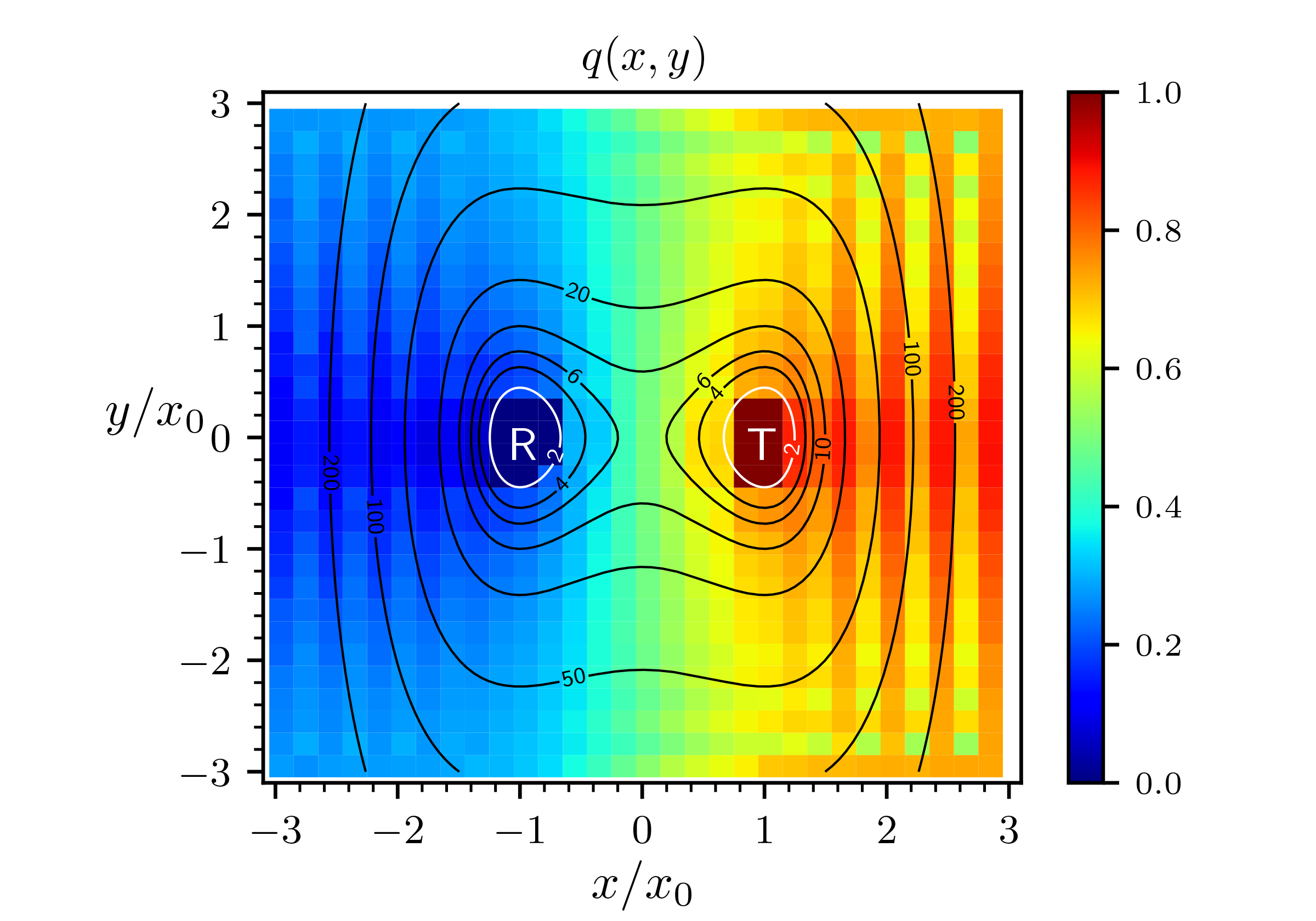}
\caption{Committor for a 2D active Brownian particle with parameters ($\text{Pe} = 4.5$, $\ell^{*} = 7$), in the double-well potential. The plot is obtained from the finite difference algorithm with grid points with double dimension along both $x$ and $y$ compared to the values used throughout the previous study, averaged along the angular component. The contour lines represent the underlying energy landscape, with the energies in units of $k_{B}T$ reported for each contour line. The white lines with $U(x,y) = 2 k_{B}T$ represent the boundaries of the R and T regions.}
\label{fig:bf_comm_double_grid}
\end{figure}
Since the numerical estimation of the committor within the grid in the finite-difference approach depends on the accuracy in determining the boundaries of the grid, we checked that the committor values on the edges of the grid were obtained with sufficient precision.
First, we verified that a different choice of the integration time step for the brute-force simulations did not introduce a systematic shift of the value of the committor on the edges, and since this was not the case, we also increased the number of simulated trajectories on the edges by a factor of $10$ to check if the statistics was sufficient.
None of these approaches provided a considerable change in the finite-difference solution, which, instead, is sensitive to the dimension of the grid-points.
A doubling in the size of the grid points, in fact, introduced an even larger error in the determination of the grid even for smaller values of the self-propulsion, see Fig.~\ref{fig:bf_comm_double_grid}.
Therefore, an increase of the self-propulsion speed requires a decrease in the dimension of the grid points to determine with high accuracy the value of the committor.
Unfortunately, the solution of the linear system with the finite difference algorithm for the grid that we employed already requires more than $6$ GB of RAM, and a doubling of the grid points along the $x$ and $y$ components is associated with an increase of the necessary memory up to more than $100$ GB, in a typical ‘‘curse of dimensionality'' fashion.

Nevertheless, the finite difference algorithm does provide a large computational advantage comparing to finding the values of the committor on the grid via brute-force simulations.
In the case of point A, determining the committor using the finite difference algorithm (also considering the time required to simulate the committor on the edges of the grid via brute force) required a time of about $0.08$ of the time needed to obtain the same result using brute-force simulations with a statistics of $10^{3}$ simulations per grid point.
This efficiency slightly changes depending on the parameters of the ABP.
For point C for instance, a higher persistence makes the particle spend a long time in surfing before reaching R or T, thus increasing the brute-force times and shifting time required for the finite-difference procedure to $0.07$ of the time needed for the brute-force simulations.

\section{Chapter conclusions}

In this chapter we extended the concept of committor function to active Brownian particles performing target-search processes.

After providing an analytic derivation of the committor function for ABPs, we exploited the fact that the committor is a solution for the Backward Kolmogorov equation to compute it numerically in an efficient way, using a finite-difference algorithm.
Subsequently, we validated our analytic derivation against brute-force simulations, and we used the finite-difference results to estimate the committor for an ABP looking for a target in the paradigmatic double-well potential.
Depending on the ABP parameters, the committor displays characteristics that differ considerably from the passive particle case, for example the fact that the shape and characteristics of the iso-committor hypersurfaces change depending on the ABP parameters.
Additionally, we found that the self-propulsion orientation plays a pivotal role in determining the success odds of the agent and it represents an advantageous feature that is precluded to passive searchers.
Finally, we observed that the counterintuitive target-search pathways found in the previous chapters are less evident from the committor characterization, suggesting how the study of this quantity in active systems needs to be complemented by the analysis of the steady-state distributions for the system.

The generalization of this framework to the active Brownian particle case paves the road for future applications of this concept to different active particle models and systems.

\afterpage{\null\newpage}

\pagestyle{plain}

\afterpage{\null\newpage}

\pagestyle{plain}

\chapter{Case study: self-propelled materials}
\pagestyle{fancy}

\section{Experimental validation of the committor function for active agents}

In the previous chapter, we generalized the concept of the committor function to the case of an active particle system, modeled through the ABP model, and we applied it to the target-search problem.
We now want to use this theory to study macroscopic objects capable of self-propelling on a water surface and performing target search, which will serve as an experimental comparison for our framework.

As we introduced in section \ref{sec:art_micr}, self-propelled macroscopic objects constitute one example of artificial self-propelling agents that are at the center of the current research in the active-matter field.
In particular, since a long time camphor scrapings have been found capable to float on a water surface and perform both directed motion and rotational diffusion~\cite{Nakata1997}, and it has been discovered that their rotational process is influenced by the shape of the object, its temperature and its surface tension.
Additionally, objects obtained by combining camphor and camphene have been found to form malleable materials capable of self-propelling with speed and trajectory shape depending on the shape of the object and on the ratio of the camphor and camphene constituents~\cite{Löffler2019}.
Furthermore, the addition of polypropylene to the object was found to improve its quality and control the dissipation of the camphor and camphene molecules, therefore improving the stability of the compound and also reducing the effect of its composition on its activity~\cite{Löffler2021}.

Here, we start by presenting the general experimental techniques used to generate and extract trajectories of these self-propelling objects, to then introduce the design of the target-search process in the experiment and the procedure employed to analyze experimental trajectories and estimate values of the committor.
We then provide a direct comparison with the committor obtained by brute-force integration of the equations of motion for an ABP, finding qualitative agreement between the curves notwithstanding the differences between the macroscopic experimental setup and the microscopic nature of ABPs.
Finally, we conclude this chapter by presenting the ongoing study of such a macroscopic particle moving on top of a submerged ‘‘energy landscape'', modeled using a rigid object with an appropriate shape.

\section{Experimental setup}
The attempt of an experimental validation of our results that we present in this chapter relies on the study of two different systems, where an object made of different camphor mixtures with other compounds is recorded using a mounted digital camera (NEX VG20EH, SONY) while floating on a water surface\footnote{All the experiments presented in this chapter were performed by Richard J.G. L\"offler}.
These systems, although displaying relevant differences to an active Brownian particle (\textit{e.g.} the translational diffusion is negligible for these macroscopic swimmers, hydrodynamics instead is not, and the speed and rotational diffusion processes are not constant in time), still share fundamental similarities with an ABP, such as the ability of self-propelling with a process displaying rotational diffusion that makes them potential candidates for a qualitative validation of our theory.

\begin{figure}[H]
\centering
\includegraphics[height=80mm]{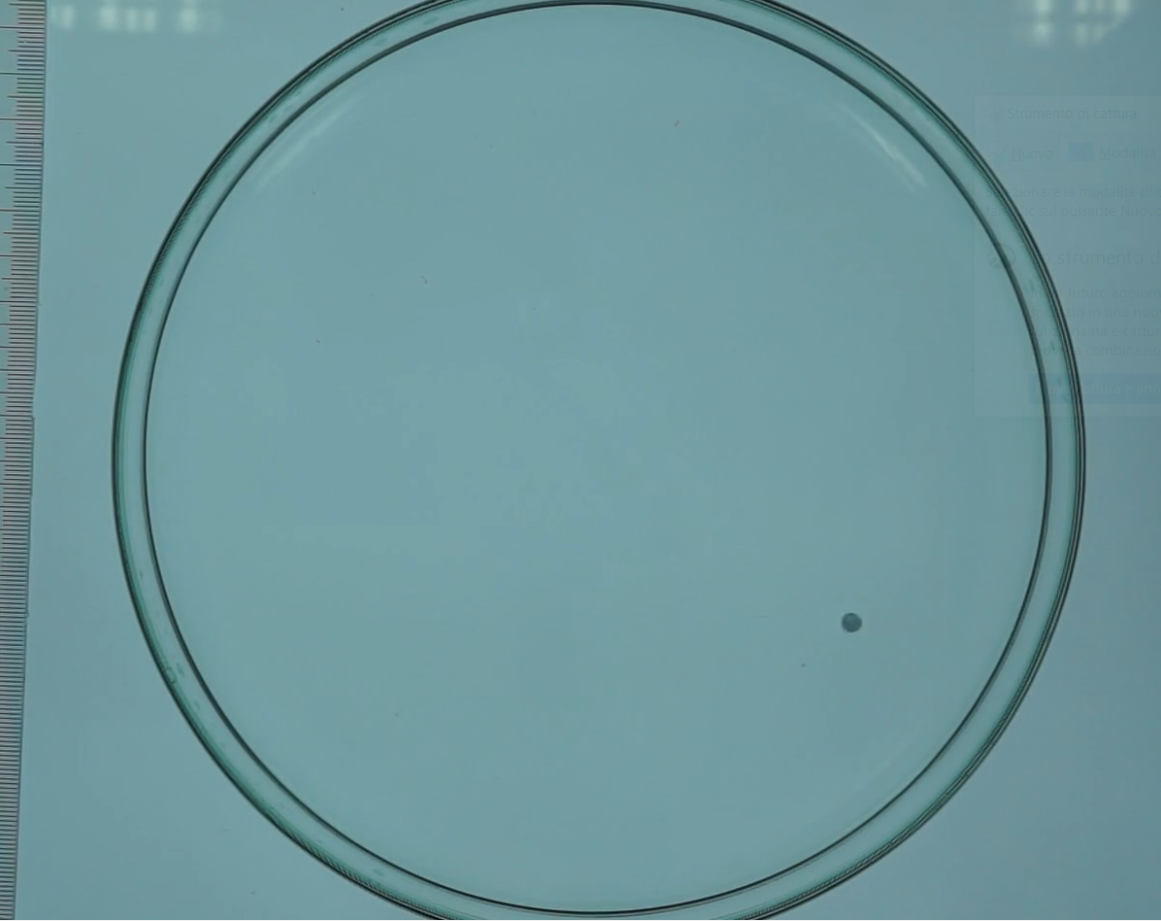}
\caption{Experimental setup of the committor evaluation during a target-search processes in a Petri dish, in a top-down view. A self-propelled camphor-camphene-polymer disk moves on a water surface within the confining boundary of a Petri dish. The picture was provided by Richard J.G. L\"offler.}
\label{fig:Petri_exp}
\end{figure}

In the first system we study, discussed in section \ref{sec:simple_target_search}, a disk-shaped object was released on a water surface within the confining boundary provided by a Petri dish (see Fig.~\ref{fig:Petri_exp}).
In these experiments, the disk composition is made in some cases of pure camphor, and in other cases of a mixture of camphor, camphene and a polymer, which is sharing an equivalent behavior to the pure camphor disk and is found more stable throughout the experiment.
In this system we proceed to design a simple configuration to study target-search processes: we select as the R state for the process a circular region located at the center of the dish and as T a circular segment of the dish.
We then select three points within the dish and we compute the committor in these configurations as a function of the self-propulsion orientation.

The second system, instead, aims at reproducing the target-search process occurring in an energy landscape, where this time the self-propelling object is a droplet obtained from a mixture of paraffin, camphor and a dye (oil red and sudan blue 2), moving on a water surface.
The outline of this research project is sketched in section \ref{sec:energy_land_targ_search}.

\section{Simple target-search processes in experiments}
\label{sec:simple_target_search}

\begin{figure}[H]
\centering
\includegraphics[height=100mm]{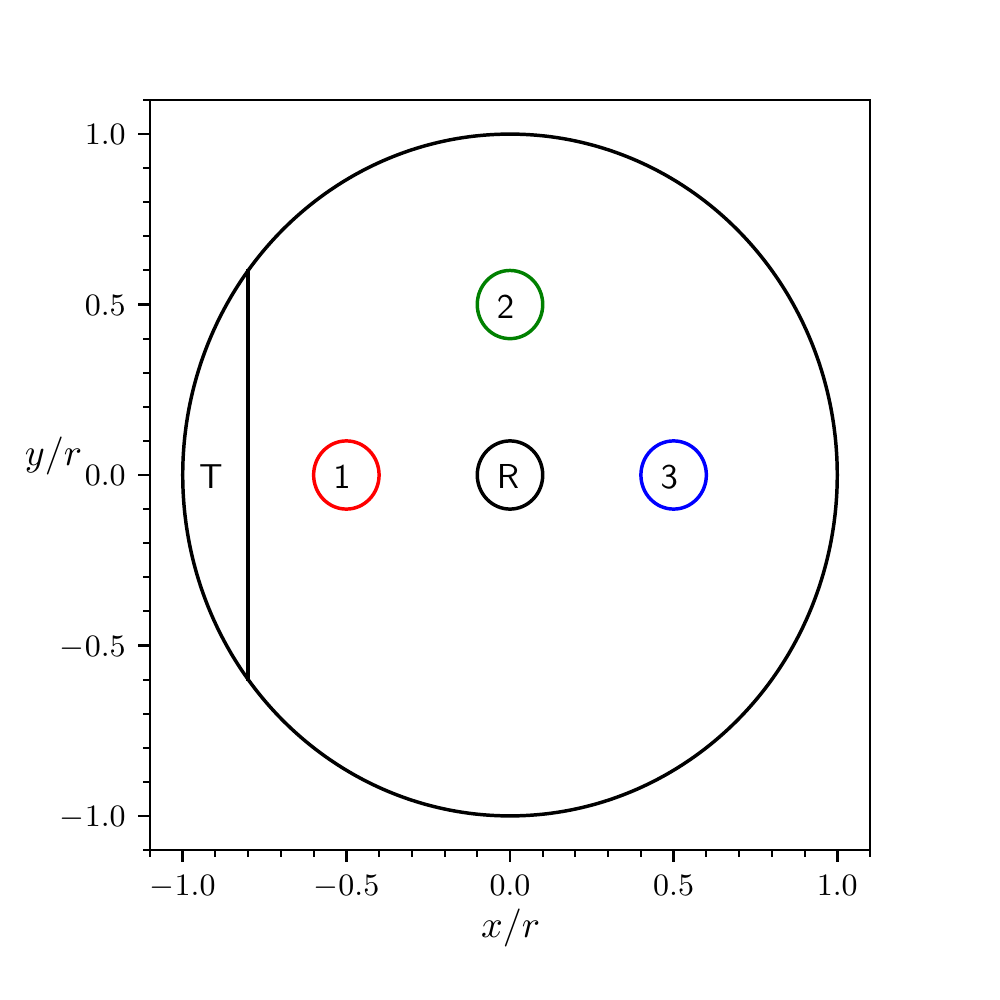}
\caption{Configuration of the experiment, in units of $r$. Region R is a circle located at the center of the dish, T is a circular segment on the left side of the dish, and points $1$, $2$, and $3$ are the regions where the committor is evaluated as an angular function.}
\label{fig:comm_sketch}
\end{figure}

In the first system we study, a disk-shaped object with a radius $r_{\text{searcher}}$ of about $2.5$ mm is released on a water surface within the confining boundary provided by a Petri dish with a radius $r$ of about $9.7$ cm (see Fig.~\ref{fig:Petri_exp}).
In this system we select as the R state for the process a circular region with center located at the center of the Petri dish and a radius of $0.1 \; r$, and as T a circular segment of the dish, characterized by having $x<-0.8 \; r$ (see Fig.~\ref{fig:comm_sketch}).
We then select three non-overlapping circles within the dish (points $1$, $2$, and $3$ in Fig.~\ref{fig:comm_sketch}, with a radius $r_{c} = 0.1 \; r$) and we compute the committor in these points as a function of the self-propulsion orientation of the agent when leaving the circles.
This orientation is estimated from the infinitesimal displacement along the $x$ and $y$ directions and is then discretized in $4$ angular slices (a slice with $-\pi/4 \! \leq \! \vartheta \! < \! \pi/4$, labeled as $0$ or $2\pi$, a slice with $\pi/4 \! \leq \! \vartheta \! < \! 3\pi/4$, labeled as $\pi/2$, a slice with $3\pi/4 \! \leq \! \vartheta \! < \! 5\pi/4$, labeled as $\pi$, and a slice with $5\pi/4 \! \leq \! \vartheta \! < \! 7\pi/4$, labeled as $3\pi/2$).

In the following discussion, we use the radius $r$ of the dish to set the length scale of the problem.

\subsection{Experimental committor and brute-force validation}

We obtain the committor values by first generating long trajectories of the self-propelled object moving in a disk, both in the case of experiments and by brute-force molecular-dynamics simulations.
Then, we extract the positions from the long trajectories and we select the relevant trajectory slices that can be used to compute the committor (so the slices leaving one of the three points and reaching R or T).
The committor is subsequently obtained for each point and angular slice as the fraction of the trajectories reaching T (starting from that point and angular slice) over the total number of trajectories (again for that point and angular slice).
Note that some trajectories might be used to compute the committor for two different points simultaneously (for example a trajectory slice starting in point 1, passing through point 2 and then entering R or T will be used to compute the committor both in point 1 and in point 2).
Trajectories exiting one of the points, reentering in it, and then visiting R or T will instead be counted only once, starting from the first time they leave the point.
We also point out that the three points are circular regions of finite extension (\textit{i.e.} not point-like, with radius $r_{c}$), so that sufficient statistics can be extracted from few long experimental trajectories (which have a variable time duration in the order of few hours for each trajectory).

\subsubsection{Experimental estimation of the committor}

After extracting the list of positions during the $5$ experiments performed, we select the trajectory slices starting from (or visiting) one of the three points and reaching R or T, obtaining a total number $n$ of about $900$ trajectory slices, which yields an approximate number of $80$ trajectories per point and angular slice, even though this number varies depending on the point and angular slice.

The committor is then estimated following the procedure outlined above, and the errors for each point and angular slice are computed as standard errors of the mean:
\begin{equation}
\sigma_{m} = \frac{\sigma}{\sqrt{n}} \; ,
\end{equation}
where $n$ is the total number of trajectory slices for that point and angular slice, and $\sigma$ is the standard deviation of the data:
\begin{equation}
\sigma = \sqrt{\frac{n_{\text{T}}(1-q)^{2} + n_{\text{R}}(0-q)^{2}}{n}} \; ,
\end{equation}
with $n_{\text{T}}$ is the number of trajectory slices that reach T for the considered point and angular slice, $n_{\text{R}}$ is the analogous number of slices that reach R, and $q$ is the value of the committor for that point and angular slice.

\subsubsection{Estimating the committor from brute-force simulations}

To reproduce the experimental results through brute-force simulations, we first tried to extract an average speed and rotational diffusion coefficient from the experimental trajectories.
However, the self-propulsion parameters are changing considerably during the experiments and are not always following the same trend, therefore we decided to simulate the active Brownian particle with some trial parameters that could yield a comparable behavior to the self-propelled disks.

Since the translational diffusion process is negligible for the self-propelled camphor object, to have a better comparison with experiments we used a reduced ABP model where the diffusion coefficient $D$ is set to $0$, yielding the following equations of motion:
\begin{subequations}
\begin{eqnarray}\label{eom}
\bm{r}_{i\!+\!1} &=& \bm{r}_{i} + v\, \bm{u}_{i} \, \Delta t - \mu \bm{\nabla} U(\bm{r}_{i}) \Delta t \;,\\ \label{eom2}
\vartheta_{i\!+\!1} &=& \vartheta_{i} + \sqrt{2D_{\vartheta}\Delta t} \, \eta_i\;.
\end{eqnarray}
\end{subequations}
Consequently, provided that $r$ is used to define the length scale, we define the time scale as $\tau = 1/D_{\vartheta}$, $\mu$ is used to fix the energy scale as $k_{B}T = r^{2}/\tau \mu$ (which is used only to determine the interaction of the particle with the boundaries), and consequently a single dimensionless parameter is left to determine the activity of the particle, which is the persistence $\ell = v \tau / r$.
The boundaries of the Petri dish are modeled using a quartic potential in the following form:
\begin{equation}
U(x,y) = (x^{2}+y^{2})^{2} \; ,
\end{equation}
which acts on the particle only if the self-propelled agent gets closer than $0.01 \; r$ to the walls.

The persistence of the particle $\ell$ was set to $0.2$, and the time step employed in the simulations was $5 \cdot 10^{-3} \tau$.
From a long brute-force simulation of such a particle about $3.4 \cdot 10^{4}$ trajectory slices were extracted, yielding an average number of about $2.9 \cdot 10^{3}$ trajectories per point and angular slice.

\subsubsection{Comparison between experiments and simulations}

The selected value for the persistence of the particle employed in the brute-force simulation allowed us to have a direct qualitative comparison of the committor function in the experiments of the self-propelled disk and in simulations of an ABP in a similar environment.

\begin{figure}[H]
\centering
\includegraphics[height=100mm]{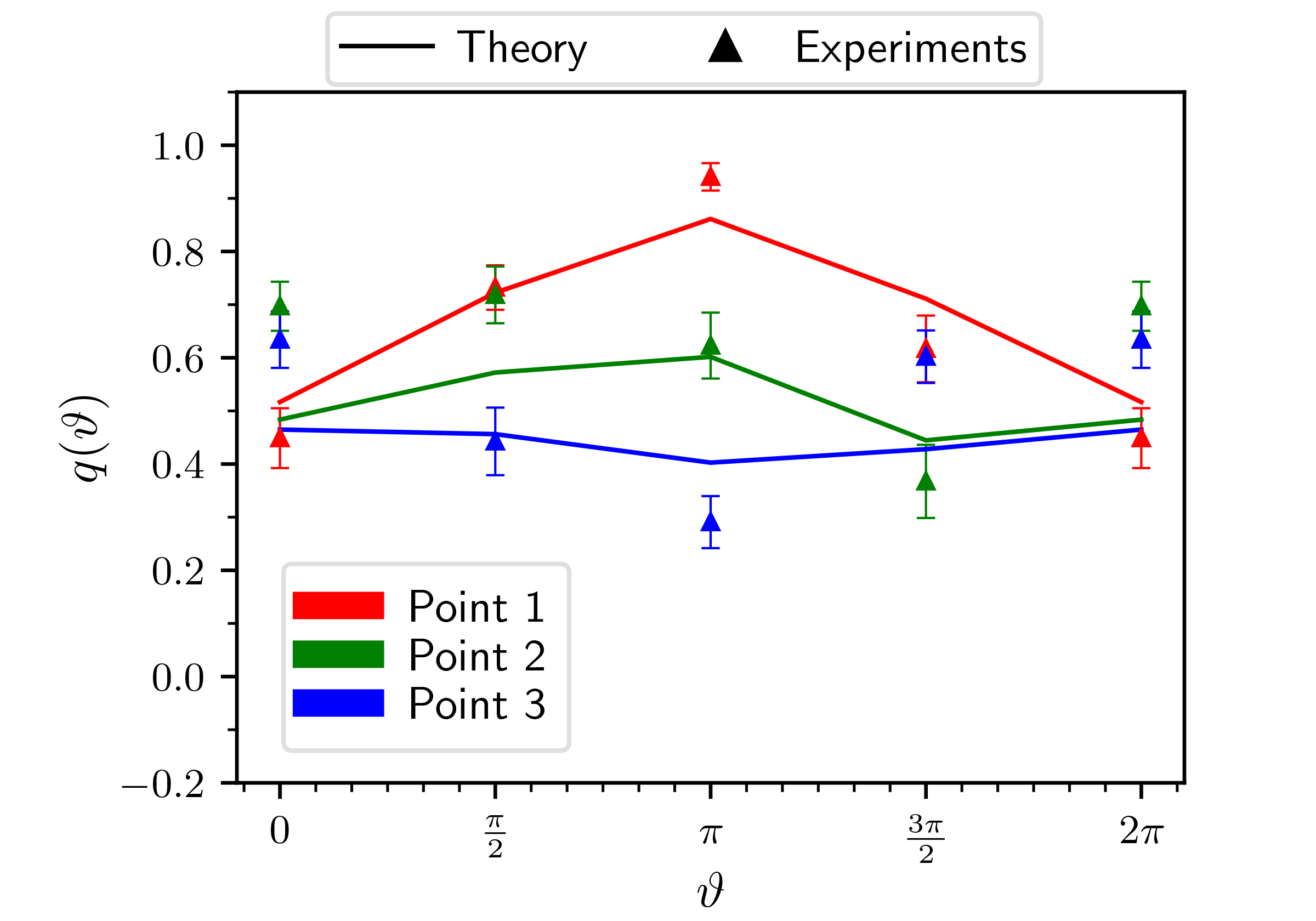}
\caption{Committor function from experiments (symbols with error bars) and simulations (continuous lines), as a function of the self-propulsion orientation of the particle when it exits the disks where the committor is computed. The three different colors indicate the three points selected in the experiments and represented in Fig.~\ref{fig:comm_sketch}.}
\label{fig:comm_comparison}
\end{figure}

The committor function obtained from the brute-force simulation of an ABP qualitatively reproduces the characteristics of the curves observed from the experiments (see Fig.~\ref{fig:comm_comparison}).
However, as can be inferred from Fig.~\ref{fig:comm_comparison}, the quantitative agreement between the curves is lacking, as expected from the considerable differences between the two systems at hand.
Nevertheless, the location of the maxima and minima for the function is correctly reproduced from the brute-force simulations, whose curves display a shape analogous to the one observed in the experiments.
In particular, this analysis highlights once again the relevance of the self-propulsion orientation in determining the success odds of the search, which is also influenced by the position in the dish.

\begin{figure}[H]
\centering
\subfigure{\label{fig:typical_traj_shuttle_sim}\includegraphics[height=50mm]{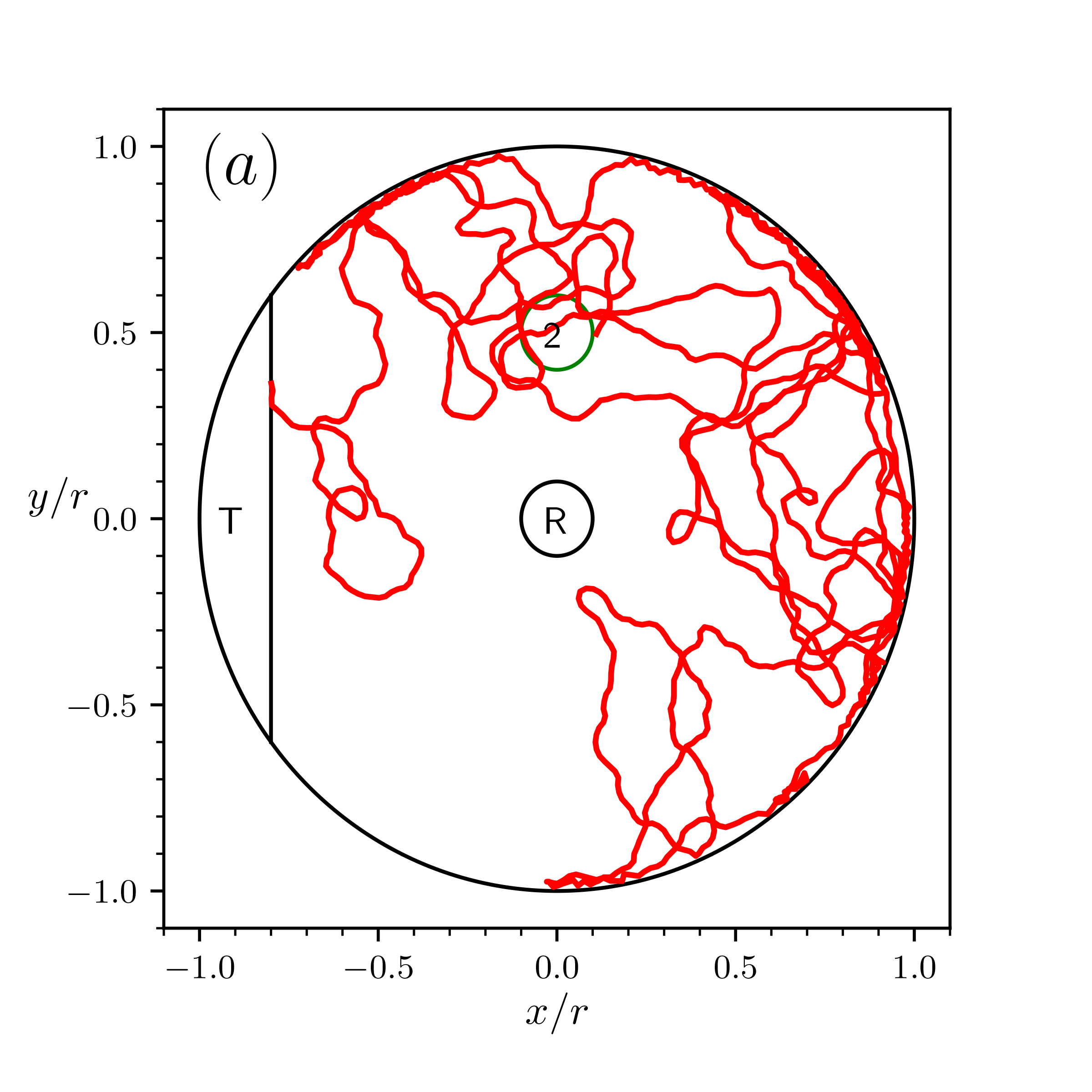}}
\subfigure{\label{fig:comm_larger_pers}\includegraphics[height=50mm]{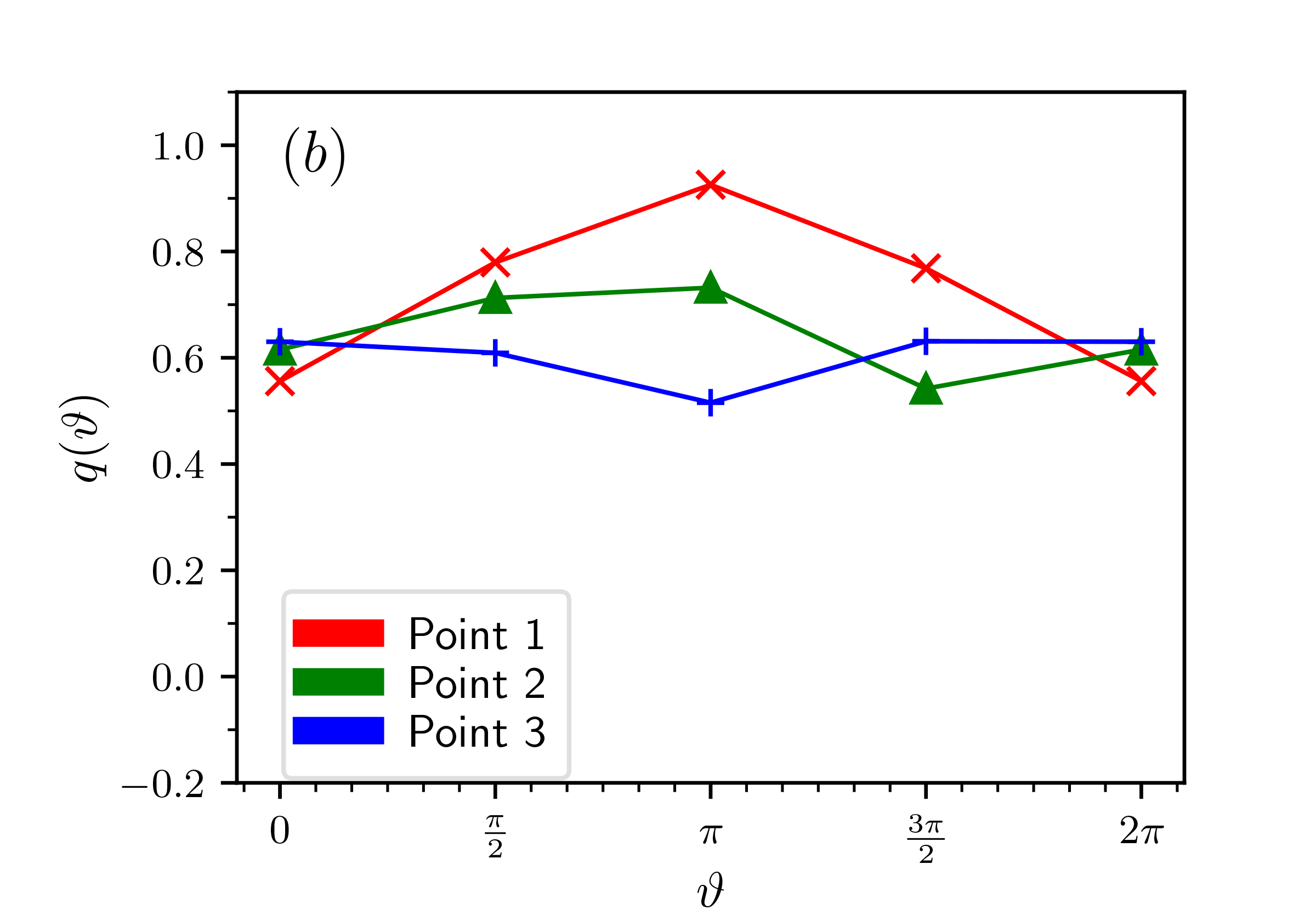}}
\caption{‘‘Shuttle'' of active Brownian particles. (a) Typical trajectory for an ABP reaching the target T, obtained from brute-force simulations. Due to the persistence of motion the particle spends large fractions of the travel time close to the boundaries. (b) The ‘‘shuttle'' effect increases with the persistence. Committor in the case with a persistence of $\ell = 0.4$.}
\label{fig:shuttle_effect}
\end{figure}

Point $1$ (Fig.~\ref{fig:comm_comparison}, red curve) represents a region located between the R and the T state (see Fig.~\ref{fig:comm_sketch}).
For particles visiting this position, the largest target-finding probability is obtained, with no surprises, when the particle leaves point $1$ with an angle of $\pi$ (\textit{i.e.} directed towards T).
Instead, the committor displays a minimum value if the particle is leaving point $1$ with an angle of $0$, or equivalently $2 \pi$ (\textit{i.e.} directed towards R).

\begin{figure}[H]
\centering
\includegraphics[height=120mm]{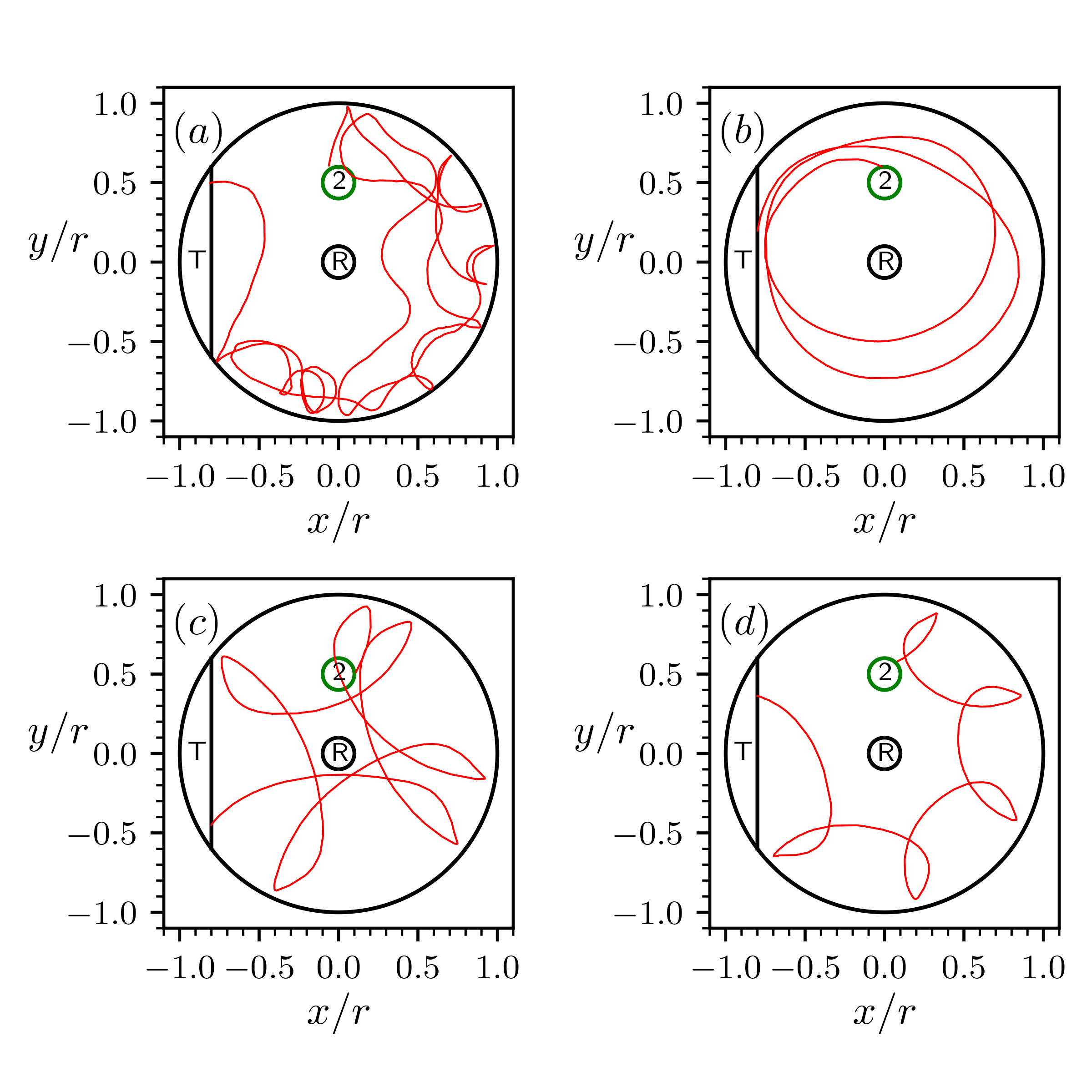}
\caption{(a-d) Representative trajectory slices obtained from experiments of a self-propelled camphor-camphene-polymer disk.}
\label{fig:exp_trajs_quarter}
\end{figure}

Point $2$, instead, shows a much more intriguing behavior.
If the particle is exiting the circle with an angle of $3\pi/2$, so directed towards R, a minimum in the committor is found due to the increased chances of meeting R at the beginning of the trajectory.
At the same time, a particle exiting from point $2$ with an angle of $\pi$ instead displays a much larger value of the committor due to its orientation pointing towards T and the larger shape of T compared to R.
However, even if the particle leaves the region with an angle of $\pi/2$ or $0$ (so, in this last case, directed on the opposite direction compared to T) the committor shows large values (in the experiments they are comparable or larger than for an angle of $\pi$, while in simulations they are slightly smaller, see Fig.~\ref{fig:comm_comparison} green curve).
In the case of the ABP employed in the simulations, this result can be explained considering the persistence of motion typical of self-propelled agents.
If the particle exits from point $2$ with an angle of $0$ or $\pi/2$, it will point towards the confining boundaries.
In this case the particle will meet the boundary, and, depending on the rotational diffusion process, some time will be required for the agent to move away and detach from the walls (which depends also on the level of convexity of the boundaries).
This leads to long times spent by the agent in following the boundaries (see Fig.~\ref{fig:shuttle_effect}(a) for a typical trajectory of an ABP with the selected persistence reaching T by spending long times close to the boundaries).
Since in this system configuration the target is located close to the confining walls, the agent will show increased chances of meeting T in its trajectory compared to meeting R, which would occur only by detaching from the walls and visiting the center of the dish.
In the following, we shall refer to this process of reaching T by following the boundaries as ‘‘shuttle'' effect.
The shuttle effect increases with the persistence of the particle, as is confirmed in Fig.~\ref{fig:shuttle_effect}(b) which shows higher values of the committor for the same system with a double persistence.

The modulation of the committor function for the active agent used in the experiments, instead, cannot be explained through the shuttle effect alone.
The trajectories observed in the experiments display a behavior that is very much dependent the conditions of the system at the time when these trajectories are observed during experiments (see Fig.~\ref{fig:exp_trajs_quarter}(a-d)).
In particular, we believe that the trajectory shapes are influenced by the remaining ‘‘fuel'' for the particle self-propulsion and by the shape of the object (small asymmetries in the shape of the disk lead to qualitatively different behaviors, similarly to what happens at the microscopic scale~\cite{bech2016}).
Notwithstanding these differences at the single-trajectory level, however, the committor obtained from the experiments shows on average a comparable behavior to the one found for an ABP.

\begin{figure}[H]
\centering
\subfigure{\label{fig:comm_pass_curves}\includegraphics[height=50mm]{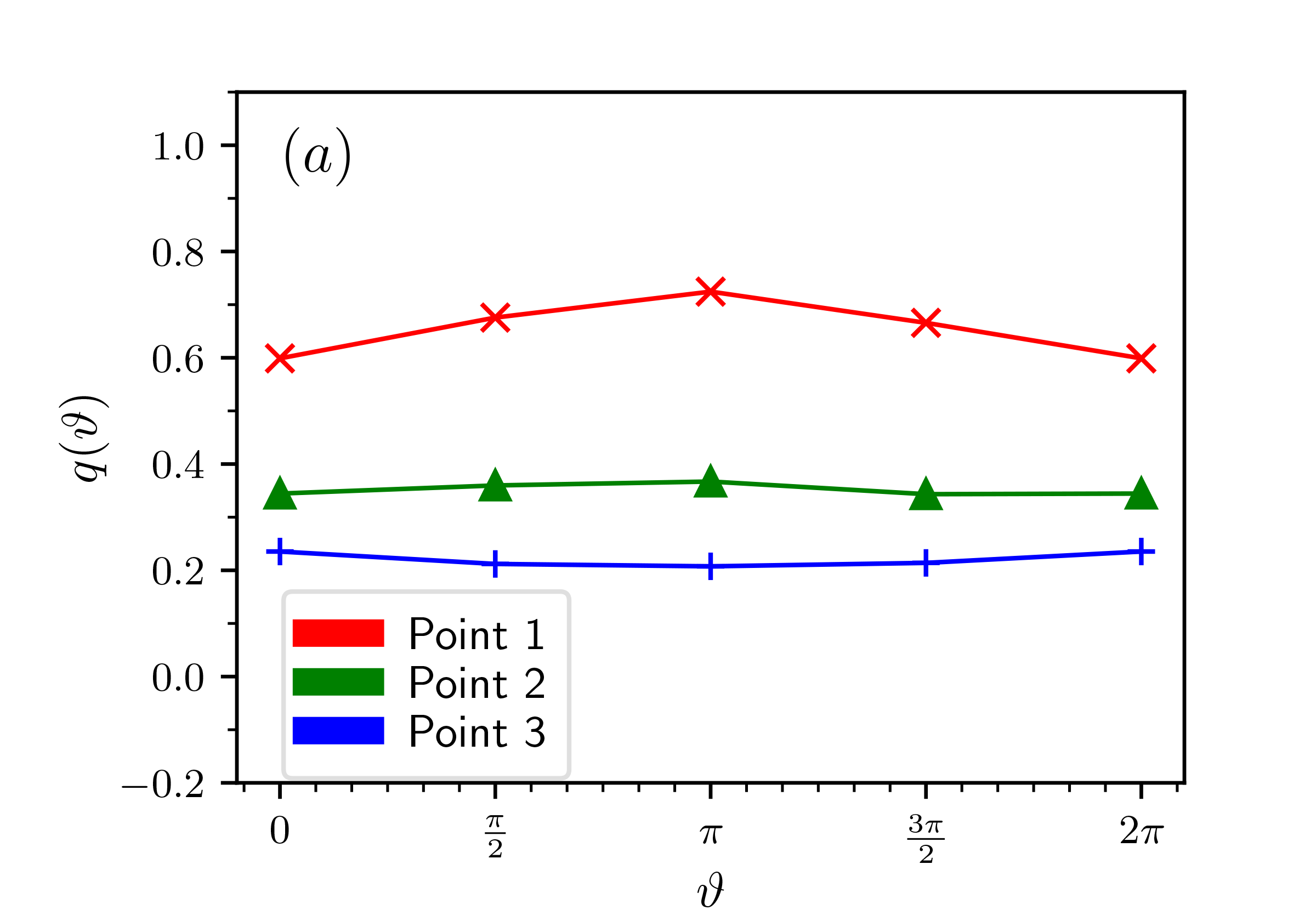}}
\subfigure{\label{fig:comm_pass_curves_rc_half}\includegraphics[height=50mm]{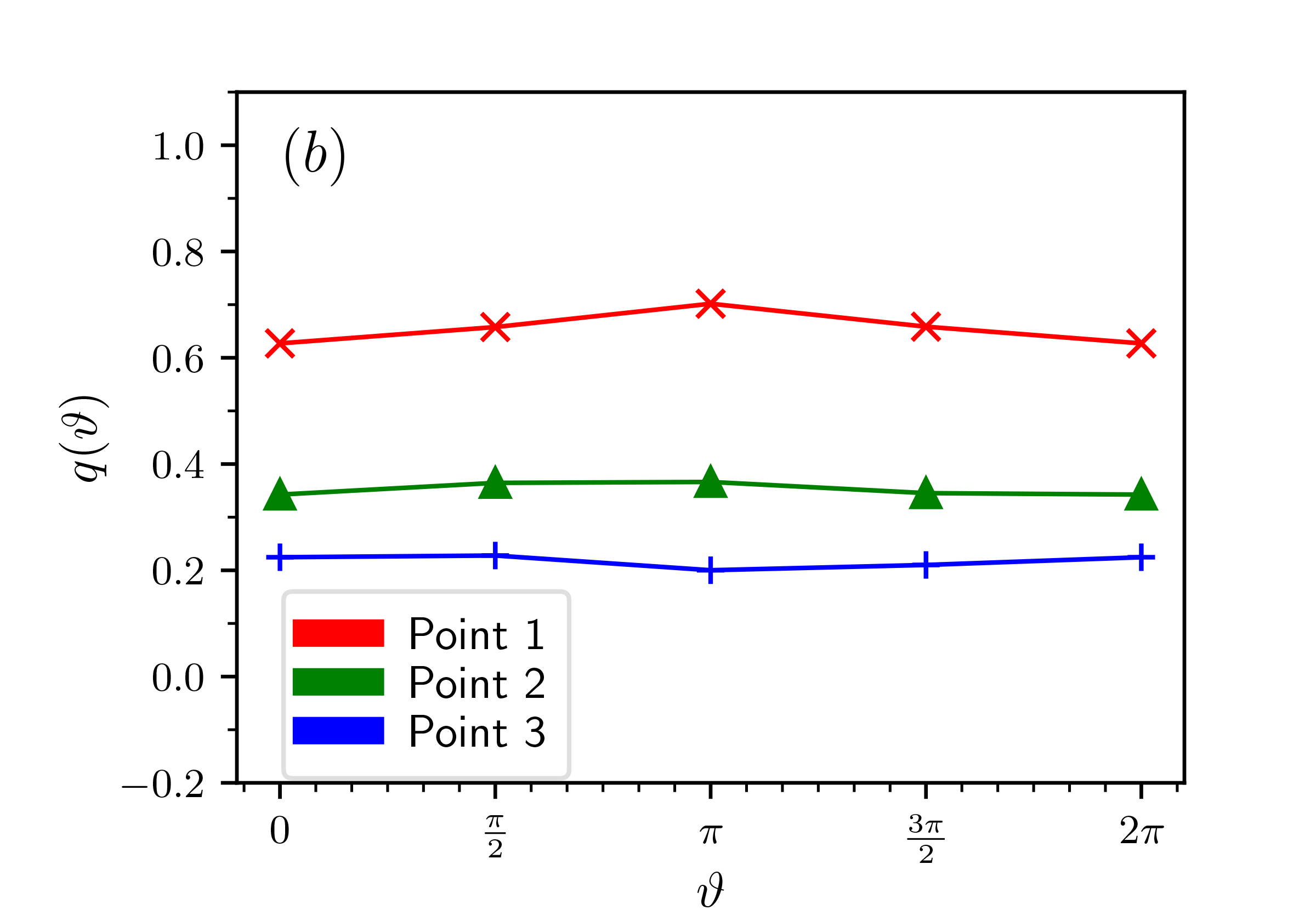}}
\caption{Committor as a function of the angle for a passive particle, obtained from brute-force simulations with $D = 10^{-3}$. Each distribution is obtained from a statistics of $10^{4}$ trajectory slices. (a) Committor obtained in the case with $r_{c} = 0.1 \; r$. (b) Committor obtained in the case with $r_{c} = 0.05 \; r$.}
\label{fig:comm_pass_curv}
\end{figure}

Finally, point $3$ shows a behavior consistent with the previous observations.
If the particle leaves the circle with an angle of $\pi$, it will be directed towards T, but at the same time it will likely meet R earlier since it is located between point $3$ and T, causing a small value for the committor.
If, instead, the particle exits with other angles, the agent will meet the boundaries and will subsequently use the ‘‘shuttle'' to reach T.
In particular, this is more evident in the case with angle $0$, which, due to the position of point $3$, has higher chances to take the particle to the walls.

The modulation observed in the committor function depending on the self-propulsion orientation of the active agent has no counterpart in the passive case (see Fig.~\ref{fig:comm_pass_curv}).
As expected, for a passive particle performing Brownian motion the committor function is independent of the angle and it is just a function of the initial position of the particle.
However, in the specific case of the considered system, there is some residual dependence on the angle as can be seen from Fig.~\ref{fig:comm_pass_curv}(a).
This is caused by how the initial angle of the self-propulsion is computed and from the finite size of the regions where the committor is measured: since the orientation is obtained from the infinitesimal displacements along $x$ and $y$, a particle exiting from the right side of a disk will have an initial angle of about $0$, while a particle exiting on the left it will have an angle of $\pi$.
Due to the finite-size effects of the system ($r_{c} \nrightarrow 0$), a particle exiting from the right side of a circular region will have a different distance from R and T compared to a particle that exits the same disk from the opposite side.
This results in a fictitious dependence of the committor on the angle, which disappears as $r_{c} \rightarrow 0$ (see Fig.~\ref{fig:comm_pass_curv}(b) for the case with a radius $r_{c}$ of $0.5$ the value used in our analysis).

We conclude this section by pointing out that, even though the microscopic behavior of the agent employed in the experiments is different from an ABP, a breaking of the symmetry in the committor is observed in both the considered scenarios of self-propelled agents compared to passive searchers.
This modulation of the committor function is generated by the persistence in the swimming mechanism of the searcher, which is absent in the case of a passive particle.
These results highlight once more the relevance of the self-propulsion mechanism in determining the target-search odds.
However, provided the considerable differences observed in the behavior of the camphor disks compared to an ABP, further validation studies will be required.
To this end, a possibility would be to include an angular drift term in the ABP equations of motion to mimic the asymmetries of the camphor disks (similarly to the Brownian circle swimmer model~\cite{bech2016}), which is one of our currently ongoing studies.

\subsection{First passage time distribution from simulations}

To close this study for a simple target-search process, we take advantage of the trajectory slices produced using the brute-force simulations to characterize the distributions of the first passage times to reach the T region (\textit{i.e.} the distributions of the times required for a particle to reach T for the first time, starting from one of the points considered during the committor calculation).

\begin{figure}[H]
\centering
\includegraphics[height=100mm]{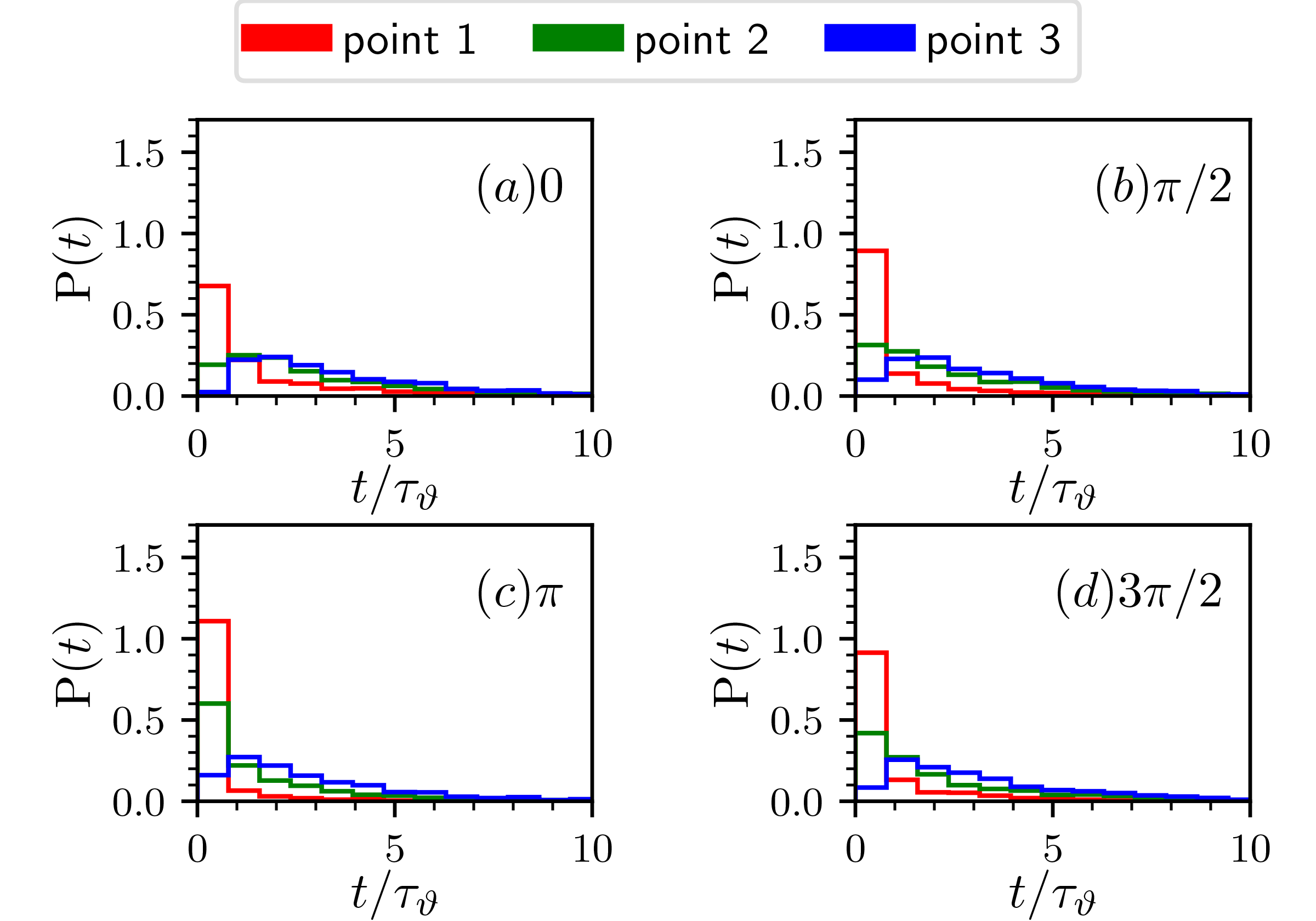}
\caption{First passage time distributions for each point and angular slice. Each distribution is obtained from a statistics of about $1.5 \cdot 10^{3}$ first passage times.}
\label{fig:fpt_distr_sim}
\end{figure}

Quite intuitively, the closer is the starting point to T, the more the distributions are peaked at short times (see Fig.~\ref{fig:fpt_distr_sim}).
Additionally, for all points it is found that an initial orientation of $\pi$ (directed towards T) ensures the largest peak at short times.
The more the orientation shifts away from $\pi$, the more the peak at short times decreases in favor of a longer tail of the distributions.

Note that these first passage time distribution are only an approximation of the real distribution for two reasons: first, they consider only trajectories that reach T without visiting R first, and second the finite dimension of the three circular regions for which the committor is computed provides a small contribution to the distributions that would not be observed in the case of point-like regions.

\section{Outlook: target-search in a landscape with submerged barriers}
\label{sec:energy_land_targ_search}

We conclude this chapter by introducing an ongoing study of a paraffin-camphor-dye droplet performing target search in an energy landscape.

\begin{figure}[H]
\centering
\subfigure{\label{fig:exp_landscape}\includegraphics[height=60mm]{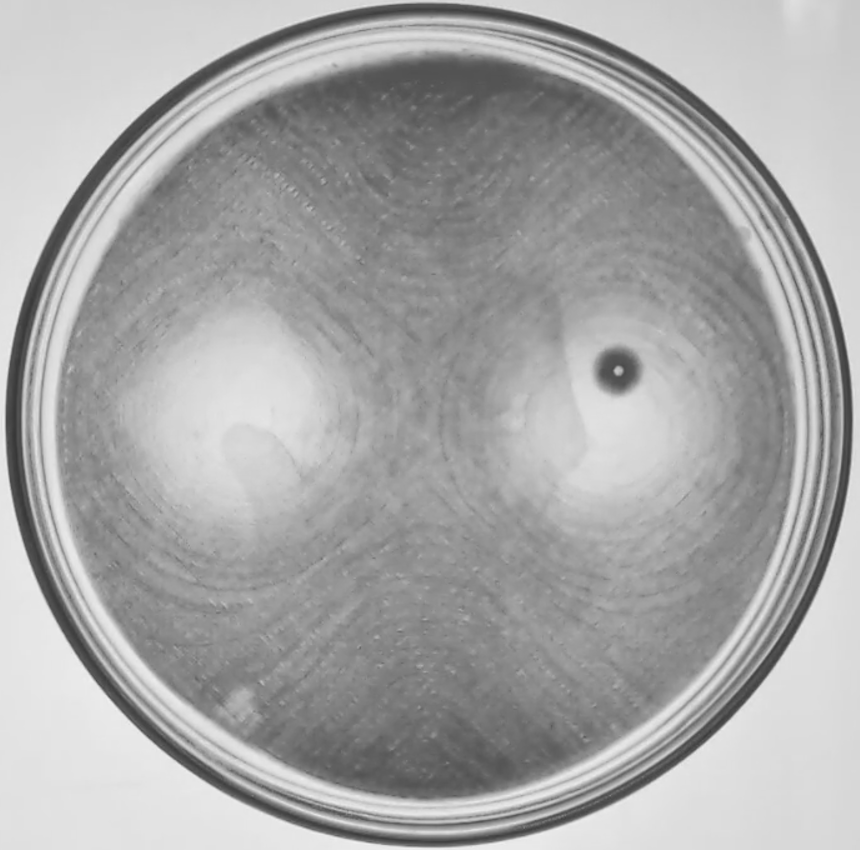}}
\subfigure{\label{fig:plastic_landscape}\includegraphics[height=60mm]{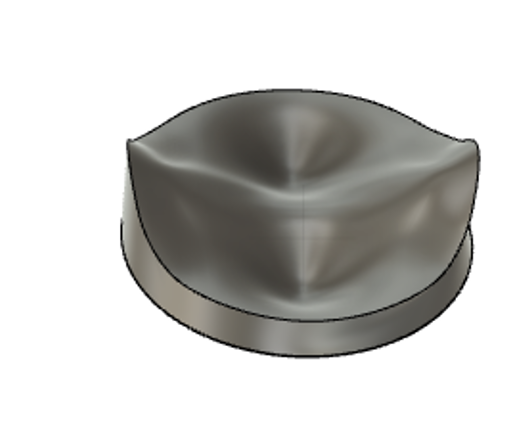}}
\caption{\textit{Left panel}: Top-down view of a paraffin-camphor-dye droplet moving in a Petri dish with a submerged plastic landscape, visible from the grey shades within the dish. \textit{Right panel}: Representation of the plastic landscape used in the experiments. The landscape is similar to a double-well potential, with two minima separated by an energy barrier. Both figures were provided by Richard J.G. L\"offler.}
\label{fig:plastic_land_exp}
\end{figure}

To find an experimental validation of the target-search pathways we found in the case of an ABP navigating in an energy landscape (discussed in chapters \ref{ch:prl_tps_abps} and \ref{ch:nav_strat}), we mimic the effects of an energy landscape by inserting a rigid plastic landscape underwater (see Fig.~\ref{fig:plastic_land_exp}).
Although the droplet is moving on a water surface, we found that, if the liquid volume located between the plastic landscape and the water surface is thin enough, the droplet will get slowed down due to hydrodynamics interactions, allowing us to reproduce a qualitatively similar behavior to the one of an ABP exploring an energy landscape.
Therefore, we design the plastic landscape to resemble a double well, and we select the R and T states as the bottom of the two wells.
In future stages of this work, we will then proceed to analyze the reactive paths obtained from experiments and we will use them for a qualitative comparison with simulations of an ABP navigating in a double-well potential inserted in the confining walls of the Petri dish.

\section{Chapter conclusions}
In this chapter we attempted an experimental validation of our results for the committor function of a self-propelled agent.

Here, we employed a self-propelled camphor object to investigate the dependence of the committor on the self-propulsion orientation.
After designing a simple experimental setup, we compared the committor function obtained from a self-propelled camphor or camphor-camphene-polymer disk exploring a Petri dish to brute-force simulations of an ABP moving in a similar environment.
From this analysis we found that, even if the microscopic and single-trajectory behavior of the two systems is significantly different, a qualitative agreement on the modulation of the committor as a function of the self-propulsion orientation is observed between the two cases.
This picture has no counterpart in the behavior of a passive particle, for which the committor is independent of the orientation of the particle, highlighting again the relevance of the self-propulsion mechanism in target-search processes.
Additionally, we found that the self-propulsion orientation also influences the distribution of the first passage times required to reach the target in this simple experimental setup.

Finally, we concluded this chapter by introducing the experimental setup of an ongoing study, which aims at reproducing at a qualitative level the counterintuitive target-search pathways characterized in the previous chapters of this thesis.

\afterpage{\null\newpage}

\pagestyle{plain}

\chapter{Conclusions}
\pagestyle{fancy}

Active particles are self-propelled objects that display a plethora of interesting phenomena in nature.
Thanks to their self-propulsion and energy-consumption mechanism, they are out-of-equilibrium systems that cannot be described with the laws of equilibrium thermodynamics.
In particular, the understanding of how these system perform target search is a fundamental query that has yet to be addressed in detail and that has fundamental implications for many processes in the nanotechnology of the $21^{\text{st}}$ century, such as targeted drug delivery or water decontamination.

The purpose of this thesis was to provide a set of tools that can be used to fully characterize the target-search process performed by self-propelled particles, and, at the same time, to study how these active agents manage to find their targets.
To this end, we resorted to the Transition Path Theory framework and we focused on the study of an active agent, described by the paradigmatic active Brownian particle model, performing target-search in a complex environment.
This environment has been modeled through an external energy landscape directly acting on the equations of motion of such a particle and characterized by the presence of high energy barriers, so to mimic the difficulties in finding the target during this processes.
The original results we presented in this thesis work are divided in two categories: the tools that we generalized to the non-equilibrium case of an ABP performing target search, which are discussed in chapters \ref{ch:prl_tps_abps} and \ref{ch:comm_function} and which provide the main theoretical contribution of this thesis, and the complete analysis of target-search behavior of active agents, discussed throughout all chapters.

The first step in this PhD project focused on developing an algorithm capable of efficiently simulating rare transitions in an out-of-equilibrium setting such as the one at hand.
To this end, we generalized the famous Transition Path Sampling algorithm to non-equilibrium settings, overcoming for the first time its limitations on the applicability to these systems.
With this result, we proved that the lack of microscopic reversibility typical of these scenarios can be circumvented by selecting any set of rules for the system dynamics backward in time, as long as this choice is correctly accounted for in the acceptance probabilities of the newly generated transition paths.
We then showed that this new version of the TPS samples the correct underlying distributions in the case of an ABP searching for a target in a double-well potential by validating the algorithm against brute-force simulations.
However, the activity of the particle, which allows the agent to climb more easily the energy barriers in the system compared to a passive searcher, together with the increased length of the target-search paths reduces the efficiency of the TPS algorithm compared to the passive case.
Notwithstanding this fact, in the considered cases the generalized TPS algorithm is found to be more efficient than plain direct integration of the equations of motion.
Finally, we showed that the efficiency of the algorithm depends both on the considered parameters for the ABP model and on the choice of the backward dynamics for the system, which suggests the need of using backward rules yielding trajectories as similar as possible to the ones obtained through the forward dynamics reverted in time.
This generalized version of TPS paves the way to an extension of the applicability of this renowned algorithm to a completely new category of systems, namely out-of-equilibrium systems.
Possible extensions of this work might include the application of the generalized TPS to other active particle models involved in rare transitions, as well as its use in the study of other out-of-equilibrium systems, which can be achieved as long as a reasonable choice of the backward dynamics is found.

Using our generalized TPS, we analyzed the target-finding process of an ABP in energy landscapes characterized by high energy barriers, observing for the first time a couterintuitive behavior of these self-propelled particles compared to passive ones.
In particular, our study of the typical Transition-Path-Theory observables (transition probability densities and currents) showed that the active agent, thanks to its self-propulsion and persistence, is found more frequently in higher energy regions of the landscape compared to the passive case.
Additionally, the active particle reaches the target by surfing along the confining energy boundaries, crossing the barrier far away from the saddle point without following the minimum energy path that is usually taken by a passive particle.
In doing so, striking long-lasting target-search pathways emerge, which exit from the back of the reactant basin R and reach the target region T from the back.

The shape and characteristics of the active target-search paths were found to change depending on the ABP parameters, which motivated us to perform a systematic study on the full parameter space of the model in search of an optimal strategy that the agent can adopt to reach the target more frequently.
This study highlighted that an optimal region in the two-dimensional parameter space for the model exists, which ensures the most effective target-finding process.
The associated strategy relies on selecting an optimal value of the agent's persistence, so to avoid passive-like behavior found at small persistence or too long detours at the system boundaries in the case of a too large persistence, and then to increase the P{\'e}clet number as much as possible, to travel quickly along the selected trajectory shapes.
Interestingly, this general strategy is found independent of the details of the landscape at hand, and the introduction of additional metastable states in the system did not considerably influence the target-search strategy in this rugged environment.

To provide a complete description of the target-search process using Transition Path Theory, we then applied the concept of committor function to our out-of-equilibrium system.
After deriving the Fokker-Planck equation for an ABP, we exploited our knowledge of the agent propagator to find an analytic expression for the committor in our target-search problem.
Subsequently, after showing that also in the active-particle case the committor is a solution to the Backward Kolmogorov equation, we efficiently computed the committor on a grid by solving this equation numerically through a finite-difference algorithm.
The pipeline that we applied to the ABP model can be easily extended to other active models and landscapes as long as the system is Markovian, it obeys a Fokker-Planck equation, and one can derive an expression for the propagator.
Unfortunately, this approach has the drawback of becoming computationally unfeasible as soon as the number of dimensions of the system increases or, more generally, when the number of the grid points required to capture the relevant features of the process increases.
We then used this framework to analyze once again the problem of an ABP searching a target in a double-well potential, providing a characterization of the committor for three relevant parameter sets of the model.
After validating our finite-differences committor against the one observed using a direct integration of the equations of motion, we found that the particle parameters strongly influence the shape of the iso-committor hypersurfaces and that the self-propulsion orientation of the agent plays a pivotal role in determining the success odds of the search.
Additionally, this analysis highlighted that the picture obtained from the committor alone is not sufficient to capture all the relevant features of the active target-search, and such a study requires to be complemented by the characterization of other observables for the process, such as the transition probability density and transition current.

Finally, we discussed how to experimentally validate our active-Brownian-particle committor.
To this end, we studied the behavior of a macroscopic camphor self-propelled object floating on water surface within the confining environment of a Petri dish, which, notwithstanding the considerable differences with respect to a microscopic ABP, still shares relevant similarities with it (\textit{e.g.} the ability of self-propelling and performing rotational diffusion).
After designing a simple experimental setup, we extracted an estimate of the Commitor as a function of the self-propulsion orientation from experimental data, and we subsequently compared it to the committor obtained for an ABP moving in a similar environment.
Although the differences between the two systems are reflected on the typical trajectories observed, the behavior of the committor in the two cases is qualitatively similar.
Also in this case, the picture emerging from the experiments highlights the relevance of the self-propulsion orientation on the odds of finding a target, marking it again as an advantageous feature typical of active agents in these processes.
This experimental framework proved to be useful to capture the qualitative behavior of active agents, and our currently ongoing research aims at finding the counterintuitive target-search patterns that we observed in the case of an ABP.

\afterpage{\null\newpage}

\pagestyle{plain}

\begin{appendices}

\afterpage{\null\newpage}

\pagestyle{plain}

\chapter{Langevin dynamics}
\label{app:LD}

In this appendix we will review the main concepts of Langevin dynamics and its related properties.

Consider a particle immersed in a thermal reservoir. The motion of such a particle can be efficiently described using Langevin dynamics, which relies on introducing a friction term and a stochastic term to mimick the interactions of the particle with the fluid molecules, without treating them explicitly.
The friction term represents the friction between the particle and the fluid molecules, which will hinder the particle motion.
The stochastic term instead represents the energy received from the collision with the bath particles, which will in turn translate into motion.
Therefore, if we consider a particle of mass $m$ immersed in a thermal bath and subject to an external force $\bm{\text{F}}$, the Langevin equation for the particle reads:
\begin{equation}
m \ddot{\bm{r}} = \bm{\text{F}} - \gamma \dot{\bm{r}} + \bm{f} \; ,
\label{eq:ULD}
\end{equation}
where $\gamma$ is the viscous drag coefficient of the fluid and $\bm{f}$ is the random force exerted by the bath molecules on the particles.
This random force is subject to some constraints, in particular these forces are Gaussian distributed and they have zero average, to reflect the fact that the ``kicks'' coming from the surrounding molecules have no preferential direction and the forces are not correlated.
Additionally, the strength of these random forces is provided by the \emph{fluctuation-dissipation theorem}, which states that the intensity of the random forces is proportional to the viscous coefficient of the surrounding fluid through the relation:
\begin{equation}
\langle f_{i}(t) f_{j} (t') \rangle = 2 \gamma k_{B}T \delta_{i,j} \delta (t-t')
\end{equation}

This formulation is usually known in the literature as \emph{underdamped Langevin dynamics}.
From the limit of the underdamped Langevin dynamics where the viscous forces dominate over inertial ones, the \emph{overdamped Langevin dynamics}  can be obtained.
In fact, in the case of a particle for which the inertial forces are much smaller than the viscous forces, Eq.~\ref{eq:ULD} becomes:
\begin{equation}
\dot{\bm{r}} = \frac{\bm{\text{F}}}{\gamma} + \frac{\bm{f}}{\gamma}
\label{eq:OLD}
\end{equation}

We point out that the mean square displacement of a particle obeying the overdamped Langevin dynamics can be computed as:
\begin{equation}
\langle \Delta \bm{r}^{2} \rangle = 2n\frac{k_{B}T}{\gamma}t \; ,
\end{equation}
where $n$ is the system dimensionality, and recalling Einstein's calculation on Brownian motion~\cite{Einstein1905} (which can be described by an overdamped Langevin equation)
\begin{equation}
\langle \Delta \bm{r}^{2} \rangle = 2 n D t \; ,
\end{equation}
with $D$ translational diffusion coefficient, the Einstein-Smoluchovski relation can be obtained:
\begin{equation}
D = \frac{k_{B}T}{\gamma} \; .
\end{equation}
For a complete derivation see for example~\cite{Zwanzig2001}.

\chapter{Fokker-Planck equation}
\label{app:FP}
In this appendix we will recall the main concepts on the Fokker-Planck (FP) equation and how it can be derived in the case of a passive particle obeying the overdamped Langevin dynamics.

The Fokker-Planck equation is an equation of motion for the distribution of a fluctuating variable, therefore, given a stochastic variable $x$ the FP equation can provide the temporal evolution of the distribution $P$ that characterizes it.~\cite{Gardiner2009}

Consider a stochastic differential equation (SDE) in the form:
\begin{equation}
dx(t) = a[x(t),t]dt + b[x(t),t]dW(t)
\label{eq:SDE1d}
\end{equation}
where $a[x(t),t]$ is a term representing all the deterministic contributions to the equation, called drift term, and $b[x(t),t]dW(t)$ is a stochastic term associated with a Wiener process.
For an SDE in this form, an associated FP equation will exist, and it will be given by:
\begin{equation}
\frac{\partial p}{\partial t} = -\frac{\partial}{\partial x}\big[a(x,t)p\big] + \frac{1}{2}\frac{\partial^{2}}{\partial x^{2}}\big[b(x,t)^{2}p\big] \; ,
\end{equation}
where $p = p(x,t)$ is the distribution for the variable $x$ at time $t$.
Additionally, a propagator for the system $p(\bar{x},t|x_{0},t_{0})$ will exist, describing the probability of observing the variable $x$ assuming a value $\bar{x}$ at time $t$ provided that $x = x_{0}$ at time $0$, and this propagator will be solution to the same FP equation.
In fact, if $t \rightarrow t_{0}$ then $p(\bar{x},t|x_{0},t_{0}) \rightarrow p(x_{0},t_{0})$, so the propagator becomes a delta function for the position $x_{0}$ at time $t_{0}$.

Now, in the more general case of a multivariate system with $N$ variables the associated system of SDEs reads:
\begin{equation}
    d\bm{x} = \bm{A}(\bm{x},t) dt + \bm{B}(\bm{x},t) d\bm{W}(t) \; ,
\end{equation}
where $d\bm{x}$ is an $N$-dimensional vector of stochastic variables, $\bm{A}(\bm{x},t)$ is an $N$-dimensional drift vector representing the deterministic terms, $\bm{B}(\bm{x},t)$ is an $N \times M$ matrix and $d\bm{W}(t)$ is an $M$-dimensional Wiener process.
Then the probability density $p(\bm{x},t)$ for $\bm{x}$ satisfies the following FP equation (here the dependence of $p$ from $\bm{x}$ and $t$ is omitted for simplicity):
\begin{equation}
    \frac{\partial p}{\partial t} = - \sum_{i=1}^{N} \frac{\partial}{\partial x_{i}} \big[A_{i}(\bm{x},t)p\big] + \sum_{i=1}^{N}\sum_{j=1}^{N} \frac{\partial^{2}}{\partial x_{i} \partial x_{j}} \big[D_{ij}(\bm{x},t)p\big]
    \label{eq:SDEND}
\end{equation}
with
\begin{equation}
    D_{ij}(\bm{x},t) = \frac{1}{2} \sum_{k=1}^{M}B_{ik}(\bm{x},t)B_{jk}(\bm{x},t)
\end{equation}
In the case of a passive particle obeying the overdamped Langevin dynamics, subject to an external potential, and moving in a two-dimensional space, the SDE system for the particle reads:
\begin{eqnarray}\label{eom:pp1}
dx &=& - D\, \beta\, \nabla_{x}U(x,y)\, dt + \sqrt{2D\, dt} \, \xi_{x}\;,\\ \label{eom:pp2}
dy &=& - D\, \beta\, \nabla_{y}U(x,y)\, dt + \sqrt{2D\, dt} \, \xi_{y}\;.
\end{eqnarray}
where $\beta = 1/k_{B}T$, $D$ is the translational diffusion coefficient, and the components of $\bm{\xi}_{i} = (\xi_{x},\xi_{y})$ are independent Gaussian random variables with zero average and unit variance.
Consequently we have:
\begin{equation}
    d\bm{x} = \begin{pmatrix} dx\\dy \end{pmatrix}\;,
\end{equation}
\begin{equation}
    \bm{A}(\bm{x},t) = \begin{pmatrix} - D\, \beta\, \nabla_{x}U(x,y) \\ - D\, \beta\, \nabla_{y}U(x,y) \end{pmatrix}\;,
\end{equation}
\begin{equation}
    \bm{B}(\bm{x},t) = \begin{pmatrix} \sqrt{2D} & 0 \\ 0 & \sqrt{2D} \end{pmatrix}\;,
\end{equation}
\begin{equation}
    d\bm{W}(t) = \begin{pmatrix} \xi_{x}\, \sqrt{dt} \\ \xi_{y}\, \sqrt{dt} \end{pmatrix}\;.
\end{equation}
Now, since $D_{ij}(\bm{x},t) \neq 0 \iff i = j$ and in particular $D_{11} = D_{22} = D$, the associate FP equation obtained from Eq.~\ref{eq:SDEND} becomes:
\begin{equation}
\frac{\partial p}{\partial t} = \frac{\partial}{\partial x} \bigg[ D\, \beta\, \nabla_{x}U(x,y) p \bigg]\, +\frac{\partial}{\partial y} \bigg[ D\, \beta\, \nabla_{y}U(x,y) p \bigg]\, + D \frac{\partial^{2}}{\partial x^{2}} p\, + D \frac{\partial^{2}}{\partial y^{2}} p\; ,
\end{equation}
which can then be rewritten in a more compact form:
\begin{equation}
    \frac{\partial p}{\partial t} = \bm{\nabla} \bigg[ D\, \beta\, \bm{\nabla}U(\bm{x}) p \bigg]\, + D\, \nabla^{2} p\;.
    \label{eq:fpp}
\end{equation}
The FP equation can be expressed both in an operator form:
\begin{equation}
\frac{\partial p}{\partial t} = - \hat{H}_{\text{FP}} \; p
\end{equation}
with $\hat{H}_{\text{FP}}$ the Fokker-Planck operator:
\begin{equation}
\hat{H}_{\text{FP}} = - D \bm{\nabla} \cdot \bigg[ \bm{\nabla} + \beta\, \bm{\nabla} U(\bm{x}) \bigg] \; ,
\end{equation}
and as a continuity equation:
\begin{equation}
    \frac{\partial p}{\partial t} = -\bm{\nabla} \cdot \bm{J} \; ,
    \label{eq:fpc}
\end{equation}
where $\bm{J} = \bm{J}(\bm{x},t)$ is the probability current, defined as:
\begin{equation}
    \bm{J} = - D [\bm{\nabla} + \beta \bm{\nabla} U(\bm{x})] p \; .
\end{equation}
The FP equation in continuity form expresses the particle's conservation in the system: if $\bm{J}(\bm{x},t)=0$ there is no probability flux and $p(\bm{x},t)$ is constant in time. If, on the other hand, $\bm{J}(\bm{x},t) \neq 0$ then $p(\bm{x},t)$ changes with time.
According to Eq.~\ref{eq:fpc}, the only way for which $p(\bm{x},t)$ changes with time is due to a non-zero probability current, and not through creation or destruction of particles since no sink or source terms are present.

\chapter{Path Integral representation of stochastic dynamics}
\label{app:PI}

In this section, we will provide a brief general introduction on the path integrals and we will derive a path-integral representation for a passive particle obeying the overdamped Langevin dynamics.

\begin{figure}[H]
\centering
\includegraphics[height=50mm]{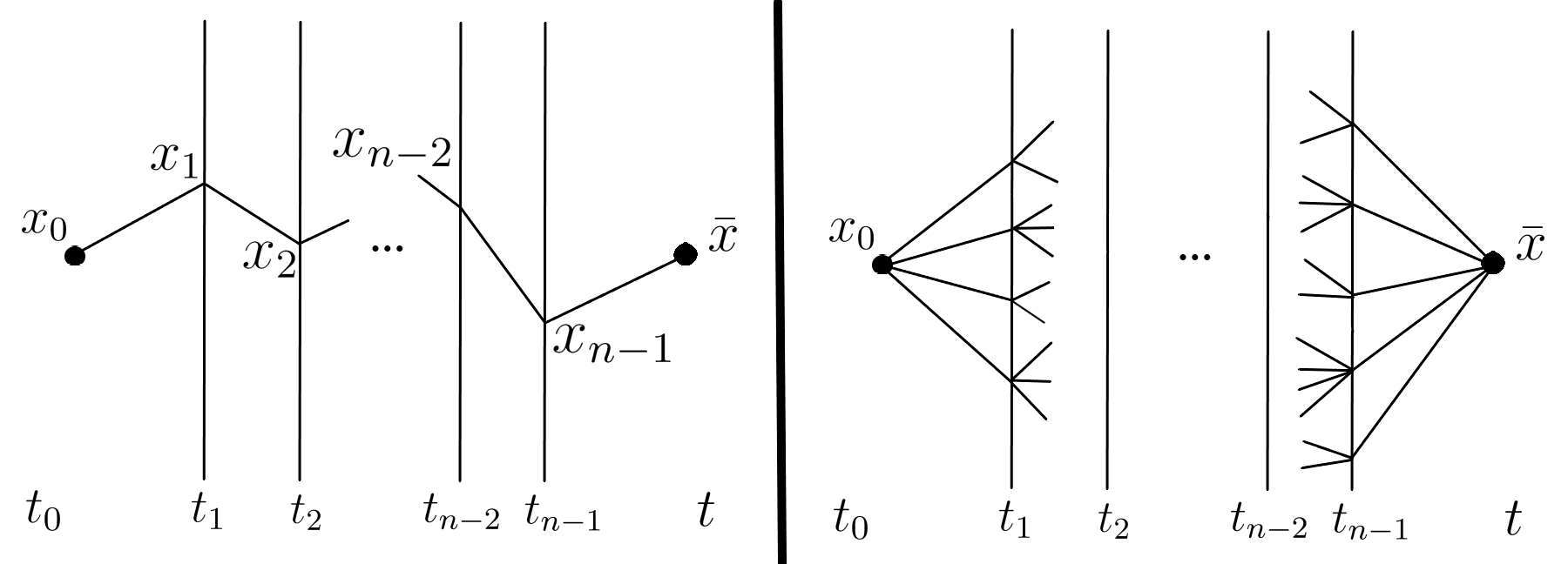}
\caption{\textit{Left panel}: discretized representation of the trajectory linking the states $(x_{0},t_{0})$ and $(\bar{x},t)$ for a deterministic system. \textit{Right panel}: discretized representation of a bundle of possible trajectories linking the initial state $(x_{0},t_{0})$ to the final state $(\bar{x},t)$ for a stochastic system.}
\label{fig:app_pi}
\end{figure}

Path integrals are a convenient tool to represent fluctuating and continuous line-like structures.~\cite{Kleinert2009}
For the purpose of this thesis, they can be useful in particular to represent the propagator probability for a stochastic system.
Consider the propagator $p(\bm{\omega}_{\text{f}},t|\bm{\omega}_{0},t_{0})$, which quantifies the probability of observing the system in a state $\bm{\omega}_{\text{f}}$ at time $t$ provided that it was in a state $\bm{\omega}_{0}$ at time $t_{0}$.
In a deterministic system, this transition from the state $\bm{\omega}_{0}$ to the state $\bm{\omega}_{\text{f}}$ in a time $t-t_{0}$ is uniquely determined by a single trajectory, which will visit a specific sequence of states $\bm{\omega}_{k}$ as time increases from $t_{0}$ to $t$.
However, if the system obeys a stochastic dynamics, the trajectory leading the system from the state $\bm{\omega}_{0}$ to the state $\bm{\omega}_{\text{f}}$ is not uniquely determined, and the transition can rather occur following a bundle of different paths, each with its own associated probability of being selected.
A visual representation of such a deterministic or stochastic path for a one-variable system can be seen in Fig.~\ref{fig:app_pi}, where we have introduced a time discretization to represent the intermediate states of the transition in a convenient way.
Therefore, in the case of a system obeying a stochastic dynamics, one must find a way to account for the probability of all possible paths that the system can take to move from the state $\bm{\omega}_{0}$ to the final state $\bm{\omega}_{\text{f}}$.
This is exactly the result that a path-integral evaluation of the process can provide.
In fact, a path-integral representation for the propagator of the system will allow to integrate the contribution to the probability coming from all the possible paths in an elegant fashion.

We will now proceed to show how a path-integral representation can be obtained in the case of a passive particle moving in two dimensions and obeying the overdamped Langevin equation.
Again, the equations of motion for such a particle can be obtained as discretized Langevin equation in It\^{o} form:
\begin{eqnarray}\label{eom:ppi1}
x_{i} &=& x_{i-1} - D\, \beta\, \nabla_{x}U(x_{i-1},y_{i-1})\, \Delta t + \Delta W_{x,i}(t)\;,\\ \label{eom:ppi2}
y_{i} &=& y_{i-1} - D\, \beta\, \nabla_{y}U(x_{i-1},y_{i-1})\, \Delta t + \Delta W_{y,i}(t)\;.
\end{eqnarray}
where $\beta = 1/k_{B}T$, $D$ is the translational diffusion coefficient, and $W_{\alpha}(t)$ are independent Wiener processes representing the translational diffusion terms.
These processes follow a Gaussian distribution with $\langle W_{\alpha} \rangle = 0$ and $\langle W_{\alpha}(t_{i}) W_{\alpha}(t_{j}) \rangle = 2 D \Delta t \delta_{i,j}$.

Note that here we adopt the indexing notation used in~\cite{Elber2000} for the noise and the positions, while this will be in general different from the one used in the main text of the manuscript and in our publications.

Since this system is Markovian, its trajectory in configuration space can be uniquely determined once the series of Gaussian random variables for the noise processes is obtained.~\cite{Elber2000}
We can easily evaluate the probability distribution of the noise terms through a straightforward product of Gaussian distributions, which for the $x$ variable reads:
\begin{equation}
P[W_{x}(t_{1}) \rightarrow W_{x}(t_{2}) \rightarrow ... \rightarrow W_{x}(t_{N})] = \left( \frac{1}{\sqrt{4\pi D\Delta t}} \right)^{N} \prod_{k=1}^{N} e^{-\frac{W_{x_{k}}^2}{4D\Delta t}}
\end{equation}
Now, through a change of variables we can obtain the probability of observing a series of visited positions $x$.
We do so by using the change of variables $W_{x,i} = x_{i}-x_{i-1} + \Delta t D \beta \nabla_{x} U(x_{i-1},y_{i-1})$, and the probability for the series becomes:
\begin{equation}
P[x(t_{1}) \rightarrow x(t_{2}) \rightarrow ... \rightarrow x(t_{N})] = \left( \frac{1}{\sqrt{4\pi D\Delta t}} \right)^{N} \prod_{k=1}^{N} e^{-\frac{\left(x_{k}-x_{k-1}+\Delta t D \beta \nabla_{x} U(x_{k-1},y_{k-1})\right)^2}{4D\Delta t}} J \left[ \frac{\partial W}{\partial x} \right] \; ,
\end{equation}
where $J \left[ \frac{\partial W}{\partial x} \right]$ is the determinant of the Jacobian matrix for the change in coordinates. It reads:
\begin{center}
$J \left[ \frac{\partial W}{\partial x} \right] = det$
\(
\begin{bmatrix}
\frac{\partial W_{1}}{\partial x_{1}} & \frac{\partial W_{1}}{\partial x_{2}} & \dots & \frac{\partial W_{1}}{\partial x_{N}} \\
\frac{\partial W_{2}}{\partial x_{1}} & \frac{\partial W_{2}}{\partial x_{2}} & \dots & \frac{\partial W_{2}}{\partial x_{N}} \\
\vdots & \vdots & \ddots & \vdots \\
\frac{\partial W_{N}}{\partial x_{1}} & \frac{\partial W_{N}}{\partial x_{2}} & \dots & \frac{\partial W_{N}}{\partial x_{N}} \\
\end{bmatrix}
\) ,
\end{center}
and therefore:
\begin{center}
$J \left[ \frac{\partial W}{\partial x} \right] = det$
\(
\begin{bmatrix}
1 & 0 & \dots & 0 \\
-1 + \Delta t D \beta \nabla_{x}^{2} U(x_{0},y_{0}) & 1 & \dots & 0 \\
\vdots & \ddots & \ddots & \vdots \\
0 & \dots & -1 + \Delta t D \beta \nabla_{x}^{2} U(x_{N-2},y_{N-2}) & 1 \\
\end{bmatrix}
\)
\end{center}
which yields:
\begin{equation}
J \left[ \frac{\partial W}{\partial x} \right] = 1^{N} \; ,
\end{equation}
and finally the probability distribution for the x positions reads:
\begin{equation}
P[x(t_{1}) \rightarrow x(t_{2}) \rightarrow ... \rightarrow x(t_{N})] = \left( \frac{1}{\sqrt{4\pi D\Delta t}} \right)^{N} \prod_{k=1}^{N} e^{-\frac{\left(x_{k}-x_{k-1}+\Delta t D \beta \nabla_{x} U(x_{k-1},y_{k-1})\right)^2}{4D\Delta t}} \; .
\end{equation}

After obtaining an analogous expression for the $y$ variable, one can calculate the total probability distribution for a specific trajectory:
\begin{equation}
\begin{aligned}[t]
& P[x(t_{1}) \rightarrow x(t_{2}) \rightarrow ... \rightarrow x(t_{N}),y(t_{1}) \rightarrow y(t_{2}) \rightarrow ... \rightarrow y(t_{N})] = \\
&= \left( \frac{1}{\sqrt{4\pi D\Delta t}} \right)^{2N} \prod_{k=1}^{N} e^{-\frac{\left(x_{k}-x_{k-1}+\Delta t D \beta \nabla_{x} U(x_{k-1},y_{k-1})\right)^2}{4D\Delta t}} e^{-\frac{\left(y_{k}-y_{k-1}+\Delta t D \beta \nabla_{y} U(x_{k-1},y_{k-1})\right)^2}{4D\Delta t}}
\end{aligned}
\end{equation}

Now that the expression for the probability of a generic trajectory is obtained, the path-integral representation of the propagator can be achieved by integrating over all possible paths going from $(\bm{\omega}_{0},t_{0})$ to $(\bm{\omega}_{\text{f}},t)$ (where $\bm{\omega} = (x,y)$), therefore:
\begin{equation}
\begin{aligned}[t]
& p(\bm{\omega}_{\text{f}},t|\bm{\omega}_{0}) = \left( \frac{1}{\sqrt{4\pi D\Delta t}} \right)^{2N} \int \mathcal{D} x \int \mathcal{D} y \cdot \\
\cdot e&^{-\frac{1}{4D}\int_{t_{0}}^{t_{f}}d\tau [(\dot{x}(\tau) +\Delta t D \beta \nabla_{x} U(x(\tau),y(\tau)))^{2}+(\dot{y}(\tau) +\Delta t D \beta \nabla_{y} U(x(\tau),y(\tau)))^{2}]} \; ,
\end{aligned}
\end{equation}
where $\mathcal{D}x$ is a shorthand notation for $\mathcal{D}x(\tau)$ and similarly for $\mathcal{D}y$. This expression can be rewritten as:
\begin{equation}
p(\bm{\omega}_{\text{f}},t|\bm{\omega}_{0}) = \mathcal{Z}^{-1} \int \mathcal{D} \bm{\omega} \; e^{-\frac{1}{4D}S_{OM}[\bm{\omega}]} \; ,
\end{equation}
where $\mathcal{Z}$ is a normalization constant and $S_{OM}[\bm{\omega}]$ is the Onsager-Machlup functional:
\begin{equation}
S_{OM}[\bm{\omega}] = \int_{t_{0}}^{t} d\tau \big[ \dot{\bm{\omega}}(\tau) + D \beta \nabla U(\bm{\omega}(\tau)) \big]^{2} \; .
\end{equation}

\chapter{Forward and backward Kolmogorov equations}
\label{app:BK}

In this section we will briefly review the concept of backward Kolmogorov equation.

Given a system obeying an overdamped Langevin dynamics, the forward Kolmogorov equation is simply the Fokker-Planck equation for the system, which can be expressed in the form:
\begin{equation}
\frac{\partial p}{\partial t} = - \hat{H}_{\text{FP}} \; p
\label{eq:app_fpb}
\end{equation}
with $\hat{H}_{\text{FP}}$ the associated Fokker-Planck operator:
\begin{equation}
\hat{H}_{\text{FP}} = - D \bm{\nabla} \cdot \bigg[ \bm{\nabla} + \beta\, \bm{\nabla}U(\bm{x}) \bigg] \; ,
\end{equation}
Now, since the system is Markovian and satisfies the forward Kolmogorov equation, the committor function satisfies the backward Kolmogorov equation:
\begin{equation}
\hat{H}^{\dag}_{\text{FP}}q(\bm{x}_{i}) = 0 \; ,
\end{equation}
where $\hat{H}^{\dag}_{\text{FP}}$ is the adjoint of the Fokker-Planck operator (backward Kolmogorov operator):
\begin{equation}
\hat{H}^{\dag}_{\text{FP}} = -D(\nabla^{2} - \beta \bm{\nabla} U(\bm{x}) \cdot \bm{\nabla}) \; .
\end{equation}
To prove this statement, it is sufficient recalling that $p^{*}_{\partial W}(\bm{x}_{f},t|\bm{x}_{0},t_{0})$ satisfies also the backward Kolmogorov equation (because it is a FP propagator solution to the FP equation Eq.~\ref{eq:app_fpb} for a Markovian system), therefore~\cite{Orland2011}:
\begin{equation}
-\frac{\partial p^{*}_{\partial W}}{\partial t_{0}} = -  \hat{H}^{\dag}_{\text{FP}} p^{*}_{\partial W} \; ,
\end{equation}
which, by assuming time homogeneity can be rewritten as:
\begin{equation}
\frac{\partial p^{*}_{\partial W}}{\partial t} = -  \hat{H}^{\dag}_{\text{FP}} p^{*}_{\partial W} \; ,
\end{equation}
and by applying the backward Kolmogorov operator to the definition of the committor we get:
\begin{equation}
\begin{aligned}[t]
   \hat{H}^{\dag}_{\text{FP}}q(\bm{x}_{i}) & = - \hat{H}^{\dag}_{\text{FP}} \int_{0}^{\infty} dt \int_{\partial \text{P}} d\bm{\sigma}' \cdot \bm{J}^{*}_{\partial \text{W}}(\bm{x}',t|\bm{x}_{i},t_{0})\\
   & = D \int_{0}^{\infty} dt \int_{\partial \text{P}} d\bm{\sigma}' \cdot [\bm{\nabla}' + \beta \bm{\nabla}' U(\bm{x}')] \hat{H}^{\dag}_{\text{FP}} p^{*}_{\partial \text{W}}(\bm{x}',t|\bm{x}_{i},t_{0}) \\
& = - D \int_{0}^{\infty} dt \frac{\partial}{\partial t} \int_{\partial \text{P}} d\bm{\sigma}' \cdot [\bm{\nabla}' + \beta \bm{\nabla}' U(\bm{x}')] p^{*}_{\partial \text{W}}(\bm{x}',t|\bm{x}_{i},t_{0})\\
   & = \int_{0}^{\infty} dt \frac{\partial}{\partial t} \int_{\partial \text{P}} d\bm{\sigma}' \cdot \bm{J}^{*}_{\partial \text{W}}(\bm{x}',t|\bm{x}_{i},t_{0})\\
   & = F(t = \infty) - F(t = 0) = 0
\end{aligned}
\end{equation}
 where we used the definition of the flux of the probability current through the surface $\partial \text{P}$:
 \begin{equation}
     F(\partial \text{P},t; \bm{x}_{i}) = \int_{\partial \text{P}} d\bm{\sigma}' \bm{J}^{*}_{\partial \text{W}}(\bm{x}',t|\bm{x}_{i},t_{0})\; ,
 \end{equation}
together with the fact that for an ergodic system if you wait long enough the system will visit all the configurations, therefore also the ones included in P, giving $ F(t = \infty) = 0$, and that if the initial configuration is at a finite distance from P it will never be able to reach it in a time $t = 0$, therefore $F(t = 0) = 0$.

\end{appendices}

\afterpage{\null\newpage}

\pagestyle{plain}

\afterpage{\null\newpage}

\pagestyle{plain}

\printbibliography[heading=bibintoc,title={Bibliography}]

\afterpage{\null\newpage}

\pagestyle{plain}

\afterpage{\null\newpage}

\pagestyle{plain}

\chapter*{Publications and Manuscripts}
\addcontentsline{toc}{chapter}{Publications and Manuscripts}
\pagestyle{fancy}

\section*{Author Contributions}

\noindent
\textsc{Luigi Zanovello, Michele Caraglio, Thomas Franosch, and Pietro Faccioli}\\
\textit{Target Search of Active Agents Crossing High Energy Barriers}\\
Physical Review Letters \textbf{126}, 018001 (2021). doi: \href{https://journals.aps.org/prl/abstract/10.1103/PhysRevLett.126.018001}{10.1103/PhysRevLett.126.018001}\\

\begin{adjustwidth}{1cm}{}
\noindent
M.C., T.F. and P.F. designed research; L.Z. and M.C. performed research\\
and analyzed data; all authors wrote the paper.
\end{adjustwidth}

\vspace{0.6cm}

\noindent
\textsc{Luigi Zanovello, Pietro Faccioli, Thomas Franosch, and Michele Caraglio}\\
\textit{Optimal navigation strategy of active Brownian particles in target-search problems}\\
The Journal of Chemical Physics \textbf{155}, 084901 (2021). doi: \href{https://aip.scitation.org/doi/10.1063/5.0064007}{10.1063/5.0064007}\\

\begin{adjustwidth}{1cm}{}
\noindent
L.Z., M.C., T.F. and P.F. designed research; L.Z. performed research and analyzed data; all authors wrote the paper.
\end{adjustwidth}

\vspace{0.6cm}

\noindent
\textsc{Luigi Zanovello, Richard J.G. L\"offler, Michele Caraglio, Thomas Franosch, Martin M. Hanczyc, and Pietro Faccioli}\\
\textit{Survival strategies of artificial active agents}\\
Scientific Reports \textbf{13}, 5616 (2023). doi: \href{https://www.nature.com/articles/s41598-023-32267-3}{10.1038/s41598-023-32267-3}\\

\begin{adjustwidth}{1cm}{}
\noindent
L.Z., R.J.G.L., M.C., T.F., M.M.H., and P.F. designed research; L.Z. performed research and analyzed data; R.J.G.L. performed experiments; all authors wrote the paper.
\end{adjustwidth}




\afterpage{\null\newpage}

\pagestyle{plain}

\chapter*{Acknowledgments}
\addcontentsline{toc}{chapter}{Acknowledgments}
\pagestyle{fancy}

So, you got to the finishing line of this thesis!
Here we are, at the end of this long and not always easy three-year journey that took me throughout my PhD experience.
However, before closing this work, a few (actually not so few) words are due to thank everyone who was part of this journey and my years as a student.

First and foremost I want to thank my parents (how original!), Carlo and Luigina.
Jokes aside, you supported me throughout my decisions and gave me the best education I could hope for.
No thanks will be enough for spending so much effort to give me the best possibilities for my future.
Big thanks to my sister Lucia as well, you are part of my life for what feels like forever.
As my ``almost-twin'' with two years difference, you were always there during the highs and the lows.
We shared most of our interests, friends, and free time, and thanks to you my existence on this rock floating in space was much more enjoyable.
A lot of appreciation is also due to my girlfriend, Leandra.
You always supported me and my choices, and during these eight years together you managed to endure my most annoying traits while we were growing together.
I am impressed by your tenacity!

Now I want to say how thankful I am to the friends that accompany me since my childhood.
Not only to have met you but also for the opportunity that we were able to forge into a friendship that can transcend space and time, like in a fairy tale.
You are my all-time favored monster hunters, D\&D players, fellow writers and/or drinkers, and I should really stop here because this list can go on forever and I still need to finish this manuscript.
Big thanks also to the friends I met only later on in life, from nearby towns to the internet world.
Whether you are physically close to me or really far, whether we are still in touch or it has been a while since the last time we spoke, I still think of you as my friends and you have a special place in my heart.
Next, I want to say thank you to all the friends that I made in this weird and interesting place that is the university.
If I hadn't met you during my journey, these years in the academic world would not have been even close to half as interesting and enjoyable as they were.
You gave me much more than a couple of laughs together, and if sometimes I just disappear or I forget when your birthdays are, please don't get mad at me, it does not reflect how much I appreciate your friendship.

This PhD would not have been possible without the help of my supervisors, Pietro, Thomas, and also Michele, who helped me achieve this result and find my path in the complex environment that is research.
You helped me grow a lot in the past few years, not only as a scientist but also as a person, and for this reason, I will always be grateful.
Additionally, I want to thank all the people of the research groups that I have been part of in Innsbruck and Trento.
Getting to know you was my pleasure, and you helped me bear the burden of academic life.
The memories I have with you from the group dinners or retreats will stay with me forever.

Finally, I want to thank all the other people who thought me something in these years, from how to craft medieval armor to how to get and stay fit.
Physical and mental health are equally important, and your advice inspired me and helped me improve during this time made difficult by the struggles of both the PhD and the pandemic.

Phew, it seems I made it!
This thesis is finished and with it also my years as a student.
I don't know what the future will bring me, but the things to do or that need to be fixed in this world are many, and I won't stay here sitting and waiting.

\afterpage{\null\newpage}

\pagestyle{plain}

\end{document}